# Fundamental Limits of Communications in Interference Networks

## Part III: Information Flow in Strong Interference Regime

Reza K. Farsani[1]

Email: reza_khosravi@alum.sharif.ir

***Abstract:*** **This paper, as the continuation of Parts I and II on Fundamental Limits of Communication in Interference Networks, is related to the study of information flow in networks with *strong interference*. First, the concept of strong interference is more broadly defined: *a network is said to be in the strong interference regime when the coding scheme achieving the capacity region is such that each receiver decodes all transmitted messages by its connected transmitters*. Next, conditions under which the strong interference regime holds are explored. The two-receiver networks are considered initially. A unified outer bound for the capacity region of these networks is established. It is shown that this outer bound can be systematically translated into simple and computable capacity outer bounds for special cases such as the two-user Classical Interference Channel (CIC) and the Broadcast Channel with Cognitive Relays (BCCR) with common information. For these channels, special cases are presented where our outer bounds are tight, which yield the exact capacity. More importantly, by using the derived outer bounds, a strong interference regime is identified for the general two-receiver interference networks with any arbitrary topology. This strong interference regime, which is represented by only two conditions, includes all previously known results for simple topologies such as the two-user CIC, the cognitive radio channel, and many others. Then, networks with arbitrary number of receivers are considered. Finding non-trivial strong interference regime for such networks, specifically for the CICs with more than two users, has been one of the open problems in network information theory. In this paper, we will give a solution to this problem. Specifically, a new approach is developed based on which one can obtain strong interference regimes not only for the multi-user CICs but also for any interference network of arbitrary large sizes. To this development, some new lemmas are proved which have a central role in the derivations. As a result, this paper establishes the first non-trivial capacity result for the multi-user classical interference channel. A general formula is also presented to derive strong interference conditions for any given network topology. In fact, for the first time in the history of network information theory, we establish a capacity result for a general single-hop communication network regardless of that how much its topology is complex.**

[1]Reza K. Farsani was with the department of electrical engineering, Sharif University of Technology. He is by now with the school of cognitive sciences, Institute for Research in Fundamental Sciences (IPM), Tehran, Iran.

**To:**

**THE MAHDI**

# Contents

# I. INTRODUCTION

As given in Part I of our multi-part papers [1], the interference network is referred to as a general single-hop communication scenario which is composed of a number of transmitters and a number of receivers without any interactive/relay node. In Part I [1], we provided a detailed review of relevant papers which deal with information theoretic capacity limits for these networks. Also, we studied in details the basic building blocks of such networks including the Multiple Access Channel (MAC), the Broadcast Channel (BC), the Classical Interference Channel (CIC), and the Cognitive Radio Channel (CRC). The study of large multi-user/multi-message networks was launched in Part II [2] where we considered degraded networks. We derived a full characterization of the sum-rate capacity for the general degraded interference networks with any arbitrary topology. Following up our systematic study, in this part, we consider the information flow in the strong interference regime.

What do we mean by the strong interference network? For the first time, Carleial [9] determined a case for the two-user CIC where interference does not reduce the capacity. In this regime, which is now called the very strong interference regime, the optimal strategy to achieve the capacity is that each receiver first decodes the interference to remove it from its received signal, and then decodes its respective signal. Later, Sato [10] developed this result and identified a regime (that includes the very strong interference regime) for the two-user Gaussian CIC where the joint decoding of both messages at both receivers is optimal and achieves the capacity. The Sato's result was also extended to the discrete channel in [11]. This regime is called "strong interference regime". Thus, for the two-user CIC in the strong interference regime, the capacity region is derived by viewing the channel as a compound MAC [12]. In [13] such a strong interference regime was also identified for some variations of the two-user CIC, specifically for the CRC. In all these papers, the strong interference regime is defined based on the conditions under which the capacity region is derived by allowing each receiver to decode all messages. In Section III, we shall demonstrate the incompleteness/inconsistency of such definitions and show that this concept is required to be more broadly defined. For this purpose, we consider the CRC and discuss the strong interference conditions derived in [13] for the general case and those derived for the specific Gaussian channel. Surprisingly, we prove that the conditions for the Gaussian channel are not consistent with those for the general case. As a result of this analysis, we identify a new class of CRCs for which the decoding of both messages at both receivers achieves the capacity; nonetheless, it is not included in the strong interference regime derived in [13] for the channel in the general case. Thus, we need to revisit the concept of strong interference to have a consistent definition so that it can be uniformly and consistently applicable to all interference networks with any topology. In Section III, we will define the strong interference regime as follows: *a network is said to be in the strong interference regime when the coding scheme achieving the capacity region is such that each receiver decodes all transmitted messages by its connected transmitters*. Indeed, this basic characteristic deserves to be considered as a criterion while referring to the strong interference.

Next, we explore conditions for general interference networks under which the strong interference regime holds. First, we consider the two-receiver networks in Section IV. We establish a general unified outer bound on the capacity region for these networks. Our outer bound is derived based on a systematic view such that all its constraints are represented by a unique expression. By this approach, we obtain very useful capacity outer bounds which are systematically adaptable to different network scenarios. To demonstrate this fact, by exploiting the derived unified outer bound, we establish simple computable capacity outer bounds for some network topologies including the two-user CIC and the Broadcast Channel with Cognitive Relays (BCCR) with common message. Indeed, our new approach to establish capacity outer bounds contains important benefits; specifically, it clearly depicts how the structure of the outer bound varies by network topology. Moreover, one can flexibly make use of the derived outer bounds to prove capacity results for networks with specific conditions. Consequently, we identify strong interference regimes for the two-user CIC and the BCCR with common message. Also, we derive less-noisy receiver conditions for these scenarios under which a successive decoding scheme is sum-rate optimal. As we will prove, our general outer bounds are indeed efficient to derive strong interference conditions for any arbitrary two-receiver network. A remarkable point is that our strong interference regime for the two-receiver networks is represented by only two conditions. Also, it includes all previously known results for simple topologies such as the two-user CIC, the cognitive interference channel and many others.

In Section V, we consider the multi-receiver interference networks. First, we show that the outer bounds derived for two-receiver networks can be extended to scenarios with arbitrary number of receivers. As a result, we establish useful capacity outer bounds for networks with arbitrary topologies which are tighter than the existing trivial cut-set bound [14]. As we will demonstrate, these outer bounds are also efficient to derive strong interference conditions for some special cases. Nonetheless, for many multi-receiver networks, the obtained outer bounds lead to trivial cases when exploring the strong interference regime. In fact, finding strong interference conditions for interference networks with more than two receivers has been a challenging problem in network information theory. Specifically, it has been an open problem [15, page 6-68] for the multi-receiver CICs. In the present part of our multi-part papers, we give a solution to this problem. We develop a new approach where one can obtain strong interference regimes not only for the multi-user CICs but also for any interference network of arbitrary size. Our methodology includes a new technique for deriving single-letter outer bounds for multi-receiver networks with specific conditions. Also, some new lemmas are proved which have a central role in our derivations. As a result, in this paper, the first non-trivial capacity result for the multi-user classical interference channels is established. We also provide a general formula to derive strong interference conditions for any given network topology. In fact, for the first time in the history of network information theory, we establish a capacity result for a general single-hop communication network regardless of that how much its topology is complex.

Finally, in Section VI, we concentrate on the BCCR with common message and discuss capacity bounds for this network. The purpose for this study is two folds. The BCCR with common message is a network which includes all basic structures presented in Part I [1] as special cases. In this section, we will provide a random coding scheme for such a network based on the systematic achievability design given in our previous paper [7, 8]. Our achievability scheme for the BCCR with common message systematically combines the Han-Kobayashi scheme for the two-user CIC [16] and the Marton's coding scheme for the two-user BC [17]; while, it employs only a simple rate splitting which is identically exploited in the Han-Kobayashi scheme. In our scheme, each message is split into only two parts; thus, reducing the computational complexity of the resulting achievable rate region. We argue that the same coding scheme can be developed for other two-receiver interference networks. The second purpose is to show that the outer bounds derived in Section IV for two-receiver networks are useful to establish exact capacity regions not only in the strong interference regime but also for many other scenarios such as more-capable and semi-deterministic networks. Hence, using the derived inner and outer bounds for the BCCR with common message, we establish the capacity region for a class of more-capable CRCs with common message[2]. In fact, we develop a framework where one can intuitively obtain capacity results for many interference networks with specific characteristics such as more-capability. In our methodology, large multi-user/multi-message networks are systematically treated based on the basic building blocks.

Preliminaries, definitions and notations are also given in the following section where we prove some new technical lemmas which have a central role in the derivations of the present part as well as Part IV [4] of our multi-part papers.

# II. PRELIMINARIES

In this part of our multi-part papers, we use the same notations and definitions as Part I [1, Sec. II]. Also, we assume the reader is familiar with the preliminaries provided in Part I [1, Sec. II.B] regarding the interference networks and their structures. Here, we briefly review the general interference network. The network model has been shown in Fig. 1.

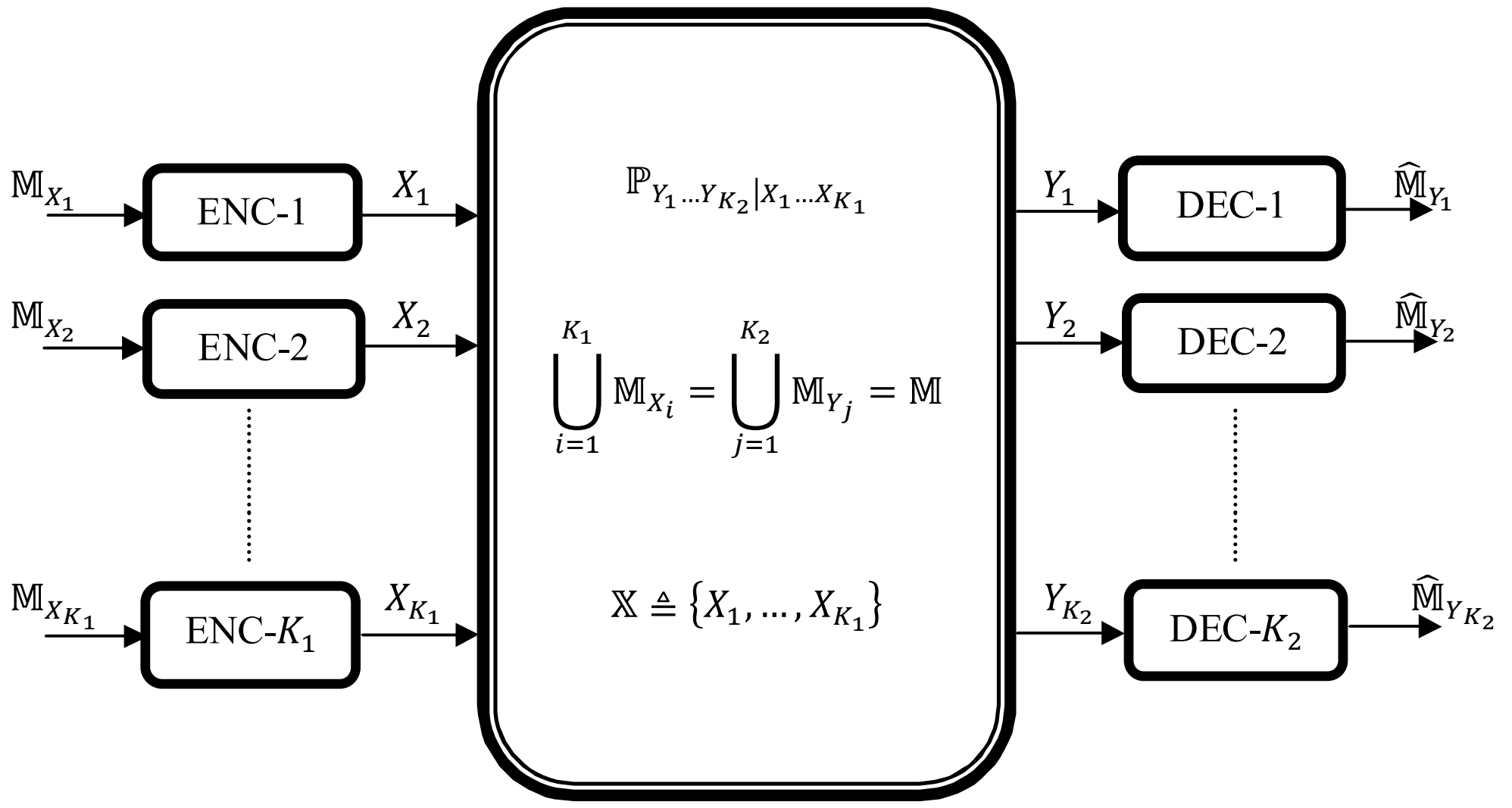

Figure 1. The General Interference Netwrok (GIN).

[2]The CRC with common message is a sub-network of the BCCR with common message. Comparing with the cognitive radio channel [13], this model includes the additional feature that the cognitive transmitter sends also a common message to both receivers.

In this scenario, $K_1$ transmitters send independent messages $\mathbb{M} \triangleq \{M_1, \dots, M_K\}$ to $K_2$ receivers; the transmitter $X_i, i = 1, \dots, K_1$ sends the messages $\mathbb{M}_{X_i}$ over the network, and the receiver $Y_j, j = 1, \dots, K_2$ decodes the messages $\mathbb{M}_{Y_j}$. Therefore, we have:

$$\bigcup_{i=1}^{K_1} \mathbb{M}_{X_i} = \bigcup_{j=1}^{K_2} \mathbb{M}_{Y_j} = \mathbb{M} \tag{1}$$

The network transition probability function $\mathbb{P}_{Y_1 \dots Y_{K_2}|X_1 \dots X_{K_1}}(y_1, \dots, y_{K_2}|x_1, \dots, x_{K_1})$ describes the relation between the inputs and the outputs. The coding procedure for the network was discussed in details in Subsection II.B of Part I [1]. Throughout the paper, we denote the set of input signals by $\mathbb{X}$:

$$\mathbb{X} \triangleq \{X_1, \dots, X_{K_1}\} \tag{2}$$

Also, the general Gaussian interference network with real-valued input and output signals is given by:

$$\begin{bmatrix} Y_1 \\ \vdots \\ Y_{K_2} \end{bmatrix} = \begin{bmatrix} a_{11} & \cdots & a_{1K_1} \\ \vdots & \ddots & \vdots \\ a_{K_2 1} & \cdots & a_{K_2 K_1} \end{bmatrix} \begin{bmatrix} X_1 \\ \vdots \\ X_{K_1} \end{bmatrix} + \begin{bmatrix} Z_1 \\ \vdots \\ Z_{K_2} \end{bmatrix} \tag{3}$$

where the parameters $\{a_{ji}\}_{\substack{j=1,\dots,K_2 \\ i=1,\dots,K_1}}$ are (fixed) real-valued numbers, $\{X_i\}_{i=1}^{K_1}$ are the input symbols, $\{Y_j\}_{j=1}^{K_2}$ are the output symbols and the noise terms $\{Z_j\}_{j=1}^{K_2}$ are zero-mean unit-variance Gaussian Random Variables (RVs). The $i^{th}$ encoder is subject to an average power constraint as: $\mathbb{E}[X_i^2] \leq P_i$, where $P_i \in \mathbb{R}_+$, $i = 1, \dots, K_1$.

Also, we recall the concept of connected and unconnected transmitters with respect to receivers in an interference network which was explicitly defined in Part I [1, Def. II.9]. Briefly, for a given network, a transmitter is connected to a receiver if its receiving signal statistically depends on the input signal with respect to that transmitter; otherwise, the transmitter is unconnected to the receiver. According to the notations of Part I [1, Def. II.9], for a given receiver $Y_j, j = 1, \dots, K_2$, the sets of connected and unconnected transmitters are represented by $\mathbb{X}_{c \to Y_j}$ and $\mathbb{X}_{c \nrightarrow Y_j}$, respectively, where $\mathbb{X}_{c \to Y_j}$ and $\mathbb{X}_{c \nrightarrow Y_j}$ are subsets of $\mathbb{X} \triangleq \{X_1, \dots, X_{K_1}\}$. As argued in [1, Lemma II.6], for any arbitrary interference, without loss of generality, one can assume that with respect to every receiver there is no message known only at its unconnected transmitters. In other words, we impose that:

$$\mathbb{M}_{Y_j} \subseteq \bigcup_{X_i \in \mathbb{X}_{c \to Y_j}} \mathbb{M}_{X_i}, \qquad j = 1, \dots, K_2 \tag{4}$$

Clearly, for each message belonging to $\mathbb{M}_{Y_j}$, there exists at least one transmitter connected to $Y_j$ which transmits that message. In the present paper (Part III) and also in future parts, we need some new definitions and preliminaries as given below.

***Definition 1:*** *Consider the general interference network in Fig. 1 with the corresponding message set $\mathbb{M}$. Also, for every subset $\Omega$ of $\mathbb{M}$, we define:*

$$\mathbb{X}_{\Omega} \triangleq \{X_i \in \mathbb{X} : \mathbb{M}_{X_i} \subseteq \Omega\} \tag{5}$$

*Therefore, the set $\mathbb{X}_{\Omega}$ is composed of all input signals that their messages lies in the set $\Omega$.*

***Definition 2:*** **Connected and Unconnected Messages**

*Consider the general interference network in Fig. 1 with the corresponding message set* $\mathbb{M}$. *For a given receiver* $Y_j$, *the set of unconnected messages* $\mathbb{M}_{c\nrightarrow Y_j}$ *is defined as follows:*

$$\mathbb{M}_{c\nrightarrow Y_j} \triangleq \left(\bigcup_{X_i\in\mathbb{X}_{c\nrightarrow Y_j}} \mathbb{M}_{X_i}\right) - \left(\bigcup_{X_i\in\mathbb{X}_{c\rightarrow Y_j}} \mathbb{M}_{X_i}\right), \qquad j=1,\dots,K_2 \tag{6}$$

*where* $\mathbb{X}_{c\rightarrow Y_j}$ *and* $\mathbb{X}_{c\nrightarrow Y_j}$ *respectively denote the set of connected and unconnected transmitters corresponding to the receiver* $Y_j$. *Also, the set of connected messages with respect to the receiver* $Y_j$ *is defined as:* $\mathbb{M}_{c\rightarrow Y_j} \triangleq \mathbb{M} - \mathbb{M}_{c\nrightarrow Y_j}$.

Based on Definition 2, one can readily deduce the following fact.

***Observation 1:*** *Consider the general interference network in Fig. 1. For each receiver* $Y_j$, $j=1,\dots,K_2$, *the unconnected messages* $\mathbb{M}_{c\nrightarrow Y_j}$ *are statistically independent of the received signal* $Y_j$.

***Definition 3:*** *Suppose* $m\in\mathbb{N}$. *Let* $\Lambda_m:\mathbb{N}^m\rightarrow\mathbb{N}$ *be a bijection. The order relation* $<_{\Lambda_m}$ *induced by* $\Lambda_m(.)$ *on the set* $\mathbb{N}^m$ *is defined as follows. For every* $(a_1,\dots,a_m)$ *and* $(b_1,\dots,b_m)$ *in* $\mathbb{N}^m$, *where* $(a_1,\dots,a_m)\neq(b_1,\dots,b_m)$, *we have:*

$$(a_1,\dots,a_m) <_{\Lambda_m} (b_1,\dots,b_m) \qquad \Leftrightarrow \qquad \Lambda_m(a_1,\dots,a_m) < \Lambda_m(b_1,\dots,b_m) \tag{7}$$

*Also, the "min" operator with respect to* $<_{\Lambda_m}$, *denoted by* $min\Lambda_m$, *is defined as follows. Let* $S$ *be a nonempty subset of* $\mathbb{N}^m$. *We have:*

$$min\Lambda_m\, S \triangleq \Lambda_m^{-1}(min\ \ \{\Lambda_m(s) : s\in S\}) \tag{8}$$

*where* $\Lambda_m^{-1}(.)$ *denotes the inverse function. The "max" operator is defined similarly.*

## ➢ New Lemmas

In what follows, we derive some useful lemmas which are extensively used throughout the paper. The following results have a central role for our derivations in the present as well as future parts of our multi-part papers.

Let $\mathcal{Y}_1,\mathcal{Y}_2,\mathcal{X}_1,\mathcal{X}_2$ be arbitrary sets and $\mathbb{P}(y_1,y_2|x_1,x_2)$ be a conditional probability function defined on the set $\mathcal{Y}_1\times\mathcal{Y}_2\times\mathcal{X}_1\times\mathcal{X}_2$. Consider the following inequality:

$$I(X_1;Y_1|X_2) \leq I(X_1;Y_2|X_2) \tag{9}$$

In Part I [1, Lemma III.1], we proved that if the inequality (9) holds for all product PDFs $P_{X_1}(x_1)P_{X_2}(x_2)$, then it also holds for all joint PDFs $P_{X_1X_2}(x_1,x_2)$. As discussed in [1, Sec. III.A], a direct consequence of this result is that dependence between the input variables does not change the strong interference conditions for the two-user CIC. Now we intend to present a generalization of this result.

***Lemma 1)*** *Let* $\mathcal{Y}_1,\mathcal{Y}_2,\mathcal{X}_1,\mathcal{X}_2,\dots,\mathcal{X}_{\mu_1},\mathcal{X}_{\mu_1+1},\dots,\mathcal{X}_{\mu_1+\mu_2}$ *be arbitrary sets, where* $\mu_1,\mu_2\in\mathbb{N}$ *are arbitrary natural numbers. Let also* $\mathbb{P}(y_1,y_2|x_1,x_2,\dots,x_{\mu_1},x_{\mu_1+1},\dots,x_{\mu_1+\mu_2})$ *be a given conditional probability distribution defined on the set* $\mathcal{Y}_1\times\mathcal{Y}_2\times\mathcal{X}_1\times\mathcal{X}_2\times\dots\times\mathcal{X}_{\mu_1}\times\mathcal{X}_{\mu_1+1}\times\dots\times\mathcal{X}_{\mu_1+\mu_2}$. *Consider the following inequality:*

$$I(X_1,\dots,X_{\mu_1};Y_1|X_{\mu_1+1},\dots,X_{\mu_1+\mu_2}) \leq I(X_1,\dots,X_{\mu_1};Y_2|X_{\mu_1+1},\dots,X_{\mu_1+\mu_2}) \tag{10}$$

*If the inequality* (10) *holds for all PDFs* $P_{X_1\dots X_{\mu_1}X_{\mu_1+1}\dots X_{\mu_1+\mu_2}}(x_1,\dots,x_{\mu_1},x_{\mu_1+1},\dots,x_{\mu_1+\mu_2})$ *with the following factorization:*

$$P_{X_1\dots X_{\mu_1}X_{\mu_1+1}\dots X_{\mu_1+\mu_2}} = P_{X_1\dots X_{\mu_1}}(x_1,\dots,x_{\mu_1})P_{X_{\mu_1+1}}(x_{\mu_1+1})P_{X_{\mu_1+2}}(x_{\mu_1+2})\dots P_{X_{\mu_1+\mu_2}}(x_{\mu_1+\mu_2}) \tag{11}$$

*then, we have:*

$$I(X_1,\dots,X_{\mu_1};Y_1|X_{\mu_1+1},\dots,X_{\mu_1+\mu_2},D) \le I(X_1,\dots,X_{\mu_1};Y_2|X_{\mu_1+1},\dots,X_{\mu_1+\mu_2},D) \tag{12}$$

*for all joint PDFs* $P_{DX_1\dots X_{\mu_1}X_{\mu_1+1}\dots X_{\mu_1+\mu_2}}(d,x_1,\dots,x_{\mu_1},x_{\mu_1+1},\dots,x_{\mu_1+\mu_2})$ *where* $D \to X_1,\dots,X_{\mu_1},X_{\mu_1+1},\dots,X_{\mu_1+\mu_2} \to Y_1,Y_2$ *forms a Markov chain.*

*Proof of Lemma 1)* First we show that (10) implies the following inequality:

$$I(X_1,\dots,X_{\mu_1};Y_1|X_{\mu_1+1},\dots,X_{\mu_1+\mu_2},W) \le I(X_1,\dots,X_{\mu_1};Y_2|X_{\mu_1+1},\dots,X_{\mu_1+\mu_2},W) \tag{13}$$

for all PDFs $P_{WX_1\dots X_{\mu_1}X_{\mu_1+1}\dots X_{\mu_1+\mu_2}}(w,x_1,\dots,x_{\mu_1},x_{\mu_1+1},\dots,x_{\mu_1+\mu_2})$ with:

$$P_{WX_1\dots X_{\mu_1}X_{\mu_1+1}\dots X_{\mu_1+\mu_2}} = P_W P_{X_1\dots X_{\mu_1}|W}P_{X_{\mu_1+1}|W}P_{X_{\mu_1+2}|W}\dots P_{X_{\mu_1+\mu_2}|W} \tag{14}$$

where $W \to X_1,\dots,X_{\mu_1},X_{\mu_1+1},\dots,X_{\mu_1+\mu_2} \to Y_1,Y_2$ forms a Markov chain. To prove this inequality, one can write:

$$\begin{aligned}
&I(X_1,\dots,X_{\mu_1};Y_1|X_{\mu_1+1},\dots,X_{\mu_1+\mu_2},W)\\
&\quad=\sum_w P_W(w)I(X_1,\dots,X_{\mu_1};Y_1|X_{\mu_1+1},\dots,X_{\mu_1+\mu_2},w)\\
&\quad=\sum_w P_W(w)I(X_1,\dots,X_{\mu_1};Y_1|X_{\mu_1+1},\dots,X_{\mu_1+\mu_2})_{\langle P_{X_1\dots X_{\mu_1}|w}\times P_{X_{\mu_1+1}|w}\times P_{X_{\mu_1+2}|w}\times\dots\times P_{X_{\mu_1+\mu_2}|w}\rangle}\\
&\quad\overset{(a)}{\le}\sum_w P_W(w)I(X_1,\dots,X_{\mu_1};Y_2|X_{\mu_1+1},\dots,X_{\mu_1+\mu_2})_{\langle P_{X_1\dots X_{\mu_1}|w}\times P_{X_{\mu_1+1}|w}\times P_{X_{\mu_1+2}|w}\times\dots\times P_{X_{\mu_1+\mu_2}|w}\rangle}\\
&\quad=\sum_w P_W(w)I(X_1,\dots,X_{\mu_1};Y_2|X_{\mu_1+1},\dots,X_{\mu_1+\mu_2},w)\\
&\quad=I(X_1,\dots,X_{\mu_1};Y_2|X_{\mu_1+1},\dots,X_{\mu_1+\mu_2},W)
\end{aligned} \tag{15}$$

where the notation $I(A;B|C)_{\langle P(.)\rangle}$ indicates that the mutual information function $I(A;B|C)$ is evaluated by the distribution $P(.)$. Note that for any given $w$, the function $P_{X_1\dots X_{\mu_1}|w}\times P_{X_{\mu_1+1}|w}\times P_{X_{\mu_1+2}|w}\times\dots\times P_{X_{\mu_1+\mu_2}|w}$ is a probability distribution defined over the set $\mathcal{X}_1\times\dots\times\mathcal{X}_{\mu_1}\times\mathcal{X}_{\mu_1+1}\times\dots\times\mathcal{X}_{\mu_1+\mu_2}$ with the factorization (11). The inequality (a) is due to (10).

Now, having at hand the inequality (13), one can substitute $W \equiv (D,X_{\mu_1+1},X_{\mu_1+2},\dots,X_{\mu_1+\mu_2})$ with an arbitrary joint distribution[3] on the set $\mathcal{D}\times\mathcal{X}_{\mu_1+1}\times\dots\times\mathcal{X}_{\mu_1+\mu_2}$ . By this substitution, we derive that (12) holds for all joint PDFs $P_{DX_{\mu_1+1}\dots X_{\mu_1+\mu_2}}P_{X_1\dots X_{\mu_1}|DX_{\mu_1+1}\dots X_{\mu_1+\mu_2}}$. The proof is thus complete. ■

***Remark 1)*** It is essential to remark that the inequality (12) holds only for those auxiliary random variables "$D$" where $D \to X_1,\dots,X_{\mu_1},X_{\mu_1+1},\dots,X_{\mu_1+\mu_2} \to Y_1,Y_2$ forms a Markov chain. For example, it is wrong to set $D \equiv Y_2$ in (12) and deduce that:

$$I(X_1,\dots,X_{\mu_1};Y_1|X_{\mu_1+1},\dots,X_{\mu_1+\mu_2},Y_2) \overset{?}{=} 0 \tag{16}$$

[3]We have this liberty because $P_W(w)$ in (14) is arbitrary.

In general, (10) *does not* imply the equality (16). For more explanation, consider a two-user BC with input $X$ and outputs $Y_1$ and $Y_2$ with transition probability function $\mathbb{P}(y_1, y_2|x)$. Consider the following condition:

$$I(X;Y_2) \leq I(X;Y_1) \qquad \textit{for all joint PDFs} \qquad P_X(x) \tag{17}$$

The condition (17) actually represents the class of *more-capable* BCs [18]. According to Lemma 1, (17) implies that:

$$I(X;Y_2|D) \leq I(X;Y_1|D) \qquad \textit{for all joint PDFs} \qquad P_{DX}(d,x) \tag{18}$$

Let us now set $D \equiv Y_1$ in (18). We obtain that $I(X;Y_2|Y_1) = 0$. It is clear that the latter equality implies that $X \to Y_1 \to Y_2$ form a Markov chain, i.e., $Y_2$ is a degraded version of $Y_1$. But we know from [18] that the more-capable BCs strictly include the degraded BCs as a subset. In other words, the condition (17) in general does not imply that $I(X;Y_2|Y_1) = 0$. The fact is that in (18) we require that $D \to X \to Y_1, Y_2$ form a Markov chain. Therefore, the choice $D \equiv Y_1$ is not admissible.

***Corollary 1)*** *Let $\mathcal{L}$ be an arbitrary subset of $\{1, \dots, \mu_1\}$. Denote $\mathbb{X}_{\mathcal{L}} \triangleq \{X_i : i \in \mathcal{L}\}$. If the inequality* (10) *holds for all joint PDFs* (11), *then we have:*

$$I\left(\{X_1, \dots, X_{\mu_1}\} - \mathbb{X}_{\mathcal{L}}; Y_1 \middle| \mathbb{X}_{\mathcal{L}}, X_{\mu_1+1}, \dots, X_{\mu_1+\mu_2}, D\right) \leq I\left(\{X_1, \dots, X_{\mu_1}\} - \mathbb{X}_{\mathcal{L}}; Y_2 \middle| \mathbb{X}_{\mathcal{L}}, X_{\mu_1+1}, \dots, X_{\mu_1+\mu_2}, D\right) \tag{19}$$

*for all joint PDFs $P_{DX_1 \dots X_{\mu_1} X_{\mu_1+1} \dots X_{\mu_1+\mu_2}}(d, x_1, \dots, x_{\mu_1}, x_{\mu_1+1}, \dots, x_{\mu_1+\mu_2})$ where $D \to X_1, \dots, X_{\mu_1}, X_{\mu_1+1}, \dots, X_{\mu_1+\mu_2} \to Y_1, Y_2$ forms a Markov chain.*

*Proof of Corollary 1)* It is sufficient to replace $D$ with $(D, \mathbb{X}_{\mathcal{L}})$ in (12). ■

Next let us consider a Gaussian transition probability function where the outputs $Y_1$ and $Y_2$ are given by:

$$\begin{cases} Y_1 \triangleq a_1 X_1 + a_2 X_2 + \dots + a_{\mu_1} X_{\mu_1} + a_{\mu_1+1} X_{\mu_1+1} + \dots + a_{\mu_1+\mu_2} X_{\mu_1+\mu_2} + Z_1 \\ Y_2 \triangleq b_1 X_1 + b_2 X_2 + \dots + b_{\mu_1} X_{\mu_1} + b_{\mu_1+1} X_{\mu_1+1} + \dots + b_{\mu_1+\mu_2} X_{\mu_1+\mu_2} + Z_2 \end{cases} \tag{20}$$

where $Z_1$ and $Z_2$ are zero-mean unit-variance Gaussian random variables; also, $X_1, X_2, \dots, X_{\mu_1}, X_{\mu_1+1}, \dots, X_{\mu_1+\mu_2}$ are real-valued power-constrained random variables independent of $(Z_1, Z_2)$ and $a_1, a_2, \dots, a_{\mu_1}, a_{\mu_1+1}, \dots, a_{\mu_1+\mu_2}$ and $b_1, b_2, \dots, b_{\mu_1}, b_{\mu_1+1}, \dots, b_{\mu_1+\mu_2}$ are fixed real numbers. Our purpose is to determine sufficient conditions such that the inequality (12) holds for all joint PDFs $P_{DX_1 \dots X_{\mu_1} X_{\mu_1+1} \dots X_{\mu_1+\mu_2}}(d, x_1, \dots, x_{\mu_1}, x_{\mu_1+1}, \dots, x_{\mu_1+\mu_2})$. A sufficient condition is given by the following lemma:

***Lemma 2)*** *Consider the Gaussian system in* (20). *If the following condition satisfies:*

$$\frac{a_1}{b_1} = \frac{a_2}{b_2} = \dots = \frac{a_{\mu_1}}{b_{\mu_1}} = \alpha, \qquad |\alpha| \leq 1 \tag{21}$$

*then, the inequality* (12) *holds for all joint PDFs $P_{DX_1 \dots X_{\mu_1} X_{\mu_1+1} \dots X_{\mu_1+\mu_2}}(d, x_1, \dots, x_{\mu_1}, x_{\mu_1+1}, \dots, x_{\mu_1+\mu_2})$ where $D$ is independent of $(Z_1, Z_2)$.*

*Proof of Lemma 2)* First note that if $D$ is independent of $(Z_1, Z_2)$, then $D \to X_1, \dots, X_{\mu_1}, X_{\mu_1+1}, \dots, X_{\mu_1+\mu_2} \to Y_1, Y_2$ forms a Markov chain. It is sufficient to prove that (10) holds. Now define:

$$\tilde{Y}_1 \triangleq \alpha Y_2 + (\alpha b_{\mu_1+1} - a_{\mu_1+1}) X_{\mu_1+1} + (\alpha b_{\mu_1+2} - a_{\mu_1+2}) X_{\mu_1+2} + \dots + (\alpha b_{\mu_1+\mu_2} - a_{\mu_1+\mu_2}) X_{\mu_1+\mu_2} + \sqrt{1-\alpha^2}\tilde{Z}_1 \tag{22}$$

where $\tilde{Z}_1$ is a Gaussian random variable with zero mean and unit variance which is independent of $(Z_1, Z_2)$. By considering (20), it is readily derived that $\tilde{Y}_1$ is statistically equivalent to $Y_1$ in the sense of:

$$\mathbb{P}(\tilde{y}_1 | x_1, \dots, x_{\mu_1}, x_{\mu_1+1}, \dots, x_{\mu_1+\mu_2}) \approx \mathbb{P}(y_1 | x_1, \dots, x_{\mu_1}, x_{\mu_1+1}, \dots, x_{\mu_1+\mu_2})$$

Therefore, for all input distributions, we have:

$$\begin{aligned} I(X_1, \dots, X_{\mu_1}; Y_1 | X_{\mu_1+1}, \dots, X_{\mu_1+\mu_2}) &= I(X_1, \dots, X_{\mu_1}; \tilde{Y}_1 | X_{\mu_1+1}, \dots, X_{\mu_1+\mu_2}) \\ &\leq I(X_1, \dots, X_{\mu_1}; \tilde{Y}_1, Y_2 | X_{\mu_1+1}, \dots, X_{\mu_1+\mu_2}) \\ &\overset{(a)}{=} I(X_1, \dots, X_{\mu_1}; Y_2 | X_{\mu_1+1}, \dots, X_{\mu_1+\mu_2}) + I(X_1, \dots, X_{\mu_1}; \tilde{Y}_1 | Y_2, X_{\mu_1+1}, \dots, X_{\mu_1+\mu_2}) \\ &= I(X_1, \dots, X_{\mu_1}; Y_2 | X_{\mu_1+1}, \dots, X_{\mu_1+\mu_2}) \end{aligned}$$

where (a) holds because $X_1, \dots, X_{\mu_1} \to Y_2, X_{\mu_1+1}, \dots, X_{\mu_1+\mu_2} \to \tilde{Y}_1$ forms a Markov chain from (22). The proof is thus complete. ■

***Remarks 2:***

1. The proof of Lemma 2 indicates that under the condition (21), given $X_{\mu_1+1}, \dots, X_{\mu_1+\mu_2}$, the signal $Y_1$ is a stochastically degraded version of $Y_2$.
2. The relation (21) is a sufficient condition under which (10) holds; however, in general, the inequality (10) may not be equivalent to (21). It is also essential to note that the condition (21) is not derived by evaluating (10) for Gaussian input distributions. Only for the case of $\mu_1 = 1$, the condition (21) can be equivalently derived by evaluating (10) for Gaussian input distributions.

The result of Lemma 1 is crucial to identify strong interference regime for the two-receiver interference networks. In the next lemma, we provide a multi-letter extension to this lemma which is necessary to identify strong interference regime for networks of arbitrary large sizes.

***Lemma 3)*** *Fix the conditional PDF* $\mathbb{P}(y_1, y_2 | x_1, \dots, x_{\mu_1}, x_{\mu_1+1}, \dots, x_{\mu_1+\mu_2})$. *Assume that the inequality* (10) *holds for all joint PDFs* (11). *For a given arbitrary natural number* $n$, *let* $\mathbb{P}(y_1^n, y_2^n | x_1^n, x_2^n, \dots, x_{\mu_1}^n, x_{\mu_1+1}^n, \dots, x_{\mu_1+\mu_2}^n)$ *be a memoryless* $n$*-tuple extension of* $\mathbb{P}(y_1, y_2 | x_1, x_2, \dots, x_{\mu_1}, x_{\mu_1+1}, \dots, x_{\mu_1+\mu_2})$, *i.e.,*

$$\mathbb{P}(y_1^n, y_2^n | x_1^n, x_2^n, \dots, x_{\mu_1}^n, x_{\mu_1+1}^n, \dots, x_{\mu_1+\mu_2}^n) = \prod_{t=1}^{n} \mathbb{P}(y_{1,t}, y_{2,t} | x_{1,t}, x_{2,t}, \dots, x_{\mu_1,t}, x_{\mu_1+1,t}, \dots, x_{\mu_1+\mu_2,t}) \tag{23}$$

*Then, the following inequality holds:*

$$I(X_1^n, \dots, X_{\mu_1}^n; Y_1^n | X_{\mu_1+1}^n, \dots, X_{\mu_1+\mu_2}^n, D) \leq I(X_1^n, \dots, X_{\mu_1}^n; Y_2^n | X_{\mu_1+1}^n, \dots, X_{\mu_1+\mu_2}^n, D) \tag{24}$$

*for all joint PDFs* $P_{DX_1^n \dots X_{\mu_1}^n X_{\mu_1+1}^n \dots X_{\mu_1+\mu_2}^n}(d, x_1^n, \dots, x_{\mu_1}^n, x_{\mu_1+1}^n, \dots, x_{\mu_1+\mu_2}^n)$ *where* $D \to X_1^n, \dots, X_{\mu_1}^n, X_{\mu_1+1}^n, \dots, X_{\mu_1+\mu_2}^n \to Y_1^n, Y_2^n$ *forms a Markov chain.*

*Proof of Lemma 3)* First note that, according to Lemma 1, since (10) holds for all joint PDFs (11), the inequality (12) also holds for all joint PDFs $P_{DX_1 \dots X_{\mu_1} X_{\mu_1+1} \dots X_{\mu_1+\mu_2}}(d, x_1, \dots, x_{\mu_1}, x_{\mu_1+1}, \dots, x_{\mu_1+\mu_2})$ where $D \to X_1, \dots, X_{\mu_1}, X_{\mu_1+1}, \dots, X_{\mu_1+\mu_2} \to Y_1, Y_2$ forms a Markov chain. Now consider the two sides of (24). For a given vector $A^n$, denote $A^{n \backslash t} \triangleq (A^{t-1}, A_{t+1}^n)$ where $t = 1, .., n$. Define:

$$\bar{\bar{D}} \triangleq (X_{\mu_1+1}^{n \backslash t}, \dots, X_{\mu_1+\mu_2}^{n \backslash t}, Y_2^{t-1}, Y_{1,t+1}^n, D) \tag{25}$$

We have:

$$I(X_1^n,\dots,X_{\mu_1}^n;Y_2^n|X_{\mu_1+1}^n,\dots,X_{\mu_1+\mu_2}^n,D)-I(X_1^n,\dots,X_{\mu_1}^n;Y_1^n|X_{\mu_1+1}^n,\dots,X_{\mu_1+\mu_2}^n,D)$$
$$=\sum_{t=1}^{n}I(X_1^n,\dots,X_{\mu_1}^n;Y_{2,t}|X_{\mu_1+1}^n,\dots,X_{\mu_1+\mu_2}^n,Y_2^{t-1},D)-\sum_{t=1}^{n}I(X_1^n,\dots,X_{\mu_1}^n;Y_{1,t}|X_{\mu_1+1}^n,\dots,X_{\mu_1+\mu_2}^n,Y_{1,t+1}^n,D)$$
$$\overset{(a)}{=}\sum_{t=1}^{n}I(X_{1,t},\dots,X_{\mu_1,t};Y_{2,t}|X_{\mu_1+1}^n,\dots,X_{\mu_1+\mu_2}^n,Y_2^{t-1},D)-\sum_{t=1}^{n}I(X_{1,t},\dots,X_{\mu_1,t};Y_{1,t}|X_{\mu_1+1}^n,\dots,X_{\mu_1+\mu_2}^n,Y_{1,t+1}^n,D)$$
$$\overset{(b)}{=}\sum_{t=1}^{n}I(X_{1,t},\dots,X_{\mu_1,t},Y_{1,t+1}^n;Y_{2,t}|X_{\mu_1+1}^n,\dots,X_{\mu_1+\mu_2}^n,Y_2^{t-1},D)-\sum_{t=1}^{n}I(X_{1,t},\dots,X_{\mu_1,t},Y_2^{t-1};Y_{1,t}|X_{\mu_1+1}^n,\dots,X_{\mu_1+\mu_2}^n,Y_{1,t+1}^n,D)$$
$$=\sum_{t=1}^{n}I(X_{1,t},\dots,X_{\mu_1,t};Y_{2,t}|X_{\mu_1+1}^n,\dots,X_{\mu_1+\mu_2}^n,Y_{1,t+1}^n,Y_2^{t-1},D)-\sum_{t=1}^{n}I(X_{1,t},\dots,X_{\mu_1,t};Y_{1,t}|X_{\mu_1+1}^n,\dots,X_{\mu_1+\mu_2}^n,Y_{1,t+1}^n,Y_2^{t-1},D)$$
$$+\sum_{t=1}^{n}I(Y_{1,t+1}^n;Y_{2,t}|X_{\mu_1+1}^n,\dots,X_{\mu_1+\mu_2}^n,Y_2^{t-1},D)-\sum_{t=1}^{n}I(Y_2^{t-1};Y_{1,t}|X_{\mu_1+1}^n,\dots,X_{\mu_1+\mu_2}^n,Y_{1,t+1}^n,D)$$
$$\overset{(c)}{=}\sum_{t=1}^{n}I(X_{1,t},\dots,X_{\mu_1,t};Y_{2,t}|X_{\mu_1+1}^n,\dots,X_{\mu_1+\mu_2}^n,Y_{1,t+1}^n,Y_2^{t-1},D)-\sum_{t=1}^{n}I(X_{1,t},\dots,X_{\mu_1,t};Y_{1,t}|X_{\mu_1+1}^n,\dots,X_{\mu_1+\mu_2}^n,Y_{1,t+1}^n,Y_2^{t-1},D)$$
$$=\sum_{t=1}^{n}\left(I(X_{1,t},\dots,X_{\mu_1,t};Y_{2,t}|X_{\mu_1+1,t},\dots,X_{\mu_1+\mu_2,t},\bar{\bar{D}})-I(X_{1,t},\dots,X_{\mu_1,t};Y_{1,t}|X_{\mu_1+1,t},\dots,X_{\mu_1+\mu_2,t},\bar{\bar{D}})\right)$$
$$\overset{(d)}{\geq}0$$
(26)

where equality (a) and (b) hold because the memorylessness property (23) implies the following Markov relations:

$$X_1^{n\backslash t},\dots,X_{\mu_1}^{n\backslash t},X_{\mu_1+1}^{n\backslash t},\dots,X_{\mu_1+\mu_2}^{n\backslash t},Y_2^{t-1},Y_{1,t+1}^n,D\to X_{1,t},\dots,X_{\mu_1,t},X_{\mu_1+1,t},\dots,X_{\mu_1+\mu_2,t}\to Y_{1,t},Y_{2,t},\qquad t=1,\dots,n$$
(27)

Also, equality (c) is due to Csiszar-Korner identity [19, Lemma 7] according which the $3^{rd}$ and the $4^{th}$ expressions in the left hand side of (c) are equal; finally, (d) is due to the inequality (12) in which $D$ is replaced by $\bar{\bar{D}}$. The proof is thus complete. ■

Let us consider the special case of $\mu_1=\mu_2=2$. In [11, Appendix], it is shown that if the following condition holds:

$$I(X_1;Y_1|X_2)\leq I(X_1;Y_2|X_2)\quad \text{for all joint PDFs}\quad P_{X_1}P_{X_2}$$
(28)

then, we have:

$$I(X_1^n;Y_1^n|X_2^n)\leq I(X_1^n;Y_2^n|X_2^n)\quad \text{for all joint PDFs}\quad P_{X_1^n}P_{X_2^n}$$
(29)

The proof of [11] is based on induction which requires establishing some sophisticated Markov chains (see [11, Appendix]). Moreover, the authors of [11] are able to derive (29) only for product distributions $P_{X_1^n}\times P_{X_2^n}$. Our proof in Lemma 3 is considerably simpler since, instead of sophisticated induction-based arguments, it is derived by a direct application of the Csiszar-Korner identity [19, Lemma 7]. Also, by using the consequence of Lemma 1, we are able to prove (29) for all arbitrary joint PDFs $P_{X_1^nX_2^n}$. As we will see throughout our multi-part papers, such extension is critical for deriving strong interference regime for large multi-user/multi-message networks.

Below, we also derive a variation of Lemma 1 which is useful to identify networks with a sequence of less noisy receivers (these networks will be studied in details in Part IV [4]).

***Lemma 4)*** *Let* $\mathcal{Y}_1,\mathcal{Y}_2,\mathcal{X}_1,\mathcal{X}_2,\ldots,\mathcal{X}_{\mu_1},\mathcal{X}_{\mu_1+1},\ldots,\mathcal{X}_{\mu_1+\mu_2}$ *be arbitrary sets, where* $\mu_1,\mu_2\in\mathbb{N}$ *are arbitrary natural numbers. Let also* $\mathbb{P}(y_1,y_2|x_1,x_2,\ldots,x_{\mu_1},x_{\mu_1+1},\ldots,x_{\mu_1+\mu_2})$ *be a given conditional probability distribution defined on the set* $\mathcal{Y}_1\times\mathcal{Y}_2\times\mathcal{X}_1\times\mathcal{X}_2\times\ldots\times\mathcal{X}_{\mu_1}\times\mathcal{X}_{\mu_1+1}\times\ldots\times\mathcal{X}_{\mu_1+\mu_2}$. *Consider the inequality below:*

$$I(U;Y_1|X_{\mu_1+1},\ldots,X_{\mu_1+\mu_2})\leq I(U;Y_2|X_{\mu_1+1},\ldots,X_{\mu_1+\mu_2}) \tag{30}$$

*If the inequality* (30) *holds for all PDFs* $P_{UX_1\ldots X_{\mu_1}X_{\mu_1+1}\ldots X_{\mu_1+\mu_2}}(x_1,\ldots,x_{\mu_1},x_{\mu_1+1},\ldots,x_{\mu_1+\mu_2})$ *with the following factorization:*

$$P_{UX_1\ldots X_{\mu_1}X_{\mu_1+1}\ldots X_{\mu_1+\mu_2}}=P_{UX_1\ldots X_{\mu_1}}(u,x_1,\ldots,x_{\mu_1})P_{X_{\mu_1+1}}(x_{\mu_1+1})P_{X_{\mu_1+2}}(x_{\mu_1+2})\ldots P_{X_{\mu_1+\mu_2}}(x_{\mu_1+\mu_2}) \tag{31}$$

*then, we have:*

$$I(U;Y_1|X_{\mu_1+1},\ldots,X_{\mu_1+\mu_2},D)\leq I(U;Y_2|X_{\mu_1+1},\ldots,X_{\mu_1+\mu_2},D) \tag{32}$$

*for all joint PDFs* $P_{DUX_1\ldots X_{\mu_1}X_{\mu_1+1}\ldots X_{\mu_1+\mu_2}}(d,u,x_1,\ldots,x_{\mu_1},x_{\mu_1+1},\ldots,x_{\mu_1+\mu_2})$ *where* $D,U\to X_1,\ldots,X_{\mu_1},X_{\mu_1+1},\ldots,X_{\mu_1+\mu_2}\to Y_1,Y_2$ *form a Markov chain.*

*Proof of Lemma 4)* The proof is rather similar to Lemma 1. First, note that (30) implies the following inequality:

$$I(U;Y_1|X_{\mu_1+1},\ldots,X_{\mu_1+\mu_2},W)\leq I(U;Y_2|X_{\mu_1+1},\ldots,X_{\mu_1+\mu_2},W) \tag{33}$$

for all PDFs $P_{WUX_1\ldots X_{\mu_1}X_{\mu_1+1}\ldots X_{\mu_1+\mu_2}}(w,u,x_1,\ldots,x_{\mu_1},x_{\mu_1+1},\ldots,x_{\mu_1+\mu_2})$ with:

$$P_{WUX_1\ldots X_{\mu_1}X_{\mu_1+1}\ldots X_{\mu_1+\mu_2}}=P_WP_{UX_1\ldots X_{\mu_1}|W}P_{X_{\mu_1+1}|W}P_{X_{\mu_1+2}|W}\ldots P_{X_{\mu_1+\mu_2}|W} \tag{34}$$

where $W,U\to X_1,\ldots,X_{\mu_1},X_{\mu_1+1},\ldots,X_{\mu_1+\mu_2}\to Y_1,Y_2$ forms a Markov chain. This can be proved by following the same lines as (15). Now, having at hand the inequality (34), one can substitute $W\equiv(X_{\mu_1+1},X_{\mu_1+2},\ldots,X_{\mu_1+\mu_2},D)$ with an arbitrary joint distribution on the set $\mathcal{X}_{\mu_1+1}\times\ldots\times\mathcal{X}_{\mu_1+\mu_2}\times\mathcal{D}$. By this substitution, we obtain that the inequality (32) holds for all joint PDFs $P_{X_{\mu_1+1}\ldots X_{\mu_1+\mu_2}D}P_{UX_1\ldots X_{\mu_1}|X_{\mu_1+1}\ldots X_{\mu_1+\mu_2}D}$. The proof is complete. ■

Finally, similar to Lemma 2, one can derive conditions under which for the Gaussian system (20), the inequality (32) holds for all joint PDFs $P_{DUX_1\ldots X_{\mu_1}X_{\mu_1+1}\ldots X_{\mu_1+\mu_2}}(d,u,x_1,\ldots,x_{\mu_1},x_{\mu_1+1},\ldots,x_{\mu_1+\mu_2})$. We present a sufficient condition in the following lemma.

***Lemma 5)*** *Consider the Gaussian system in* (20). *If* (21) *holds, then the inequality* (32) *is satisfied for all joint PDFs* $P_{DUX_1\ldots X_{\mu_1}X_{\mu_1+1}\ldots X_{\mu_1+\mu_2}}(d,u,x_1,\ldots,x_{\mu_1},x_{\mu_1+1},\ldots,x_{\mu_1+\mu_2})$, *where* $(D,U)$ *is independent of* $(Z_1,Z_2)$.

*Proof of Lemma 5)* The proof is actually similar to Lemma 2. In essence, if the condition (21) holds, given $X_{\mu_1+1},\ldots,X_{\mu_1+\mu_2}$, the signal $Y_1$ is a stochastically degraded version of $Y_2$. This fact was previously discussed in Remark 2. Therefore, (32) is always satisfied for all joint PDFs $P_{DUX_1\ldots X_{\mu_1}X_{\mu_1+1}\ldots X_{\mu_1+\mu_2}}(d,u,x_1,\ldots,x_{\mu_1},x_{\mu_1+1},\ldots,x_{\mu_1+\mu_2})$. ■

***Remark 3)*** Lemma 5 provides only a sufficient condition for the Gaussian system (20) to satisfy the inequality (32) for all input distributions. However, it is not clear if these conditions are also necessary.

By these preliminaries, we are now ready to develop our results in the subsequent sections.

# III. A GENERAL DEFINITION FOR STRONG INTERFERENCE REGIME

As discussed in introduction, the main sections in this part of our multi-part papers are concerned with the behavior of information flow in general interference networks with strong interference. Before all, it is required to provide an exact definition of the concept of "strong interference". This issue is of great importance, specifically, when considering large multi-message networks. In the following, we first briefly discuss the well-known strong interference regime for some basic models. We demonstrate, by examples, the incompleteness/inconsistency of the usual definitions for this regime. Accordingly, we provide a more precise definition for the concept of strong interference based on the fundamental characteristic of the optimal coding scheme in this regime.

Consider first the two-user CIC as shown in Fig. 2. As one of the main building blocks of all interference networks, this channel was studied in details in Part I [1].

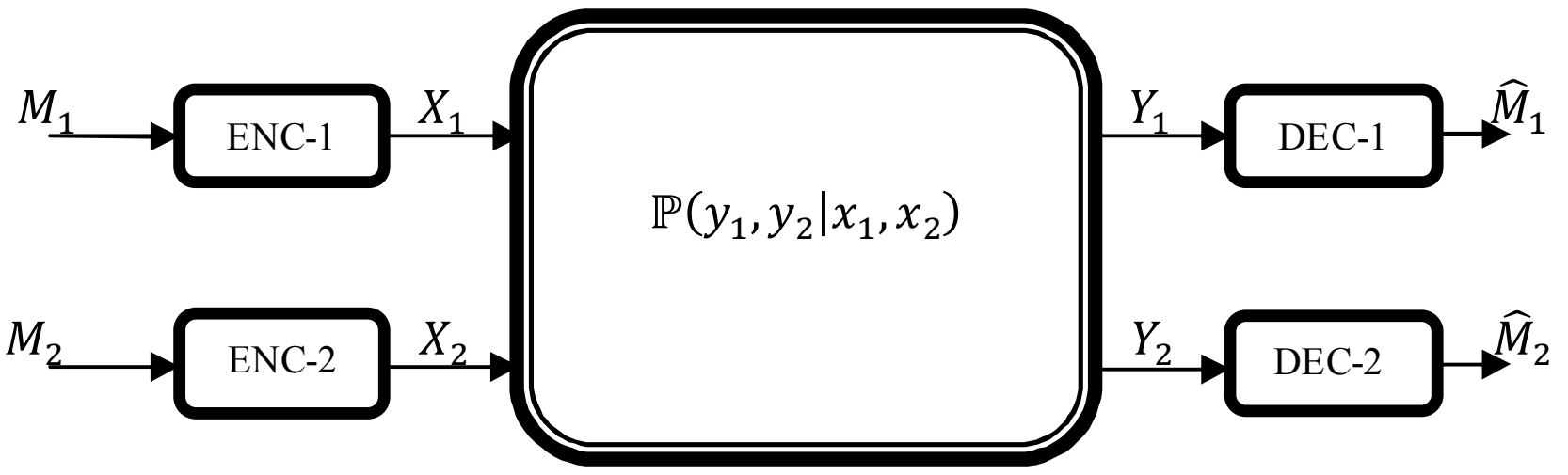

Figure 2. The two-user Classical Interference Channel (CIC).

The Gaussian channel is formulated as follows:

$$\begin{cases} Y_1 = X_1 + aX_2 + Z_1 \\ Y_2 = bX_1 + X_2 + Z_2 \end{cases} \tag{35}$$

where $Z_1, Z_2$ are zero-mean unit-variance Gaussian RVs; also, the inputs are subject to the power constraints $\mathbb{E}[X_i^2] \leq P_i, i = 1,2$.

During attempts to establish capacity results for this channel, the "very strong interference[4]" regime was first distinguished in [9] as the case where interference does not reduce capacity. This regime is given by holding the following inequalities:

$$\begin{cases} I(X_1;Y_1|X_2) \leq I(X_1;Y_2) \\ I(X_2;Y_2|X_1) \leq I(X_2;Y_1) \end{cases} \quad \text{for all } P_{X_1}(x_1)P_{X_2}(x_2) \tag{36}$$

In this regime, the interference level is so strong that each receiver is able to first decode the interference and remove it from its received signal (without any rate penalty) and then decode its own signal. Later, Sato [10] determined conditions for the Gaussian channel where joint decoding of both messages at each receiver is optimal and achieves the capacity. Sato's condition for the Gaussian channel (35) is given below:

$$|a| \geq 1, \qquad |b| \geq 1 \tag{37}$$

He proved that in this case the capacity region is derived by viewing the channel as a compound MAC [12] in which both messages are required at both receivers. In other words, the interference level is so strong (but possibly weaker than the very strong interference case) that it is optimal to jointly decode both the signal and the interference at each receiver. Afterwards, the condition (37) is called the "strong interference" regime for the Gaussian channel. Sato also conjectured [10] that for any two-user CIC satisfying the following conditions:

$$\begin{cases} I(X_1;Y_1|X_2) \leq I(X_1;Y_2|X_2) \\ I(X_2;Y_2|X_1) \leq I(X_2;Y_1|X_1) \end{cases} \quad \text{for all joint PDFs } P_{X_1}(x_1)P_{X_2}(x_2) \tag{38}$$

[4]In fact, the condition (36) is coined as the "strong interference" regime in [9]. After identifying the strong interference conditions in the sense of (37) and (38) in [10, 11], the condition (36) is commonly referred to as the "very strong interference" regime.

the capacity region is achieved by decoding both messages at each receiver. Note that by evaluating (38) with Gaussian input distributions, we directly derive (37). Later, Costa and El Gamal in [11] proved that Sato's conjecture is indeed true. Hence, the strong interference channel is referred to as a channel satisfying the conditions (38). It is also clear that the CICs with strong interference (38) contain those with very strong interference (36) as a subset.

As we proved in Part I [1, Lemma III.2], the strong interference conditions (37) and (38) are equivalent. This means that for the Gaussian channel if the conditions (38) hold for the Gaussian input distributions (equivalently (37) holds), they also hold for arbitrary input distributions. Therefore, for the Gaussian CIC, the conditions (37) and (38) are consistent and represent the same situation.

Now, the question is that whether the definition of the strong interference regime based on the conditions (38) always has the desired completeness? The response is negative: an immediate counterexample is the one-sided CIC (the ZIC). Consider the ZIC for which the channel transition probability $\mathbb{P}(y_1, y_2|x_1, x_2)$ is factorized as:

$$\mathbb{P}(y_1, y_2|x_1, x_2) = \mathbb{P}(y_1|x_1, x_2)\mathbb{P}(y_2|x_2) \tag{39}$$

It is clear that for this channel, the first inequality of (38) never holds. Thus, regarding the ZIC, one cannot refer to the condition (38) as a valid definition for the strong interference regime. Nonetheless, it is still possible to define a meaningful strong interference regime for this channel. Consider the ZIC which satisfies the following condition:

$$I(X_2; Y_2) \leq I(X_2; Y_1|X_1) \tag{40}$$

for all joint PDFs $P_{X_1}(x_1)P_{X_2}(x_2)$. As we showed in Part I [1, Proposition III. 21], under the condition (40), the optimal coding strategy is that the receiver $Y_1$ decodes both messages (both the interference and its signal) and the receiver $Y_2$ decodes only its respective message. Note that since the received signal at the receiver $Y_2$ is not really interfered by any other signal, this receiver decodes only its own signal. Therefore, the condition (40) represents a strong interference regime for the ZIC.

Next let us consider the CRC as shown in Fig. 3. This channel was also investigated in details in Part I [1] as one of the basic structures for the interference networks.

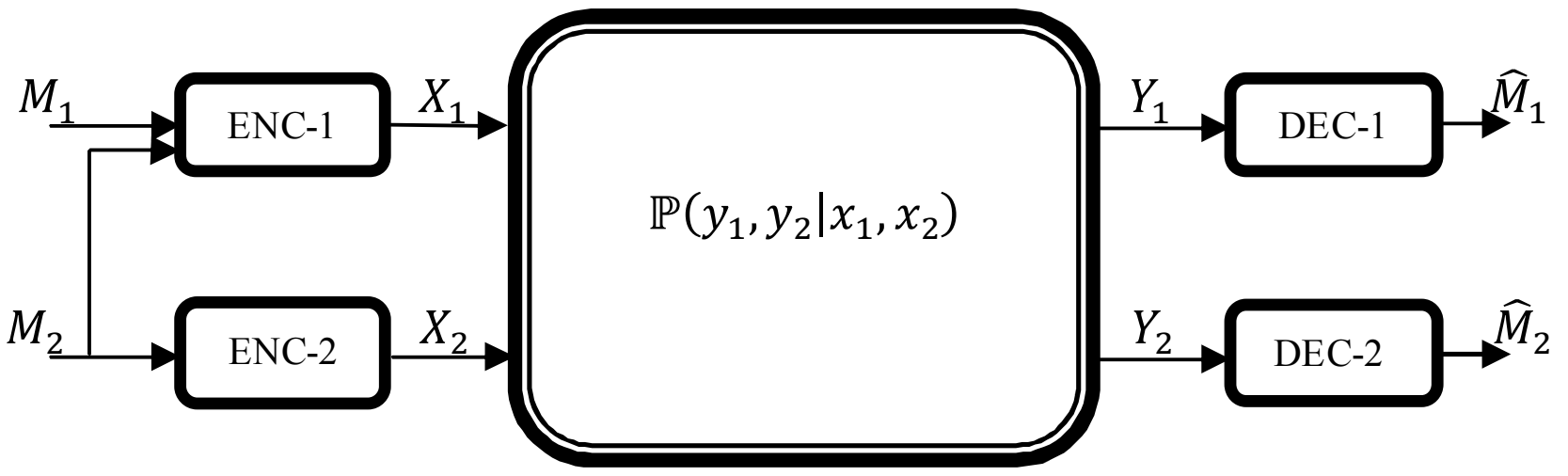

Figure 3. The Cognitive Radio Channel (CRC).

For the CRC, a strong interference regime was defined in [13] as described below:

$$\begin{cases} I(X_1, X_2; Y_2) \leq I(X_1, X_2; Y_1) \\ I(X_1; Y_1|X_2) \leq I(X_1; Y_2|X_2) \end{cases} \quad \text{for all joint PDFs } P_{X_1X_2}(x_1, x_2) \tag{41}$$

As we see, this condition is very similar to the condition (38) representing the strong interference regime for the two-user CIC. The authors in [13] proved that the capacity region of the CRC satisfying (41) is achieved by decoding both messages at each receiver. Let us consider the Gaussian channel. The Gaussian CRC is also formulated similar to (35). By evaluating the condition (41) for Gaussian input distributions, we derive the following:

$$\begin{cases} |1 + a\alpha| \geq |b + \alpha| \\ |1 - a\alpha| \geq |b - \alpha| \\ |b| \geq 1 \end{cases} \tag{42}$$

It is interesting to note that the conditions (42) are *not consistent* with those in (41). Clearly, unlike the Gaussian CIC for which the strong interference conditions (37) are equivalent to those in (38), for the Gaussian CRC, the conditions (42) are not equivalent to (41). As mentioned before, the conditions (42) are obtained by evaluating (41) for Gaussian input distributions; nonetheless, (42) do not imply that the inequalities (41) hold for all joint PDFs $P_{X_1X_2}(x_1,x_2)$. To show this point, consider (41). According to Lemma 1 and Corollary 1, if the first inequality of (41) holds for all joint PDFs $P_{X_1X_2}(x_1,x_2)$, then we have:

$$I(X_2;Y_2|X_1) \leq I(X_2;Y_1|X_1) \qquad \text{for all } P_{X_1X_2}(x_1,x_2) \tag{43}$$

Therefore, the conditions (41) are equivalent to:

$$\begin{cases} I(X_1,X_2;Y_2) \leq I(X_1,X_2;Y_1) \\ I(X_1;Y_1|X_2) = I(X_1;Y_2|X_2) \end{cases} \qquad \text{for all } P_{X_1X_2}(x_1,x_2) \tag{44}$$

One can readily see if the Gaussian channel satisfies (44), then it is required that $|b| = 1$. But for satisfying the conditions (42), one may set $|b| > 1$. In other words, unlike (44), the conditions (42) do not necessarily imply that $|b|$ is equal to 1. The reason behind this inconsistency is that the conditions (42) imply the inequalities (41) hold for *Gaussian input distributions*, however, they do not imply that these inequalities hold for *all joint PDFs* $P_{X_1X_2}(x_1,x_2)$.

Let us look at this problem by another view. As the authors proved in [13], we know for a CRC if both inequalities (41) are satisfied for all joint PDFs $P_{X_1X_2}(x_1,x_2)$, then it is optimal to decode both messages at each receiver. Now, inspired by the above argument, we intend to show that a similar result still holds under conditions strictly weaker than (41). We present these new conditions in the following theorem;

***Theorem 1)*** *Consider the CRC as depicted in Fig. 3. Assume that the channel transition probability function* $\mathbb{P}(y_1,y_2|x_1,x_2)$ *satisfies the following conditions:*

1. *The following inequality holds for all product PDFs* $P_{X_1}(x_1)P_{X_2}(x_2)$*:*

$$I(X_1;Y_1|X_2) \leq I(X_1;Y_2|X_2) \tag{45}$$

2. *The following two rate regions are equal:*

$$\bigcup_{P_{X_1X_2}(x_1,x_2)} \left\{ \begin{array}{l} (R_1,R_2) \in \mathbb{R}_+^2 : \\ R_1 \leq I(X_1;Y_1|X_2) \\ R_1 + R_2 \leq \begin{Bmatrix} I(X_1,X_2;Y_1), \\ I(X_1,X_2;Y_2) \end{Bmatrix} \end{array} \right\} = \bigcup_{P_{X_1X_2}(x_1,x_2)} \left\{ \begin{array}{l} (R_1,R_2) \in \mathbb{R}_+^2 : \\ R_1 \leq I(X_1;Y_1|X_2) \\ R_1 + R_2 \leq I(X_1,X_2;Y_2) \end{array} \right\} \tag{46}$$

*Then the capacity region is achieved by decoding both messages at each receiver and is given by:*

$$\bigcup_{P_{X_1X_2}(x_1,x_2)} \left\{ \begin{array}{l} (R_1,R_2) \in \mathbb{R}_+^2 : \\ R_1 \leq I(X_1;Y_1|X_2) \\ R_1 + R_2 \leq I(X_1,X_2;Y_2) \end{array} \right\} \tag{47}$$

*Proof of Theorem 1)* First note that, according to Lemma 1, if the inequality (45) holds for all product PDFs $P_{X_1}(x_1)P_{X_2}(x_2)$, then it also holds for all joint PDFs $P_{X_1X_2}(x_1,x_2)$. By applying a simple superposition coding scheme at the transmitters and also decoding both messages at both receivers, we obtain the following achievable rate region:

$$\bigcup_{P_{X_1X_2}(x_1,x_2)} \left\{ \begin{array}{l} (R_1,R_2) \in \mathbb{R}_+^2 : \\ R_1 \leq \min \begin{Bmatrix} I(X_1;Y_1|X_2), \\ I(X_1;Y_2|X_2) \end{Bmatrix} \\ R_1 + R_2 \leq \begin{Bmatrix} I(X_1,X_2;Y_1), \\ I(X_1,X_2;Y_2) \end{Bmatrix} \end{array} \right\} \tag{48}$$

Considering the conditions (45) and (46), we deduce that (48) is equal to (47). The converse part is derived using the UV-outer bound for the CRC given in [1, Corollary III.2]. The bound on the rate $R_1$ is obvious. To derive the bound on the sum rate, note that the inequality (45) implies the following

$$I(X_1;Y_1|V,X_2) \leq I(X_1;Y_2|V,X_2) \tag{49}$$

for all $P_{VX_1X_2}(v,x_1,x_2)$. Now by using the UV-outer bound in [1, Corollary III.2], we have:

$$\begin{aligned} R_1+R_2 &\leq I(X_1;Y_1|V,X_2)+I(V,X_2;Y_2) \\ &\leq I(X_1;Y_2|V,X_2)+I(V,X_2;Y_2) \\ &= I(X_1,X_2;Y_2) \end{aligned}$$

The proof is thus complete. ■

***Remarks 4:***

1. It should be noted that the equality (46) to hold, it is sufficient to have:

$$I(X_1,X_2;Y_2) \leq I(X_1,X_2;Y_1), \tag{50}$$

for those PDFs which achieve the boundary points for each of the rate regions in (46). But it is not required that the inequality (50) holds for all joint PDFs $P_{X_1X_2}(x_1,x_2)$. For example, consider the Gaussian channel. According to Lemma 2, if $|b| \geq 1$, then the condition (45) is satisfied. Moreover, it is clear that all mutual information functions in (46) are maximized for Gaussian distributions. Thereby, for the equality (46) to hold, it is sufficient that (50) is valid for Gaussian input distributions. In other words, for the Gaussian CRC, the conditions (42) are equivalent to the conditions of Theorem 1. As a direct consequence, one can state that the class of CRCs satisfying the conditions (45) and (46) strictly contains the channels satisfying (41).
2. Based on the previous remark, Theorem 1 establishes the capacity region for a new class of CRCs which is not included in any other class with known capacity. Even the more-capable CRCs, for which we derived the capacity region in Part I of our multi-part papers [1, Th. III.8], do not contain the channels satisfying (45) and (46). Because for the more-capable channel considered in [1, Th. III.8], it is required that the inequality (50) holds for all joint PDFs.
3. The conditions (45) and (46) can be referred to as a new strong interference regime for the CRC.

The above discussion demonstrates the incompleteness/inconsistency of defining the strong interference regime based on conditions such as (38) or (41), albeit a common practice in the literature. Therefore, we need to look at this problem by a broader perspective. In interference networks, each receiver perceives its respective messages as the information and the non-respective messages which are transmitted by its connected (or by both its connected and unconnected) transmitters as the interference. Now consider the conditions (38), (41) and (45)-(46). These conditions represent certain relations among the inputs and the outputs for various network topologies which yield the optimality of decoding both information and interference at all receivers. In other words, different interference networks in the strong interference regime have a unique characteristic from the viewpoint of optimal coding scheme. This characteristic implies that the strategy of *treating the interference as information* at all receivers achieves the capacity. Therefore, instead of conditions such as (38), (41), or (45)-(46), it is better to define the strong interference regime based on this common feature as given below.

***Definition 4) Strong Interference Regime for the General Interference Networks***

*The general interference network in Fig. 1 is said to be in the strong interference regime when the coding scheme achieving the capacity region is such that "each receiver decodes all messages transmitted by its connected (or by both its connected and unconnected) transmitters".*

***Remark 5)*** Using Definition 2, one could state that in the strong interference regime, the optimal coding strategy is that each receiver decodes all its connected messages.

We note that Definition 4 is uniformly applicable to all interference networks. It represents the unique nature of all interference networks in the strong interference regime. Based on this definition, conditions such as (38), (41), or (45)-(46) constitute only sufficient conditions for the corresponding network to be in the strong interference regime. It would not be surprising if one derives other conditions under which the network lies in this regime.

# IV. General Interference Networks with Two Receivers

In this section, we concentrate on the General Interference Networks with Two Receivers (GINTR). First we derive unified outer bounds for the capacity region of such networks. We then investigate some specific network scenarios in details to demonstrate the efficiency of these outer bounds for deriving capacity results. Using the derived outer bounds, we next obtain a strong interference regime for any two-receiver interference network. In Part IV of our multi-part papers [4], we will demonstrate that our outer bounds are also useful to study networks with less-noisy receivers.

## IV.A) Unified Outer Bounds

Recall that in Part I [1], we established capacity outer bounds based on a common framework for the basic structures of the interference networks, i.e., the two-user BC, the two-user CIC, and the CRC. Now, we intend to build such outer bounds for all networks of arbitrary large sizes. Specifically, here we consider the interference networks with two receivers. Our result is given in the next theorem. To present our theorem, we need to define a few more notations.

***Definition 5:*** *Let $\{R_1,\dots,R_K\}$ be a $K$-tuple of non-negative real numbers, where $K$ is a natural number. Let also $\mathbb{M}=\{M_1,\dots,M_K\}$ be a set of $K$ indexed elements. Assume that $\Omega$ is an arbitrary subset of $\mathbb{M}$. The partial-sum $\boldsymbol{R}_{\Sigma_\Omega}$ with respect to $\Omega$ is defined as follows:*

$$\boldsymbol{R}_{\Sigma_\Omega} \triangleq \sum_{l\in \underline{id}_\Omega} R_l \tag{51}$$

Note that the identification of the set $\Omega$, $\underline{id}_\Omega$, was defined in Part I [1, Def. II.1].

***Theorem 2)*** **Unified Outer Bound for the Capacity Region of the GINTRs**

*Consider the general interference network with two receivers that is derived by setting $K_2=2$ in the network shown in Fig. 1. Define the rate region $\mathfrak{R}_{o:(1)}^{GINTR}$ as follows:*

$$\mathfrak{R}_{o:(1)}^{GINTR} \triangleq \bigcup_{\mathcal{P}_o^{GINTR}} \left\{ \begin{array}{l} (R_1,\dots,R_K)\in\mathbb{R}_+^K: \\ \forall\,\mu\in\mathbb{N},\quad \forall\,\Omega_1^{Y_1},\Omega_2^{Y_1},\dots,\Omega_\mu^{Y_1}\subseteq\mathbb{M}_{Y_1},\quad \forall\,\Omega_1^{Y_2},\Omega_2^{Y_2}\dots,\Omega_\mu^{Y_2}\subseteq\mathbb{M}_{Y_2}: \\ \quad with\ \ \langle\forall\, l_1,l_2\in[1:\mu], l_1\neq l_2:\ \ \Omega_{l_1}^{Y_1}\cap\Omega_{l_2}^{Y_1}=\Omega_{l_1}^{Y_2}\cap\Omega_{l_2}^{Y_2}=\Omega_{l_1}^{Y_1}\cap\Omega_{l_2}^{Y_2}=\emptyset\rangle \\ \forall\ \Omega_s\subseteq\mathbb{M}-\left(\Omega_1^{Y_1}\cup\Omega_2^{Y_1}\cup\dots\cup\Omega_\mu^{Y_1}\cup\Omega_1^{Y_2}\cup\Omega_2^{Y_2}\cup\dots\cup\Omega_\mu^{Y_2}\right) \\ \qquad\qquad\qquad\qquad\qquad\qquad , \\ \displaystyle\sum_{l=1}^{\mu}\boldsymbol{R}_{\Sigma_{\Omega_l^{Y_1}}}+\boldsymbol{R}_{\Sigma_{\Omega_l^{Y_2}}}\le\sum_{l=1}^{\mu}\left(\begin{array}{l} I\left(\Omega_l^{Y_1};Y_1\middle|Z,\Omega_1^{Y_1},\dots,\Omega_{l-1}^{Y_1},\Omega_1^{Y_2},\dots,\Omega_{l-1}^{Y_2},\Omega_l^{Y_2},\Omega_s,Q\right) \\ +I\left(\Omega_l^{Y_2};Y_2\middle|Z,\Omega_1^{Y_1},\dots,\Omega_{l-1}^{Y_1},\Omega_1^{Y_2},\dots,\Omega_{l-1}^{Y_2},\Omega_s,Q\right) \end{array}\right) \\ \qquad\qquad\qquad +\min\{I(Z;Y_1|\Omega_s,Q),I(Z;Y_2|\Omega_s,Q)\} \end{array} \right\} \tag{52}$$

*where* $\mathcal{P}_o^{GINTR}$ *denotes the set of all joint PDFs* $P_{QM_1\ldots M_KZX_1\ldots X_{K_1}}(q,m_1,\ldots,m_K,z,x_1,\ldots,x_{K_1})$ *satisfying:*

$$P_{QM_1\ldots M_KZX_1\ldots X_{K_1}}=P_Q\times P_{M_1}\times\ldots\times P_{M_K}\times P_{Z|M_1\ldots M_KQ}\times P_{X_1|\mathbb{M}_{X_1},Q}\times\ldots\times P_{X_{K_1}|\mathbb{M}_{X_{K_1}},Q} \tag{53}$$

*Also, the PDFs* $P_{M_l}, l=1,\ldots,K$, *are uniformly distributed, and* $P_{X_i|\mathbb{M}_{X_i},Q}\in\{0,1\}$ *for* $i=1,\ldots,K_1$. *The set* $\mathfrak{R}_{o:(1)}^{GINTR}$ *constitutes an outer bound for the capacity region.*

*Proof of Theorem 2)* Refer to Appendix. ■

***Remarks 6:***

1. The unified outer bound $\mathfrak{R}_{o:(1)}^{GINTR}$ consists of the following parameters:
   - The RVs representing the receiver signals, $Y_1,Y_2$.
   - The $K$ auxiliary random variables $M_1,\ldots,M_K$ which actually represent the messages.
   - The time-sharing random variable $Q$.
   - The auxiliary random variable $Z$.

   Such characterization of the outer bounds is indeed very useful to understand the nature of information flow in large multi-message networks. This fact will be verified later.
2. For each $\mu\in\mathbb{N}$ where $\mu\le\max\{\|\mathbb{M}_{Y_1}\|,\|\mathbb{M}_{Y_2}\|\}$, the constraint given in (52) contains at most $2(\mu+1)$ mutual information terms, where the random variable $Z$ appears after the conditioning operator in all except the last two terms, i.e., $I(Z;Y_1|\Omega_s,Q)$ and $I(Z;Y_2|\Omega_s,Q)$.
3. Although the outer bound $\mathfrak{R}_{o:(1)}^{GINTR}$ includes many different constraints, the proof style and also the description given in (52) show that all these constraints have a "*unique nature*" in derivation.

The unified outer bound $\mathfrak{R}_{o:(1)}^{GINTR}$ in (52) is fairly general. It indeed includes all possible constraints one can establish on the communication rates $R_1,\ldots,R_K$ provided that only the mutual information functions containing the random variables $Y_1,Y_2,M_1,\ldots,M_K,Z,Q$ are allowed. In general, however, the evaluation of the outer bound seems to be rather complex, but it is very useful to establish tight capacity outer bounds with satisfactory performance for different interference networks. This eligibility is demonstrated in the sequel. Moreover, it should be noted that the complexity of evaluation of this outer bound is essentially due to the large number of its constraints and not on its auxiliary random variables; because, the auxiliary random variables are restricted to be $M_1,\ldots,M_K,Z$ as well as a time-sharing parameter $Q$. In other words, while the outer bound is described with numerous constraints, it has a well-formed structure. Based on this structure, given a certain network, we can always select specific constraints of the bound and derive an efficient computable outer bound as will be shown by examples below. It is also remarkable to note that for the two-user BC without common message, one can easily check that the outer bound $\mathfrak{R}_{o:(1)}^{GINTR}$ in (52) is reduced to the one we derived in Part I of our multi-part papers [1, Th. III.1]. As discussed in [1, Remark III.3], this bound may be strictly tighter than the UV-bound [1, Eq. III~13], which is known as the best capacity outer bound for the two-user BC without common message [20].

In the following corollary, by relaxing some constraints of (52), we first present another outer bound which is possibly weaker than $\mathfrak{R}_{o:(1)}^{GINTR}$; nonetheless, it has a more convenient description. It can be used to intuitively establish capacity outer bounds with satisfactory performance (that are optimal in many important cases) for different network topologies.

***Corollary 2)*** *Define the rate region* $\mathfrak{R}_{o:(2)}^{GINTR}$ *as follows:*

$$\mathfrak{R}_{o:(2)}^{GINTR} \triangleq \bigcup_{\mathcal{P}_o^{GINTR}} \left\{ \begin{array}{l} (R_1,\dots,R_K) \in \mathbb{R}_+^K: \\ \forall\, \Omega_0,\Omega_1,\Omega_2 : \ \Omega_0 \subseteq \mathbb{M}_{Y_1} \cap \mathbb{M}_{Y_2}, \Omega_1 \subseteq \mathbb{M}_{Y_1}, \Omega_2 \subseteq \mathbb{M}_{Y_2}, \\ \qquad \text{with } \ \langle \Omega_1 \cap \Omega_2 = \Omega_0 \cap \Omega_1 = \Omega_0 \cap \Omega_2 = \emptyset \rangle \\ \qquad , \\ \forall\, \Omega_s : \ \Omega_s \subseteq \mathbb{M} - (\Omega_0 \cup \Omega_1 \cup \Omega_2) \\ \qquad , \\ \langle 1 \rangle : \boldsymbol{R}_{\Sigma\Omega_0} + \boldsymbol{R}_{\Sigma\Omega_1} \le I(Z,\Omega_0,\Omega_1;Y_1|\Omega_s,Q) \\ \langle 2 \rangle : \boldsymbol{R}_{\Sigma\Omega_0} + \boldsymbol{R}_{\Sigma\Omega_2} \le I(Z,\Omega_0,\Omega_2;Y_2|\Omega_s,Q) \\ \langle 3 \rangle : \boldsymbol{R}_{\Sigma\Omega_0} + \boldsymbol{R}_{\Sigma\Omega_1} + \boldsymbol{R}_{\Sigma\Omega_2} \le I(\Omega_1;Y_1|Z,\Omega_0,\Omega_2,\Omega_s,Q) + I(Z,\Omega_0,\Omega_2;Y_2|\Omega_s,Q) \\ \langle 4 \rangle : \boldsymbol{R}_{\Sigma\Omega_0} + \boldsymbol{R}_{\Sigma\Omega_1} + \boldsymbol{R}_{\Sigma\Omega_2} \le I(\Omega_2;Y_2|Z,\Omega_0,\Omega_1,\Omega_s,Q) + I(Z,\Omega_0,\Omega_1;Y_1|\Omega_s,Q) \\ \qquad , \\ \langle 5 \rangle : \boldsymbol{R}_{\Sigma\Omega_0} + \boldsymbol{R}_{\Sigma\Omega_1} + \boldsymbol{R}_{\Sigma\Omega_2} \le I(\Omega_1;Y_1|Z,\Omega_0,\Omega_2,\Omega_s,Q) + I(\Omega_2;Y_2|Z,\Omega_0,\Omega_s,Q) \\ \qquad\qquad + I(Z,\Omega_0;Y_1|\Omega_s,Q) \\ \langle 6 \rangle : \boldsymbol{R}_{\Sigma\Omega_0} + \boldsymbol{R}_{\Sigma\Omega_1} + \boldsymbol{R}_{\Sigma\Omega_2} \le I(\Omega_2;Y_2|Z,\Omega_0,\Omega_1,\Omega_s,Q) + I(\Omega_1;Y_1|Z,\Omega_0,\Omega_s,Q) \\ \qquad\qquad + I(Z,\Omega_0;Y_2|\Omega_s,Q) \end{array} \right\} \tag{54}$$

*where $\mathcal{P}_o^{GINTR}$ is given as in Theorem 2, (see (53)). The set $\mathfrak{R}_{o:(2)}^{GINTR}$ constitutes an outer bound for the capacity region of the general interference network with two receivers.*

***Remarks 7:***

1. For the sake of simplicity, hereafter, we use $\mathfrak{R}_{o:(2)}^{GINTR}\langle i \rangle, i = 1,\dots,6$, to denote the $i^{th}$ expression (inequality) of the outer bound $\mathfrak{R}_{o:(2)}^{GINTR}$; for example,

$$\mathfrak{R}_{o:(2)}^{GINTR}\langle 3 \rangle : \ \boldsymbol{R}_{\Sigma\Omega_0} + \boldsymbol{R}_{\Sigma\Omega_1} + \boldsymbol{R}_{\Sigma\Omega_2} \le I(\Omega_1;Y_1|Z,\Omega_0,\Omega_2,\Omega_s,Q) + I(Z,\Omega_0,\Omega_2;Y_2|\Omega_s,Q)$$

2. Note that in the outer bound $\mathfrak{R}_{o:(2)}^{GINTR}$ in (54), each constraint contains at most three mutual information functions. In fact, for a wide range of models, specially for the strong and mixed interference networks, the first four constraints of $\mathfrak{R}_{o:(2)}^{GINTR}$ are sufficient to explicitly derive the capacity results. The last two constraints of $\mathfrak{R}_{o:(2)}^{GINTR}$ are useful to establish the capacity region in some semi-deterministic channels such as the semi-deterministic BC with common message. Note that the $WUV$-outer bound (see [1, Proposition III.5]) on the capacity region of the two-user BC with common message, which is optimal for the semi-deterministic channel, can be directly derived as a special case of $\mathfrak{R}_{o:(2)}^{GINTR}$.

*Proof of Corollary 2)* Consider the unified outer bound $\mathfrak{R}_{o:(1)}^{GINTR}$ given in (52) for the network. The outer bound $\mathfrak{R}_{o:(2)}^{GINTR}$ in (54) is configured by only certain constraints of $\mathfrak{R}_{o:(1)}^{GINTR}$. Let $\Omega_0,\Omega_1,\Omega_2,\Omega_s$ be pairwise disjoint subsets of the messages $\mathbb{M}$, where $\Omega_0 \subseteq \mathbb{M}_{Y_1} \cap \mathbb{M}_{Y_2}, \Omega_1 \subseteq \mathbb{M}_{Y_1}, \Omega_2 \subseteq \mathbb{M}_{Y_2}$ and $\Omega_s \subseteq \mathbb{M} - (\Omega_0 \cup \Omega_1 \cup \Omega_2)$. From (52), we have:

- ✓ *Select:* $\mu = 1,\ \Omega_1^{Y_1} = \Omega_0 \cup \Omega_1,\ \Omega_1^{Y_2} = \emptyset,\ \Omega_s = \Omega_s \Rightarrow$

$$\begin{aligned} \boldsymbol{R}_{\Sigma\Omega_1^{Y_1}} = \boldsymbol{R}_{\Sigma\Omega_0} + \boldsymbol{R}_{\Sigma\Omega_1} &\le I(\Omega_0,\Omega_1;Y_1|Z,\Omega_s,Q) + \min\{I(Z;Y_1|\Omega_s,Q), I(Z;Y_2|\Omega_s,Q)\} \\ &\le I(Z,\Omega_0,\Omega_1;Y_1|\Omega_s,Q) \end{aligned} \tag{55}$$

- ✓ *Select:* $\mu = 1,\ \Omega_1^{Y_1} = \emptyset,\ \Omega_1^{Y_2} = \Omega_0 \cup \Omega_2,\ \Omega_s = \Omega_s \Rightarrow$

$$\begin{aligned} \boldsymbol{R}_{\Sigma\Omega_1^{Y_2}} = \boldsymbol{R}_{\Sigma\Omega_0} + \boldsymbol{R}_{\Sigma\Omega_2} &\le I(\Omega_0,\Omega_2;Y_2|Z,\Omega_s,Q) + \min\{I(Z;Y_1|\Omega_s,Q), I(Z;Y_2|\Omega_s,Q)\} \\ &\le I(Z,\Omega_0,\Omega_2;Y_2|\Omega_s,Q) \end{aligned} \tag{56}$$

✓ *Select:* $\mu = 1,\ \Omega_1^{Y_1} = \Omega_1,\ \Omega_1^{Y_2} = \Omega_0 \cup \Omega_2,\ \Omega_s = \Omega_s\ \Rightarrow$

$$\begin{aligned}\boldsymbol{R}_{\Sigma_{\Omega_1^{Y_1}}} + \boldsymbol{R}_{\Sigma_{\Omega_1^{Y_2}}} &= \boldsymbol{R}_{\Sigma\Omega_0} + \boldsymbol{R}_{\Sigma\Omega_1} + \boldsymbol{R}_{\Sigma\Omega_2}\\ &\le I(\Omega_1;Y_1|Z,\Omega_0,\Omega_2,\Omega_s,Q) + I(\Omega_0,\Omega_2;Y_2|Z,\Omega_s,Q) + \min\{I(Z;Y_1|\Omega_s,Q), I(Z;Y_2|\Omega_s,Q)\}\\ &\le I(\Omega_1;Y_1|Z,\Omega_0,\Omega_2,\Omega_s,Q) + I(Z,\Omega_0,\Omega_2;Y_2|\Omega_s,Q)\end{aligned} \tag{57}$$

✓ *Select:* $\mu = 2,\ \Omega_1^{Y_1} = \Omega_0 \cup \Omega_1,\ \Omega_2^{Y_1} = \emptyset,\ \Omega_1^{Y_2} = \emptyset,\ \Omega_2^{Y_2} = \Omega_2,\ \Omega_s = \Omega_s\ \Rightarrow$

$$\begin{aligned}\boldsymbol{R}_{\Sigma_{\Omega_1^{Y_1}}} + \boldsymbol{R}_{\Sigma_{\Omega_1^{Y_2}}} + \boldsymbol{R}_{\Sigma_{\Omega_2^{Y_1}}} + \boldsymbol{R}_{\Sigma_{\Omega_2^{Y_2}}} &= \boldsymbol{R}_{\Sigma\Omega_0} + \boldsymbol{R}_{\Sigma\Omega_1} + \boldsymbol{R}_{\Sigma\Omega_2}\\ &\le I(\Omega_0,\Omega_1;Y_1|Z,\Omega_s,Q) + I(\Omega_2;Y_2|Z,\Omega_0,\Omega_1,\Omega_s,Q) + \min\{I(Z;Y_1|\Omega_s,Q), I(Z;Y_2|\Omega_s,Q)\}\\ &\le I(\Omega_2;Y_2|Z,\Omega_0,\Omega_1,\Omega_s,Q) + I(Z,\Omega_0,\Omega_1;Y_1|\Omega_s,Q)\end{aligned} \tag{58}$$

✓ *Select:* $\mu = 2,\ \Omega_1^{Y_1} = \Omega_0,\ \Omega_2^{Y_1} = \Omega_1,\ \Omega_1^{Y_2} = \emptyset,\ \Omega_2^{Y_2} = \Omega_2,\ \Omega_s = \Omega_s\ \Rightarrow$

$$\begin{aligned}\boldsymbol{R}_{\Sigma_{\Omega_1^{Y_1}}} + \boldsymbol{R}_{\Sigma_{\Omega_1^{Y_2}}} + \boldsymbol{R}_{\Sigma_{\Omega_2^{Y_1}}} + \boldsymbol{R}_{\Sigma_{\Omega_2^{Y_2}}} &= \boldsymbol{R}_{\Sigma\Omega_0} + \boldsymbol{R}_{\Sigma\Omega_1} + \boldsymbol{R}_{\Sigma\Omega_2}\\ &\le I(\Omega_0;Y_1|Z,\Omega_s,Q) + I(\Omega_1;Y_1|Z,\Omega_0,\Omega_2,\Omega_s,Q) + I(\Omega_2;Y_2|Z,\Omega_0,\Omega_s,Q) + \min\{I(Z;Y_1|\Omega_s,Q), I(Z;Y_2|\Omega_s,Q)\}\\ &\le I(\Omega_1;Y_1|Z,\Omega_0,\Omega_2,\Omega_s,Q) + I(\Omega_2;Y_2|Z,\Omega_0,\Omega_s,Q) + I(Z,\Omega_0;Y_1|\Omega_s,Q)\end{aligned} \tag{59}$$

✓ *Select:* $\mu = 3,\ \Omega_1^{Y_1} = \emptyset,\ \Omega_2^{Y_1} = \Omega_1,\ \Omega_3^{Y_1} = \emptyset,\ \Omega_1^{Y_2} = \Omega_0,\ \Omega_2^{Y_2} = \emptyset,\ \Omega_3^{Y_2} = \Omega_2,\ \Omega_s = \Omega_s\ \Rightarrow$

$$\begin{aligned}\boldsymbol{R}_{\Sigma_{\Omega_1^{Y_1}}} + \boldsymbol{R}_{\Sigma_{\Omega_1^{Y_2}}} + \boldsymbol{R}_{\Sigma_{\Omega_2^{Y_1}}} + \boldsymbol{R}_{\Sigma_{\Omega_2^{Y_2}}} &= \boldsymbol{R}_{\Sigma\Omega_0} + \boldsymbol{R}_{\Sigma\Omega_1} + \boldsymbol{R}_{\Sigma\Omega_2}\\ &\le I(\Omega_0;Y_2|Z,\Omega_s,Q) + I(\Omega_1;Y_1|Z,\Omega_0,\Omega_s,Q) + I(\Omega_2;Y_2|Z,\Omega_0,\Omega_1,\Omega_s,Q) + \min\{I(Z;Y_1|\Omega_s,Q), I(Z;Y_2|\Omega_s,Q)\}\\ &\le I(\Omega_2;Y_2|Z,\Omega_0,\Omega_1,\Omega_s,Q) + I(\Omega_1;Y_1|Z,\Omega_0,\Omega_s,Q) + I(Z,\Omega_0;Y_2|\Omega_s,Q)\end{aligned} \tag{60}$$

By collecting (55)-(60), we derive (54) and the proof is thus complete. ■

In the following subsections, we show that the unified outer bound $\mathfrak{R}_{o:(2)}^{GINTR}$ in (54) can be easily translated to computable outer bounds for different networks which are optimal for important special cases as well. To this end, we consider two networks: 1) the two-user CIC with common message, and 2) the BCCR with common message.

## IV.A.1) Two-User Classical Interference Channel with Common Message

First, consider the two-user CIC with common information wherein two transmitters send three messages over the channel. Each transmitter sends a private message to its respective receiver; also, both transmitters cooperatively send a common message to both receivers. Each receiver decodes its respective private message and also the common message. The channel model is depicted in Fig. 4. This channel was also previously studied in [21-23].

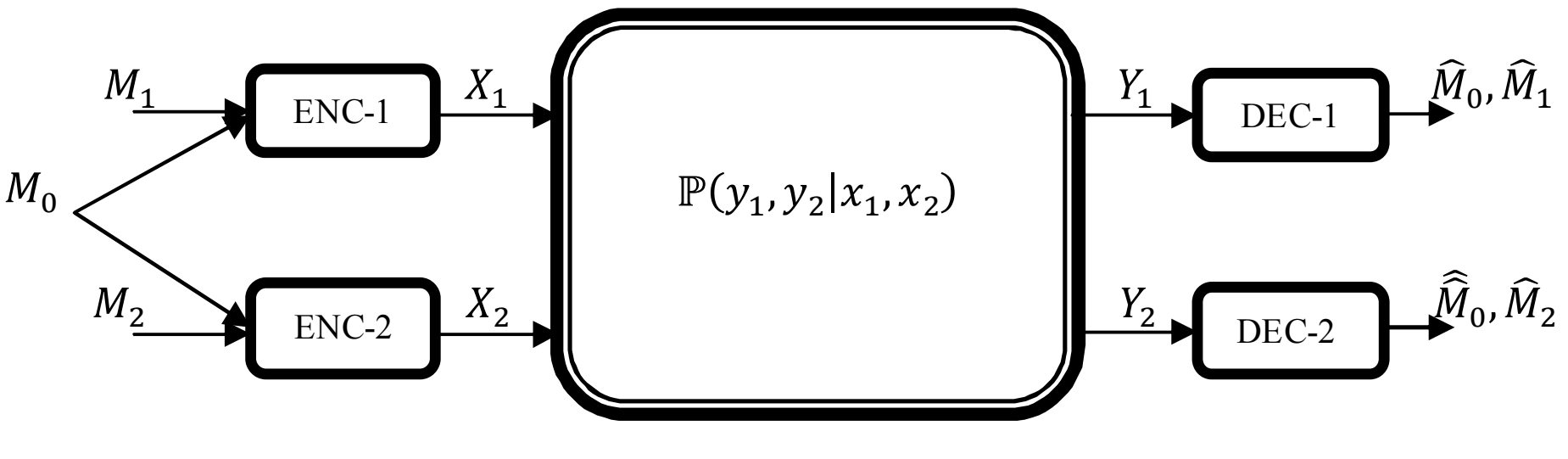

Figure 4. The two-user CIC with common message.

The two-user CIC with common message is derived from the general interference network defined in Part I of our multi-part papers [1, Sec. II] (see also Fig. 1) by setting $K_1 = K_2 = 2$, $\mathbb{M} = \{M_0, M_1, M_2\}$, $\mathbb{M}_{X_1} = \mathbb{M}_{Y_1} = \{M_0, M_1\}$, and $\mathbb{M}_{X_2} = \mathbb{M}_{Y_2} = \{M_0, M_2\}$. The Gaussian channel is formulated as follows:

$$\begin{cases} Y_1 = X_1 + aX_2 + Z_1 \\ Y_2 = bX_1 + X_2 + Z_2 \end{cases} \tag{61}$$

where $Z_1$ and $Z_2$ are zero-mean and unit-variance Gaussian RVs; also, the inputs are subject to the power constraints $\mathbb{E}[X_i^2] \leq P_i, i = 1,2$.

In the following, we first derive a computable outer bound on the capacity region of this channel using the unified outer bound $\mathfrak{R}_{o:(2)}^{GINTR}$ given in (54). Then, we present special cases where this outer bound is optimal and yields the capacity.

***Theorem 3)*** *Define the rate region* $\mathfrak{R}_o^{CIC_{cm}}$ *as follows:*

$$\mathfrak{R}_o^{CIC_{cm}} \triangleq \bigcup_{\mathcal{P}_o^{CIC_{cm}}} \left\{ \begin{array}{l} (R_0, R_1, R_2) \in \mathbb{R}_+^3 : \\ R_1 \leq \min \left\{ \begin{array}{c} I(X_1;Y_1|X_2,W), I(U,X_1;Y_1|W) \\ I(X_1;Y_1|X_2,V,W) + I(V;Y_2|X_2,W) \end{array} \right\} \\ R_2 \leq \min \left\{ \begin{array}{c} I(X_2;Y_2|X_1,W), I(V,X_2;Y_2|W) \\ I(X_2;Y_2|X_1,U,W) + I(U;Y_1|X_1,W) \end{array} \right\} \\ R_0 + R_1 \leq I(W,U,X_1;Y_1) \\ R_0 + R_2 \leq I(W,V,X_2;Y_2) \\ R_1 + R_2 \leq I(X_1;Y_1|X_2,V,W) + I(V,X_2;Y_2|W) \\ R_1 + R_2 \leq I(X_2;Y_2|X_1,U,W) + I(U,X_1;Y_1|W) \\ R_0 + R_1 + R_2 \leq I(X_1;Y_1|X_2,V,W) + I(W,V,X_2;Y_2) \\ R_0 + R_1 + R_2 \leq I(X_2;Y_2|X_1,U,W) + I(W,U,X_1;Y_1) \end{array} \right\} \tag{62}$$

*where* $\mathcal{P}_o^{CIC_{cm}}$ *denotes the set of all joint PDFs* $P_{WUVX_1X_2}(w,u,v,x_1,x_2)$ *satisfying:*

$$P_{WUVX_1X_2} = P_W P_{X_1|W} P_{X_2|W} P_{UV|X_1X_2W} \tag{63}$$

*Also,* $P_{X_1|UW}(x_1|u,w) \in \{0,1\}$ *and* $P_{X_2|VW}(x_2|v,w) \in \{0,1\}$*. The set* $\mathfrak{R}_o^{CIC_{cm}}$ *constitutes an outer bound for the capacity region of the two-user CIC with common message in Fig. 4.*

*Proof of Theorem 3)* Consider the unified outer bound $\mathfrak{R}_{o:(2)}^{GINTR}$ given in (54) for the general interference network with two receivers. For the two-user CIC with common message, we have: $\mathbb{M}_{Y_1} = \{M_0, M_1\}$, and $\mathbb{M}_{Y_2} = \{M_0, M_2\}$. In addition, $X_1$ is a deterministic function of $(M_0, M_1, Q)$ and $X_2$ is a deterministic function of $(M_0, M_2, Q)$. Also, note that $(M_0, M_1, M_2, Z, Q) \to X_1, X_2 \to Y_1, Y_2$ forms a Markov chain. Define the random variables $U, V$ and $W$ as follows:

$$U \triangleq (Z, M_1), \qquad V \triangleq (Z, M_2), \qquad W \triangleq (M_0, Q) \tag{64}$$

Therefore, we can write:

- ✓ *Select:* $\Omega_0 = \emptyset,\ \Omega_1 = \{M_1\},\ \Omega_2 = \emptyset,\ \Omega_s = \{M_2, M_0\} \Rightarrow$

$$\begin{aligned} \mathfrak{R}_{o:(2)}^{GINTR}\langle 1 \rangle : \ R_1 &\leq I(Z, M_1; Y_1|M_2, M_0, Q) = I(Z, M_1, X_1; Y_1|X_2, M_2, M_0, Q) \\ &\leq I(X_1; Y_1|X_2, M_0, Q) = I(X_1; Y_1|X_2, W) \end{aligned} \tag{65}$$

$$\begin{aligned} \mathfrak{R}_{o:(2)}^{GINTR}\langle 3 \rangle : \ R_1 &\leq I(M_1; Y_1|Z, M_2, M_0, Q) + I(Z; Y_2|M_2, M_0, Q) \\ &= I(M_1, X_1; Y_1|X_2, Z, M_2, M_0, Q) + I(Z; Y_2|X_2, M_2, M_0, Q) \\ &\leq I(X_1; Y_1|X_2, V, W) + I(V; Y_2|X_2, W) \end{aligned} \tag{66}$$

✓ *Select:* $\Omega_0 = \emptyset,\ \Omega_1 = \{M_1\},\ \Omega_2 = \emptyset,\ \Omega_s = \{M_0\}\ \Rightarrow$

$$\mathfrak{R}_{o:(2)}^{GINTR}\langle 1\rangle:\ R_1 \leq I(Z,M_1;Y_1|M_0,Q) = I(Z,M_1,X_1;Y_1|M_0,Q) = I(U,X_1;Y_1|W) \tag{67}$$

✓ *Select:* $\Omega_0 = \emptyset,\ \Omega_1 = \emptyset,\ \Omega_2 = \{M_2\},\ \Omega_s = \{M_1,M_0\}\ \Rightarrow$

$$\begin{aligned}\mathfrak{R}_{o:(2)}^{GINTR}\langle 2\rangle:\ R_2 &\leq I(Z,M_2;Y_2|M_1,M_0,Q) = I(Z,M_2,X_2;Y_2|X_1,M_1,M_0,Q)\\ &\leq I(X_2;Y_2|X_1,M_0,Q) = I(X_2;Y_2|X_1,W)\end{aligned} \tag{68}$$

$$\begin{aligned}\mathfrak{R}_{o:(2)}^{GINTR}\langle 4\rangle:\ R_2 &\leq I(M_2;Y_2|Z,M_1,M_0,Q) + I(Z;Y_1|M_1,M_0,Q)\\ &= I(M_2,X_2;Y_2|X_1,Z,M_1,M_0,Q) + I(Z;Y_1|X_1,M_1,M_0,Q)\\ &\leq I(X_2;Y_2|X_1,U,W) + I(U;Y_1|X_1,W)\end{aligned} \tag{69}$$

✓ *Select:* $\Omega_0 = \emptyset,\ \Omega_1 = \emptyset,\ \Omega_2 = \{M_2\},\ \Omega_s = \{M_0\}\ \Rightarrow$

$$\mathfrak{R}_{o:(2)}^{GINTR}\langle 2\rangle:\ R_2 \leq I(Z,M_2;Y_2|M_0,Q) = I(Z,M_2,X_2;Y_2|M_0,Q) = I(V,X_2;Y_2|W) \tag{70}$$

✓ *Select:* $\Omega_0 = \{M_0\},\ \Omega_1 = \{M_1\},\ \Omega_2 = \{M_2\},\ \Omega_s = \emptyset\ \Rightarrow$

$$\begin{aligned}\mathfrak{R}_{o:(2)}^{GINTR}\langle 1\rangle:\ R_0 + R_1 &\leq I(Z,M_0,M_1;Y_1|Q) = I(Z,M_0,M_1,X_1;Y_1|Q)\\ &\leq I(Q,Z,M_0,M_1,X_1;Y_1) = I(W,U,X_1;Y_1)\end{aligned} \tag{71}$$

$$\begin{aligned}\mathfrak{R}_{o:(2)}^{GINTR}\langle 2\rangle:\ R_0 + R_2 &\leq I(Z,M_0,M_2;Y_2|Q) = I(Z,M_0,M_2,X_2;Y_2|Q)\\ &\leq I(Q,Z,M_0,M_2,X_2;Y_2) = I(W,V,X_2;Y_2)\end{aligned} \tag{72}$$

✓ *Select:* $\Omega_0 = \emptyset,\ \Omega_1 = \{M_1\},\ \Omega_2 = \{M_2\},\ \Omega_s = \{M_0\}\ \Rightarrow$

$$\begin{aligned}\mathfrak{R}_{o:(2)}^{GINTR}\langle 3\rangle:\ R_1 + R_2 &\leq I(M_1;Y_1|Z,M_2,M_0,Q) + I(Z,M_2;Y_2|M_0,Q)\\ &= I(M_1,X_1;Y_1|X_2,Z,M_2,M_0,Q) + I(Z,M_2,X_2;Y_2|M_0,Q)\\ &= I(X_1;Y_1|X_2,V,W) + I(V,X_2;Y_2|W)\end{aligned} \tag{73}$$

$$\begin{aligned}\mathfrak{R}_{o:(2)}^{GINTR}\langle 4\rangle:\ R_1 + R_2 &\leq I(M_2;Y_2|Z,M_1,M_0,Q) + I(Z,M_1;Y_1|M_0,Q)\\ &= I(M_2,X_2;Y_2|X_1,Z,M_1,M_0,Q) + I(Z,M_1,X_1;Y_1|M_0,Q)\\ &= I(X_2;Y_2|X_1,U,W) + I(U,X_1;Y_1|W)\end{aligned} \tag{74}$$

✓ *Select:* $\Omega_0 = \{M_0\},\ \Omega_1 = \{M_1\},\ \Omega_2 = \{M_2\},\ \Omega_s = \emptyset\ \Rightarrow$

$$\begin{aligned}\mathfrak{R}_{o:(2)}^{GINTR}\langle 3\rangle:\ R_0 + R_1 + R_2 &\leq I(M_1;Y_1|Z,M_0,M_2,Q) + I(Z,M_0,M_2;Y_2|Q)\\ &\leq I(M_1,X_1;Y_1|X_2,Z,M_0,M_2,Q) + I(Q,Z,M_0,M_2,X_2;Y_2)\\ &= I(X_1;Y_1|X_2,V,W) + I(W,V,X_2;Y_2)\end{aligned} \tag{75}$$

$$\begin{aligned}\mathfrak{R}_{o:(2)}^{GINTR}\langle 4\rangle:\ R_0 + R_1 + R_2 &\leq I(M_2;Y_2|Z,M_0,M_1,Q) + I(Z,M_0,M_1;Y_1|Q)\\ &\leq I(M_2,X_2;Y_2|X_1,Z,M_0,M_1,Q) + I(Q,Z,M_0,M_1,X_1;Y_1)\\ &= I(X_2;Y_2|X_1,U,W) + I(W,U,X_1;Y_1)\end{aligned} \tag{76}$$

By collecting (65)-(76), we obtain the outer bound $\mathfrak{R}_o^{CICcm}$ given in (62). The proof is thus complete. ■

***Remarks 8:***

1. Note that the outer bound $\mathfrak{R}_o^{CICcm}$ in (62) alongside the appropriate input power constraints, i.e., $\mathbb{E}[X_i^2] \leq P_i, i = 1,2$, is valid also for the Gaussian channel (61). By using the Entropy Power Inequality (EPI), one can prove that for the Gaussian channel this outer bound is optimized for Gaussian distributions. Thereby, an explicit characterization of the bound can be derived for the Gaussian channel. However, we omit the derivations due to space considerations.
2. The rate region $\mathfrak{R}_o^{CICcm}$ in (62) is the first computable non-trivial outer bound for the CIC with common message that holds for both Gaussian and discrete channels.
3. Note that for the case of $M_1 \equiv \emptyset$ (or $M_2 \equiv \emptyset$), the network in Fig. 4 is reduced to the cognitive interference channel studied in [24]. By setting $R_1 = 0$ in the outer bound $\mathfrak{R}_o^{CICcm}$ in (62), one can readily show that the remaining region is optimized for $W \equiv X_1$ which coincides with the capacity region of the cognitive interference channel [24, Th. 4]. Thus, our outer bound is optimal for the cases where either $R_1 = 0$ or $R_2 = 0$.

For the two-user CIC with common message in Fig. 4 a strong interference regime was identified in [13]. In that paper, the authors showed that if the channel satisfies the following conditions:

$$\begin{cases} I(X_1;Y_1|X_2) \leq I(X_1;Y_2|X_2) \\ I(X_2;Y_2|X_1) \leq I(X_2;Y_1|X_1) \end{cases} \quad \text{for all joint PDFs} \quad P_{X_1}(x_1)P_{X_2}(x_2) \tag{77}$$

then the channel has strong interference. Note that for the Gaussian channel in (61) the conditions (77) are equivalent to $|a| \geq 1$ and $|b| \geq 1$. These conditions are actually identical to the ones in (38) and (37) which were previously discerned in [10] and [11] as the strong interference regime for the channel without common message. The style of the proof presented in [13] for the strong interference regime (77) was based on the techniques devised in [10, 11]. In the next theorem, we prove that the outer bound $\mathfrak{R}_o^{CICcm}$ in (62) is optimal under the conditions (77); therefore, establishing a new proof for the problem.

***Proposition 1)*** *Consider the two-user CIC with common message in Fig. 4. If the channel satisfies the conditions* (77)*, then it lies in the strong interference regime. In this case, the capacity region is given by:*

$$\bigcup_{P_W P_{X_1|W} P_{X_2|W}} \left\{ \begin{array}{l} (R_0, R_1, R_2) \in \mathbb{R}_+^3 : \\ R_1 \leq I(X_1;Y_1|X_2,W) \\ R_2 \leq I(X_2;Y_2|X_1,W) \\ R_1 + R_2 \leq \min\begin{Bmatrix} I(X_1,X_2;Y_1|W), \\ I(X_1,X_2;Y_2|W) \end{Bmatrix} \\ R_0 + R_1 + R_2 \leq \min\begin{Bmatrix} I(X_1,X_2;Y_1), \\ I(X_1,X_2;Y_2) \end{Bmatrix} \end{array} \right\} \tag{78}$$

*This result also holds for the Gaussian channel in* (61) *if* $|a| \geq 1$ *and* $|b| \geq 1$.

*Proof of Proposition 1)* Note that based on Lemma 1, the strong interference conditions (77) imply the following:

$$\begin{cases} I(X_1;Y_1|X_2,D) \leq I(X_1;Y_2|X_2,D) \\ I(X_2;Y_2|X_1,D) \leq I(X_2;Y_1|X_1,D)' \end{cases} \quad \text{for all joint PDFs} \quad P_{DX_1X_2}(d,x_1,x_2) \tag{79}$$

The achievability of (78) is readily derived by requiring that each receiver decodes all messages. To prove the optimality of this scheme, using the outer bound $\mathfrak{R}_o^{CICcm}$ in (62), we can write:

$$\begin{aligned} R_1 + R_2 &\leq I(X_1;Y_1|X_2,V,W) + I(V,X_2;Y_2|W) \\ &\overset{(a)}{\leq} I(X_1;Y_2|X_2,V,W) + I(V,X_2;Y_2|W) = I(X_1,X_2;Y_2|W) \end{aligned}$$

The inequality (a) is due to the first condition of (79) where $D$ is replaced by $(V,W)$. Similarly,

$$R_1 + R_2 \le I(X_2;Y_2|X_1,U,W) + I(U,X_1;Y_1|W)$$
$$\overset{(b)}{\le} I(X_2;Y_1|X_1,U,W) + I(U,X_1;Y_1|W) = I(X_1,X_2;Y_1|W)$$

The inequality (b) is due to the second condition of (79) in which $D$ is replaced by $(U,W)$. Also, we have:

$$R_0 + R_1 + R_2 \le I(X_1;Y_1|X_2,V,W) + I(W,V,X_2;Y_2)$$
$$\le I(X_1;Y_1|X_2,V,W) + I(W,V,X_2;Y_2) = I(X_1,X_2;Y_2)$$

$$R_0 + R_1 + R_2 \le I(X_2;Y_2|X_1,U,W) + I(W,U,X_1;Y_1)$$
$$\le I(X_2;Y_1|X_1,U,W) + I(W,U,X_1;Y_1) = I(X_1,X_2;Y_1)$$

Moreover, the first two constraints of (78) are directly found in the outer bound $\mathfrak{R}_o^{CICcm}$. For the Gaussian channel in (61), it is sufficient to know that the conditions (77) are equivalent to $|a| \ge 1$ and $|b| \ge 1$. ■

In Part I [1, Sec. III.C], we identified a class of less-noisy CICs (without common message) for which the successive decoding scheme is sum-rate optimal. Now, using the outer bound $\mathfrak{R}_o^{CICcm}$ in (62), we prove a similar result for the channel with common message. This result is given in the following theorem.

***Theorem 4)*** *Consider the two-user CIC with common message in Fig. 4. Assume that the channel satisfies the following less-noisy condition:*

$$\begin{cases} I(V;Y_2|X_2) \le I(V;Y_1|X_2) \\ I(X_2;Y_2|X_1) \le I(X_2;Y_1|X_1) \end{cases} \quad \text{for all PDFs} \quad P_{VX_1}(v,x_1)P_{X_2}(x_2) \tag{80}$$

*In this case, the successive decoding scheme achieves the sum-rate capacity. It is given by:*

$$\max_{P_{X_1X_2}} \begin{cases} I(X_1;Y_1|X_2) + I(X_2;Y_2), \\ I(X_1,X_2;Y_1) \end{cases} \tag{81}$$

*Also, for the Gaussian channel (61) the conditions (80) are equivalent to $|b| \le 1 \le |a|$ and the sum-rate capacity is given by:*

$$\max_{0\le\rho\le1} \begin{cases} \psi\left((1-\rho^2)P_1\right) + \psi\left(\dfrac{\rho^2b^2P_1 + P_2 + 2b\rho\sqrt{P_1P_2}}{1+(1-\rho^2)b^2P_1}\right), \\ \psi\left(P_1 + a^2P_2 + 2a\rho\sqrt{P_1P_2}\right) \end{cases} \tag{82}$$

*Proof of Theorem 4)* To achieve the sum-rate (82), the message $M_2$ is withdrawn from the transmission scheme. The common message $M_0$ is encoded by a codeword constructed by the random variable $X_2$ based on $P_{X_2}(x_2)$. The private message $M_1$ is encoded using a codeword constructed by $X_1$. This codeword is superimposed upon the common message codeword $X_2(M_0)$ according to $P_{X_1|X_2}(x_1|x_2)$. The first and the second transmitters send $X_1(M_0,M_1)$ and $X_2(M_0)$, respectively over the channel. The second receiver decodes only its corresponding codeword $X_2(M_0)$. The first receiver applies a successive decoding scheme; it first decodes $X_2(M_0)$ and then decodes $X_1(M_0,M_1)$. One can easily show that this strategy leads to the sum-rate (82). To prove the converse part, first note that according to Lemma 4 the first inequality of (80) readily extends to the following:

$$I(V;Y_2|X_2) \le I(V;Y_1|X_2), \qquad \text{for all PDFs} \quad P_{VX_1X_2}(v,x_1,x_2) \tag{83}$$

Also, according to Lemma 1 the second inequality in (80) extends to:

$$I(X_2;Y_2|X_1,D) \le I(X_2;Y_1|X_1,D), \qquad \text{for all PDFs} \quad P_{DX_1X_2}(d,x_1,x_2) \tag{84}$$

Now consider the outer bound $\mathfrak{R}_o^{CICcm}$ in (62). For the sum-rate we have:

$$\begin{aligned} R_0+R_1+R_2 &\leq I(X_1;Y_1|X_2,V,W)+I(W,V,X_2;Y_2) \\ &= I(X_1;Y_1|X_2,V,W)+I(X_2;Y_2)+I(W,V;Y_2|X_2) \\ &\overset{(a)}{\leq} I(X_1;Y_1|X_2,V,W)+I(X_2;Y_2)+I(W,V;Y_1|X_2) \\ &= I(X_1;Y_2|X_2)+I(X_2;Y_2) \end{aligned} \tag{85}$$

where inequality (a) is due to (83) in which $V$ is replaced by $(W,V)$. Also,

$$\begin{aligned} R_0+R_1+R_2 &\leq I(X_2;Y_2|X_1,U,W)+I(W,U,X_1;Y_1) \\ &\overset{(b)}{\leq} I(X_2;Y_1|X_1,U,W)+I(W,U,X_1;Y_1) \\ &= I(X_1,X_2;Y_1) \end{aligned} \tag{86}$$

where inequality (b) is due to (84) in which $D$ is replaced by $(U,W)$. The bounds (85) and (86) should be considered over all PDFs $P_W P_{X_1|W} P_{X_2|W}$. Since $W$ does not appear in the last expressions in (85) and (86), the optimum value for $W$ is equal to $X_1$ or $X_2$. This completes the proof of the converse part. For the Gaussian channel in (61), one can readily show that the conditions (80) are equivalent to $|b| \leq 1 \leq |a|$. Moreover, using the EPI, it is not difficult to show that (81) is optimized for Gaussian input distributions. ■

***Remark 9:*** Unlike the channel without common message (see Part I [1, Th. III.6]), to achieve the sum-rate (81), the message $M_2$ is not included in the transmission scheme. In fact, to achieve the maximum sum-rate, the transmitter $X_2$ tries to transmit information for the less noisy (stronger) receiver $Y_1$. For the channel without common message, the transmitter $X_2$ is not able to contribute in transmitting information for the stronger receiver; perforce, it transmits the message $M_2$ to its respective receiver. Indeed, in the networks with less noisy receivers to achieve the sum-rate, each transmitter tries, if possible, to contribute in sending information for the stronger receivers and avoid sending messages with respect to the more noisy receivers. This characteristic of the less noisy networks is similar to the degraded ones; see our detailed discussion in Part II [2].

Next, let us examine the one-sided channel. Consider a channel for which the transition probability function is decomposed as follows:

$$\mathbb{P}(y_1,y_2|x_1,x_2)=\mathbb{P}(y_1|x_1,x_2)\mathbb{P}(y_2|x_2) \tag{87}$$

Note that for the channel in (87), the strong interference conditions (77) cannot be satisfied simultaneously because these conditions imply that the interference at both receivers is strong while for the one-sided channel (87), the receiver $Y_2$ experiences no interference. Nonetheless, one can still derive a strong interference regime for this channel. This result is given in the next proposition.

***Proposition 2)*** *Consider the one-sided two-user CIC (87) with common message in Fig. 4. If the channel satisfies the following condition:*

$$I(X_2;Y_2) \leq I(X_2;Y_1|X_1) \quad \textit{for all joint PDFs} \quad P_{X_1}(x_1)P_{X_2}(x_2) \tag{88}$$

*then it has strong interference. The capacity region is given by:*

$$\bigcup_{P_W P_{X_1|W} P_{X_2|W}} \left\{ \begin{array}{l} (R_0,R_1,R_2) \in \mathbb{R}_+^3 : \\ R_1 \leq I(X_1;Y_1|X_2,W) \\ R_2 \leq I(X_2;Y_2|W) \\ R_0+R_2 \leq I(X_2;Y_2) \\ R_1+R_2 \leq I(X_1,X_2;Y_1|W) \\ R_0+R_1+R_2 \leq I(X_1,X_2;Y_1) \end{array} \right\} \tag{89}$$

*This result also holds for the one-sided Gaussian channel in (61) with and* $b=0$ *if* $|a| \geq 1$.

*Proof of Proposition 2)* First note that for the one-sided channel in (87), the condition (88) can be re-formulated as:

$$I(X_2;Y_2) \overset{(a)}{=} I(X_2;Y_2|X_1) \leq I(X_2;Y_1|X_1) \quad \textit{for all joint PDFs} \quad P_{X_1}(x_1)P_{X_2}(x_2) \tag{90}$$

where equality (a) is due to (87). Therefore, according to Lemma 1, we have:

$$I(X_2;Y_2|X_1,D) \leq I(X_2;Y_1|X_1,D) \quad \text{for all joint PDFs} \quad P_{DX_1X_2}(d,x_1,x_2) \tag{91}$$

To obtain the achievability of (89), the messages $M_0, M_1$ and $M_2$ are encoded exactly similar to a MAC with common message. First, the message $M_0$ is encoded by a codeword constructed by $W$ based on $P_W(w)$. The private message $M_i, i = 1,2$, is then encoded by a codeword constructed by $X_i$ based on $P_{X_i|W}$; in other words, the codeword $X_i(M_i, M_0)$ is superimposed on $W(M_0)$. The $i^{th}$ transmitter, $i = 1,2$, then sends $X_i(M_i, M_0)$ over the channel. The receiver $Y_1$ jointly decodes all the codewords $W(M_0), X_1(M_1, M_0)$ and $X_2(M_2, M_0)$. The receiver $Y_2$ jointly decodes $W(M_0)$ and $X_2(M_2, M_0)$. A simple analysis of this scheme leads to the following achievable rate region:

$$\begin{cases} R_2 \leq I(X_2;Y_2|W) \\ R_0 + R_2 \leq I(W,X_2;Y_2) = I(X_2;Y_2) \\ R_1 \leq I(X_1;Y_1|X_2,W) \\ R_2 \leq I(X_2;Y_1|X_1,W) \\ R_1 + R_2 \leq I(X_1,X_2;Y_1|W) \\ R_0 + R_1 + R_2 \leq I(X_1,X_2;Y_1) \end{cases} , \quad \text{for all} \quad P_W P_{X_1|W} P_{X_2|W} \tag{92}$$

The constraint $R_2 \leq I(X_2;Y_1|X_1,W)$ in (92) is actually redundant because we have:

$$R_2 \leq I(X_2;Y_2|W) \overset{(a)}{=} I(X_2;Y_2|X_1,W) \overset{(b)}{\leq} I(X_2;Y_1|X_1,W) \tag{93}$$

where equality (a) is due to (87) and inequality (b) is derived by setting $D \equiv W$ in (91) for all joint PDFs $P_W P_{X_1|W} P_{X_2|W}$. Thereby, we derive the achievability of (89). For the converse part, consider the outer bound $\mathfrak{R}_o^{CICcm}$ in (62). The first constraint of (89) is directly given in (62); also, we have:

$$R_2 \leq I(X_2;Y_2|X_1,W) \overset{(a)}{=} I(X_2;Y_2|W)$$

$$R_0 + R_2 \leq I(W,V,X_2;Y_2) \overset{(b)}{=} I(X_2;Y_2)$$

where equalities (a) and (b) are due to (87). Moreover, one can derive:

$$\begin{aligned} R_1 + R_2 &\leq I(X_2;Y_2|X_1,U,W) + I(U,X_1;Y_1|W) \\ &\overset{(a)}{\leq} I(X_2;Y_1|X_1,U,W) + I(U,X_1;Y_1|W) \\ &= I(X_1,X_2;Y_1|W) \end{aligned}$$

$$\begin{aligned} R_0 + R_1 + R_2 &\leq I(X_2;Y_2|X_1,U,W) + I(W,U,X_1;Y_1) \\ &\overset{(b)}{\leq} I(X_2;Y_1|X_1,U,W) + I(W,U,X_1;Y_1) \\ &= I(W,X_1,X_2;Y_1) = I(X_1,X_2;Y_1) \end{aligned}$$

where inequalities (a) and (b) are due to (91). For the Gaussian channel with $b = 0$, it is sufficient to note that the condition (88) is reduced to $|a| \geq 1$. The proof is thus complete. ■

***Remark 10:*** According to (90), the strong interference condition (88) is nothing more than the second inequality of (77) which is reduced to (88) for the one-sided channel (87). Therefore, one may represent the strong interference regime for the one-sided channel by the condition (90). The latter representation is helpful for describing the strong interference regime for large networks where some of the receivers may be unconnected to certain transmitters.

## IV.A.2) Broadcast Channel with Cognitive Relays (BCCR)

Now let us consider the BCCR with common message. This interference network is composed of three transmitters and two receivers. One transmitter (broadcasting node) sends three messages $M_0, M_1, M_2$ to the receivers while being assisted by two relay transmitters. The first relay has access to the message $M_1$ and the second relay to the message $M_2$. The messages $M_0$ and $M_1$ are decoded at the first receiver and the messages $M_0$ and $M_2$ at the second receiver. Fig. 5 depicts the channel model.

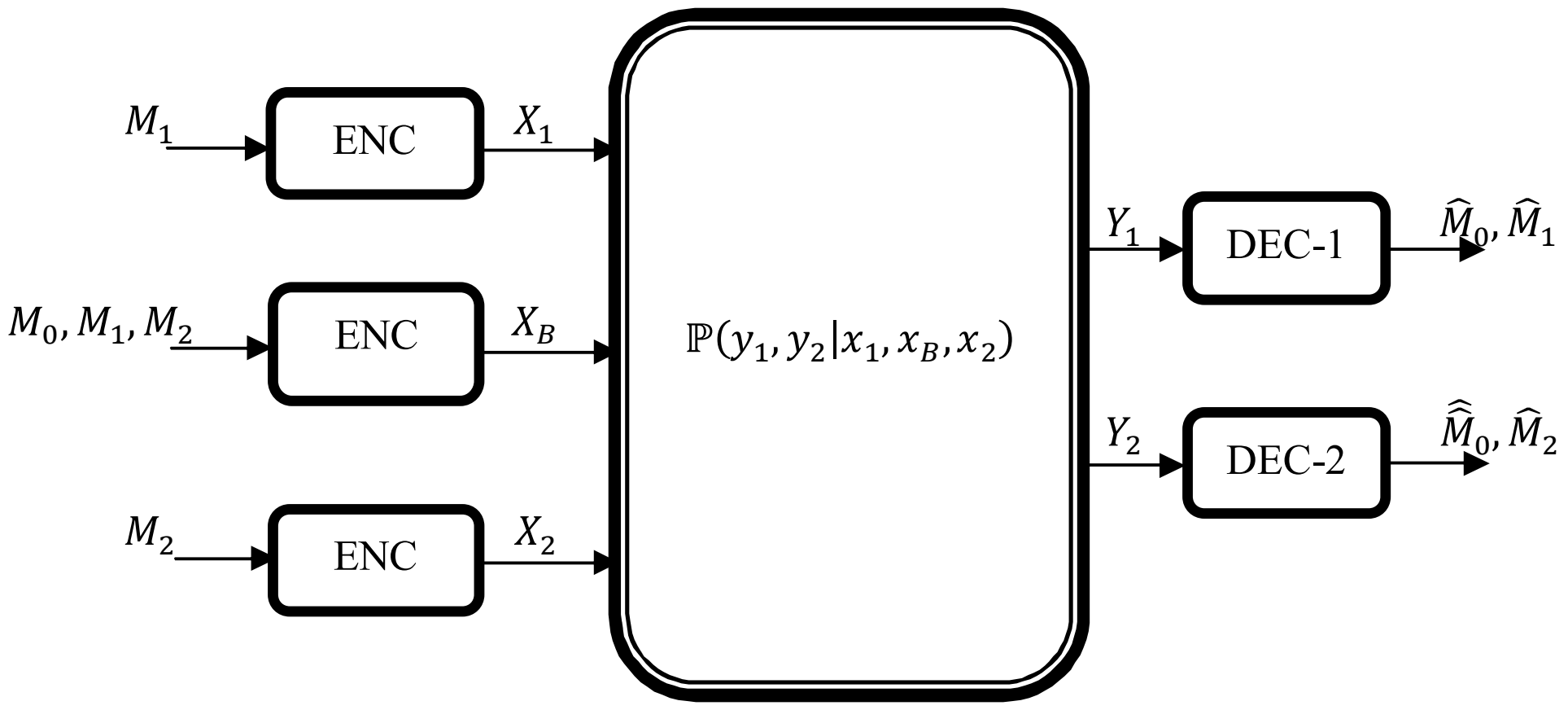

Figure 5. The Broadcast Channel with Cognitive Relays (BCCR).

Note that the BCCR contains the two-user CIC, the CRC and the two-user BC with common message as special cases. Besides, this model is reduced to the one considered in [25] when there is no common message, i.e., $M_0 \equiv \emptyset$. The BCCR is obtained from the general interference network defined in Part I [1, Sec. II.B] (see also Fig. 1) by setting $K_1 = 3$, $K_2 = 2$, $\mathbb{M} = \{M_0, M_1, M_2\}$, $\mathbb{M}_{X_1} = \{M_1\}$, $\mathbb{M}_{X_2} = \{M_2\}$, $\mathbb{M}_{X_B} = \{M_0, M_1, M_2\}$, $\mathbb{M}_{Y_1} = \{M_0, M_1\}$ and $\mathbb{M}_{Y_2} = \{M_0, M_2\}$. The Gaussian channel is formulated as follows:

$$\begin{cases} Y_1 = a_B X_B + a_1 X_1 + a_2 X_2 + Z_1 \\ Y_2 = b_B X_B + b_1 X_1 + b_2 X_2 + Z_2 \end{cases} \tag{94}$$

where $Z_1$ and $Z_2$ are zero-mean unit-variance Gaussian RVs and the inputs are subject to power constraints $\mathbb{E}[X_i^2] \leq P_i, i = B, 1,2$.

In the following, we derive a computable capacity outer bound for this channel using the unified outer bound $\mathfrak{R}_{o:(2)}^{GINTR}$ given in (54). Then, we utilize this outer bound to establish capacity results for some special cases. This network and its related models will be further studied in Subsection VI.

***Theorem 5)*** *Define the rate region* $\mathfrak{R}_o^{BCCR}$ *as follows:*

$$\mathfrak{R}_o^{BCCR} \triangleq \bigcup_{\mathcal{P}_o^{BCCR}} \left\{ \begin{array}{l} (R_0, R_1, R_2) \in \mathbb{R}_+^3 : \\ R_0 \leq \min\limits_{i\in\{1,2\}} \min \left\{ \begin{array}{c} I(U, W; Y_i|X_1, Q), I(V, W; Y_i|X_2, Q), \\ I(W; Y_i), I(X_B; Y_i|X_1, X_2, Q) \end{array} \right\}, \\ R_0 + R_1 \leq \min \left\{ \begin{array}{c} I(X_1, X_B; Y_1|X_2, Q), I(U, W, X_1; Y_1|Q), \\ I(U, X_1; Y_1|W, Q) + I(W; Y_2|Q) \\ I(X_1, X_B; Y_1|X_2, V, W, Q) + I(V, W; Y_2|X_2, Q) \end{array} \right\}, \\ R_0 + R_2 \leq \min \left\{ \begin{array}{c} I(X_2, X_B; Y_2|X_1, Q), I(V, W, X_2; Y_2|Q), \\ I(V, X_2; Y_2|W, Q) + I(W; Y_1|Q) \\ I(X_2, X_B; Y_2|X_1, U, W, Q) + I(U, W; Y_1|X_1, Q) \end{array} \right\}, \\ R_0 + R_1 + R_2 \leq I(X_1, X_B; Y_1|X_2, V, W, Q) + I(V, W, X_2; Y_2|Q) \\ R_0 + R_1 + R_2 \leq I(X_2, X_B; Y_2|X_1, U, W, Q) + I(U, W, X_1; Y_1|Q) \\ R_0 + R_1 + R_2 \leq I(X_1, X_B; Y_1|X_2, V, W, Q) + I(V, X_2; Y_2|W, Q) + I(W; Y_1|Q) \\ R_0 + R_1 + R_2 \leq I(X_2, X_B; Y_2|X_1, U, W, Q) + I(U, X_1; Y_1|W, Q) + I(W; Y_2|Q) \end{array} \right\} \tag{95}$$

*where* $\mathcal{P}_o^{BCCR}$ *denotes the set of all joint PDFs* $P_{QX_1X_2X_BUV}(q,x_1,x_2,x_B,u,v,w)$ *satisfying:*

$$P_{QX_1X_2X_BUV}(q,x_1,x_2,x_B,u,v) = P_Q(q)P_{X_1|Q}(x_1|q)P_{X_2|Q}(x_2|q)P_{X_B|X_1X_2Q}(x_B|x_1,x_2,q)P_{UVW|X_BX_1X_2Q}(u,v,w|x_1,x_2,x_B,q) \quad (96)$$

*as well as,* $P_{X_1|UQ}(x_1|u,q) \in \{0,1\}$, $P_{X_2|VQ}(x_2|v,q) \in \{0,1\}$ *and* $P_{X_B|UVQ}(x_B|u,v,q) \in \{0,1\}$. *The rate region* $\mathfrak{R}_o^{BCCR}$ *constitutes an outer bound for the capacity region of the BCCR.*

*Proof of Theorem 5)* Consider the unified outer bound $\mathfrak{R}_{o:(2)}^{GINTR}$ given in (54) for the general interference network with two receivers. For the BCCR in Fig. 5, we have: $\mathbb{M}_{Y_1} = \{M_0, M_1\}$ and $\mathbb{M}_{Y_2} = \{M_0, M_2\}$. Moreover, $X_1$ is a deterministic function of $(M_1, Q)$, $X_2$ is a deterministic function of $(M_2, Q)$, and $X_B$ is a deterministic function of $(M_0, M_1, M_2, Q)$. Also, note that $(M_0, M_1, M_2, Z, Q) \to X_1, X_2, X_B \to Y_1, Y_2$ forms a Markov chain. Define the following new random variables:

$$U \triangleq (Z, M_0, M_1), \qquad V \triangleq (Z, M_0, M_2), \qquad W \triangleq (Z, M_0) \quad (97)$$

Therefore, we can write:

- ✓ *Select:* $\Omega_0 = \{M_0\}$, $\Omega_1 = \emptyset$, $\Omega_2 = \emptyset$, $\Omega_s = \{M_1\}$ $\Rightarrow$ for $i = 1,2$:

$$\begin{aligned}\mathfrak{R}_{o:(2)}^{GINTR}\langle i\rangle : \; R_0 &\leq I(Z, M_0; Y_i|M_1, Q) = I(Z, M_0; Y_i|X_1, M_1, Q)\\ &\leq I(Z, M_0, M_1; Y_i|X_1, Q) = I(U, W; Y_i|X_1, Q)\end{aligned} \quad (98)$$

- ✓ *Select:* $\Omega_0 = \{M_0\}$, $\Omega_1 = \emptyset$, $\Omega_2 = \emptyset$, $\Omega_s = \{M_2\}$ $\Rightarrow$ for $i = 1,2$:

$$\begin{aligned}\mathfrak{R}_{o:(2)}^{GINTR}\langle i\rangle : \; R_0 &\leq I(Z, M_0; Y_i|M_2, Q) = I(Z, M_0; Y_i|X_2, M_2, Q)\\ &\leq I(Z, M_0, M_2; Y_i|X_2, Q) = I(V, W; Y_i|X_2, Q)\end{aligned} \quad (99)$$

- ✓ *Select:* $\Omega_0 = \{M_0\}$, $\Omega_1 = \emptyset$, $\Omega_2 = \emptyset$, $\Omega_s = \{M_1, M_2\}$ $\Rightarrow$ for $i = 1,2$:

$$\mathfrak{R}_{o:(2)}^{GINTR}\langle i\rangle : \; R_0 \leq I(Z, M_0; Y_i|M_1, M_2, Q) = I(X_B, Z, M_0; Y_i|X_1, X_2, M_1, M_2, Q) \leq I(X_B; Y_i|X_1, X_2, Q) \quad (100)$$

- ✓ *Select:* $\Omega_0 = \{M_0\}$, $\Omega_1 = \emptyset$, $\Omega_2 = \emptyset$, $\Omega_s = \emptyset$ $\Rightarrow$ for $i = 1,2$:

$$\mathfrak{R}_{o:(2)}^{GINTR}\langle i\rangle : \; R_0 \leq I(Z, M_0; Y_i|Q) = I(W; Y_i|Q) \quad (101)$$

- ✓ *Select:* $\Omega_0 = \{M_0\}$, $\Omega_1 = \{M_1\}$, $\Omega_2 = \emptyset$, $\Omega_s = \{M_2\}$ $\Rightarrow$

$$\mathfrak{R}_{o:(2)}^{GINTR}\langle 1\rangle : \; R_0 + R_1 \leq I(Z, M_0, M_1; Y_1|M_2, Q) = I(X_1, X_B, Z, M_0, M_1; Y_1|X_2, M_2, Q) \leq I(X_1, X_B; Y_1|X_2, Q) \quad (102)$$

- ✓ *Select:* $\Omega_0 = \{M_0\}$, $\Omega_1 = \{M_1\}$, $\Omega_2 = \emptyset$, $\Omega_s = \emptyset$ $\Rightarrow$

$$\mathfrak{R}_{o:(2)}^{GINTR}\langle 3\rangle : \; R_0 + R_1 \leq I(M_1; Y_1|Z, M_0, Q) + I(Z, M_0; Y_2|Q) = I(U, X_1; Y_1|W, Q) + I(W; Y_2|Q) \quad (103)$$

- ✓ *Select:* $\Omega_0 = \{M_0\}$, $\Omega_1 = \emptyset$, $\Omega_2 = \{M_2\}$, $\Omega_s = \{M_1\}$ $\Rightarrow$

$$\mathfrak{R}_{o:(2)}^{GINTR}\langle 2\rangle : \; R_0 + R_2 \leq I(Z, M_0, M_2; Y_2|M_1, Q) = I(X_2, X_B, Z, M_0, M_2; Y_2|X_1, M_1, Q) \leq I(X_2, X_B; Y_2|X_1, Q) \quad (104)$$

✓ *Select:* $\Omega_0 = \{M_0\},\ \Omega_1 = \emptyset,\ \Omega_2 = \{M_2\},\ \Omega_s = \emptyset\ \Rightarrow$

$$\mathfrak{R}_{o:(2)}^{GINTR}\langle 4\rangle: \ R_0 + R_2 \leq I(M_2;Y_2|Z,M_0,Q) + I(Z,M_0;Y_1|Q) = I(V,X_2;Y_2|W,Q) + I(W;Y_1|Q) \tag{105}$$

✓ *Select:* $\Omega_0 = \{M_0\},\ \Omega_1 = \{M_1\},\ \Omega_2 = \emptyset,\ \Omega_s = \{M_2\}\ \Rightarrow$

$$\begin{aligned}\mathfrak{R}_{o:(2)}^{GINTR}\langle 3\rangle: \ R_0 + R_1 &\leq I(M_1;Y_1|Z,M_0,M_2,Q) + I(Z,M_0;Y_2|M_2,Q)\\ &= I(X_1,X_B,M_1;Y_1|X_2,Z,M_0,M_2,Q) + I(Z,M_0;Y_2|X_2,M_2,Q)\\ &\leq I(X_1,X_B;Y_1|X_2,Z,M_0,M_2,Q) + I(Z,M_0,M_2;Y_2|X_2,Q)\\ &= I(X_1,X_B;Y_1|X_2,V,W,Q) + I(V,W;Y_2|X_2,Q)\end{aligned} \tag{106}$$

✓ *Select:* $\Omega_0 = \{M_0\},\ \Omega_1 = \emptyset,\ \Omega_2 = \{M_2\},\ \Omega_s = \{M_1\}\ \Rightarrow$

$$\begin{aligned}\mathfrak{R}_{o:(2)}^{GINTR}\langle 4\rangle: \ R_0 + R_2 &\leq I(M_2;Y_2|Z,M_0,M_1,Q) + I(Z,M_0;Y_1|M_1,Q)\\ &= I(X_2,X_B,M_2;Y_2|X_1,Z,M_0,M_1,Q) + I(Z,M_0;Y_1|X_1,M_1,Q)\\ &\leq I(X_2,X_B;Y_2|X_1,Z,M_0,M_1,Q) + I(Z,M_0,M_1;Y_1|X_1,Q)\\ &= I(X_2,X_B;Y_2|X_1,U,W,Q) + I(U,W;Y_1|X_1,Q)\end{aligned} \tag{107}$$

✓ *Select:* $\Omega_0 = \{M_0\},\ \Omega_1 = \{M_1\},\ \Omega_2 = \{M_2\},\ \Omega_s = \emptyset\ \Rightarrow$

$$\mathfrak{R}_{o:(2)}^{GINTR}\langle 1\rangle: \ R_0 + R_1 \leq I(Z,M_0,M_1;Y_1|Q) = I(Z,M_0,M_1,X_1;Y_1|Q) = I(U,W,X_1;Y_1|Q) \tag{108}$$

$$\mathfrak{R}_{o:(2)}^{GINTR}\langle 2\rangle: \ R_0 + R_2 \leq I(Z,M_0,M_2;Y_2|Q) = I(Z,M_0,M_2,X_2;Y_2|Q) = I(V,W,X_2;Y_2|Q) \tag{109}$$

$$\begin{aligned}\mathfrak{R}_{o:(2)}^{GINTR}\langle 3\rangle: \ R_0 + R_1 + R_2 &\leq I(M_1;Y_1|Z,M_0,M_2,Q) + I(Z,M_0,M_2;Y_2|Q)\\ &= I(X_1,X_B,M_1;Y_1|X_2,Z,M_0,M_2,Q) + I(Z,M_0,M_2,X_2;Y_2|Q)\\ &= I(X_1,X_B;Y_1|X_2,Z,M_0,M_2,Q) + I(Z,M_0,M_2,X_2;Y_2|Q)\\ &= I(X_1,X_B;Y_1|X_2,V,W,Q) + I(V,W,X_2;Y_2|Q)\end{aligned} \tag{110}$$

$$\begin{aligned}\mathfrak{R}_{o:(2)}^{GINTR}\langle 4\rangle: \ R_0 + R_1 + R_2 &\leq I(M_2;Y_2|Z,M_0,M_1,Q) + I(Z,M_0,M_1;Y_1|Q)\\ &= I(X_2,X_B,M_2;Y_2|X_1,Z,M_0,M_1,Q) + I(Z,M_0,M_1,X_1;Y_1|Q)\\ &= I(X_2,X_B;Y_2|X_1,Z,M_0,M_1,Q) + I(Z,M_0,M_1,X_1;Y_1|Q)\\ &= I(X_2,X_B;Y_2|X_1,U,W,Q) + I(U,W,X_1;Y_1|Q)\end{aligned} \tag{111}$$

$$\begin{aligned}\mathfrak{R}_{o:(2)}^{GINTR}\langle 5\rangle: \ R_0 + R_1 + R_2 &\leq I(M_1;Y_1|Z,M_0,M_2,Q) + I(M_2;Y_2|Z,M_0,Q) + I(Z,M_0;Y_1|Q)\\ &= I(X_1,X_B,M_1;Y_1|X_2,Z,M_0,M_2,Q) + I(M_2,X_2;Y_2|Z,M_0,Q) + I(Z,M_0;Y_1|Q)\\ &= I(X_1,X_B;Y_1|X_2,Z,M_0,M_2,Q) + I(M_2,X_2;Y_2|Z,M_0,Q) + I(Z,M_0;Y_1|Q)\\ &= I(X_1,X_B;Y_1|X_2,V,W,Q) + I(V,X_2;Y_2|W,Q) + I(W;Y_1|Q)\end{aligned} \tag{112}$$

$$\begin{aligned}\mathfrak{R}_{o:(2)}^{GINTR}\langle 6\rangle: \ R_0 + R_1 + R_2 &\leq I(M_2;Y_2|Z,M_0,M_1,Q) + I(M_1;Y_1|Z,M_0,Q) + I(Z,M_0;Y_2|Q)\\ &= I(X_2,X_B,M_2;Y_2|X_1,Z,M_0,M_1,Q) + I(M_1,X_1;Y_1|Z,M_0,Q) + I(Z,M_0;Y_2|Q)\\ &= I(X_2,X_B;Y_2|X_1,Z,M_0,M_1,Q) + I(M_1,X_1;Y_1|Z,M_0,Q) + I(Z,M_0;Y_2|Q)\\ &= I(X_2,X_B;Y_2|X_1,U,W,Q) + I(U,X_1;Y_1|W,Q) + I(W;Y_2|Q)\end{aligned} \tag{113}$$

Now by collecting (98)-(113), we obtain the outer bound $\mathfrak{R}_o^{BCCR}$ given in (95). ■

***Remark 11:*** Comparing the outer bound $\mathfrak{R}_o^{CICcm}$ in (62) and the outer bound $\mathfrak{R}_o^{BCCR}$ in (95), we can deduce how the characteristics of the unified bound $\mathfrak{R}_{o:(2)}^{GINTR}$ in (54) vary by network topology. For example, the auxiliary random variable $W$ in $\mathfrak{R}_o^{CICcm}$ is reminiscent of the variable conveying the common message in the two-user MAC (see Part II [2, Sec. II.A] for details), while in $\mathfrak{R}_o^{BCCR}$ it is reminiscent of the variable $W$ in the WUV-outer bound for the two-user BC with common message (see Part I [1, Proposition III.5]). This insight can be adapted to extract useful capacity outer bounds for other network topologies from the unified bound $\mathfrak{R}_{o:(2)}^{GINTR}$ in (54).

Now, we consider special cases for which the outer bound $\mathfrak{R}_o^{BCCR}$ given in (95) is optimal and yields the capacity. First consider a channel which satisfies the following conditions:

$$\begin{cases} I(X_1,X_B;Y_1|X_2) \le I(X_1,X_B;Y_2|X_2), & \text{for all PDFs} \quad P_{X_1X_B}(x_1,x_B)P_{X_2}(x_2) \\ I(X_2,X_B;Y_2|X_1) \le I(X_2,X_B;Y_1|X_1), & \text{for all PDFs} \quad P_{X_1}(x_1)P_{X_2X_B}(x_2,x_B) \end{cases} \tag{114}$$

In the next theorem, we prove that such a channel lies in the strong interference regime.

***Proposition 3)*** *The BCCR depicted in Fig. 5 with the conditions (114) has strong interference regime. The capacity region is given by:*

$$\bigcup_{P_QP_{X_1|Q}P_{X_2|Q}P_{X_B|X_1X_2Q}} \left\{ \begin{array}{l} (R_0,R_1,R_2) \in \mathbb{R}_+^3: \\ R_0 \le \min\left\{ \begin{array}{l} I(X_B;Y_1|X_1,X_2,Q), \\ I(X_B;Y_2|X_1,X_2,Q) \end{array} \right\} \\ R_0 + R_1 \le I(X_1,X_B;Y_1|X_2,Q) \\ R_0 + R_2 \le I(X_2,X_B;Y_2|X_1,Q) \\ R_0 + R_1 + R_2 \le I(X_1,X_2,X_B;Y_1|Q) \\ R_0 + R_1 + R_2 \le I(X_1,X_2,X_B;Y_2|Q) \end{array} \right\} \tag{115}$$

*Proof of Proposition 3)* Note that according to Lemma 1 the conditions (114) extend to:

$$\begin{array}{c} I(X_1,X_B;Y_1|X_2,D) \le I(X_1,X_B;Y_2|X_2,D) \\ I(X_2,X_B;Y_2|X_1,D) \le I(X_2,X_B;Y_1|X_1,D) \end{array} \tag{116}$$

for all PDFs $P_{DX_1X_2X_B}(d,x_1,x_2,x_B)$. The achievability is derived by requiring that each receiver decodes all three messages. The message $M_i, i = 1,2$, is encoded by a codeword constructed by $X_i$ based on $P_{X_i}(x_i)$. The message $M_0$ is encoded by a codeword constructed by $X_B$ which is superimposed upon $(X_1(M_1), X_2(M_2))$ in accordance with $P_{X_B|X_1X_2}(x_B|x_1,x_2)$. Each receiver decodes all the codewrods $X_1(M_1), X_2(M_2)$ and $X_B(M_0,M_1,M_2)$. One can readily show that, under the conditions (116), this scheme achieves the rate region (115) for $Q \equiv \emptyset$, and therefore, a time-shared version of the scheme achieves the entire region. For the converse part consider the outer bound $\mathfrak{R}_o^{BCCR}$ given in (95). The bounds on $R_0$, $R_0 + R_1$ and $R_0 + R_2$ are directly found in $\mathfrak{R}_o^{BCCR}$. Moreover, we have:

$$\begin{aligned} R_0 + R_1 + R_2 &\le I(X_1,X_B;Y_1|X_2,V,W,Q) + I(V,W,X_2;Y_2|Q) \\ &\overset{(a)}{\le} I(X_1,X_B;Y_2|X_2,V,W,Q) + I(V,W,X_2;Y_2|Q) \\ &= I(X_1,X_2,X_B;Y_2|Q) \end{aligned}$$

where (a) is due to the first inequality of (116); also,

$$\begin{aligned} R_0 + R_1 + R_2 &\le I(X_2,X_B;Y_2|X_1,U,W,Q) + I(U,W,X_1;Y_1|Q) \\ &\overset{(b)}{\le} I(X_2,X_B;Y_1|X_1,U,W,Q) + I(U,W,X_1;Y_1|Q) \\ &= I(X_1,X_2,X_B;Y_1|Q) \end{aligned}$$

where (b) is due to the second inequality of (116). The proof is thus complete. ■

Now, let us examine the Gaussian BCCR in (94). It should be noted that the conditions (114) also represent a strong interference regime for the Gaussian channel. Unfortunately, it is difficult to transform these conditions into non-trivial constraints on the channel gains $a_1, a_2, a_B, b_1, b_2, b_B$. By applying the result of Lemma 2, we derive the trivial case where $a_1 = a_2 = a_B = b_1 = b_2 = b_B$. One might think that by evaluation of the inequalities (114) for Gaussian input distributions a strong interference regime is derived for the Gaussian channel. But indeed this is not the case. The fact is that for the Gaussian BCCR, the constraints which are obtained by evaluation of (114) for Gaussian distributions never imply that these inequalities hold for all arbitrary distributions as given in (116). Therefore, by simple evaluation of the constraints (114) one cannot derive an explicit characterization of a strong interference regime for the Gaussian channel (94).

We also identify a class of BCCRs with less noisy receivers. Specifically, assume that the channel satisfies the following conditions:

$$\begin{cases} I(V;Y_2|X_2) \leq I(V;Y_1|X_2), & \text{for all PDFs} \quad P_{VX_1X_B}P_{X_2} \\ I(X_2,X_B;Y_2|X_1) \leq I(X_2,X_B;Y_1|X_1), & \text{for all PDFs} \quad P_{X_1}P_{X_2X_B} \end{cases} \tag{117}$$

In the next theorem, using the outer bound $\mathfrak{R}_o^{BCCR}$ in (95), we establish the sum-rate capacity for this case.

***Theorem 6)*** *Consider the BCCR in Fig. 5 with the conditions (117). In this case, the receiver $Y_1$ is less noisy than the receiver $Y_2$ and the sum-rate capacity is given by:*

$$\max_{P_QP_{X_1|Q}P_{X_2|Q}P_{X_B|X_1X_2Q}} \begin{Bmatrix} I(X_1,X_B;Y_1|X_2,Q) + I(X_2;Y_2|Q) \\ I(X_1,X_2,X_B;Y_1|Q) \end{Bmatrix} \tag{118}$$

*Proof of Theorem 6)* The achievability is derived by a successive decoding scheme. The message $M_0$ is withdrawn from the transmission scheme. The private message $M_i, i = 1,2$, is encoded by a codeword constructed by $X_i$ based on $P_{X_i}(x_i)$. The $3^{rd}$ transmitter also generates a codeword $X_B(M_1,M_2)$ superimposed upon the codewords $X_1(M_1)$ and $X_2(M_2)$ according to $P_{X_B|X_1X_2}(x_B|x_1,x_2)$. The first receiver first decodes $X_2(M_2)$ and removes its interference effect and then jointly decodes the signals $X_1(M_1), X_B(M_1,M_2)$. The second receiver decodes only the signal $X_2(M_2)$. The resulting achievable rate region is described by the following constraints:

$$\begin{cases} R_1 \leq I(X_1,X_B;Y_1|X_2) \\ R_2 \leq \min\big(I(X_2;Y_2), I(X_2;Y_1)\big) \end{cases} \tag{119}$$

for all joint PDFs $P_{X_1}(x_1)P_{X_2}(x_2)P_{X_B|X_1X_2}(x_B|x_1,x_2)$. This scheme achieves the sum-rate (118) for the case of $Q \equiv \emptyset$, and thereby a time-shared version of the scheme achieves the entire region. To prove the converse part, note that the conditions (117) imply that:

$$\begin{gathered} I(V,W;Y_2|X_2,Q) \leq I(V,W;Y_1|X_2,Q) \\ I(X_2,X_B;Y_2|X_1,U,W,Q) \leq I(X_2,X_B;Y_1|X_1,U,W,Q) \end{gathered} \tag{120}$$

for all joint PDFs $P_QP_{X_1|Q}P_{X_2|Q}P_{X_B|X_1X_2Q}P_{UVW|X_BX_1X_2Q}$. This extension is actually a consequence of Lemmas 1 and 4. Now according to the outer bound $\mathfrak{R}_o^{BCCR}$ in (95) we have:

$$\begin{aligned} R_0 + R_1 + R_2 &\leq I(X_1,X_B;Y_1|X_2,V,W,Q) + I(V,W,X_2;Y_2|Q) \\ &= I(X_1,X_B;Y_1|X_2,V,W,Q) + I(X_2;Y_2|Q) + I(V,W;Y_2|X_2,Q) \\ &\overset{(a)}{\leq} I(X_1,X_B;Y_1|X_2,V,W,Q) + I(X_2;Y_2|Q) + I(V,W;Y_1|X_2,Q) \\ &= I(X_1,X_B;Y_1|X_2,Q) + I(X_2;Y_2|Q) \end{aligned}$$

where (a) is due to the first inequality of (120); also,

$$\begin{aligned} R_0 + R_1 + R_2 &\leq I(X_2,X_B;Y_2|X_1,U,W,Q) + I(U,W,X_1;Y_1|Q) \\ &\overset{(b)}{\leq} I(X_2,X_B;Y_1|X_1,U,W,Q) + I(U,W,X_1;Y_1|Q) \\ &= I(X_1,X_2,X_B;Y_1|Q) \end{aligned}$$

where (b) is due to the second inequality of (120). The proof is thus complete. ■

***Remarks 12:*** The result of Theorem 6 holds also for the Gaussian channel (94). For this channel, using Lemma 2, one can readily show that if the following relation holds:

$$\frac{b_1}{a_1}=\frac{b_2}{a_2}=\frac{b_B}{a_B}=\alpha, \qquad \text{with} \qquad |\alpha|\leq 1 \tag{121}$$

then, the inequalities (117) are satisfied. In fact, the Gaussian BCCR (94) with the condition (121) is (stochastically) degraded in the sense that $Y_2$ is statistically equivalent to a noisy version of $Y_1$. Therefore, its sum-rate capacity is also given in Part II [2].

Next, let us deal with the one-sided BCCR. Consider a channel where the probability transition function is decomposed as follows:

$$\mathbb{P}(y_1,y_2|x_1,x_B,x_2)=\mathbb{P}(y_1|x_1,x_B,x_2)\mathbb{P}(y_2|x_2) \tag{122}$$

In this scenario, the receiver $Y_2$ experiences no interference. In the next proposition, we present a strong interference regime for this network (note that the conditions (114) cannot be satisfied simultaneously for the one-sided channel (122)).

***Proposition 4)*** *Consider the one-sided BCCR in (122). If the channel satisfies the following condition:*

$$I(X_2;Y_2)\leq I(X_2,X_B;Y_1|X_1), \qquad \text{for all PDFs} \qquad P_{X_1}(x_1)P_{X_2X_B}(x_1,x_B) \tag{123}$$

*then, it is in the strong interference regime. The capacity region is given by:*

$$\bigcup_{P_QP_{X_1|Q}P_{X_2|Q}P_{X_B|X_1X_2Q}}\left\{\begin{array}{l}(R_0,R_1,R_2)\in\mathbb{R}_+^3: \quad R_0=0\\ R_1\leq I(X_1,X_B;Y_1|X_2,Q)\\ R_2\leq I(X_2;Y_2|Q)\\ R_1+R_2\leq I(X_1,X_2,X_B;Y_1|Q)\end{array}\right\} \tag{124}$$

*Proof of Proposition 4)* First note that the condition (123) beside (122) implies that:

$$I(X_2;Y_2)=I(X_2,X_B;Y_2|X_1)\leq I(X_2,X_B;Y_1|X_1), \qquad \text{for all PDFs} \qquad P_{X_1}(x_1)P_{X_2X_B}(x_1,x_B) \tag{125}$$

Therefore, according to Lemma 1, we have:

$$I(X_2,X_B;Y_2|X_1,D)\leq I(X_2,X_B;Y_1|X_1,D), \qquad \text{for all PDFs} \qquad P_{DX_1X_2X_B}(d,x_1,x_2,x_B) \tag{126}$$

For the one-sided BCCR, according to Lemma II.6 of Part I [1, Sec. II], no positive rate can be achieved for the message $M_0$ because this message should be decoded at a receiver which is not connected to it. This is also seen from the outer bound $\mathfrak{R}_o^{BCCR}$ in (95) if the condition (122) is considered. To derive the achievability, the messages $M_1$ and $M_2$ are encoded by the codewords $X_1(M_1),X_2(M_2)$ and $X_B(M_1,M_2)$ at the corresponding transmitters; these codewords are generated according to $P_{X_1}(x_1),P_{X_2}(x_2)$ and $P_{X_B|X_1X_2}(x_B|x_1,x_2)$, respectively. The receiver $Y_1$ jointly decodes all three codewords $X_1(M_1),X_2(M_2)$ and $X_B(M_1,M_2)$. The receiver $Y_2$ decodes only the codeword $X_2(M_2)$. The resulting achievable rate region is described by:

$$\left\{\begin{array}{l}R_2\leq I(X_2;Y_2)\\ R_1\leq I(X_1,X_B;Y_1|X_2)\\ R_2\leq I(X_2,X_B;Y_1|X_1)\\ R_1+R_2\leq I(X_1,X_2,X_B;Y_1)\end{array}\right. , \qquad \text{for all PDFs} \quad P_{X_1}P_{X_2}P_{X_B|X_1X_2} \tag{127}$$

The third constraint of (127) is actually redundant because we have:

$$R_2\leq I(X_2;Y_2)=I(X_2,X_B;Y_2|X_1)\overset{(a)}{\leq} I(X_2,X_B;Y_1|X_1), \quad \text{for all PDFs} \quad P_{X_1}P_{X_2}P_{X_B|X_1X_2} \tag{128}$$

where (a) is due to (126) that is considered for $D \equiv \emptyset$ and the joint PDFs of the form $P_{X_1}P_{X_2}P_{X_B|X_1X_2}$. Therefore, the coding scheme achieves the rate region (124) for $Q \equiv \emptyset$, and a time-shared version of it achieves the entire region. To prove the converse part, using the outer bound $\mathfrak{R}_o^{BCCR}$ in (95) we obtain:

$$
\begin{aligned}
R_1 &\leq I(X_1, X_B; Y_1|X_2, Q) \\
R_2 &\leq I(X_2, X_B; Y_2|X_1, Q) = I(X_2; Y_2|Q) \\
R_1 + R_2 &\leq I(X_2, X_B; Y_2|X_1, U, W, Q) + I(U, W, X_1; Y_1|Q) \\
&\overset{(a)}{\leq} I(X_2, X_B; Y_1|X_1, U, W, Q) + I(U, W, X_1; Y_1|Q) \\
&= I(X_1, X_2, X_B; Y_1|Q)
\end{aligned}
$$

where (a) is due to (126). The proof is complete. ■

***Remark 13:*** Similar to the two-sided strong interference BCCR in (122), for the one-sided Gaussian channel which is derived by setting $b_2 = b_B = 0$ in (94), it is difficult to transform the condition (123) into non-trivial constraints in term of the parameters $a_1, b_1, b_2, b_B$. The simple evaluation of (123) for Gaussian distributions does not yield equivalent constraints.

## IV.B) Information Flow in Strong Interference Regime

In previous subsections, we derived unified outer bounds on the capacity region of the general interference networks with two receivers. By considering the two-user CIC with common message and the BCCR, we showed that the derived outer bounds are efficient to identify strong interference regimes for these channels. Now, we intend to demonstrate this capability for all interference networks with two-receivers. To this aim, the procedure we followed for the two-user CIC or the BCCR in previous subsection is not efficient because for large multi-user/multi-message networks one involve to handle numerous constraints while describing rate regions. Thus, we need to develop a new methodology which is given below.

Let us once more consider the two-user CIC with common message in Fig. 4. As derived in Subsection IV.A.1, this channel under the conditions (77) lies in the strong interference regime. To identify this regime, one may proceed as follows. First, an achievable rate region is derived by viewing the channel as a compound MAC [12] with a common message. In this scheme, the messages are encoded at the transmitters exactly similar to the two-user MAC with a common message [26]; also, each receiver jointly decodes all messages. The resulting rate region is given in the following:

$$
\bigcup_{P_W P_{X_1|W} P_{X_2|W}} \left\{
\begin{array}{l}
(R_0, R_1, R_2) \in \mathbb{R}_+^3 : \\
R_1 \leq \min\left\{\begin{array}{l} I(X_1; Y_1|X_2, W), \\ I(X_1; Y_2|X_2, W) \end{array}\right\} \\
R_2 \leq \min\left\{\begin{array}{l} I(X_2; Y_2|X_1, W), \\ I(X_2; Y_1|X_1, W) \end{array}\right\} \\
R_1 + R_2 \leq \min\left\{\begin{array}{l} I(X_1, X_2; Y_1|W), \\ I(X_1, X_2; Y_2|W) \end{array}\right\} \\
R_0 + R_1 + R_2 \leq \min\left\{\begin{array}{l} I(X_1, X_2; Y_1), \\ I(X_1, X_2; Y_2) \end{array}\right\}
\end{array}
\right\}
\tag{129}
$$

In the second step, we explore for the conditions by which this rate region is optimal. These conditions are indeed given by (77). Let us carefully examine the procedure of proving the optimality of (129) under the strong interference conditions. As a direct consequence of these conditions, the following two constraints can be relaxed from (129):

$$
\left\{\begin{array}{l} R_1 \leq I(X_1; Y_2|X_2, W) \\ R_2 \leq I(X_2; Y_1|X_1, W) \end{array}\right.
\tag{130}
$$

These constraints are redundant in characterization of (129) if the conditions (77) hold. Nevertheless, let us indicate an interesting characteristic regarding the description of capacity region in the strong interference regime. Clearly, consider the following constraints of the outer bound $\mathfrak{R}_o^{CICcm}$ in (62):

$$\begin{cases} R_1 \leq I(X_1;Y_1|X_2,V,W) + I(V;Y_2|X_2,W) \\ R_2 \leq I(X_2;Y_2|X_1,U,W) + I(U;Y_1|X_1,W) \end{cases} \tag{131}$$

Note that the strong interference conditions (77) yield (based on Lemma 1):

$$\begin{cases} I(X_1;Y_1|X_2,V,W) \leq I(X_1;Y_2|X_2,V,W) \\ I(X_2;Y_2|X_1,U,W) \leq I(X_2;Y_1|X_1,U,W) \end{cases} \tag{132}$$

Now by substituting (132) in (131), we obtain the constraints (130). This fact reveals that for deriving the capacity region in the strong interference regime, it is not required to apply the strong interference conditions to simplify the achievable rate region obtained by requiring all receivers decode all messages. Instead, all the constraints of this latter achievable rate region could be derived using the outer bound $\mathfrak{R}_o^{CICcm}$ in (62). The two other constraints on the individual rates $R_1$ and $R_2$ are directly given in the outer bound $\mathfrak{R}_o^{CICcm}$. It remains to prove the bounds on the sum of rates. By inspecting our method given in Subsection IV.A.1 for the derivation of the desired constraints on $R_1+R_2$ and $R_0+R_1+R_2$ (see also the derivation of the outer bound $\mathfrak{R}_o^{CICcm}$ in (62) from $\mathfrak{R}_{o:(2)}^{GINTR}$ in (54)), one can easily perceive that we made use of the following constraints of the general outer bound $\mathfrak{R}_{o:(2)}^{GINTR}$ in (54):

$$\begin{cases} \mathfrak{R}_{o:(2)}^{GINTR}\langle 3\rangle : \boldsymbol{R}_{\Sigma\Omega_0} + \boldsymbol{R}_{\Sigma\Omega_1} + \boldsymbol{R}_{\Sigma\Omega_2} \leq I(\Omega_1;Y_1|Z,\Omega_0,\Omega_2,\Omega_s,Q) + I(Z,\Omega_0,\Omega_2;Y_2|\Omega_s,Q) \\ \mathfrak{R}_{o:(2)}^{GINTR}\langle 4\rangle : \boldsymbol{R}_{\Sigma\Omega_0} + \boldsymbol{R}_{\Sigma\Omega_1} + \boldsymbol{R}_{\Sigma\Omega_2} \leq I(\Omega_2;Y_2|Z,\Omega_0,\Omega_1,\Omega_s,Q) + I(Z,\Omega_0,\Omega_1;Y_1|\Omega_s,Q) \end{cases} \tag{133}$$

In Subsection IV.A.1, we first translated the constraints (133) into suitable bounds given in $\mathfrak{R}_o^{CICcm}$ in (62) and then made use of them to prove the desired constraints on $R_1+R_2$ and $R_0+R_1+R_2$ for the strong interference regime. Here, we intend to directly obtain one of the constraints on the sum-rate in (129) using (133). The new proof style sheds light on our future derivations in this subsection. Let us choose $\Omega_0 = M_0$, $\Omega_1 = M_1$, $\Omega_2 = M_2$ and $\Omega_s = \emptyset$. From $\mathfrak{R}_{o:(2)}^{GINTR}\langle 3\rangle$ we have:

$$R_0 + R_1 + R_2 \leq I(M_1;Y_1|Z,M_0,M_2,Q) + I(Z,M_0,M_2;Y_2|Q) \tag{134}$$

For the two-user CIC with common message, the constraint (134) should be evaluated over all joint PDFs $P_Q P_{M_0} P_{M_1} P_{M_2} P_{X_1|M_1M_0Q} P_{X_2|M_2M_0Q}$ where $X_i$ is a deterministic function of $(M_i, M_0, Q)$, $i=1,2$. Now, we claim that the strong interference conditions (77) imply the following inequality:

$$I(M_1;Y_1|Z,M_0,M_2,Q) \leq I(M_1;Y_2|Z,M_0,M_2,Q) \tag{135}$$

First note that we have:

$$\begin{cases} I(M_1;Y_1|Z,M_0,M_2,Q) = I(X_1;Y_1|X_2,Z,M_0,M_2,Q) \\ \\ I(M_1;Y_2|Z,M_0,M_2,Q) = I(X_1;Y_2|X_2,Z,M_0,M_2,Q) \end{cases} \tag{136}$$

The equalities in (136) are indeed due the fact that $X_i$ is a deterministic function of $(M_i, M_0, Q)$, $i=1,2$ and also the fact that, given the inputs $X_1, X_2$, the outputs $Y_1, Y_2$ are independent of other random variables. Now consider the strong interference conditions (77). The remarkable consequence of Lemma 1 is that it extends the strong interference conditions (77) to:

$$\begin{cases} I(X_1;Y_1|X_2,D) \leq I(X_1;Y_2|X_2,D) \\ I(X_1;Y_2|X_2,D) \leq I(X_1;Y_1|X_2,D) \end{cases}, \quad \text{for all } P_{DX_1X_2}(d,x_1,x_2) \tag{137}$$

Based on (137), one can directly write:

$$I(X_1;Y_1|X_2,Z,M_0,M_2,Q) \leq I(X_1;Y_2|X_2,Z,M_0,M_2,Q) \tag{138}$$

In fact, (138) holds regardless of the distribution of $Q, M_0, M_1, M_2, Z, X_1, X_2$ because (137) holds for all $P_{DX_1X_2}(d, x_1, x_2)$. Considering (136) and (138), we obtain (135).

The above approach in fact leads us to an alternate characterization of the capacity region in the strong interference regime. Here, we refer the reader to our detailed discussion in Part II of our multi-part papers [2, Sec. II.A and IV] regarding different characterizations for the capacity region of the MACs with common messages. Hereafter, we assume that the reader is familiar with the prerequisites developed there. As discussed before, the capacity region of the compound MAC with a common message (this network is similar to the two-user CIC with common message in Fig. 4 except that each receiver is required to decode both messages) is given by (129). Such characterization of the capacity region, which is due to Slepian and Wolf [26], is derived using the superposition coding technique. The respective coding scheme was discussed in details in Part II [2, Sec. II.A]. Briefly, the common message is encoded using a codeword served as the cloud center and the private messages are encoded by codewords (satellites) which are superimposed on the common message codeword. On the other side, each receiver jointly decodes all the messages. An alternative characterization of the capacity region is given below:

$$\bigcup_{\substack{P_QP_{M_0}P_{M_1}P_{M_2} \\ X_i=f_i(M_0,M_i,Q),i=1,2}} \left\{ \begin{array}{l} (R_0,R_1,R_2) \in \mathbb{R}_+^3 : \\ R_0 \leq \min\{I(M_0;Y_1|M_1,M_2,Q), I(M_0;Y_2|M_1,M_2,Q)\} \\ R_1 \leq \min\{I(M_1;Y_1|M_0,M_2,Q), I(M_1;Y_2|M_0,M_2,Q)\} \\ R_2 \leq \min\{I(M_2;Y_1|M_0,M_1,Q), I(M_2;Y_2|M_0,M_1,Q)\} \\ \qquad\qquad , \\ R_0+R_1 \leq \min\{I(M_0,M_1;Y_1|M_2,Q), I(M_0,M_1;Y_2|M_2,Q)\} \\ R_0+R_2 \leq \min\{I(M_0,M_2;Y_1|M_1,Q), I(M_0,M_2;Y_2|M_1,Q)\} \\ R_1+R_2 \leq \min\{I(M_1,M_2;Y_1|M_0,Q), I(M_1,M_2;Y_1|M_0,Q)\} \\ \qquad\qquad , \\ R_0+R_1+R_2 \leq \min\{I(M_0,M_1,M_2;Y_1|Q), I(M_0,M_1,M_2;Y_2|Q)\} \end{array} \right\} \tag{139}$$

This characterization, which is due to Han [27] is more complex than that in (129); nevertheless, it represents a very simple coding scheme for the corresponding network. It states that to achieve the capacity region of the MACCM, it is sufficient to encode the messages separately using independent codewords (without any binning or superposition coding); in other words, the message $M_i$ is encoded using a codeword $M_i^n$ generated based on $P_{M_i}(m_i)$, $i = 0,1,2$. A time-sharing codeword $Q^n$ is also generated independently based on $P_Q(q)$ which is revealed to all parties. The transmitter $X_i, i = 1,2$ then generates its codeword $X_i^n$ as $X_i^n = f_i(M_0^n, M_i^n, Q^n)$[5], where $f_i(.)$ is an arbitrary deterministic function, and sends it over the channel. At each receiver, all the codewords $M_0^n, M_1^n, M_2^n$ are jointly decoded. A simple analysis of this achievability scheme leads to the rate region (139). Another important benefit of this approach is that one can simply describe the capacity region for the general MAC with arbitrary distribution of messages among transmitters. See Part II [2] for a detailed discussion. It should be noted the rate region (139) is equivalent to (129) because both are coincide with the capacity region of the compound MAC with common messages. Thereby, we can take advantage of (139) as an alternate characterization of the capacity region for the two-user CIC with common message in Fig. 4 in the strong interference regime (77).

Now let us investigate the rate region (139). As we demonstrated, having at hand the strong interference conditions (77), one can derive all the constraints of the rate region (129) using the outer bound $\mathfrak{R}_{o:(2)}^{GINTR}$ in (54); even a constraint can be directly relaxed by the conditions (77). Now the question is that if the same conclusion holds for the characterization (139)? Strikingly, this is indeed the case. In fact, if the strong interference conditions (77) hold, one can derive all the constraints of the rate region (139) by using only the general bounds in (133). We examine some nontrivial cases. Specifically, consider the following constraints:

$$\begin{array}{lr} R_0 \leq I(M_0;Y_1|M_1,M_2,Q) & \text{(i)} \\ R_0+R_1 \leq I(M_0,M_1;Y_2|M_2,Q) & \text{(ii)} \end{array} \tag{140}$$

To derive (140~i), by choosing $\Omega_0 = M_0, \Omega_1 = \emptyset, \Omega_2 = \emptyset$ and $\Omega_s = \{M_1, M_2\}$ from $\mathfrak{R}_{o:(2)}^{GINTR}\langle 4\rangle$, we have:

$$R_0 \leq I(M_0,Z;Y_1|M_1,M_2,Q) \overset{(a)}{=} I(M_0;Y_1|M_1,M_2,Q) \tag{141}$$

[5]Here, the notation $X_i^n = f_i(M_0^n, M_i^n, Q^n)$ actually indicates $X_{i,t} = f_i(M_{0,t}, M_{i,t}, Q_t), t = 1, \dots, n$.

where (a) holds because $X_1$ and $X_2$ are given by deterministic functions of $(M_0, M_1, M_2, Q)$, and hence $Z \to M_0, M_1, M_2, Q \to Y_1, Y_2$ forms a Markov chain. To derive (140~ii), by choosing $\Omega_0 = M_0$, $\Omega_1 = M_1$, $\Omega_2 = \emptyset$ and $\Omega_s = \{M_2\}$, from $\mathfrak{R}_{o:(2)}^{GINTR}\langle 3\rangle$ we have:

$$\begin{aligned} R_0 + R_1 &\leq I(M_1; Y_1|Z, M_0, M_2, Q) + I(Z, M_0; Y_2|M_2, Q) \\ &\overset{(b)}{\leq} I(M_1; Y_2|Z, M_0, M_2, Q) + I(Z, M_0; Y_2|M_2, Q) \\ &= I(Z, M_0, M_1; Y_2|M_2, Q) \overset{(c)}{=} I(M_0, M_1; Y_2|M_2, Q) \end{aligned} \tag{142}$$

where inequality (b) is obtained from (135); also, equality (c) holds because $Z \to M_0, M_1, M_2, Q \to Y_1, Y_2$ form a Markov chain. All the other constraints of the characterization (139) could be derived from the bounds in (133) by following the same arguments, regardless of that whether a constraint is redundant or not.

The above theory can be developed for other network scenarios. For example, the capacity region of the BCCR in Fig. 5 in the strong interference regime (114) can be re-described by the same approach. Strikingly, it can be described by a rate region identical to that of (139) except that the union should be taken over all PDFs of the following form:

$$P_Q P_{M_0} P_{M_1} P_{M_2} P_{X_1|M_1Q} P_{X_2|M_2Q} P_{X_3|M_0M_1M_2Q} \quad \text{with} \quad P_{X_1|M_1Q}, P_{X_2|M_2Q}, P_{X_3|M_0M_1M_2Q} \in \{0,1\} \tag{143}$$

Now, we are ready to state our main result in this subsection which is given in the next theorem.

***Theorem 7)*** **Two-Receiver Interference Networks: Information Flow in Strong Interference Regime**

*Consider a general interference network with two receivers which is derived from the general network in Fig. 1 by setting $K_2 = 2$. If the transition probability function of the network satisfies the following conditions:*

$$\boxed{\begin{array}{ll} I\left(\mathbb{X} - \mathbb{X}_{\mathbb{M}_{Y_2}}; Y_1 \middle| \mathbb{X}_{\mathbb{M}_{Y_2}}\right) \leq I\left(\mathbb{X} - \mathbb{X}_{\mathbb{M}_{Y_2}}; Y_2 \middle| \mathbb{X}_{\mathbb{M}_{Y_2}}\right) & \text{for all joint PDFs} \quad P_{\mathbb{X}-\mathbb{X}_{\mathbb{M}_{Y_2}}} \prod_{X_i \in \mathbb{X}_{\mathbb{M}_{Y_2}}} P_{X_i} \\ I\left(\mathbb{X} - \mathbb{X}_{\mathbb{M}_{Y_1}}; Y_2 \middle| \mathbb{X}_{\mathbb{M}_{Y_1}}\right) \leq I\left(\mathbb{X} - \mathbb{X}_{\mathbb{M}_{Y_1}}; Y_1 \middle| \mathbb{X}_{\mathbb{M}_{Y_1}}\right) & \text{for all joint PDFs} \quad P_{\mathbb{X}-\mathbb{X}_{\mathbb{M}_{Y_1}}} \prod_{X_i \in \mathbb{X}_{\mathbb{M}_{Y_1}}} P_{X_i} \end{array}} \tag{144}$$

*then the network has strong interference, i.e., the decoding of all messages at each receiver achieves the capacity region. Specifically, the outer bound $\mathfrak{R}_{o:(2)}^{GINTR}$ in (54) is optimal for such a network.*

*Proof of Theorem 7)* First note that, according to Lemma 1, if the inequalities in (144) hold for the given PDFs, then they also hold for any arbitrary joint PDF $P_{\mathbb{X}}$ on the input signals $\mathbb{X} = \{X_1, \dots, X_{K_1}\}$. Now, consider the rate region below:

$$\bigcup_{\substack{P_Q P_{M_1} P_{M_2} \dots P_{M_K} \\ \times \prod_{i=1}^{K_1} P_{X_i|\mathbb{M}_{X_i}, Q}}} \left\{ \begin{array}{l} (R_1, R_2, \dots, R_K): \\ \forall\, \Omega \subseteq \mathbb{M} = \{M_1, \dots, M_K\}: \\ \boldsymbol{R}_{\Sigma\Omega} \leq \min \left\{ \begin{array}{l} I(\Omega; Y_1|\mathbb{M} - \Omega, Q), \\ I(\Omega; Y_2|\mathbb{M} - \Omega, Q) \end{array} \right\} \end{array} \right\} \tag{145}$$

The rate region (145) is achievable for the network. It is actually the capacity region if both receivers are required to decode all messages. Let us briefly describe the corresponding achievability scheme. Each message $M_i$ belonging to $\{M_1, \dots, M_K\}$ is encoded using a codeword $M_i^n$ generated based on the distribution $P_{M_i}(m_i), i = 1, \dots, K$. A time-sharing codeword $Q^n$ is also generated independently based on $P_Q(q)$; this codeword is revealed to all parties. The transmitter $X_i, i = 1, \dots, K_1$, then generates its codeword $X_i^n$ based on $P_{X_i|\mathbb{M}_{X_i}, Q}$ and sends it over the network. At each receiver, all the codewords $M_1^n, M_2^n, \dots, M_K^n$ are jointly decoded. By a simple analysis, we derive the achievable rate region (145). Next we prove that, under the conditions (144), this achievable region is optimal.

Let $\Omega$ be an arbitrary subset of $\{M_1,\dots,M_K\}$. Define the subsets $\Omega_{Y_1}$ and $\Omega_{Y_2}$ as follows:

$$\Omega_{Y_1} \triangleq \left(\mathbb{M}_{Y_1} - \mathbb{M}_{Y_2}\right)\cap\Omega, \qquad \Omega_{Y_2} \triangleq \mathbb{M}_{Y_2}\cap\Omega \tag{146}$$

Note that, $\Omega_{Y_1}$ and $\Omega_{Y_2}$ constitutes a partition for the set $\Omega$. Consider the general outer bound $\mathfrak{R}_{o:(2)}^{GINTR}$ in (54). We need to apply the constraints $\mathfrak{R}_{o:(2)}^{GINTR}\langle 3\rangle$ and $\mathfrak{R}_{o:(2)}^{GINTR}\langle 4\rangle$. Specifically, by substituting $\Omega_0=\emptyset$, $\Omega_1=\Omega_{Y_1}$, $\Omega_2=\Omega_{Y_2}$ and $\Omega_s=\mathbb{M}-\Omega$ in $\mathfrak{R}_{o:(2)}^{GINTR}\langle 3\rangle$, we obtain:

$$\boldsymbol{R}_{\sum\Omega}=\boldsymbol{R}_{\sum\Omega_1}+\boldsymbol{R}_{\sum\Omega_2}\leq I\left(\Omega_{Y_1};Y_1\middle|Z,\Omega_{Y_2},\mathbb{M}-\Omega,Q\right)+I\left(Z,\Omega_{Y_2};Y_2\middle|\mathbb{M}-\Omega,Q\right) \tag{147}$$

Consider the first mutual information function in (147). Note that from (146) we can write:

$$\Omega_{Y_2}\cup(\mathbb{M}-\Omega)=\mathbb{M}-\Omega_{Y_1} \tag{148}$$

Also, since $\Omega_{Y_1}\cup\left(\Omega_{Y_2}\cup\mathbb{M}-\Omega\right)=\mathbb{M}$ and $\mathbb{M},Q\to\mathbb{X}\to Y_1,Y_2$ form a Markov chain, we have:

$$I\left(\Omega_{Y_1};Y_1\middle|Z,\Omega_{Y_2},\mathbb{M}-\Omega,Q\right)=I\left(\mathbb{X}-\mathbb{X}_{\mathbb{M}-\Omega_{Y_1}};Y_1\middle|\mathbb{X}_{\mathbb{M}-\Omega_{Y_1}},Z,\Omega_{Y_2},\mathbb{M}-\Omega,Q\right) \tag{149}$$

Note that we have $\mathbb{M}_{Y_2}\subseteq\mathbb{M}-\Omega_{Y_1}$; thereby, $\mathbb{X}_{\mathbb{M}_{Y_2}}$ is a subset of $\mathbb{X}_{\mathbb{M}-\Omega_{Y_1}}$ and thus we derive:

$$\mathbb{X}-\mathbb{X}_{\mathbb{M}-\Omega_{Y_1}}\subseteq\mathbb{X}-\mathbb{X}_{\mathbb{M}_{Y_2}} \tag{150}$$

Now consider the first inequality in (144). According to Lemma 1, this condition implies that:

$$I\left(\mathbb{X}-\mathbb{X}_{\mathbb{M}_{Y_2}};Y_1\middle|\mathbb{X}_{\mathbb{M}_{Y_2}},D\right)\leq I\left(\mathbb{X}-\mathbb{X}_{\mathbb{M}_{Y_2}};Y_2\middle|\mathbb{X}_{\mathbb{M}_{Y_2}},D\right)\qquad\text{for all joint PDFs } P_{D\mathbb{X}} \tag{151}$$

Also, since the inclusion (150) holds, according to the Corollary 1, one can write:

$$I\left(\mathbb{X}-\mathbb{X}_{\mathbb{M}-\Omega_{Y_1}};Y_1\middle|\mathbb{X}_{\mathbb{M}-\Omega_{Y_1}},D\right)\leq I\left(\mathbb{X}-\mathbb{X}_{\mathbb{M}-\Omega_{Y_1}};Y_2\middle|\mathbb{X}_{\mathbb{M}-\Omega_{Y_1}},D\right)\qquad\text{for all joint PDFs } P_{D\mathbb{X}} \tag{152}$$

Hence, from (152) we directly obtain:

$$I\left(\mathbb{X}-\mathbb{X}_{\mathbb{M}-\Omega_{Y_1}};Y_1\middle|\mathbb{X}_{\mathbb{M}-\Omega_{Y_1}},Z,\Omega_{Y_2},\mathbb{M}-\Omega,Q\right)\leq I\left(\mathbb{X}-\mathbb{X}_{\mathbb{M}-\Omega_{Y_1}};Y_2\middle|\mathbb{X}_{\mathbb{M}-\Omega_{Y_1}},Z,\Omega_{Y_2},\mathbb{M}-\Omega,Q\right) \tag{153}$$

By substituting (153) into (149), we obtain:

$$\begin{aligned}
I\left(\Omega_{Y_1};Y_1\middle|Z,\Omega_{Y_2},\mathbb{M}-\Omega,Q\right)&\leq I\left(\mathbb{X}-\mathbb{X}_{\mathbb{M}-\Omega_{Y_1}};Y_2\middle|\mathbb{X}_{\mathbb{M}-\Omega_{Y_1}},Z,\Omega_{Y_2},\mathbb{M}-\Omega,Q\right)\\
&\overset{(a)}{=} I\left(\mathbb{X}-\mathbb{X}_{\mathbb{M}-\Omega_{Y_1}},\Omega_{Y_1};Y_2\middle|\mathbb{X}_{\mathbb{M}-\Omega_{Y_1}},Z,\Omega_{Y_2},\mathbb{M}-\Omega,Q\right)\\
&\overset{(b)}{=} I\left(\Omega_{Y_1};Y_2\middle|Z,\Omega_{Y_2},\mathbb{M}-\Omega,Q\right)
\end{aligned} \tag{154}$$

where equality (a) holds because $\mathbb{M},Q\to\mathbb{X}\to Y_2$ form a Markov chain; (b) holds because the inputs $\mathbb{X}-\mathbb{X}_{\mathbb{M}-\Omega_{Y_1}}$ are given by deterministic functions of $(\mathbb{M},Q)$ and the inputs $\mathbb{X}_{\mathbb{M}-\Omega_{Y_1}}$ by deterministic functions of $\left(\Omega_{Y_2},\mathbb{M}-\Omega,Q\right)$.

Finally, by substituting (154) into (147), we derive:

$$\boldsymbol{R}_{\sum\Omega} \leq I(\Omega_{Y_1};Y_2|Z,\Omega_{Y_2},\mathbb{M}-\Omega,Q)+I(Z,\Omega_{Y_2};Y_2|\mathbb{M}-\Omega,Q)$$

$$=I(Z,\Omega_{Y_1},\Omega_{Y_2};Y_2|\mathbb{M}-\Omega,Q)=I(Z,\Omega;Y_2|\mathbb{M}-\Omega,Q)\overset{(a)}{=}I(\Omega;Y_2|\mathbb{M}-\Omega,Q)$$

where equality (a) holds because $Z \to \mathbb{M},Q \to Y_2$ form a Markov chain. Symmetrically, using the second inequality of (144) and the bound $\mathfrak{R}_{o:(2)}^{GINTR}\langle 4\rangle$, one can derive the other constraint in (145). The proof is complete. ■

***Remarks 14:***

1. The first condition in (144) implies the strong interference at the receiver $Y_2$ and the second condition implies the strong interference at the receiver $Y_1$.
2. Consider a two-receiver network in which there is no transmitter that broadcasts private messages to both receivers. In these networks, each message set $\mathbb{M}_{X_i}, i=1,\dots,K_1$, is a subset of at least one of the sets $\mathbb{M}_{Y_1}$ and $\mathbb{M}_{Y_2}$. Hence, we have:

$$\begin{cases}\mathbb{X}-\mathbb{X}_{\mathbb{M}_{Y_2}}=\mathbb{X}_{\mathbb{M}_{Y_1}}-\mathbb{X}_{\mathbb{M}_{Y_2}}\\ \mathbb{X}-\mathbb{X}_{\mathbb{M}_{Y_1}}=\mathbb{X}_{\mathbb{M}_{Y_2}}-\mathbb{X}_{\mathbb{M}_{Y_1}}\end{cases}$$

   Thus, for such networks the strong interference regime (144) can be re-written as follows:

$$\begin{cases}I\left(\mathbb{X}_{\mathbb{M}_{Y_1}};Y_1\middle|\mathbb{X}_{\mathbb{M}_{Y_2}}\right)\leq I\left(\mathbb{X}_{\mathbb{M}_{Y_1}};Y_2\middle|\mathbb{X}_{\mathbb{M}_{Y_2}}\right) & \text{for all } P_{\mathbb{X}_{\mathbb{M}_{Y_1}}-\mathbb{X}_{\mathbb{M}_{Y_2}}}\prod_{X_i\in\mathbb{X}_{\mathbb{M}_{Y_2}}}P_{X_i}\\ I\left(\mathbb{X}_{\mathbb{M}_{Y_2}};Y_2\middle|\mathbb{X}_{\mathbb{M}_{Y_1}}\right)\leq I\left(\mathbb{X}_{\mathbb{M}_{Y_2}};Y_1\middle|\mathbb{X}_{\mathbb{M}_{Y_1}}\right) & \text{for all } P_{\mathbb{X}_{\mathbb{M}_{Y_2}}-\mathbb{X}_{\mathbb{M}_{Y_1}}}\prod_{X_i\in\mathbb{X}_{\mathbb{M}_{Y_1}}}P_{X_i}\end{cases} \quad (155)$$

   These conditions expressively represent a fact regarding the strong interference regime: the amount of information flowing in the undesired directions is greater than that flowing in the desired directions.
3. Consider a two-receiver interference network where one receiver is required to decode all messages, e.g., $\mathbb{M}_{Y_2}=\mathbb{M}$. For such a network, from the viewpoint of the receiver $Y_2$, all input signals contain only information and no interference is experienced at this receiver. Let us examine our conditions in (144) for this case. Since $\mathbb{M}_{Y_2}=\mathbb{M}$, we have $\mathbb{X}_{\mathbb{M}_{Y_2}}=\mathbb{X}$. Thereby, the two sides of the first inequality in (144) are both zero. In fact, for this network the first condition of (144), which was implying the strong interference at the receiver $Y_2$, vanishes and the strong interference regime is given by only the second inequality of (144), as was expected.
4. The result of Theorem 7 also holds for the Gaussian networks (3). Thus, any two-receiver Gaussian interference network (3) satisfying the conditions of (144) lies in the strong interference regime. For these networks, by applying Lemma 2, one can readily derive a strong interference regime described by explicit constraints on the network gain matrix. Nonetheless, it is critical to note that evaluating the conditions (144) using Gaussian input distributions, does not yield a valid strong interference regime in general.
5. As shown in the proof of Theorem 7, to derive the strong interference regime (144) for the general two-receiver interference network, our result in Lemma 1 and its consequence in Corollary 1 have a central role.
6. Although, the characterization (145) is useful to derive the strong interference regime (144) for the general two-receiver interference networks, but in fact it involves numerous constraints (rapidly growing with the number of users) where many of them are redundant. Using our graphical illustrations developed in Part II [2] called *MACCM Plan* of messages, one can simplify the rate region (145) by removing its redundant constraints. A detailed discussion in this regard will be given in Subsection V.B.4.
7. One can readily see that for the two-user CIC with common message in Fig. 4 and the BCCR in Fig. 5, the conditions (144) are reduced to (77) and (114), respectively.

Let us provide an example to better appreciate the significance of in Theorem 7.

***Example 1:*** **The Multiple-Access Interference Network (MAIN)**

Let us discuss a network composed of two interfering MACs. Consider a two-receiver MAIN (introduced in Part II [2, Sec. IV]) where two groups of transmitters (each group with an arbitrary size) communicate with two receivers via a common media: each group of transmitters send information to their respective receiver while causing interference to the other receiver. The network is depicted in Fig. 6.

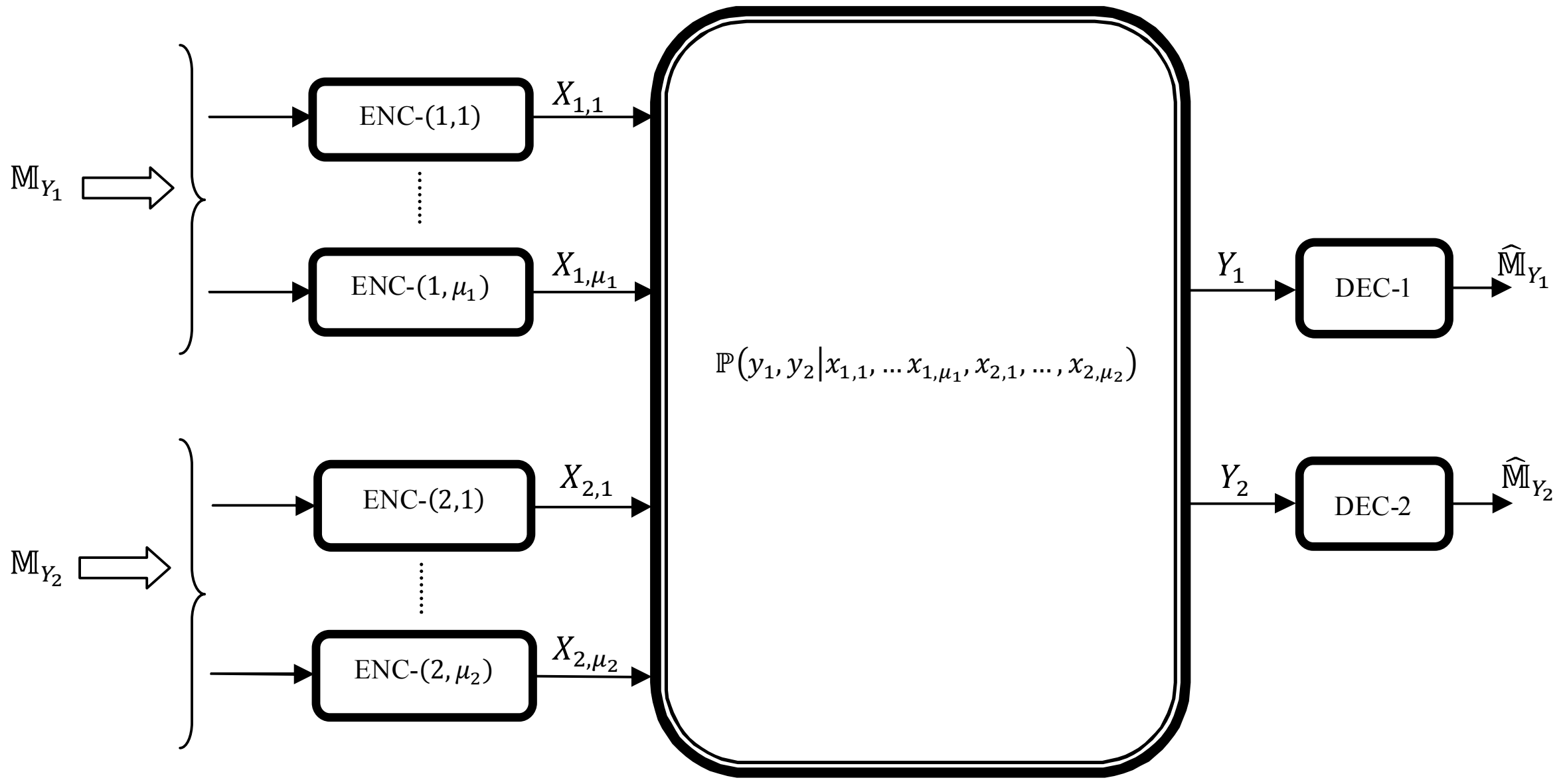

Figure 6. The two-receiver Multiple-Access-Interference Network (MAIN): two groups of transmitters $\mathbb{X}_1 \triangleq \{X_{1,1}, \dots, X_{1,\mu_1}\}$ and $\mathbb{X}_2 \triangleq \{X_{2,1}, \dots, X_{2,\mu_2}\}$ send the message sets $\mathbb{M}_{Y_1}$ and $\mathbb{M}_{Y_2}$ to the receivers $Y_1$ and $Y_2$, respectively. The arrangement of messages $\mathbb{M}_{Y_1}$ and $\mathbb{M}_{Y_2}$ among the corresponding transmitters is arbitrary.

This network is composed of two groups of transmitters $\mathbb{X}_1 \triangleq \{X_{1,1}, \dots, X_{1,\mu_1}\}$ and $\mathbb{X}_2 \triangleq \{X_{2,1}, \dots, X_{2,\mu_2}\}$ which transmit the message sets $\mathbb{M}_{Y_1}$ and $\mathbb{M}_{Y_2}$ to the receivers $Y_1$ and $Y_2$, respectively. The parameters $\mu_1$ and $\mu_2$ are arbitrary natural numbers. Also, the arrangement of messages $\mathbb{M}_{Y_1}$ and $\mathbb{M}_{Y_2}$ among the corresponding transmitters is arbitrary. Note that the sets $\mathbb{M}_{Y_1}$ and $\mathbb{M}_{Y_2}$ are disjoint.

According to Theorem 7, if the following conditions hold:

$$\begin{cases} I(X_{1,1}, \dots, X_{1,\mu_1}; Y_1 | X_{2,1}, \dots, X_{2,\mu_2}) \leq I(X_{1,1}, \dots, X_{1,\mu_1}; Y_2 | X_{2,1}, \dots, X_{2,\mu_2}) & \text{for all} \quad P_{X_{1,1} \dots X_{1,\mu_1}} \times \prod_{i \in [1:\mu_2]} P_{X_{2,i}} \\ I(X_{2,1}, \dots, X_{2,\mu_2}; Y_2 | X_{1,1}, \dots, X_{1,\mu_1}) \leq I(X_{2,1}, \dots, X_{2,\mu_2}; Y_1 | X_{1,1}, \dots, X_{1,\mu_1}) & \text{for all} \quad P_{X_{2,1} \dots X_{2,\mu_2}} \times \prod_{i \in [1:\mu_1]} P_{X_{1,i}} \end{cases} \tag{156}$$

then, the network fulfills the strong interference criterion, i.e., the decoding of all messages at each receiver is optimal and achieves the capacity. A remarkable point is that the strong interference regime (156) does not depend on the arrangement of messages $\mathbb{M}_{Y_1}$ and $\mathbb{M}_{Y_2}$ among the corresponding transmitters. In general, one can easily deduce the following fact.

***Note:*** *Consider the general two-receiver interference network which is derived from the general network in Fig. 1 by setting $K_2 = 2$. For a given set of messages $\mathbb{M}_A$ with $\mathbb{M}_{Y_1}^A$ and $\mathbb{M}_{Y_2}^A$ for the receivers $Y_1$ and $Y_2$, respectively, assume that the network satisfies the strong interference conditions in* (144). *For any other message set $\mathbb{M}_B$ with $\mathbb{M}_{Y_1}^B$ and $\mathbb{M}_{Y_2}^B$ for the receivers $Y_1$ and $Y_2$, respectively, if the following condition holds:*

$$\mathbb{X}_{\mathbb{M}_{Y_1}^A} = \mathbb{X}_{\mathbb{M}_{Y_1}^B}, \qquad \mathbb{X}_{\mathbb{M}_{Y_2}^A} = \mathbb{X}_{\mathbb{M}_{Y_2}^B} \tag{157}$$

*then, the network is still in the strong interference regime.*

For example, the conditions (156) represent a strong interference regime even if each subset of the transmitters $\{X_{1,1}, \dots, X_{1,\mu_1}\}$ and each subset of the transmitters $\{X_{2,1}, \dots, X_{2,\mu_2}\}$ cooperatively send a common message for both receivers.

Next let us consider the two-receiver Gaussian MAIN. It is formulated below:

$$\begin{cases} Y_1 \triangleq a_{1,1}X_{1,1} + a_{1,2}X_{1,2} + \dots + a_{1,\mu_1}X_{1,\mu_1} + a_{2,1}X_{2,1} + \dots + a_{2,\mu_2}X_{2,\mu_2} + Z_1 \\ Y_2 \triangleq b_{1,1}X_{1,1} + b_{1,2}X_{1,2} + \dots + b_{1,\mu_1}X_{1,\mu_1} + b_{2,1}X_{2,1} + \dots + b_{2,\mu_2}X_{2,\mu_2} + Z_2 \end{cases} \tag{158}$$

where $Z_1$ and $Z_2$ are zero-mean unit-variance Gaussian noises; $a_{1,1},\dots,a_{1,\mu_1},a_{2,1},\dots,a_{2,\mu_2}$ and $b_{1,1},\dots,b_{1,\mu_1},b_{2,1},\dots,b_{2,\mu_2}$ are fixed real numbers representing the network gains. Also, the inputs are power-constrained. Considering the conditions (156) and using Lemma 2, we directly obtain that if the network gains satisfy the following:

$$\begin{cases}\dfrac{a_{1,1}}{b_{1,1}}=\dfrac{a_{1,2}}{b_{1,2}}=\cdots=\dfrac{a_{1,\mu_1}}{b_{1,\mu_1}}=\alpha \\ \qquad\qquad , \\ \dfrac{b_{2,1}}{a_{2,1}}=\dfrac{b_{2,2}}{a_{2,2}}=\cdots=\dfrac{b_{2,\mu_2}}{a_{2,\mu_2}}=\beta\end{cases} \quad , \qquad |\alpha|,|\beta|\leq 1 \tag{159}$$

then the network lies in the strong interference regime regardless of the arrangement of messages among transmitters. Note that (159) represents only sufficient conditions for this purpose.

Once more consider the strong interference regime in (144). For the fully connected networks, i.e., those in which each receiver is connected to all transmitters, one can consistently obtain a strong interference regime based on the conditions (144). However, for networks in which some transmitters are not connected to some receivers these conditions cannot be simultaneously satisfied. For example, consider a MAIN as in Fig. 6 for which the transition probability function is factorized as follows:

$$\mathbb{P}(y_1,y_2|\mathbb{X}_1,\mathbb{X}_2)=\mathbb{P}(y_1|\mathbb{X}_1,\mathbb{X}_2)\mathbb{P}(y_2|\mathbb{X}_2) \tag{160}$$

where $\mathbb{X}_1\triangleq\{X_{1,1},\dots,X_{1,\mu_1}\}$ and $\mathbb{X}_2\triangleq\{X_{2,1},\dots,X_{2,\mu_2}\}$. For such a network the right side of the first inequality in (144) is always zero. Thereby, the first condition of (144) cannot be satisfied. In fact, in this scenario the receiver $Y_2$ experiences no interference and it is meaningless to discuss the strong interference at this receiver while the conditions (144) are regarded to the strong interference at both receivers. Therefore, to derive strong interference regime for such networks, a modified version of the conditions (144) is required. This observation enables us to strengthen the strong interference conditions in (144) as given in the next theorem.

***Theorem 8)*** *Consider a general interference network with two receivers that is derived from the general network in Fig. 1 by setting $K_2=2$. If the transition probability function of the network satisfies the following conditions:*

$$\begin{cases} I\left(\mathbb{X}-\mathbb{X}_{\mathbb{M}_{Y_2}\cup\mathbb{M}_{c\nrightarrow Y_2}};Y_1\middle|\mathbb{X}_{\mathbb{M}_{Y_2}\cup\mathbb{M}_{c\nrightarrow Y_2}}\right)\leq I\left(\mathbb{X}-\mathbb{X}_{\mathbb{M}_{Y_2}\cup\mathbb{M}_{c\nrightarrow Y_2}};Y_2\middle|\mathbb{X}_{\mathbb{M}_{Y_2}\cup\mathbb{M}_{c\nrightarrow Y_2}}\right) & \text{for all } P_{\mathbb{X}-\mathbb{X}_{\mathbb{M}_{Y_2}\cup\mathbb{M}_{c\nrightarrow Y_2}}}\prod_{X_i\in\mathbb{X}_{\mathbb{M}_{Y_2}\cup\mathbb{M}_{c\nrightarrow Y_2}}}P_{X_i} \\ I\left(\mathbb{X}-\mathbb{X}_{\mathbb{M}_{Y_1}\cup\mathbb{M}_{c\nrightarrow Y_1}};Y_2\middle|\mathbb{X}_{\mathbb{M}_{Y_1}\cup\mathbb{M}_{c\nrightarrow Y_1}}\right)\leq I\left(\mathbb{X}-\mathbb{X}_{\mathbb{M}_{Y_1}\cup\mathbb{M}_{c\nrightarrow Y_1}};Y_1\middle|\mathbb{X}_{\mathbb{M}_{Y_1}\cup\mathbb{M}_{c\nrightarrow Y_1}}\right) & \text{for all } P_{\mathbb{X}-\mathbb{X}_{\mathbb{M}_{Y_1}\cup\mathbb{M}_{c\nrightarrow Y_1}}}\prod_{X_i\in\mathbb{X}_{\mathbb{M}_{Y_1}\cup\mathbb{M}_{c\nrightarrow Y_1}}}P_{X_i}\end{cases} \tag{161}$$

*Then the network has strong interference. Moreover, the outer bound $\mathfrak{R}_{o:(2)}^{GINTR}$ given in (54) is optimal for such a network.*

*Proof of Theorem 8)* Before all, let us remark that since $\mathbb{X}_{\mathbb{M}_{Y_2}}$ is a subset of $\mathbb{X}_{\mathbb{M}_{Y_2}\cup\mathbb{M}_{c\nrightarrow Y_2}}$, according to Corollary 1, the conditions (161) imply those in (144). Therefore, Theorem 8 is more general Theorem 7. In fact, Theorem 8 itself is a special case of our general result which will be proved later in Subsection V.B.3. Hence, we only outline the proof with emphasis on showing that the outer bound $\mathfrak{R}_{o:(2)}^{GINTR}$ in (54) under the conditions (161) is optimal.

First note that, according to the Definition 4, in the strong interference regime each receiver decodes all the messages which are transmitted by its connected transmitters. Based on Definition 2, one can state that in this regime each receiver indeed decodes its connected messages. Now, consider the following rate region:

$$\bigcup_{\substack{P_QP_{M_1}P_{M_2}\dots P_{M_K}\\ \times\prod_{i=1}^{K_1}P_{X_i|\mathbb{M}_{X_i},Q}}}\begin{Bmatrix}(R_1,R_2,\dots,R_K): \\ \forall\,\Omega\subseteq\mathbb{M}_{c\to Y_1}:\ \boldsymbol{R}_{\Sigma\Omega}\leq I\left(\Omega;Y_1\middle|\mathbb{M}_{c\to Y_1}-\Omega,Q\right) \\ \forall\,\Omega\subseteq\mathbb{M}_{c\to Y_2}:\ \boldsymbol{R}_{\Sigma\Omega}\leq I\left(\Omega;Y_2\middle|\mathbb{M}_{c\to Y_2}-\Omega,Q\right)\end{Bmatrix} \tag{162}$$

This rate region is achievable for the network. To achieve this region, the messages are separately encoded using independent codewords. Also, each receiver decodes its connected messages using a jointly typical decoder. It remains to show that under the conditions (161) the rate region (162) is optimal. Consider the second constraint of (162). Let $\Omega$ be a subset of $\mathbb{M}_{c\to Y_2}$. We know that the messages $\mathbb{M}_{c\nrightarrow Y_2}$ are independent of $Y_2$ and also $\mathbb{M}_{c\to Y_2}$. Thereby, we have:

$$\boldsymbol{R}_{\sum\Omega} \leq I\left(\Omega; Y_2 \middle| \mathbb{M}_{c\to Y_2} - \Omega, Q\right) = I\left(\Omega; Y_2 \middle| \mathbb{M}_{c\to Y_2} - \Omega, \mathbb{M}_{c\nrightarrow Y_2}, Q\right) = I(\Omega; Y_2|\mathbb{M} - \Omega, Q) \tag{163}$$

Now to derive (163), one can follow exactly the same lines as in the proof of Theorem 7. However, here we know that $\Omega_{Y_1}$ (which is given similar to (146)) is a subset of $\mathbb{M}_{c\to Y_2}$; thereby, $\mathbb{M}_{c\nrightarrow Y_2}$ is a subset of $\mathbb{M} - \Omega_{Y_1}$ and thus we have: $\mathbb{M}_{Y_2} \cup \mathbb{M}_{c\nrightarrow Y_2} \subseteq \mathbb{M} - \Omega_{Y_1}$. Consequently, $\mathbb{X}_{\mathbb{M}_{Y_2} \cup \mathbb{M}_{c\nrightarrow Y_2}}$ is a subset of $\mathbb{X}_{\mathbb{M}-\Omega_{Y_1}}$ and we obtain:

$$\mathbb{X} - \mathbb{X}_{\mathbb{M}-\Omega_{Y_1}} \subseteq \mathbb{X} - \mathbb{X}_{\mathbb{M}_{Y_2} \cup \mathbb{M}_{c\nrightarrow Y_2}} \tag{164}$$

Therefore, to derive (153), instead of (151), we need to have:

$$I\left(\mathbb{X} - \mathbb{X}_{\mathbb{M}_{Y_2} \cup \mathbb{M}_{c\nrightarrow Y_2}}; Y_1 \middle| \mathbb{X}_{\mathbb{M}_{Y_2} \cup \mathbb{M}_{c\nrightarrow Y_2}}, D\right) \leq I\left(\mathbb{X} - \mathbb{X}_{\mathbb{M}_{Y_2} \cup \mathbb{M}_{c\nrightarrow Y_2}}; Y_2 \middle| \mathbb{X}_{\mathbb{M}_{Y_2} \cup \mathbb{M}_{c\nrightarrow Y_2}}, D\right) \quad \text{for all joint PDFs } P_{D\mathbb{X}} \tag{165}$$

which is guaranteed by the first condition in (161). This completes the proof. ■

Now, consider the MAIN in Fig. 6 with the transition probability in (160). For this network we have: $\mathbb{M}_{Y_2} \cup \mathbb{M}_{c\nrightarrow Y_2} = \mathbb{M}$; thereby, both sides of the first inequality in (161) are always zero. Thus, the network has strong interference if the second condition in (161) holds. Also, one can easily check that for the one-sided two-user CIC in (87) and for the one-sided BCCR in (122), the conditions (161) are respectively reduced to those of (88) and (123), (we recall that (88) is equivalent to (90) and (123) to (125)).

In addition to strong interference regime, the general outer bound $\mathfrak{R}_{o:(2)}^{GINTR}$ in (54) is a very useful tool to identify networks with less-noisy receivers for which a successive decoding scheme is sum-rate optimal. We proved this capability for the two-user CIC and the BCCR in Subsections IV.A.1 and IV.A.2. The same result will be derived for other two-receiver interference networks in Part IV of our multi-part papers [4].

# V. General Multi-Receiver Interference Networks

Thus far, we have only considered the interference networks with two receivers. For these networks, we obtained useful capacity outer bounds based on a unified framework. Also, we demonstrated the capability of these derived outer bounds to prove important capacity results. Now we intend to develop the theory for arbitrary multi-receiver networks. In general, such networks are far less understood and there exist a very few cases for which an exact characterization of the capacity region is available [15, p. 6-64]. Nonetheless, in this paper we develop many new results for these networks. In Part II of our multi-part papers [2], we derived a full characterization of the sum-rate capacity for degraded networks. Now, we intend to develop a theory for the strong interference regime. In Subsection V.A, we show that the outer bounds derived in Subsection IV.A for the two-receiver networks can be flexibly extended to interference networks of arbitrary large sizes. As a result, we obtain useful capacity outer bounds for multi-receiver interference networks which are tighter than the existing cut-set outer bound [14]. The main results of this section are addressed in Subsection V.B, where we develop a new approach which enables us to derive strong interference regime for interference networks with arbitrary configurations. Multi-receiver networks with a sequence of less noisy receivers will be studied in details in Part IV [4].

## V.A) Unified Outer Bounds

As said before, one of the main difficulties in deriving capacity bounds for multi-receiver interference networks are the lack of outer bounds with satisfactory performance. Nonetheless, in this subsection, we provide a subtle generalization of the outer bounds derived in Subsection IV.A for the networks with two receivers to the case of arbitrary number of receivers. The key idea behind this extension is that for each two arbitrary distinct subsets of receivers, we can provide constraints similar to the ones given in the outer bound $\mathfrak{R}_{o:(1)}^{GINTR}$ in (52). As a result, we derive a collection of constraints on the communication rates. Picking out desired constraints of this collection, one can establish very useful outer bounds on the capacity region of different interference networks with more than two receivers. In addition, these outer bounds are optimal for some special scenarios, as will be demonstrated in the sequel.

Consider the general interference network with $K_1$ transmitters and $K_2$ receivers shown in Fig. 1 in which the messages $\mathbb{M}=\{M_1,\dots,M_K\}$ are transmitted over the channel. Let $J$ be an arbitrary subset of $[1:K_2]$. Given the network outputs $Y_1,\dots,Y_{K_2}$, define:

$$\begin{cases} Y_J \triangleq \{Y_j : j\in J\} \\ \mathbb{M}_{Y_J} \triangleq \bigcup_{j\in J} \mathbb{M}_{Y_j} \end{cases} \tag{166}$$

Also, for two distinct subsets of $[1:K_2]$, e.g., $J_1$ and $J_2$, define:

$$\begin{cases} \mathbb{M}_{Y_{J_1}\cap Y_{J_2}} \triangleq \mathbb{M}_{Y_{J_1}} \cap \mathbb{M}_{Y_{J_2}} \\ \mathbb{M}_{Y_{J_1}-Y_{J_2}} \triangleq \mathbb{M}_{Y_{J_1}} - \mathbb{M}_{Y_{J_2}} \end{cases} \tag{167}$$

Our unified outer bound on the capacity region of the general interference networks is presented in the following theorem.

***Theorem 9) Unified Outer Bound for the General Interference Networks***

*Define the rate region $\mathfrak{R}_{o:(1)}^{GIN}$ as follows:*

$$\mathfrak{R}_{o:(1)}^{GIN} \triangleq \bigcup_{\mathcal{P}_o^{GIN}} \left\{ \begin{array}{l} (R_1,\dots,R_K)\in\mathbb{R}_+^K : \\ \forall J_1,J_2\subseteq[1:K_2], J_1\cap J_2=\emptyset : \\ \forall \mu\in\mathbb{N},\quad \forall \Omega_1^{Y_{J_1}},\Omega_2^{Y_{J_1}},\dots,\Omega_\mu^{Y_{J_1}}\subseteq\mathbb{M}_{Y_{J_1}},\quad \forall \Omega_1^{Y_{J_2}},\Omega_2^{Y_{J_2}}\dots,\Omega_\mu^{Y_{J_2}}\subseteq\mathbb{M}_{Y_{J_2}}: \\ \quad\text{with } \langle \forall\, l_1,l_2\in[1:\mu], l_1\neq l_2:\ \Omega_{l_1}^{Y_{J_1}}\cap\Omega_{l_2}^{Y_{J_1}} = \Omega_{l_1}^{Y_{J_2}}\cap\Omega_{l_2}^{Y_{J_2}} = \Omega_{l_1}^{Y_{J_1}}\cap\Omega_{l_2}^{Y_{J_2}}=\emptyset\rangle \\ \forall\, \Omega_s\subseteq\mathbb{M}-\left(\Omega_1^{Y_{J_1}}\cup\Omega_2^{Y_{J_1}}\cup\dots\cup\Omega_\mu^{Y_{J_1}}\cup\Omega_1^{Y_{J_2}}\cup\Omega_2^{Y_{J_2}}\cup\dots\cup\Omega_\mu^{Y_{J_2}}\right) \\ \qquad , \\ \sum_{l=1}^{\mu}\boldsymbol{R}_{\Sigma_{\Omega_l^{Y_{J_1}}}}+\boldsymbol{R}_{\Sigma_{\Omega_l^{Y_{J_2}}}} \leq \sum_{l=1}^{\mu}\left(\begin{array}{l} I\left(\Omega_l^{Y_{J_1}};Y_{J_1}\middle|Z_{J_1,J_2},\Omega_1^{Y_{J_1}},\dots,\Omega_{l-1}^{Y_{J_1}},\Omega_1^{Y_{J_2}},\dots,\Omega_{l-1}^{Y_{J_2}},\Omega_l^{Y_{J_2}},\Omega_s,Q\right) \\ +I\left(\Omega_l^{Y_{J_2}};Y_{J_2}\middle|Z_{J_1,J_2},\Omega_1^{Y_{J_1}},\dots,\Omega_{l-1}^{Y_{J_1}},\Omega_1^{Y_{J_2}},\dots,\Omega_{l-1}^{Y_{J_2}},\Omega_s,Q\right)\end{array}\right) \\ \qquad\qquad +\min\left\{I\left(Z_{J_1,J_2};Y_{J_1}\middle|\Omega_s,Q\right), I\left(Z_{J_1,J_2};Y_{J_2}\middle|\Omega_s,Q\right)\right\} \end{array}\right\} \tag{168}$$

*where $\mathcal{P}_o^{GIN}$ denotes the set of all joint PDFs $P_{QM_1\dots M_K\underbrace{\dots Z_{J_1,J_2}\dots}_{\substack{J_1,J_2\subseteq[1:K_2],\\ J_1\cap J_2=\emptyset}}X_1\dots X_{K_1}}\left(q,m_1,\dots,m_K,\underbrace{\dots,z_{J_1,J_2},\dots}_{\substack{J_1,J_2\subseteq[1:K_2],\\ J_1\cap J_2=\emptyset}},x_1,\dots,x_{K_1}\right)$ satisfying:*

$$P_{QM_1\dots M_K\underbrace{\dots Z_{J_1,J_2}\dots}_{\substack{J_1,J_2\subseteq[1:K_2],\\ J_1\cap J_2=\emptyset}}X_1\dots X_{K_1}} = P_Q\times P_{M_1}\times\dots\times P_{M_K}\times \underbrace{P_{\dots Z_{J_1,J_2}\dots|M_1\dots M_K Q}}_{\substack{J_1,J_2\subseteq[1:K_2],\\ J_1\cap J_2=\emptyset}}\times P_{X_1|\mathbb{M}_{X_1},Q}\times\dots\times P_{X_{K_1}|\mathbb{M}_{X_{K_1}},Q} \tag{169}$$

*Also, the PDFs $P_{M_l}, l = 1, \dots, K$, are uniformly distributed and $P_{X_i|\mathbb{M}_{X_i},Q} \in \{0,1\}$ for $i = 1, \dots, K_1$. The set $\mathfrak{R}_{o:(1)}^{GIN}$ constitutes an outer bound for the capacity region of the general interference network in Fig. 1.*

*Proof of Theorem 9)* Let $J_1, J_2 \subseteq [1:K_2]$, where $J_1 \cap J_2 = \emptyset$. By the same lines as the proof of Theorem 2 for the receivers $Y_{J_1}$ and $Y_{J_2}$, one can obtain the constraint in (168). Note that in this scenario for a length-$n$ code, we have:

$$\frac{1}{n}H\left(\mathbb{M}_{Y_{J_1}}\middle|Y_{J_1}^n\right) = \frac{1}{n}H\left(\bigcup_{j\in J_1}\mathbb{M}_{Y_j}\middle|\left\{Y_j^n\right\}_{j\in J_1}\right) \leq \sum_{j\in J_1}\frac{1}{n}H\left(\mathbb{M}_{Y_j}\middle|Y_j^n\right) \overset{(a)}{\leq} \epsilon_{J_1,n}$$
$$\frac{1}{n}H\left(\mathbb{M}_{Y_{J_2}}\middle|Y_{J_2}^n\right) = \frac{1}{n}H\left(\bigcup_{j\in J_2}\mathbb{M}_{Y_j}\middle|\left\{Y_j^n\right\}_{j\in J_2}\right) \leq \sum_{j\in J_2}\frac{1}{n}H\left(\mathbb{M}_{Y_j}\middle|Y_j^n\right) \overset{(b)}{\leq} \epsilon_{J_2,n} \tag{170}$$

where $\epsilon_{J_1,n} \to 0$ as $n \to \infty$. Also, (a) and (b) hold by Fano's inequality. Therefore, all the steps in the proof of Theorem 2 can be directly applied here; it is sufficient to replace $Y_1$ by $Y_{J_1}$ and $Y_2$ by $Y_{J_2}$ everywhere. Accordingly, the auxiliary random variables $Z_{J_1,J_2,t}, t = 1, \dots, n$, are defined as follows:

$$Z_{J_1,J_2,t} \triangleq \left(Y_{J_1}^{t-1}, Y_{J_2,t+1}^n\right) \tag{171}$$

Moreover, as indicted in the characterization (168), the side information messages $\Omega_s$ can be selected from the set $\mathbb{M} - \left(\Omega_1^{Y_{J_1}} \cup \dots \cup \Omega_\mu^{Y_{J_1}} \cup \Omega_1^{Y_{J_2}} \cup \dots \cup \Omega_\mu^{Y_{J_2}}\right)$ which necessarily contains some other messages other than $\mathbb{M}_{Y_{J_1}} \cup \mathbb{M}_{Y_{J_2}}$. The proof is thus complete. ■

***Remarks 14:***

1. The parameter $Q$ in (168) is a time-sharing random variable.
2. In the characterization (168), once we select $J_1, J_2 \subseteq [1:K_2]$ with $J_1 \cap J_2 = \emptyset$, it is not required to consider the alternative option, i.e., the option of $\acute{J}_1$ and $\acute{J}_2$ where $\acute{J}_1 = J_2$ and $\acute{J}_2 = J_1$.

Naturally, parallel to Corollary 2, one can also derive a unified outer bound with a more convenient description for the network, as given below. The main benefit of this characterization is that one can easily make use of it to derive computable capacity outer bounds for different scenarios intuitively.

***Corollary 3)*** *Define the rate region $\mathfrak{R}_{o:(2)}^{GIN}$ as follows:*

$$\mathfrak{R}_{o:(2)}^{GIN} \triangleq \bigcup_{\mathcal{P}_o^{GIN}} \left\{\begin{array}{l}(R_1, \dots, R_K) \in \mathbb{R}_+^K : \\ \forall J_1, J_2 \subseteq [1:K_2], J_1 \cap J_2 = \emptyset : \\ \forall \Omega_0, \Omega_1, \Omega_2 : \Omega_0 \subseteq \mathbb{M}_{Y_{J_1} \cap Y_{J_2}}, \quad \Omega_1 \subseteq \mathbb{M}_{Y_{J_1}}, \quad \Omega_2 \subseteq \mathbb{M}_{Y_{J_2}}, \\ \qquad \text{with } \langle \Omega_0 \cap \Omega_1 = \Omega_0 \cap \Omega_2 = \Omega_1 \cap \Omega_2 = \emptyset \rangle \\ \forall \Omega_s : \Omega_s \subseteq (\Omega_0 \cup \Omega_1 \cup \Omega_2)^c \subseteq \mathbb{M} = \{M_1, \dots, M_K\} \\ \qquad , \\ \langle 1 \rangle : \boldsymbol{R}_{\Sigma\Omega_0} + \boldsymbol{R}_{\Sigma\Omega_1} \leq I(Z_{J_1,J_2}, \Omega_0, \Omega_1; Y_{J_1}|\Omega_s, Q) \\ \langle 2 \rangle : \boldsymbol{R}_{\Sigma\Omega_0} + \boldsymbol{R}_{\Sigma\Omega_2} \leq I(Z_{J_1,J_2}, \Omega_0, \Omega_2; Y_{J_2}|\Omega_s, Q) \\ \langle 3 \rangle : \boldsymbol{R}_{\Sigma\Omega_0} + \boldsymbol{R}_{\Sigma\Omega_1} + \boldsymbol{R}_{\Sigma\Omega_2} \leq I(\Omega_1; Y_{J_1}|Z_{J_1,J_2}, \Omega_0, \Omega_2, \Omega_s, Q) + I(Z_{J_1,J_2}, \Omega_0, \Omega_2; Y_{J_2}|\Omega_s, Q) \\ \langle 4 \rangle : \boldsymbol{R}_{\Sigma\Omega_0} + \boldsymbol{R}_{\Sigma\Omega_1} + \boldsymbol{R}_{\Sigma\Omega_2} \leq I(\Omega_2; Y_{J_2}|Z_{J_1,J_2}, \Omega_0, \Omega_1, \Omega_s, Q) + I(Z_{J_1,J_2}, \Omega_0, \Omega_1; Y_{J_1}|\Omega_s, Q) \\ \qquad , \\ \langle 5 \rangle : \boldsymbol{R}_{\Sigma\Omega_0} + \boldsymbol{R}_{\Sigma\Omega_1} + \boldsymbol{R}_{\Sigma\Omega_2} \leq I(\Omega_1; Y_{J_1}|Z_{J_1,J_2}, \Omega_0, \Omega_2, \Omega_s, Q) + I(\Omega_2; Y_{J_2}|Z_{J_1,J_2}, \Omega_0, \Omega_s, Q) \\ \qquad + I(Z_{J_1,J_2}, \Omega_0; Y_{J_1}|\Omega_s, Q) \\ \langle 6 \rangle : \boldsymbol{R}_{\Sigma\Omega_0} + \boldsymbol{R}_{\Sigma\Omega_1} + \boldsymbol{R}_{\Sigma\Omega_2} \leq I(\Omega_2; Y_{J_2}|Z_{J_1,J_2}, \Omega_0, \Omega_1, \Omega_s, Q) + I(\Omega_1; Y_{J_1}|Z_{J_1,J_2}, \Omega_0, \Omega_s, Q) \\ \qquad + I(Z_{J_1,J_2}, \Omega_0; Y_{J_2}|\Omega_s, Q)\end{array}\right\} \tag{172}$$

*where $\mathcal{P}_o^{GIN}$ is given similar to Theorem 9 (see (169)). The rate region $\mathfrak{R}_{o:(2)}^{GIN}$ constitutes an outer bound on the capacity region of the general interference network in Fig. 1.*

*Proof of Corollary 3)* This outer bound is derived from $\mathfrak{R}_{o:(1)}^{GIN}$ in (168), by following the same lines as the derivation of $\mathfrak{R}_{o:(2)}^{GINTR}$ in (54) from $\mathfrak{R}_{o:(1)}^{GINTR}$ in (52). ■

In what follows, to exemplify our result, we derive a computable capacity outer bound for the three-user CIC using the unified outer bound $\mathfrak{R}_{o:(2)}^{GIN}$ in (172). Moreover, we demonstrate the utility of the derived outer bounds for establishing explicit capacity results for some specials cases.

***Example 2:*** **The Three-User CIC**

Consider the three-user CIC where three transmitters send separately independent messages to their respective receivers. Fig. 7 depicts the channel model.

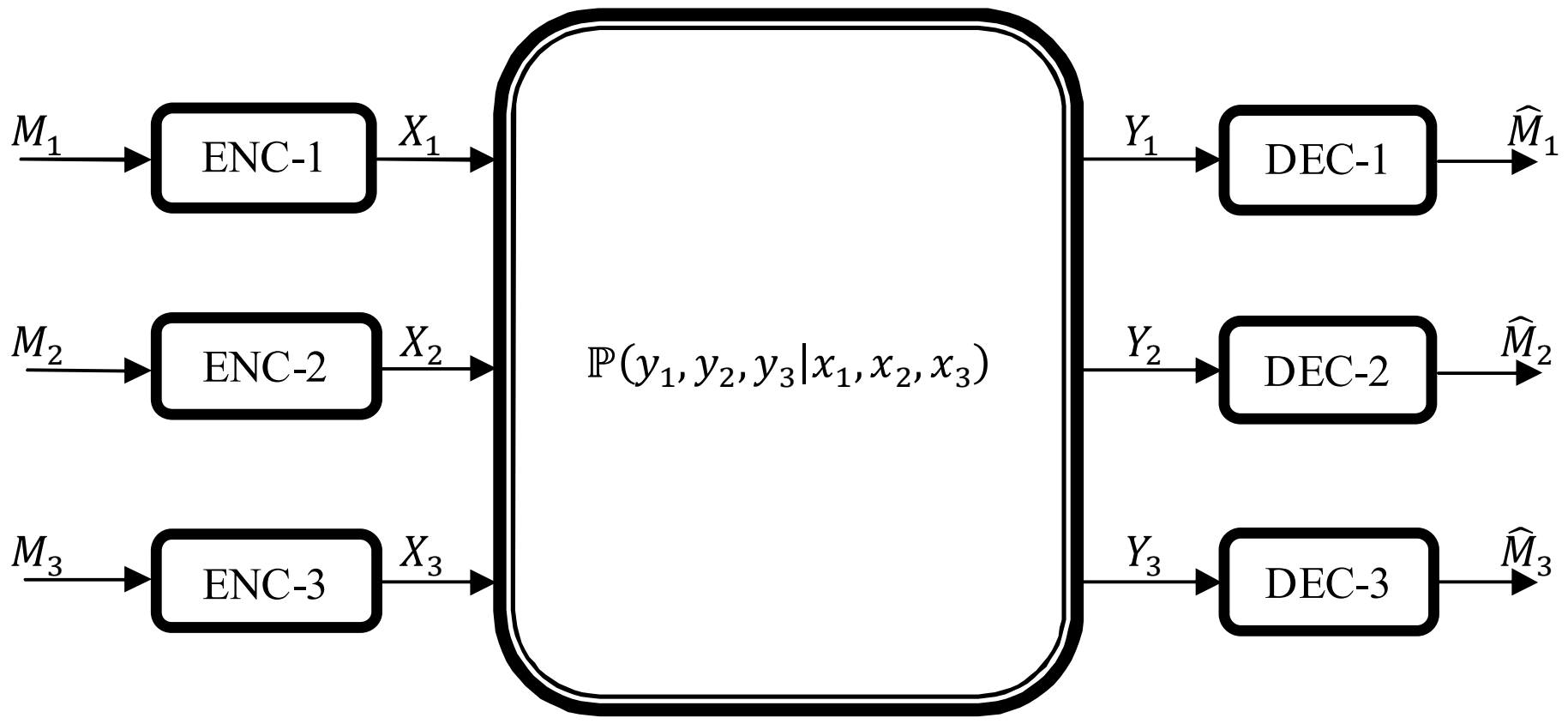

Figure 7. The three-user Classical Interference Channel (CIC).

This scenario is obtained from the general interference network in Fig. 1 by setting $K_1 = K_2 = 3$, $\mathbb{M} = \{M_1, M_2, M_3\}$, $\mathbb{M}_{X_i} = \{M_i\}, i = 1,2,3$, and $\mathbb{M}_{Y_j} = \{M_j\}, j = 1,2,3$. Also, the Gaussian channel is formulated as:

$$\begin{cases} Y_1 = X_1 + a_{12}X_2 + a_{13}X_3 + Z_1 \\ Y_2 = a_{21}X_1 + X_2 + a_{23}X_3 + Z_2 \\ Y_3 = a_{31}X_1 + a_{32}X_2 + X_3 + Z_3 \end{cases} \tag{173}$$

where $Z_1$, $Z_2$, and $Z_3$ are zero-mean unit-variance Gaussian RVs and the inputs are subject to power constraints $\mathbb{E}[X_i^2] \leq P_i, i = 1,2,3$. In the next proposition, we present a capacity outer bound for the channel.

***Proposition 5)*** *Define the rate region* $\mathfrak{R}_o^{CIC\to 3-user}$ *as follows:*

$$\mathfrak{R}_o^{CIC\to 3-user} \triangleq \bigcup_{\mathcal{P}_o^{CIC\to 3-user}} \left\{ \begin{array}{l} (R_1,R_2,R_3)\in\mathbb{R}_+^3 : \\ R_1 \le \min\left\{ \begin{array}{c} I(X_1;Y_1|X_2,X_3,Q), I(U_{1,23},X_1;Y_1|Q) \\ I(U_1,X_1;Y_1|X_3,Q), I(U_3,X_1;Y_1|X_2,Q) \end{array} \right\}, \\ R_2 \le \min\left\{ \begin{array}{c} I(X_2;Y_2|X_1,X_3,Q), I(U_{2,13},X_2;Y_2|Q) \\ I(V_1,X_2;Y_2|X_3,Q), I(U_2,X_2;Y_2|X_1,Q) \end{array} \right\}, \\ R_3 \le \min\left\{ \begin{array}{c} I(X_3;Y_3|X_1,X_2,Q), I(U_{3,12},X_3;Y_3|Q) \\ I(V_2,X_3;Y_3|X_1,Q), I(V_3,X_3;Y_3|X_2,Q) \end{array} \right\}, \\ R_1+R_2 \le \min\left\{ \begin{array}{c} I(V_{3,12},X_1,X_2;Y_1,Y_2|Q) \\ I(X_1;Y_1|X_2,X_3,V_1,Q)+I(V_1,X_2;Y_2|X_3,Q) \\ I(X_2;Y_2|X_1,X_3,U_1,Q)+I(U_1,X_1;Y_1|X_3,Q) \end{array} \right\}, \\ R_2+R_3 \le \min\left\{ \begin{array}{c} I(V_{1,23},X_2,X_3;Y_2,Y_3|Q) \\ I(X_2;Y_2|X_1,X_3,V_2,Q)+I(V_2,X_3;Y_3|X_1,Q) \\ I(X_3;Y_3|X_1,X_2,U_2,Q)+I(U_2,X_2;Y_2|X_1,Q) \end{array} \right\}, \\ R_1+R_3 \le \min\left\{ \begin{array}{c} I(V_{2,13},X_1,X_3;Y_1,Y_3|Q) \\ I(X_1;Y_1|X_2,X_3,V_3,Q)+I(V_3,X_3;Y_3|X_2,Q) \\ I(X_3;Y_3|X_1,X_2,U_3,Q)+I(U_3,X_1;Y_1|X_2,Q) \end{array} \right\}, \\ R_1+R_2+R_3 \le \min\left\{ \begin{array}{c} I(X_1;Y_1|X_2,X_3,V_{1,23},Q)+I(V_{1,23},X_2,X_3;Y_2,Y_3|Q) \\ I(X_2,X_3;Y_2,Y_3|X_1,U_{1,23},Q)+I(U_{1,23},X_1;Y_1|Q) \\ I(X_2;Y_2|X_1,X_3,V_{2,13},Q)+I(V_{2,13},X_1,X_3;Y_1,Y_3|Q) \\ I(X_1,X_3;Y_1,Y_3|X_2,U_{2,13},Q)+I(U_{2,13},X_2;Y_2|Q) \\ I(X_3;Y_3|X_1,X_2,V_{3,12},Q)+I(V_{3,12},X_1,X_2;Y_1,Y_2|Q) \\ I(X_1,X_2;Y_1,Y_2|X_3,U_{3,12},Q)+I(U_{3,12},X_3;Y_3|Q) \end{array} \right\} \end{array} \right\}$$

(174)

*where* $\mathcal{P}_o^{CIC\to 3-user}$ *denotes the set of all joint PDFs* $P_{QX_1X_2X_3U_1U_2U_3U_{1,23}U_{2,13}U_{3,12}V_1V_2V_3V_{1,23}V_{2,13}V_{3,12}}$ *satisfying:*

$$P_{QX_1X_2X_3U_1U_2U_3U_{1,23}U_{2,13}U_{3,12}V_1V_2V_3V_{1,23}V_{2,13}V_{3,12}} = P_Q P_{X_1|Q} P_{X_2|Q} P_{X_3|Q} P_{U_1U_2U_3U_{1,23}U_{2,13}U_{3,12}V_1V_2V_3V_{1,23}V_{2,13}V_{3,12}|X_3X_1X_2Q}$$

(175)

*The rate region* $\mathfrak{R}_o^{CIC\to 3-user}$ *constitutes an outer bound on the capacity region of the three-user CIC in Fig. 7.*

*Proof of Proposition 5)* Refer to Appendix. ■

***Remarks 15:***

1. The outer bound $\mathfrak{R}_o^{CIC\to 3-user}$ given in (174) for the three-user CIC is a generalization of the UV-outer bound derived in Part I [1, Corollary III.1] for the two-user case. The extension to the K-user case is straightforward.
2. It is clear that the outer bound $\mathfrak{R}_o^{CIC\to 3-user}$ is tighter than the cut-set bound [14].

It should be noted that the outer bound $\mathfrak{R}_o^{CIC\to 3-user}$ in (174) with the input power constraints $\mathbb{E}[X_i^2]\le P_i, i=1,2,3$ is also valid for the Gaussian network in (173). For this case, by dividing the problem into different interference regimes and using the EPI, one can derive explicit characterization of the bound. However, this paper does not deal with this problem due to space limitation.

Now, we present a special case for which the outer bound $\mathfrak{R}_o^{CIC\to 3-user}$ in (174) becomes identical to the capacity region. Consider a network where its transition probability function satisfies the following factorization:

$$\mathbb{P}(y_1,y_2,y_3|x_1,x_2,x_3) = \mathbb{P}(y_1|x_1,x_2,x_3)\mathbb{P}(y_2|x_2)\mathbb{P}(y_3|x_3)$$

(176)

In this network, only the first receiver experiences interference. This network is called "*many-to-one CIC*" [28]. In the following we present a strong interference regime for this scenario.

***Theorem 10)*** *Consider the three-user CIC in* (176). *If the channel satisfies the following condition:*

$$I(X_2,X_3;Y_2,Y_3) \leq I(X_3,X_2;Y_1|X_1) \quad \text{for all joint PDFs} \quad P_{X_1}P_{X_2X_3} \tag{177}$$

*then the channel lies in the strong interference regime. Specifically, the outer bound* $\mathfrak{R}_o^{CIC\to 3-user}$ *given in* (174) *is tight. The capacity region is given by:*

$$\bigcup_{P_QP_{X_1|Q}P_{X_2|Q}P_{X_3|Q}} \left\{\begin{array}{l}(R_1,R_2,R_3): \\ R_1 \leq I(X_1;Y_1|X_2,X_3,Q) \\ R_2 \leq I(X_2;Y_2|Q) \\ R_3 \leq I(X_3;Y_3|Q) \\ R_1+R_2 \leq I(X_1,X_2;Y_1|X_3,Q) \\ R_1+R_3 \leq I(X_1,X_3;Y_1|X_2,Q) \\ R_1+R_2+R_3 \leq I(X_1,X_2,X_3;Y_1|Q)\end{array}\right\} \tag{178}$$

*Proof of Theorem 10)* First note that the condition (177) can be extended as:

$$I(X_2,X_3;Y_2,Y_3|X_1,D) \leq I(X_3,X_2;Y_1|X_1,D) \quad \text{for all joint PDFs} \quad P_{DX_1X_2X_3} \tag{179}$$

Now consider the achievability scheme in which the first receiver decodes all messages and the second and the third receivers decode their own messages. The resulting rate region is similar to (178) but with the following additional constraint:

$$R_2+R_3 \leq I(X_2,X_3;Y_1|X_1,Q) \tag{180}$$

The constraint (180) is redundant because for a joint PDF of the form $P_QP_{X_1|Q}P_{X_2|Q}P_{X_3|Q}$, the condition (179) implies that:

$$R_2+R_3 \leq I(X_2;Y_2|Q)+I(X_3;Y_3|Q) \leq I(X_3,X_2;Y_1|X_1,Q) \tag{181}$$

To derive the converse part, we show that the outer bound $\mathfrak{R}_o^{CIC\to 3-user}$ in (174) is optimal. The constraints on the individual rates are trivial. We have:

$$\begin{aligned} R_1+R_2 &\leq I(X_2;Y_2|X_1,X_3,U_1,Q)+I(U_1,X_1;Y_1|X_3,Q) \\ &\overset{(a)}{\leq} I(X_2;Y_1|X_1,X_3,U_1,Q)+I(U_1,X_1;Y_1|X_3,Q) = I(X_1,X_2;Y_1|X_3,Q) \end{aligned}$$

To verify inequality (a), by setting $D = (X_3,U_1,Q)$ in (179), we obtain:

$$I(X_2;Y_2,Y_3|X_1,X_3,U_1,Q) \leq I(X_2;Y_1|X_1,X_3,U_1,Q)$$

which is stricter than (a). Similarly, we can derive:

$$\begin{aligned} R_1+R_3 &\leq I(X_3;Y_3|X_1,X_2,U_3,Q)+I(U_3,X_1;Y_1|X_2,Q) \\ &\leq I(X_3;Y_1|X_1,X_2,U_3,Q)+I(U_3,X_1;Y_1|X_2,Q) = I(X_1,X_3;Y_1|X_2,Q) \end{aligned}$$

Finally, we have:

$$\begin{aligned} R_1+R_2+R_3 &\leq I\left(X_2,X_3;Y_2,Y_3\middle|X_1,U_{1,23},Q\right)+I\left(U_{1,23},X_1;Y_1\middle|Q\right) \\ &\overset{(b)}{\leq} I\left(X_2,X_3;Y_1\middle|X_1,U_{1,23},Q\right)+I\left(U_{1,23},X_1;Y_1\middle|Q\right) = I(X_1,X_2,X_3;Y_1|Q) \end{aligned}$$

where inequality (b) is derived by setting $D \equiv \left(U_{1,23},Q\right)$ in (179). The proof is complete. ■

***Remark 16:*** A straightforward extension of the outer bound (174) for the K-user case is optimal for the K-user many-to-one CIC with strong interference.

## V.B) Information Flow in Strong Interference Regime

The outer bounds developed in previous subsection are useful to obtain strong interference regime for special scenarios but not for all networks of arbitrary large sizes. For many networks with more than two-receivers, these outer bounds lead to trivial cases when exploring the strong interference regime. In fact, the problem of determining strong interference regime for interference networks with more than two receivers, specifically for the multi-user CICs, has been an open problem in network information theory [15, page 6-68]. In this subsection, we provide a response to this problem. Clearly, we derive a strong interference regime not only for the CICs with arbitrary number of users but also for the general interference networks with arbitrary configurations. In what follows, we first address the main idea behind our derivations. Then, we present the result for the multi-user CICs. Lastly, we develop the theory for the general networks. Also, some additional examples are given to demonstrate the power of our results.

Let us first discuss an interesting related scenario. A part of the ideas for our approach in fact was derived during our attempts to obtain the capacity region of the multi-user BC with a sequence of more-capable receivers. Let consider a multi-user BC as shown in Fig. 8.

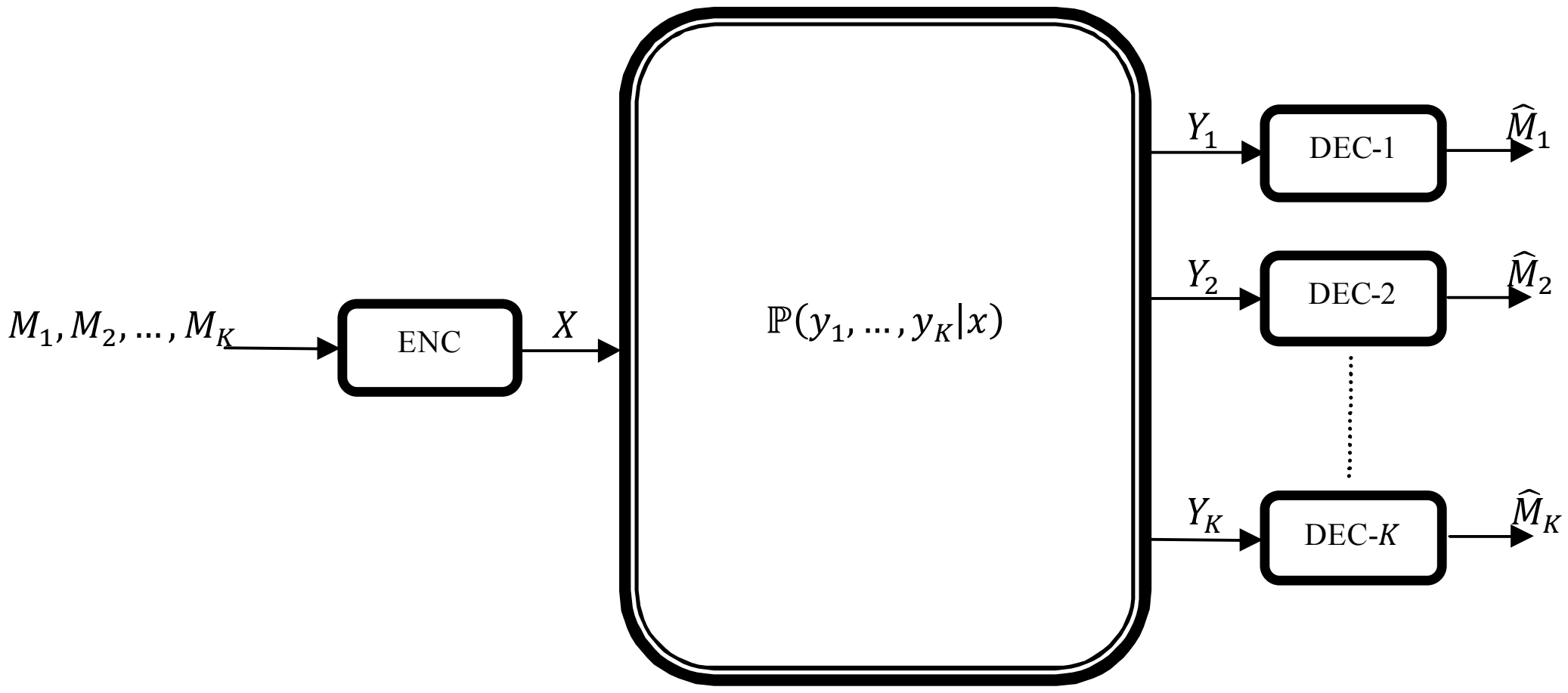

Figure 8. The K-user Broadcast Channel (BC).

The $K$-user BC is derived from the general interference network in Fig. 1 by setting $\mathbb{M} = \mathbb{M}_X = \{M_1, M_2, \dots, M_K\}$, $K_1 = 1$, $K_2 = K$, and $\mathbb{M}_{Y_j} = \{M_j\}$, $j = 1, \dots, K$, (note that since for the BC there is only one transmitter, we omit the subscripts from $\mathcal{X}_1, X_1, x_1$). Finding the capacity region for the broadcast networks is one the most difficult problems in network information theory, specifically in the case of more than two receivers. One of important classes for which the capacity region is known for the two-user case is the more-capable channels [18]. A two-user BC with receivers $Y_1$ and $Y_2$ is said to be more-capable if the condition (17) holds. For this channel, the superposition coding scheme achieves the capacity region, as given in Part I [1, Proposition III.3]. As a natural generalization, one may consider the multi-user BC with a sequence of more-capable receivers. Precisely, inspired by the two-user more-capable channel in (17), one may define a multi-user BC with a sequence of more-capable receivers to be a channel for which the following condition holds:

$$I(X;Y_K) < I(X;Y_{K-1}) < I(X;Y_{K-2}) < \cdots < I(X;Y_1) \quad \text{for all joint PDFs} \quad P_X \tag{182}$$

Unfortunately, the capacity result for the two-user more-capable BC does not seem to be straightforwardly extended to the multi-user case in (182). Nevertheless, we could find an insightful result in this regard. Clearly, we can find the sum-rate capacity for this network. Let us describe the procedure of the derivation in details. We claim that for the more-capable BC in (182), the sum-rate capacity is given below:

$$\mathcal{C}_{sum}^{more-capable} = \max_{P_X} I(X;Y_1) \tag{183}$$

The proof of achievability is trivial: the transmitter sends the message $M_1$ for the first user (stronger user) at its capacity rate and ignores the other messages. In fact, for our purposes, the proof of the converse part is important. Consider a length-$n$ code with vanishing average error probability for the network. First note that, according to Lemma 3, the condition (182) implies that:

$$I(X^n;Y_K^n|D) < I(X^n;Y_{K-1}^n|D) < I(X^n;Y_{K-2}^n|D) < \cdots < I(X^n;Y_1^n|D) \quad \text{for all joint PDFs} \quad P_{DX^n} \tag{184}$$

Now using the Fano's inequality we can write:

$$\begin{aligned}
n\sum_{l=1}^{K} R_l &\leq I(M_K;Y_K^n) + I(M_{K-1};Y_{K-1}^n) + I(M_{K-2};Y_{K-2}^n) + \cdots + I(M_1;Y_1^n) + nK\epsilon_n \\
&\leq I(M_K;Y_K^n,M_{K-1},M_{K-2},\dots,M_1) + I(M_{K-1};Y_{K-1}^n,M_{K-2},\dots,M_1) + I(M_{K-2};Y_{K-2}^n,M_{K-3},\dots,M_1) + \cdots + I(M_1;Y_1^n) + nK\epsilon_n \\
&\leq I(M_K;Y_K^n|M_{K-1},M_{K-2},\dots,M_1) + I(M_{K-1};Y_{K-1}^n|M_{K-2},\dots,M_1) + I(M_{K-2};Y_{K-2}^n|M_{K-3},\dots,M_1) + \cdots + I(M_1;Y_1^n) + nK\epsilon_n \\
&\overset{(a)}{=} I(X^n;Y_K^n|M_{K-1},M_{K-2},\dots,M_1) + I(M_{K-1};Y_{K-1}^n|M_{K-2},\dots,M_1) + I(M_{K-2};Y_{K-2}^n|M_{K-3},\dots,M_1) + \cdots + I(M_1;Y_1^n) + nK\epsilon_n \\
&\overset{(b)}{\leq} I(X^n;Y_{K-1}^n|M_{K-1},M_{K-2},\dots,M_1) + I(M_{K-1};Y_{K-1}^n|M_{K-2},\dots,M_1) + I(M_{K-2};Y_{K-2}^n|M_{K-3},\dots,M_1) + \cdots + I(M_1;Y_1^n) + nK\epsilon_n \\
&= I(X^n,M_{K-1};Y_{K-1}^n|M_{K-2},\dots,M_1) + I(M_{K-2};Y_{K-2}^n|M_{K-3},\dots,M_1) + \cdots + I(M_1;Y_1^n) + nK\epsilon_n \\
&\overset{(c)}{=} I(X^n;Y_{K-1}^n|M_{K-2},\dots,M_1) + I(M_{K-2};Y_{K-2}^n|M_{K-3},\dots,M_1) + \cdots + I(M_1;Y_1^n) + nK\epsilon_n \\
&\overset{(d)}{\leq} I(X^n;Y_{K-2}^n|M_{K-2},\dots,M_1) + I(M_{K-2};Y_{K-2}^n|M_{K-3},\dots,M_1) + \cdots + I(M_1;Y_1^n) + nK\epsilon_n \\
&= I(X^n;Y_{K-2}^n|M_{K-3},\dots,M_1) + \cdots + I(M_1;Y_1^n) + nK\epsilon_n \leq \cdots \leq I(X^n;Y_1^n) + nK\epsilon_n \leq nI(X;Y_1) + nK\epsilon_n
\end{aligned} \tag{185}$$

where $\epsilon_n \to 0$ as $n \to 0$, the equality (a) holds because $X^n$ is given by a deterministic function of $(M_1,M_2,\dots,M_K)$, inequality (b) is due to (184) where $D$ is replaced by $(M_{K-1},M_{K-2},\dots,M_1)$, equality (c) holds because, given the input sequence $X^n$, the output sequence $Y_{K-1}^n$ is independent of messages, inequality (d) is due to (184) and *etc*.

Let us review the philosophy behind the derivations. The first inequality in (185) is a direct consequence of Fano's inequality. The second one is derived by providing (virtual) side information to the receivers. In fact, the messages are sequentially given as side information to the non-respective receivers in a degraded order: $(M_{K-1},M_{K-2},\dots,M_1)$ to $Y_K$, $(M_{K-2},\dots,M_1)$ to $Y_{K-1}$ and so forth. Then, the resulting mutual information functions are successively manipulated (combined) using the more-capable condition in (182) and its extension in Lemma 3, until to arrive at a single mutual information function. The last mutual information function has a desirable property: it is composed of the input signal and one of the outputs (the stronger one in (182)). This is one of the ideas for our derivations in the following subsections.

### V.B.1) Three-User Classical Interference Channel

Consider the three-user CIC shown in Fig. 7. We intend to derive a strong interference regime for this network. We remark that the following derivations which are presented for the three-user CIC can be developed for other interference networks with any arbitrary topology, as will be given in Subsection V.B.3.

First note that according to our definition of the strong interference regime each receiver decodes all messages. The resulting achievable rate region by this scheme is given by:

$$\bigcup_{P_QP_{X_1|Q}P_{X_2|Q}P_{X_3|Q}} \left\{\begin{array}{l}
(R_1,R_2,R_3): \\
R_1 \leq \min\{I(X_1;Y_1|X_2,X_3,Q),I(X_1;Y_2|X_2,X_3,Q),I(X_1;Y_3|X_2,X_3,Q)\} \\
R_2 \leq \min\{I(X_2;Y_1|X_1,X_3,Q),I(X_2;Y_2|X_1,X_3,Q),I(X_2;Y_3|X_1,X_3,Q)\} \\
R_3 \leq \min\{I(X_3;Y_1|X_1,X_2,Q),I(X_3;Y_2|X_1,X_2,Q),I(X_3;Y_3|X_1,X_2,Q)\} \\
\\
R_1+R_2 \leq \min\{I(X_1,X_2;Y_1|X_3,Q),I(X_1,X_2;Y_2|X_3,Q),I(X_1,X_2;Y_3|X_3,Q)\} \\
R_2+R_3 \leq \min\{I(X_2,X_3;Y_1|X_1,Q),I(X_2,X_3;Y_2|X_1,Q),I(X_2,X_3;Y_3|X_1,Q)\} \\
R_1+R_3 \leq \min\{I(X_1,X_3;Y_1|X_2,Q),I(X_1,X_3;Y_2|X_2,Q),I(X_1,X_3;Y_3|X_2,Q)\} \\
\\
R_1+R_2+R_3 \leq \min\{I(X_1,X_2,X_3;Y_1|Q),I(X_1,X_2,X_3;Y_2|Q),I(X_1,X_2,X_3;Y_3|Q)\}
\end{array}\right\} \tag{186}$$

We need to derive conditions under which this rate region is optimal. Consider a length-$n$ block code for the network with vanishing error probability.

*Claim:* if the network transition probability function satisfies the following conditions:

$$\begin{cases} I(X_2;Y_2|X_1,X_3) \leq I(X_2;Y_3|X_1,X_3) & \text{for all joint PDFs} \quad P_{X_1}P_{X_2}P_{X_3} \\ I(X_2,X_3;Y_3|X_1) \leq I(X_2,X_3;Y_1|X_1) & \text{for all joint PDFs} \quad P_{X_1}P_{X_2X_3} \end{cases} \tag{187}$$

then, we have:

$$n(R_1+R_2+R_3) \leq I(X_1^n,X_2^n,X_3^n;Y_1^n)+n\epsilon_n \leq \sum_{t=1}^{n} I\left(X_{1,t},X_{2,t},X_{3,t};Y_{1,t}\right)+n\epsilon_n \tag{188}$$

where $\epsilon_n \to 0$ as $n \to 0$.

*Proof of Claim:* Based on the Fano's inequality one can write:

$$\begin{aligned} n(R_1+R_2+R_3) &\leq I(M_2;Y_2^n)+I(M_3;Y_3^n)+I(M_1;Y_1^n)+n\epsilon_n \\ &\leq I(M_2;Y_2^n|M_1,M_3)+I(M_3;Y_3^n|M_1)+I(M_1;Y_1^n)+n\epsilon_n \\ &\overset{(a)}{=} I(X_2^n;Y_2^n|X_1^n,X_3^n,M_1,M_3)+I(X_3^n,M_3;Y_3^n|X_1^n,M_1)+I(X_1^n,M_1;Y_1^n)+n\epsilon_n \\ &\overset{(b)}{\leq} I(X_2^n;Y_3^n|X_1^n,X_3^n,M_1,M_3)+I(X_3^n,M_3;Y_3^n|X_1^n,M_1)+I(X_1^n,M_1;Y_1^n)+n\epsilon_n \\ &= I(X_2^n,X_3^n;Y_3^n|X_1^n,M_1)+I(X_1^n,M_1;Y_1^n)+n\epsilon_n \\ &\overset{(c)}{\leq} I(X_2^n,X_3^n;Y_1^n|X_1^n,M_1)+I(X_1^n,M_1;Y_1^n)+n\epsilon_n \\ &= I(X_1^n,X_2^n,X_3^n;Y_1^n)+n\epsilon_n \leq \sum_{t=1}^{n} I\left(X_{1,t},X_{2,t},X_{3,t};Y_{1,t}\right)+n\epsilon_n \end{aligned} \tag{189}$$

where equality (a) holds because the input sequence $X_i^n$ is given by a deterministic function of the message $M_i, i=1,2,3$, inequality (b) is due to the first condition in (187) and its $n$-tuple extension in Lemma 3, and inequality (c) is due to the second condition in (187) and its $n$-tuple extension in Lemma 3.

Therefore, under the conditions (187), we derived one of the desired constraints on the sum-rate capacity in (186). Indeed, by the conditions (187), one can achieve further results. Clearly, these conditions imply that decoding all messages at the first receiver is optimal. Let us prove this result. Consider the constraints on the partial sum rates. First note that, according to Corollary 1, the second condition of (187) implies that:

$$\begin{cases} I(X_2;Y_3|X_1,X_3) \leq I(X_2;Y_1|X_1,X_3) \\ I(X_3;Y_3|X_1,X_2) \leq I(X_3;Y_1|X_1,X_2) \end{cases} \tag{190}$$

Comparing the first condition of (187) and the first condition of (190), we also obtain:

$$I(X_2;Y_2|X_1,X_3) \leq I(X_2;Y_1|X_1,X_3) \tag{191}$$

Now, we have:

$$\begin{aligned} n(R_1+R_2) &\leq I(M_2;Y_2^n)+I(M_1;Y_1^n)+n\epsilon_n \\ &\leq I(M_2;Y_2^n|M_1,M_3)+I(M_1;Y_1^n|M_3)+n\epsilon_n \\ &= I(X_2^n;Y_2^n|X_1^n,X_3^n,M_1,M_3)+I(X_1^n,M_1;Y_1^n|X_3^n,M_3)+n\epsilon_n \\ &\overset{(a)}{\leq} I(X_2^n;Y_1^n|X_1^n,X_3^n,M_1,M_3)+I(X_1^n,M_1;Y_1^n|X_3^n,M_3)+n\epsilon_n \end{aligned}$$

$$= I(X_1^n, X_2^n; Y_1^n | X_3^n, M_3) + n\epsilon_n \leq \sum_{t=1}^n I(X_{1,t}, X_{2,t}; Y_{1,t} | X_{3,t}) + n\epsilon_n \tag{192}$$

where inequality (a) is due to (191) and its $n$-tuple extension in Lemma 3. Also, by following the same lines as (189), one can derive:

$$R_2 + R_3 \leq I(X_2^n, X_3^n; Y_1^n | X_1^n, M_1) + n\epsilon_n \leq \sum_{t=1}^n I(X_{2,t}, X_{3,t}; Y_{1,t} | X_{1,t}) + n\epsilon_n \tag{193}$$

Note that (193) is actually the inequality (c) of (189) where the term $I(X_1^n, M_1; Y_1^n)$ from the right side and the corresponding rate $R_1$ from the left side are removed. Lastly, we have:

$$\begin{aligned} n(R_1 + R_3) &\leq I(M_3; Y_3^n) + I(M_1; Y_1^n) + n\epsilon_n \\ &\leq I(X_3^n; Y_3^n | X_1^n, X_2^n, M_1, M_2) + I(X_1^n, M_1; Y_1^n | X_2^n, M_2) + n\epsilon_n \\ &\overset{(a)}{\leq} I(X_3^n; Y_1^n | X_1^n, X_2^n, M_1, M_2) + I(X_1^n, M_1; Y_1^n | X_2^n, M_2) + n\epsilon_n \\ &= I(X_1^n, X_3^n; Y_1^n | X_2^n, M_2) + n\epsilon_n \leq \sum_{t=1}^n I(X_{1,t}, X_{3,t}; Y_{1,t} | X_{2,t}) + n\epsilon_n \end{aligned} \tag{194}$$

where inequality (a) is due to the second condition in (190). Finally, the desired constraints on the individual rates can be easily derived using the second condition of (190) and the condition (191). Thus, if (187) holds, then it is optimal to decode all messages at the first receiver. It is clear that we can follow the same procedure for the other receivers to derive conditions under which the strong interference criterion, i.e., the optimality of decoding all messages, is satisfied. For example, one can verify that if the following conditions hold:

$$\begin{cases} I(X_3; Y_3 | X_1, X_2) \leq I(X_3; Y_1 | X_1, X_2) & \text{for all joint PDFs} \quad P_{X_1} P_{X_2} P_{X_3} \\ I(X_1, X_3; Y_1 | X_2) \leq I(X_1, X_3; Y_2 | X_2) & \text{for all joint PDFs} \quad P_{X_1 X_3} P_{X_2} \end{cases} \tag{195}$$

the second receiver experiences strong interference; also, if the following hold:

$$\begin{cases} I(X_1; Y_1 | X_2, X_3) \leq I(X_1; Y_2 | X_2, X_3) & \text{for all joint PDFs} \quad P_{X_1} P_{X_2} P_{X_3} \\ I(X_1, X_2; Y_2 | X_3) \leq I(X_1, X_2; Y_3 | X_3) & \text{for all joint PDFs} \quad P_{X_1 X_2} P_{X_3} \end{cases} \tag{196}$$

the third receiver experiences strong interference. Therefore, the collection of the conditions (187), (195) and (196) constitutes a strong interference regime for the three-user CIC in Fig. 7. A remarkable point is that the necessary conditions for deriving the desired constraints on the sum-rate such as (188) are indeed sufficient to prove the optimality of decoding all messages at the receivers. In other words, once we derived the desired constraints on the sum-rate capacity using certain conditions, no additional condition is required to be introduced to prove the desired constraints on the partial sum-rates.

Let us concentrate on this collection. According to Corollary 1, the second condition of (187) implies the first condition of (195), the second condition of (195) implies the first condition in (196) and the second condition of (196) implies the first condition of (187). Therefore, a strong interference regime for the three-user is given as follows:

**A strong interference regime for the 3-user interference channel**

$$\begin{aligned} I(X_2, X_3; Y_3 | X_1) &\leq I(X_2, X_3; Y_1 | X_1) && \text{for all joint PDFs} \quad P_{X_1} P_{X_2 X_3} \\ I(X_1, X_3; Y_1 | X_2) &\leq I(X_1, X_3; Y_2 | X_2) && \text{for all joint PDFs} \quad P_{X_1 X_3} P_{X_2} \\ I(X_1, X_2; Y_2 | X_3) &\leq I(X_1, X_2; Y_3 | X_3) && \text{for all joint PDFs} \quad P_{X_1 X_2} P_{X_3} \end{aligned} \tag{197}$$

The conditions (197) to some extent represents a fact regarding the CICs that is the signal of each receiver is impaired by the joint effect of interference from all non-corresponding transmitters rather than by each transmitter's signal separately. Note that the terms in the right side of the inequalities in (197) indeed measure the amount of interference experienced by the receivers.

It should be noted that (similar to the two-receiver networks) using the conditions (197) we are able to derive all the constraints in the rate region (186); nevertheless, some of these constraints are actually redundant. In fact, if the conditions (197) hold, the rate region (186) is simplified below:

$$\bigcup_{P_Q P_{X_1|Q} P_{X_2|Q} P_{X_3|Q}} \left\{\begin{array}{l} (R_1, R_2, R_3): \\ R_1 \leq I(X_1; Y_1|X_2, X_3, Q) \\ R_2 \leq I(X_2; Y_2|X_1, X_3, Q) \\ R_3 \leq I(X_3; Y_3|X_1, X_2, Q) \\ \\ R_1 + R_2 \leq \min\{I(X_1, X_2; Y_1|X_3, Q), I(X_1, X_2; Y_2|X_3, Q)\} \\ R_2 + R_3 \leq \min\{I(X_2, X_3; Y_2|X_1, Q), I(X_2, X_3; Y_3|X_1, Q)\} \\ R_1 + R_3 \leq \min\{I(X_1, X_3; Y_1|X_2, Q), I(X_1, X_3; Y_3|X_2, Q)\} \\ \\ R_1 + R_2 + R_3 \leq \min\left\{\begin{array}{l} I(X_1, X_2, X_3; Y_1|Q), \\ I(X_1, X_2, X_3; Y_2|Q), \\ I(X_1, X_2, X_3; Y_3|Q) \end{array}\right\} \end{array}\right\}$$

(198)

The other constraints of (186) are relaxed by the conditions in (197). Let us now consider the three-user Gaussian CIC as formulated in the standard form in (173). The gain matrix of the network is as follows:

$$\begin{bmatrix} 1 & a_{12} & a_{13} \\ a_{21} & 1 & a_{23} \\ a_{31} & a_{32} & 1 \end{bmatrix}$$

(199)

Using Lemma 2, one can derive explicit constraints on the network gains under which the strong interference regime (197) holds, as given below:

$$\begin{cases} |a_{13}| \geq 1, \qquad |a_{21}| \geq 1, \qquad |a_{32}| \geq 1 \\ \qquad\qquad\qquad\qquad , \\ a_{12} = a_{13}a_{32}, \qquad a_{31} = a_{21}a_{32}, \qquad a_{23} = a_{21}a_{13} \end{cases}$$

(200)

Let examine the conditions (200). Among the six parameters in the matrix (199), the parameters $a_{13}, a_{21}$ and $a_{32}$, which no pair of them lies in either a same row or a same column, are given by arbitrary real numbers greater than one and the other parameters are dependent on these parameters.

Here, we return to the calculations (189) where we derived the constraint (188) on the sum-rate using the conditions (187). If we review these calculations, we observe that by exchanging the order of manipulating mutual information functions, one can derive conditions other than those in (187) under which decoding of all messages at the first receiver is optimal. Consider the following conditions:

$$\begin{cases} I(X_3; Y_3|X_1, X_2) \leq I(X_3; Y_2|X_1, X_2) & \text{for all joint PDFs} \quad P_{X_1}P_{X_2}P_{X_3} \\ I(X_2, X_3; Y_2|X_1) \leq I(X_2, X_3; Y_1|X_1) & \text{for all joint PDFs} \quad P_{X_1}P_{X_2X_3} \end{cases}$$

(201)

The conditions (201) are obtained by exchanging the indices "2" and "3" in (187). One can readily verify the conditions (201) also imply that the decoding of all messages at the first receiver is optimal. The derivation is similar to (189)-(194) except that the indices "2" and "3" are exchanged everywhere. Therefore, the collection of (201), (195) and (196) is also a strong interference regime for the three-user CIC. This new collection of strong interference conditions can be represented in the following form:

$$\begin{cases} I(X_3;Y_3|X_1,X_2) \leq I(X_3;Y_2|X_1,X_2) & \text{for all joint PDFs} \quad P_{X_1}P_{X_2}P_{X_3} \\ I(X_2,X_3;Y_2|X_1) \leq I(X_2,X_3;Y_1|X_1) & \text{for all joint PDFs} \quad P_{X_1}P_{X_2X_3} \\ I(X_1,X_3;Y_1|X_2) \leq I(X_1,X_3;Y_2|X_2) & \text{for all joint PDFs} \quad P_{X_1X_3}P_{X_2} \\ I(X_1,X_2;Y_2|X_3) \leq I(X_1,X_2;Y_3|X_3) & \text{for all joint PDFs} \quad P_{X_1X_2}P_{X_3} \end{cases} \tag{202}$$

For the Gaussian network in (173) Lemma 2 implies that if the following constraints hold:

$$\begin{cases} 1 \leq |a_{23}| \\ , \\ a_{21} = \dfrac{1}{a_{12}} = \dfrac{a_{23}}{a_{13}} = \alpha, \qquad |\alpha| = 1 \\ , \\ \dfrac{a_{21}}{a_{31}} = \dfrac{1}{a_{32}} = \beta, \qquad |\beta| \leq 1 \end{cases} \tag{203}$$

then, the conditions (202) are satisfied.

In fact, by following the same procedure, one can derive $2^3 = 8$ different strong interference regimes for the three-user CIC. Among these regimes, 2 ($(3-1)!$) of them are more significant: the regime in (197) and the regime that is derived by exchanging 1 by 3 in (197). The other regimes, such as the one given in (202), lead to rather trivial situations, specifically for the Gaussian networks.

For certain scenarios in which some receivers are not connected to specific transmitters, the above strong interference conditions cannot be satisfied; for example, consider a network where its transition probability function satisfies the factorization below:

$$\mathbb{P}(y_1,y_2,y_3|x_1,x_2,x_3) = \mathbb{P}(y_1|x_1,x_2)\mathbb{P}(y_2|x_2,x_3)\mathbb{P}(y_3|x_1,x_3) \tag{204}$$

In this network, each receiver experiences interference from only transmitter. This network, which was previously studied in [29], is called "*cyclic Z-interference channel*". Let us examine the first constraint of (197). For the cyclic Z-interference channel in (204), this constraint can be re-written as:

$$I(X_3;Y_3|X_1) \overset{?}{\leq} I(X_2;Y_1|X_1) \qquad \text{for all joint PDFs} \qquad P_{X_1}P_{X_2X_3} \tag{205}$$

Now consider a joint PDF on the input signals such that $X_1$ and $X_2$ are degenerate. The right side of (205) is zero while the left side may be positive. Therefore, (205) cannot be satisfied for all joint PDFs $P_{X_1}P_{X_2X_3}$. In fact, in such scenarios there is no need to prove that the decoding of all the messages at all receivers is optimal. According to Definition 4, in strong interference regime, each receiver decodes all its connected messages. Specially, for the cyclic Z-interference channel (204), the first receiver decodes the messages $M_1$ and $M_2$, the second receiver decodes $M_2$ and $M_3$ and the third receiver decodes $M_1$ and $M_3$. Thus, for this channel it is not required to derive constraints such as (188). Based on this fact, one can derive a consistent strong interference regime for this model as well. Specifically, it is not difficult to show that if the channel satisfies the following conditions:

$$\begin{cases} I(X_2;Y_2|X_1,X_3) \leq I(X_2;Y_1|X_1,X_3) \\ I(X_3;Y_3|X_1,X_2) \leq I(X_3;Y_2|X_1,X_2) \\ I(X_1;Y_1|X_2,X_3) \leq I(X_1;Y_3|X_2,X_3) \end{cases} \qquad \text{for all joint PDFs} \qquad P_{X_1}P_{X_2}P_{X_3} \tag{206}$$

then, it lies in the strong interference regime (this is a consequence of our general result in Subsection V.B.4). The Gaussian cyclic Z-interference channel is derived by setting $a_{13} = a_{21} = a_{32} = 0$ in (173). For this case, the authors in [29] showed that if $|a_{12}|, |a_{23}|, |a_{31}| \geq 1$, then the channel has strong interference. The latter condition can also be derived by evaluating (206).

Let us again consider the achievable rate region (186). We proved that under the strong interference conditions (197), this rate region is reduced to the one in (198) that is also the capacity region. Now, concentrate on the rate region (198). A remarkable point is that this rate region by itself is achievable for the general three-user CIC regardless of the conditions (197). Compare the rate region (186) with that in (198). We recall that the rate region (186) is derived by requiring that each receiver *correctly decodes all transmitted messages*. Among the constraints of this achievable rate region, those not given in (198) are actually required to correctly decode non-corresponding messages at the receivers. In other words, with the same coding scheme, each receiver can still decode *its own message* correctly if the communication rates do not satisfy those constraints of (186) which are not given in (198). Thus, the rate region (198) is really achievable for the general three-user CIC. Later, in Subsection V.B.5, we discuss this achievability scheme in detail. In general, the rate region (186) is a subset of that in (198). However, under the conditions (197), these rate regions both coincide to the capacity region. We remark that, if the conditions (197) (or similar strong interference conditions such as (202)) hold, then we are able to derive all the constraints of the achievable rate region (198) using Fano's inequality. Nonetheless, for some certain scenarios, it is not required to prove all the constraints of (198). Precisely, there exist scenarios where some of the constraints of the rate region (198) are actually redundant. Let us provide an example. Consider a three-user CIC for which the following condition holds:

$$I(X_3;Y_3|X_1,X_2,Q) \leq I(X_3;Y_1|Q) \tag{207}$$

for all joint PDFs $P_Q P_{X_1|Q} P_{X_2|Q} P_{X_3|Q}$ which achieve the boundary points of the rate region (208) given below. In this case, one can readily show that the achievable rate region (198) is simplified as follows:

$$\bigcup_{P_Q P_{X_1|Q} P_{X_2|Q} P_{X_3|Q}} \left\{ \begin{array}{l} (R_1,R_2,R_3): \\ R_1 \leq I(X_1;Y_1|X_2,X_3,Q) \\ R_2 \leq I(X_2;Y_2|X_1,X_3,Q) \\ R_3 \leq I(X_3;Y_3|X_1,X_2,Q) \\ R_1+R_2 \leq \min\{I(X_1,X_2;Y_1|X_3,Q), I(X_1,X_2;Y_2|X_3,Q)\} \\ R_2+R_3 \leq \min\{I(X_2,X_3;Y_2|X_1,Q), I(X_2,X_3;Y_3|X_1,Q)\} \\ R_1+R_3 \leq I(X_1,X_3;Y_3|X_2,Q) \\ R_1+R_2+R_3 \leq \min\left\{ \begin{array}{l} I(X_1,X_2,X_3;Y_2|Q), \\ I(X_1,X_2,X_3;Y_3|Q) \end{array} \right\} \end{array} \right\} \tag{208}$$

In fact, if the conditions (207) are satisfied, then we have:

$$\begin{array}{l} I(X_1;Y_1|X_2,X_3,Q)+I(X_3;Y_3|X_1,X_2,Q) \leq I(X_1;Y_1|X_2,X_3,Q)+I(X_3;Y_1|Q) \leq I(X_1,X_3;Y_1|X_2,Q) \\ I(X_1,X_2;Y_1|X_3,Q)+I(X_3;Y_3|X_1,X_2,Q) \leq I(X_1,X_2,X_3;Y_1|Q) \end{array} \tag{209}$$

Thus, one of the constraints on the sum-rate and also one of the constraints on the partial sum-rate $R_1+R_3$ in the achievable rate region (198) are redundant and can be relaxed which yield the rate region (208). Now, let us explore conditions under which the rate region (208) is optimal. Note that for the rate region (208), we do not need to prove the constraint (188). For the receiver $Y_1$, it is only required to derive the following constraints:

$$\begin{array}{c} R_1 \leq I(X_1;Y_1|X_2,X_3,Q) \\ R_1+R_2 \leq I(X_1,X_2;Y_1|X_3,Q) \end{array} \tag{210}$$

The first constraint of (210) is trivial. Also, according to the relations in (192), if the following condition holds:

$$I(X_2;Y_2|X_1,X_3) \leq I(X_2;Y_1|X_1,X_3) \quad \text{for all joint PDFs} \quad P_{X_1}P_{X_2}P_{X_3} \tag{211}$$

then, we can derive the second constraint of (210). Moreover, to derive the other constraints of (208), the conditions (195) and (196) are sufficient. Thereby, in addition to (207), if the collection of the conditions (211), (195) and (196) holds, then the rate region (208) is optimal and coincides with the capacity region. These conditions can be collectively represented as follows:

$$\left\{ \begin{array}{lll} I(X_3;Y_3|X_1,X_2,Q) \leq I(X_3;Y_1|Q) & \text{for all joint PDFs } P_Q P_{X_1|Q} P_{X_2|Q} P_{X_3|Q} \text{ achieving the boundary points of (208)} & \\ I(X_2;Y_2|X_1,X_3) \leq I(X_2;Y_1|X_1,X_3) & \text{for all joint PDFs} & P_{X_1}P_{X_2}P_{X_3} \\ I(X_3;Y_3|X_1,X_2) \leq I(X_3;Y_1|X_1,X_2) & \text{for all joint PDFs} & P_{X_1}P_{X_2}P_{X_3} \\ I(X_1,X_3;Y_1|X_2) \leq I(X_1,X_3;Y_2|X_2) & \text{for all joint PDFs} & P_{X_1X_3}P_{X_2} \\ I(X_1,X_2;Y_2|X_3) \leq I(X_1,X_2;Y_3|X_3) & \text{for all joint PDFs} & P_{X_1X_2}P_{X_3} \end{array} \right. \tag{212}$$

It should be remarked that under the conditions (212), the achievable rat region (198) is optimal, however, it is not clear whether the rate region (186) is also optimal. The reason is that from the conditions (212) we can not necessarily derive the constraint (193).

In fact, by following the same approach described above, one can derive many other cases for which the achievable rate region (198) is optimal. However, for these scenarios the achievable rate region (186) may be sub-optimal. We refer to such scenarios as *networks with almost decodable interference*. Later, in Subsection V.B.5, we present an exact definition for the almost-decodable interference regime.

Let us discuss the conditions (212) for the Gaussian network in (173). First note that the boundary points of the rate region (208) are achieved by Gaussian distributions without time-sharing ($Q \equiv \emptyset$). Therefore, the first condition (212) holds if:

$$\sqrt{P_1 + a_{12}^2 P_2 + 1} \leq |a_{13}| \tag{213}$$

Also, using Lemma 2, one can verify that if the following constraints hold, then the other conditions of (212) are satisfied:

$$\begin{cases} |a_{12}| \geq 1, \quad |a_{13}| \geq 1, \quad |a_{21}| \geq 1, \quad |a_{32}| \geq 1 \\ a_{31} = a_{21}a_{32}, \quad a_{23} = a_{21}a_{13} \end{cases}, \tag{214}$$

In general, for Gaussian networks, conditions such as the first condition of (212) lead to constraints on both powers and gains, unlike the other conditions of (212) which lead to constraints on only the network gains.

Thus, for the three-user Gaussian CIC (173) if the collection of the conditions (213) and (214) is satisfied, the rate region (208) coincides with the capacity region. This collection can also be described as follows:

$$\begin{cases} |a_{12}| \geq 1, \quad |a_{13}| \geq \sqrt{P_1 + a_{12}^2 P_2 + 1}, \quad |a_{21}| \geq 1, \quad |a_{32}| \geq 1 \\ a_{31} = a_{21}a_{32}, \quad a_{23} = a_{21}a_{13} \end{cases}, \tag{215}$$

## V.B.2) K-user Classical Interference Channel

Now, let us examine the CIC with arbitrary number of users as shown in Fig. 9.

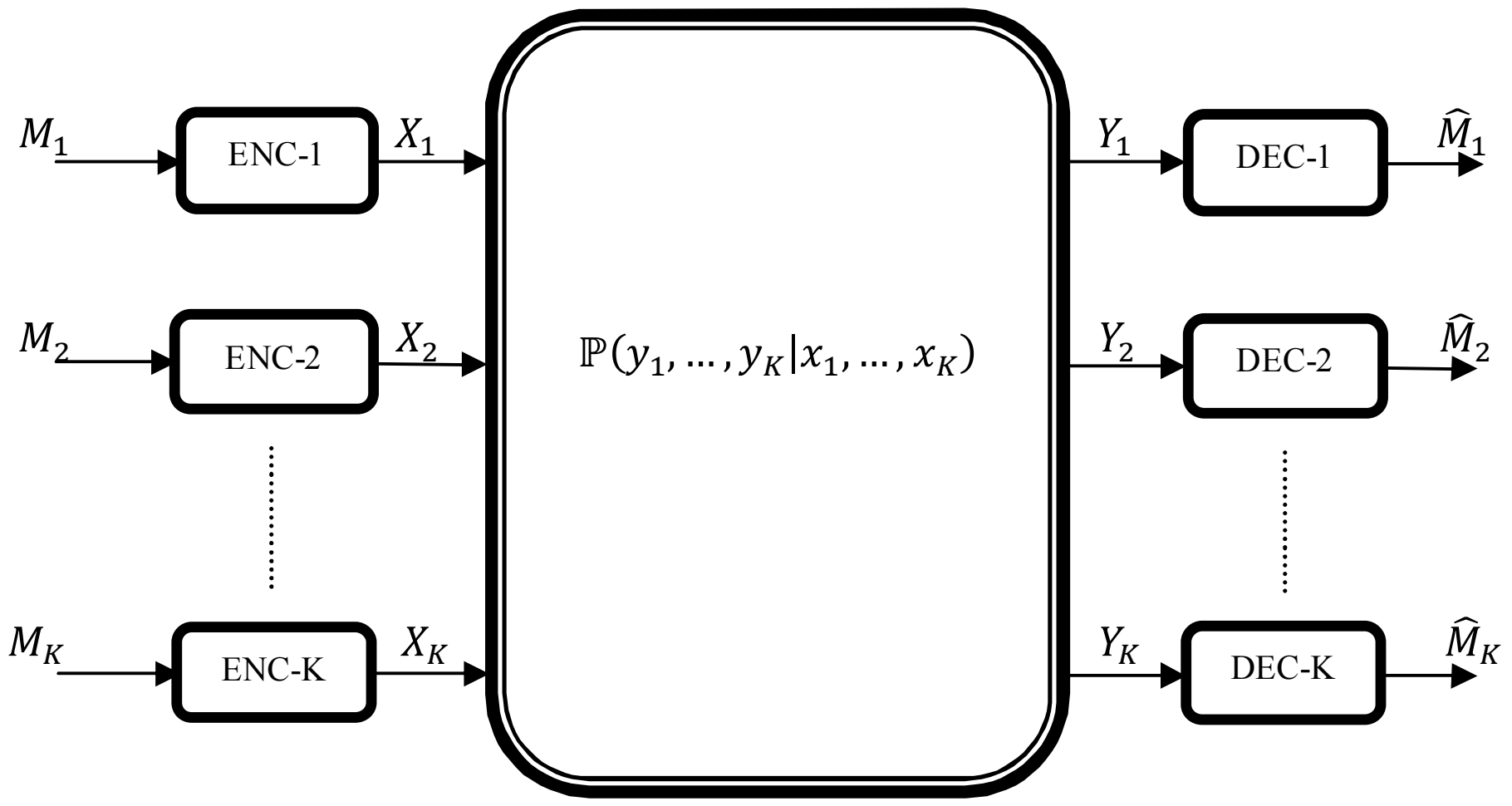

Figure 9. The K-user Classical Interference Channel (CIC).

The K-user CIC is derived from the general interference network given in Fig. 1 by setting $\mathbb{M}=\{M_1,M_2,\dots,M_K\}$, $K_1=K_2=K$, $\mathbb{M}_{X_i}=\{M_i\}$, $i=1,\dots,K$, and $\mathbb{M}_{Y_j}=\{M_j\}$, $j=1,\dots,K$. An open problem in network information theory has been to determine a strong interference regime for this network [15, page 6-68]. We now present a solution to this problem. Specifically, by following the same approach as the three-user channel, one can derive the following strong interference regime for the K-user CIC:

**A Strong Interference Regime for the K-User Interference Channel**

$$I(X_2,X_3,X_4,\dots,X_K;Y_K|X_1)\le I(X_2,X_3,X_4,\dots,X_K;Y_1|X_1) \quad \text{for all joint PDFs} \quad P_{X_2X_3X_4\dots X_K}P_{X_1}$$

$$I(X_1,X_3,X_4,\dots,X_K;Y_1|X_2)\le I(X_1,X_3,X_4,\dots,X_K;Y_2|X_2) \quad \text{for all joint PDFs} \quad P_{X_1X_3X_4\dots X_K}P_{X_2}$$

$$I(X_1,X_2,X_4,\dots,X_K;Y_2|X_3)\le I(X_1,X_2,X_4,\dots,X_K;Y_3|X_3) \quad \text{for all joint PDFs} \quad P_{X_1X_2X_4\dots X_K}P_{X_3}$$

$$\vdots$$

$$I(X_1,X_2,\dots,X_{K-1};Y_{K-1}|X_K)\le I(X_1,X_2,\dots,X_{K-1};Y_K|X_K) \quad \text{for all joint PDFs} \quad P_{X_1X_2\dots X_{K-1}}P_{X_K}$$

(216)

Note that the regime (216) is described by $K$ inequalities. In fact, for the K-user CIC by following the same lines as the three-user CIC, one can derive $\big((K-1)!\big)^K$ different strong interference regime. However, among these regimes, $(K-1)!$ of them are more significant, which are derived by exchanging the indices $1,2,3,\dots,K$, with $\vartheta(1),\vartheta(2),\dots,\vartheta(K)$, respectively in (216), where $\vartheta(.)$ is a *cyclic permutation* of the elements of the set $\{1,2,3,\dots,K\}$. It is worth noting that this is the first time where a full characterization of the capacity region is derived for a general multi-user CIC.

Let us consider the Gaussian network which is formulated below:

$$\begin{cases} Y_1=X_1+a_{12}X_2+a_{13}X_3+\cdots+a_{1K}X_K+Z_1 \\ Y_2=a_{21}X_1+X_2+a_{23}X_3+\cdots+a_{2K}X_K+Z_2 \\ \vdots \\ Y_K=a_{K1}X_1+a_{K2}X_2+\cdots+a_{KK-1}X_{K-1}+X_K+Z_K \end{cases} \tag{217}$$

where $Z_1,\dots,Z_K$ are zero-mean unit-variance Gaussian noises and the inputs are subject to power constraints $\mathbb{E}[X_i^2]\le P_i,\ i=1,\dots,K$. Using Lemma 2, one can show that if the following conditions are satisfied:

$$\begin{cases} \frac{a_{K2}}{a_{12}}=\frac{a_{K3}}{a_{13}}=\cdots=\frac{a_{KK-1}}{a_{1K-1}}=\frac{1}{a_{1K}}=\alpha_1 \\ \frac{1}{a_{21}}=\frac{a_{13}}{a_{23}}=\frac{a_{14}}{a_{24}}=\cdots=\frac{a_{1K}}{a_{2K}}=\alpha_2 \\ \frac{a_{21}}{a_{31}}=\frac{1}{a_{32}}=\frac{a_{24}}{a_{34}}=\cdots=\frac{a_{2K}}{a_{3K}}=\alpha_3 \\ \vdots \\ \frac{a_{K-1,1}}{a_{K1}}=\frac{a_{K-1,2}}{a_{K2}}=\frac{a_{K-1,3}}{a_{K3}}=\cdots=\frac{1}{a_{KK-1}}=\alpha_K \end{cases} ,\quad |\alpha_i|\le 1,\quad i=1,2,\dots,K \tag{218}$$

then, the strong interference regime (216) holds. According to the conditions (218), the parameters $a_{1K},a_{21},a_{32},\dots,a_{KK-1}$ (which no pair of them lies in either a same row or a same column of the gain matrix) are given by arbitrary real numbers greater than one and the other parameters are dependent on these parameters. It is clear that $(K-1)!$ other strong interference regimes are derived by exchanging the indices $1,2,3,\dots,K$, with $\vartheta(1),\vartheta(2),\dots,\vartheta(K)$, respectively in (218), where $\vartheta(.)$ is a cyclic permutation of the elements of the set $\{1,2,3,\dots,K\}$.

## V.B.3) General Interference Network

In this subsection, we intend to develop the approach given in previous subsections to derive strong interference regime for the general interference network with any arbitrary topology. First, we need to define the concept of *nonempty partitioning* of a given set.

***Definition 6:*** *Let $\Lambda$ be a given arbitrary set. Let also $\Lambda_1,\Lambda_2,\dots,\Lambda_P$ be nonempty subsets of $\Lambda$ such that:*

$$\begin{cases}\Lambda_{l_1}\cap\Lambda_{l_2}=\emptyset,\ \ l_1\neq l_2 \\ \bigcup_{l\in[1:P]}\Lambda_l=\Lambda\end{cases},\tag{219}$$

*The collection $\Lambda_1,\Lambda_2,\dots,\Lambda_P$ is called a nonempty partitioning of $\Lambda$.*

In the following, we also use of the notations given in (166). Our main result is given in the next theorem.

***Theorem 11)*** **General Interference Networks: Information flow in Strong Interference Regime**

Consider the general interference network with $K_1$ transmitters and $K_2$ receivers as shown in Fig. 1. For each $j\in[1:K_2]$, assume that there exist a natural number $\lambda_j$, subsets $j(1),j(2),\dots,j(\lambda_j)$ of $\{1,\dots,K_2\}-\{j\}$ and a nonempty partitioning $\mathbb{M}_{j(1)},\mathbb{M}_{j(2)}\dots,\mathbb{M}_{j(\lambda_j)}$ of the messages $\mathbb{M}_{c\to Y_j}-\mathbb{M}_{Y_j}$ such that:

$$\mathbb{M}_{j(l)}\subseteq\mathbb{M}_{Y_{j(l)}},\qquad l=1,\dots,\lambda_j\tag{220}$$

Moreover, assume that the network transition probability function satisfies the following inequalities ($l=1,\dots,\lambda_j-1$):

$$I\left(\mathbb{X}-\mathbb{X}_{\bigcup_{\theta=l+1}^{\lambda_j}\mathbb{M}_{j(\theta)}\cup\mathbb{M}_{Y_j}\cup\mathbb{M}_{c\nrightarrow Y_j}};Y_{j(l)}\middle|\mathbb{X}_{\bigcup_{\theta=l+1}^{\lambda_j}\mathbb{M}_{j(\theta)}\cup\mathbb{M}_{Y_j}\cup\mathbb{M}_{c\nrightarrow Y_j}}\right)\leq I\left(\mathbb{X}-\mathbb{X}_{\bigcup_{\theta=l+1}^{\lambda_j}\mathbb{M}_{j(\theta)}\cup\mathbb{M}_{Y_j}\cup\mathbb{M}_{c\nrightarrow Y_j}};Y_{j(l+1)}\middle|\mathbb{X}_{\bigcup_{\theta=l+1}^{\lambda_j}\mathbb{M}_{j(\theta)}\cup\mathbb{M}_{Y_j}\cup\mathbb{M}_{c\nrightarrow Y_j}}\right)\tag{221}$$

for all joint PDFs $P_{\mathbb{X}-\mathbb{X}_{\bigcup_{\theta=l+1}^{\lambda_j}\mathbb{M}_{j(\theta)}\cup\mathbb{M}_{Y_j}\cup\mathbb{M}_{c\nrightarrow Y_j}}}\prod_{X_i\in\mathbb{X}_{\bigcup_{\theta=l+1}^{\lambda_j}\mathbb{M}_{j(\theta)}\cup\mathbb{M}_{Y_j}\cup\mathbb{M}_{c\nrightarrow Y_j}}}P_{X_i}$.

also, it satisfies:

$$I\left(\mathbb{X}-\mathbb{X}_{\mathbb{M}_{Y_j}\cup\mathbb{M}_{c\nrightarrow Y_j}};Y_{j(\lambda_j)}\middle|\mathbb{X}_{\mathbb{M}_{Y_j}\cup\mathbb{M}_{c\nrightarrow Y_j}}\right)\leq I\left(\mathbb{X}-\mathbb{X}_{\mathbb{M}_{Y_j}\cup\mathbb{M}_{c\nrightarrow Y_j}};Y_j\middle|\mathbb{X}_{\mathbb{M}_{Y_j}\cup\mathbb{M}_{c\nrightarrow Y_j}}\right)\tag{222}$$

for all joint PDFs $P_{\mathbb{X}-\mathbb{X}_{\mathbb{M}_{Y_j}\cup\mathbb{M}_{c\nrightarrow Y_j}}}\prod_{X_i\in\mathbb{X}_{\mathbb{M}_{Y_j}\cup\mathbb{M}_{c\nrightarrow Y_j}}}P_{X_i}$.

Under these conditions, the network has strong interference, i.e., the optimal coding strategy to achieve the capacity is that each receiver decodes all its connected messages.

*Proof of Theorem 11)* Consider the rate region below:

$$\bigcup_{\substack{P_Q P_{M_1} P_{M_2} \dots P_{M_K} \\ \times \prod_{i=1}^{K_1} P_{X_i|\mathbb{M}_{X_i},Q}}} \left\{ \begin{array}{l} (R_1, R_2, \dots, R_K): \\ \forall j \in [1:K_2]: \\ \forall \Omega \subseteq \mathbb{M}_{c\to Y_j}: \;\; \boldsymbol{R}_{\Sigma\Omega} \leq I\left(\Omega; Y_j \middle| \mathbb{M}_{c\to Y_j} - \Omega, Q\right) \end{array} \right\}$$

(223)

The rate region (223) is achievable for the network. The achievability scheme is that all the messages are separately encoded using independent codewords. Also, each receiver decodes its connected messages using a jointly typical decoder. We now prove that under the conditions (221) and (222), this rate region is optimal. Consider a receiver $Y_j$ where $j \in [1:K_2]$. Let $\Omega$ be an arbitrary subset of $\mathbb{M}_{c\to Y_j}$. According to Observation 1, the messages $\mathbb{M}_{c\nrightarrow Y_j}$ are independent of the signal $Y_j$. They are also independent of both $\mathbb{M}_{c\to Y_j}$ and $Q$, (all messages are independent of the time-sharing parameter). Thus, we have:

$$\boldsymbol{R}_{\Sigma\Omega} \leq I\left(\Omega; Y_j \middle| \mathbb{M}_{c\to Y_j} - \Omega, Q\right) = I\left(\Omega; Y_j \middle| \mathbb{M}_{c\to Y_j} - \Omega, \mathbb{M}_{c\nrightarrow Y_j}, Q\right) = I(\Omega; Y_j | \mathbb{M} - \Omega, Q)$$

(224)

Next, we prove (224). Define:

$$\Omega_{Y_j} \triangleq \mathbb{M}_{Y_j} \cap \Omega, \qquad \Omega_{j(l)} \triangleq \mathbb{M}_{j(l)} \cap \Omega, \qquad l = 1, \dots, \lambda_j$$

(225)

Note that as $\mathbb{M}_{j(1)}, \mathbb{M}_{j(2)}, \dots, \mathbb{M}_{j(\lambda_j)}$ is a partitioning for the set $\mathbb{M}_{c\to Y_j} - \mathbb{M}_{Y_j}$, the sets $\mathbb{M}_{Y_j}, \mathbb{M}_{j(1)}, \mathbb{M}_{j(2)}, \dots, \mathbb{M}_{j(\lambda_j)}$ constitute also a partitioning for $\mathbb{M}_{c\to Y_j}$. Therefore, $\Omega_{Y_j}, \Omega_{j(1)}, \Omega_{j(2)} \dots, \Omega_{j(\lambda_j)}$ in (225) make up a partitioning for the set $\Omega$. In other words,

$$\Omega_{Y_j} \cup \Omega_{j(1)} \cup \Omega_{j(2)} \cup \dots \cup \Omega_{j(\lambda_j)} = \Omega$$

(226)

Moreover, the sets $\Omega_{Y_j}, \Omega_{j(1)}, \Omega_{j(2)}, \dots, \Omega_{j(\lambda_j)}$ are pairwise disjoint. Now consider a length-$n$ code for the network with vanishing average error probability. Using Fano's inequality, we have:

$$\frac{1}{n} H\left(\mathbb{M}_{Y_j} \middle| Y_j^n\right) \leq \epsilon_{j,n}, \qquad j = 1,2, \dots, K_2$$

(227)

where $\epsilon_{j,n} \to 0$ as $n \to \infty$. Based on (220), (225) and (227), we derive:

$$\frac{1}{n} H\left(\Omega_{Y_j} \middle| Y_j^n\right) \leq \epsilon_{j,n}, \qquad \frac{1}{n} H\left(\Omega_{j(l)} \middle| Y_{j(l)}^n\right) \leq \epsilon_{j(l),n}, \qquad l = 1, \dots, \lambda_j$$

(228)

where $\epsilon_{j(l),n} \to 0$ as $n \to \infty$. Therefore, we have:

$$\begin{aligned}
& n\,\boldsymbol{R}_{\Sigma\Omega} - n\left(\textstyle\sum_{l\in[1:\lambda_j]} \epsilon_{j(l),n} + \epsilon_{j,n}\right) \\
&\leq I\left(\Omega_{j(1)}; Y_{j(1)}^n\right) + I\left(\Omega_{j(2)}; Y_{j(2)}^n\right) + I\left(\Omega_{j(3)}; Y_{j(3)}^n\right) + \dots + I\left(\Omega_{j(\lambda_j)}; Y_{j(\lambda_j)}^n\right) + I\left(\Omega_{Y_j}; Y_j^n\right) \\
&\leq I\left(\Omega_{j(1)}; Y_{j(1)}^n, \Omega_{j(2)}, \Omega_{j(3)} \dots, \Omega_{j(\lambda_j)}, \mathbb{M} - \Omega\right) + I\left(\Omega_{j(2)}; Y_{j(2)}^n, \Omega_{j(3)} \dots, \Omega_{j(\lambda_j)}, \mathbb{M} - \Omega\right) \\
&\qquad + I\left(\Omega_{j(3)}; Y_{j(3)}^n, \Omega_{j(4)}, \dots, \Omega_{j(\lambda_j)}, \mathbb{M} - \Omega\right) + \dots + I\left(\Omega_{j(\lambda_j)}; Y_{j(\lambda_j)}^n, \Omega_{Y_j}, \mathbb{M} - \Omega\right) + I\left(\Omega_{Y_j}; Y_j^n, \mathbb{M} - \Omega\right) \\
&= I\left(\Omega_{j(1)}; Y_{j(1)}^n \middle| \Omega_{j(2)}, \Omega_{j(3)} \dots, \Omega_{j(\lambda_j)}, \mathbb{M} - \Omega\right) + I\left(\Omega_{j(2)}; Y_{j(2)}^n \middle| \Omega_{j(3)} \dots, \Omega_{j(\lambda_j)}, \mathbb{M} - \Omega\right) \\
&\qquad + I\left(\Omega_{j(3)}; Y_{j(3)}^n \middle| \Omega_{j(4)}, \dots, \Omega_{j(\lambda_j)}, \mathbb{M} - \Omega\right) + \dots + I\left(\Omega_{j(\lambda_j)}; Y_{j(\lambda_j)}^n \middle| \Omega_{Y_j}, \mathbb{M} - \Omega\right) + I\left(\Omega_{Y_j}; Y_j^n \middle| \mathbb{M} - \Omega\right) \\
&\overset{(a)}{=} I\left(\Omega_{j(1)}; Y_{j(1)}^n \middle| \mathbb{M} - \Omega_{j(1)}\right) + I\left(\Omega_{j(2)}; Y_{j(2)}^n \middle| \mathbb{M} - \cup_{\theta=1}^{2} \Omega_{j(\theta)}\right)
\end{aligned}$$

$$+I\left(\Omega_{j(3)};Y_{j(3)}^n\middle|\mathbb{M}-\cup_{\theta=1}^{3}\Omega_{j(\theta)}\right)+\cdots+I\left(\Omega_{j(\lambda_j)};Y_{j(\lambda_j)}^n\middle|\mathbb{M}-\cup_{\theta=1}^{\lambda_j}\Omega_{j(\theta)}\right)+I\left(\Omega_{Y_j};Y_j^n\middle|\mathbb{M}-\Omega\right)$$

(229)

where equality (a) is derived by definitions (225). Now consider the first term of (229). Based on the definition (225), one derive that $\cup_{\theta=2}^{\lambda_j}\mathbb{M}_{j(\theta)}\cup\mathbb{M}_{Y_j}\cup\mathbb{M}_{c\nrightarrow Y_j}$ is a subset of $\mathbb{M}-\Omega_{j(1)}$. Let $\mathbb{X}^n$ denotes the set of input codewords, i.e., $\{X_i^n : X_i\in\mathbb{X}\}$. We know that $\mathbb{X}^n$ are given by deterministic functions of $\mathbb{M}$. Moreover, the Markov chain $\mathbb{M}\to\mathbb{X}^n\to Y_1^n,Y_2^n,\dots,Y_{K_2}^n$ holds. Thus, we have:

$$\begin{aligned}
I\left(\Omega_{j(1)};Y_{j(1)}^n\middle|\mathbb{M}-\Omega_{j(1)}\right)&=I\left(\mathbb{M};Y_{j(1)}^n\middle|\mathbb{M}-\Omega_{j(1)}\right)\\
&=I\left(\mathbb{X}^n;Y_{j(1)}^n\middle|\mathbb{X}^n_{\cup_{\theta=2}^{\lambda_j}\mathbb{M}_{j(\theta)}\cup\mathbb{M}_{Y_j}\cup\mathbb{M}_{c\nrightarrow Y_j}},\mathbb{M}-\Omega_{j(1)}\right)\\
&=I\left(\mathbb{X}^n-\mathbb{X}^n_{\cup_{\theta=2}^{\lambda_j}\mathbb{M}_{j(\theta)}\cup\mathbb{M}_{Y_j}\cup\mathbb{M}_{c\nrightarrow Y_j}};Y_{j(1)}^n\middle|\mathbb{X}^n_{\cup_{\theta=2}^{\lambda_j}\mathbb{M}_{j(\theta)}\cup\mathbb{M}_{Y_j}\cup\mathbb{M}_{c\nrightarrow Y_j}},\mathbb{M}-\Omega_{j(1)}\right)\\
&\overset{(a)}{\leq}I\left(\mathbb{X}^n-\mathbb{X}^n_{\cup_{\theta=2}^{\lambda_j}\mathbb{M}_{j(\theta)}\cup\mathbb{M}_{Y_j}\cup\mathbb{M}_{c\nrightarrow Y_j}};Y_{j(2)}^n\middle|\mathbb{X}^n_{\cup_{\theta=2}^{\lambda_j}\mathbb{M}_{j(\theta)}\cup\mathbb{M}_{Y_j}\cup\mathbb{M}_{c\nrightarrow Y_j}},\mathbb{M}-\Omega_{j(1)}\right)\\
&=I\left(\mathbb{X}^n;Y_{j(2)}^n\middle|\mathbb{X}^n_{\cup_{\theta=2}^{\lambda_j}\mathbb{M}_{j(\theta)}\cup\mathbb{M}_{Y_j}\cup\mathbb{M}_{c\nrightarrow Y_j}},\mathbb{M}-\Omega_{j(1)}\right)\\
&=I\left(\mathbb{X}^n,\mathbb{M};Y_{j(2)}^n\middle|\mathbb{X}^n_{\cup_{\theta=2}^{\lambda_j}\mathbb{M}_{j(\theta)}\cup\mathbb{M}_{Y_j}\cup\mathbb{M}_{c\nrightarrow Y_j}},\mathbb{M}-\Omega_{j(1)}\right)\\
&=I\left(\mathbb{M};Y_{j(2)}^n\middle|\mathbb{M}-\Omega_{j(1)}\right)\\
&=I\left(\Omega_{j(1)};Y_{j(2)}^n\middle|\mathbb{M}-\Omega_{j(1)}\right)
\end{aligned}$$

(230)

where inequality (a) is due to (221) and its $n$-tuple extension in Lemma 3. Note that according to Lemma 3, (221) is extended as:

$$I\left(\mathbb{X}^n-\mathbb{X}^n_{\cup_{\theta=2}^{\lambda_j}\mathbb{M}_{j(\theta)}\cup\mathbb{M}_{Y_j}\cup\mathbb{M}_{c\nrightarrow Y_j}};Y_{j(1)}^n\middle|\mathbb{X}^n_{\cup_{\theta=2}^{\lambda_j}\mathbb{M}_{j(\theta)}\cup\mathbb{M}_{Y_j}\cup\mathbb{M}_{c\nrightarrow Y_j}},D\right)\leq I\left(\mathbb{X}^n-\mathbb{X}^n_{\cup_{\theta=2}^{\lambda_j}\mathbb{M}_{j(\theta)}\cup\mathbb{M}_{Y_j}\cup\mathbb{M}_{c\nrightarrow Y_j}};Y_{j(2)}^n\middle|\mathbb{X}^n_{\cup_{\theta=2}^{\lambda_j}\mathbb{M}_{j(\theta)}\cup\mathbb{M}_{Y_j}\cup\mathbb{M}_{c\nrightarrow Y_j}},D\right)$$

(231)

for any arbitrary joint PDF on $P_{D\mathbb{X}^n}$. The inequality (a) in (230) is given by setting $D\equiv\left(\mathbb{M}-\Omega_{j(1)}\right)$ in (231). By substituting (230) in (229), we obtain:

$$\begin{aligned}
&n\,\boldsymbol{R}_{\Sigma\Omega}-n\left(\textstyle\sum_{l\in[1:\lambda_j]}\epsilon_{j(l),n}+\epsilon_{j,n}\right)\\
&\leq I\left(\Omega_{j(1)};Y_{j(2)}^n\middle|\mathbb{M}-\Omega_{j(1)}\right)+I\left(\Omega_{j(2)};Y_{j(2)}^n\middle|\mathbb{M}-\cup_{\theta=1}^{2}\Omega_{j(\theta)}\right)\\
&\qquad+I\left(\Omega_{j(3)};Y_{j(3)}^n\middle|\mathbb{M}-\cup_{\theta=1}^{3}\Omega_{j(\theta)}\right)+\cdots+I\left(\Omega_{j(\lambda_j)};Y_{j(\lambda_j)}^n\middle|\mathbb{M}-\cup_{\theta=1}^{\lambda_j}\Omega_{j(\theta)}\right)+I\left(\Omega_{Y_j};Y_j^n\middle|\mathbb{M}-\Omega\right)\\
&=I\left(\Omega_{j(1)},\Omega_{j(2)};Y_{j(2)}^n\middle|\mathbb{M}-\cup_{\theta=1}^{2}\Omega_{j(\theta)}\right)+I\left(\Omega_{j(3)};Y_{j(3)}^n\middle|\mathbb{M}-\cup_{\theta=1}^{3}\Omega_{j(\theta)}\right)\\
&\qquad+\cdots+I\left(\Omega_{j(\lambda_j)};Y_{j(\lambda_j)}^n\middle|\mathbb{M}-\cup_{\theta=1}^{\lambda_j}\Omega_{j(\theta)}\right)+I\left(\Omega_{Y_j};Y_j^n\middle|\mathbb{M}-\Omega\right)
\end{aligned}$$

(232)

By following similar steps successively, using the conditions (221) and (222) one can derive:

$$\begin{aligned}
n\,\boldsymbol{R}_{\Sigma\Omega}-n\left(\textstyle\sum_{l\in[1:\lambda_j]}\epsilon_{j(l),n}+\epsilon_{j,n}\right)&\leq I\left(\Omega_{j(1)},\Omega_{j(2)},\dots,\Omega_{Y_j},\Omega_{j(\lambda_j)},\Omega_{Y_j};Y_j^n\middle|\mathbb{M}-\Omega\right)\\
&=I\left(\Omega;Y_j^n\middle|\mathbb{M}-\Omega\right)\leq\sum_{t=1}^{n}I\left(\Omega;Y_{j,t}\middle|\mathbb{M}-\Omega\right)
\end{aligned}$$

(233)

Applying a standard time-sharing argument, we obtain the desired constraint in (224). The proof is now complete. ■

***Remarks 17:***

1. Note that for each receiver $Y_j, j = 1, \dots, K_2$, the nonempty partitioning $\mathbb{M}_{j(1)}, \mathbb{M}_{j(2)}, \dots, \mathbb{M}_{j(\lambda_j)}$ of the messages $\mathbb{M}_{c\to Y_j} - \mathbb{M}_{Y_j}$ is always available provided that $\mathbb{M}_{c\to Y_j} - \mathbb{M}_{Y_j}$ is nonempty. Moreover, if $\mathbb{M}_{c\to Y_j} - \mathbb{M}_{Y_j}$ is empty, then we have $\mathbb{M}_{c\to Y_j} = \mathbb{M}_{Y_j}$ which implies that the receiver $Y_j$ decodes all its connected messages and there is no need to impose any strong interference condition at this receiver; it indeed receives no interference. Hence, using the general formula given in Theorem 11, we are able to derive strong interference regime for any network with arbitrary topology. After deriving the set of strong interference conditions for each receiver using the general formulas given in Theorem 11, the collection of these conditions constitutes a strong interference regime for the network. Similar to the three-user CIC, by collecting strong interference conditions for different receivers, many of them can usually be removed because they are implied by other stricter conditions.

2. Let us describe the procedure of the derivation of the conditions (221)-(222). Consider a receiver $Y_j, j = 1, \dots, K_2$. According to Definition 4, in the strong interference the receiver is required to decode the messages $\mathbb{M}_{c\to Y_j}$. Among these messages those in $\mathbb{M}_{Y_j}$ are naturally decoded. It remains to consider those in $\mathbb{M}_{c\to Y_j} - \mathbb{M}_{Y_j}$. We find a sequence of receivers (more precisely, a sequence of groups of receivers) $Y_{j(1)}, Y_{j(2)}, \dots, Y_{j(\lambda_j)}$ which decode these messages, i.e., each message in $\mathbb{M}_{c\to Y_j} - \mathbb{M}_{Y_j}$ belongs to one of $\mathbb{M}_{Y_{j(1)}}, \mathbb{M}_{Y_{j(2)}}, \dots, \mathbb{M}_{Y_{j(\lambda_j)}}$. The messages $\mathbb{M}_{c\to Y_j} - \mathbb{M}_{Y_j}$ are partitioned into the subsets $\mathbb{M}_{j(1)}, \mathbb{M}_{j(2)} \dots, \mathbb{M}_{j(\lambda_j)}$ which satisfy the inclusion (220). Now, using Fano's inequality we can write:

$$
\begin{aligned}
& n\,\boldsymbol{R}_{\Sigma\mathbb{M}_{c\to Y_j}} - n\left(\textstyle\sum_{l\in[1:\lambda_j]} \epsilon_{j(l),n} + \epsilon_{j,n}\right) \\
&\quad \le I\left(\mathbb{M}_{j(1)}; Y_{j(1)}^n\right) + I\left(\mathbb{M}_{j(2)}; Y_{j(2)}^n\right) + \cdots + I\left(\mathbb{M}_{j(\lambda_j)}; Y_{j(\lambda_j)}^n\right) + I\left(\mathbb{M}_{Y_j}; Y_j^n\right) \\
&\quad \le I\left(\mathbb{M}_{j(1)}; Y_{j(1)}^n \middle| \mathbb{M}_{c\nrightarrow Y_j}\right) + I\left(\mathbb{M}_{j(2)}; Y_{j(2)}^n \middle| \mathbb{M}_{c\nrightarrow Y_j}\right) + \cdots + I\left(\mathbb{M}_{j(\lambda_j)}; Y_{j(\lambda_j)}^n \middle| \mathbb{M}_{c\nrightarrow Y_j}\right) + I\left(\mathbb{M}_{Y_j}; Y_j^n \middle| \mathbb{M}_{c\nrightarrow Y_j}\right)
\end{aligned}
\tag{234}
$$

Note that the second inequality in (234) is derived by adding the messages $\mathbb{M}_{c\nrightarrow Y_j}$ as virtual side information to all the receivers $Y_{j(1)}, Y_{j(2)}, \dots, Y_{j(\lambda_j)}$ and $Y_j$. Using the procedure described before for the multi-receiver more-capable BC in (185) and also for the three-user CIC in (189), we impose suitable conditions of the form (221)-(222) so that we are able to successively combine the mutual information functions in (234) with the purpose of establishing:

$$
n\,\boldsymbol{R}_{\Sigma\mathbb{M}_{c\to Y_j}} - n\left(\textstyle\sum_{l\in[1:\lambda_j]} \epsilon_{j(l),n} + \epsilon_{j,n}\right) \le I\left(\mathbb{M}_{j(1)}, \dots, \mathbb{M}_{j(\lambda_j)}, \mathbb{M}_{Y_j}; Y_j^n \middle| \mathbb{M}_{c\nrightarrow Y_j}\right) = I\left(\mathbb{M}_{c\to Y_j}; Y_j^n \middle| \mathbb{M}_{c\nrightarrow Y_j}\right)
\tag{235}
$$

After deriving the desired constraint on the sum-rate $\boldsymbol{R}_{\Sigma\mathbb{M}_{c\to Y_j}}$ as given in (235), one can prove all the desired constraints on the partial sum-rates $\boldsymbol{R}_{\Sigma\Omega}$, where $\Omega$ is a subset of $\mathbb{M}_{c\to Y_j}$, without being introduced any new condition. The latter point was also previously demonstrated for two-receiver networks in Theorems 7 and for the three-user CIC in Subsection V.B.1.

3. One can readily show that for the two-receiver networks, the consequence of Theorem 11 is indeed reduced to that of Theorem 8. Although, the proof style of Theorem 8 is different. Recall that for two-receiver networks, we derived the strong interference regime (161) using the general outer bound $\mathfrak{R}_{o:(2)}^{GINTR}$ in (54).

4. The derivations in Theorems 8 and 11 are similar in essence. Theorem 8 was proved using the general outer bound $\mathfrak{R}_{o:(2)}^{GINTR}$ in (54) which itself was derived by the Csiszar-Korner identity. Theorem 11 was derived essentially based on our result in Lemma 3 and this lemma was also proved using the Csiszar-Korner identity.

5. The strong interference regime (177), which was derived in Theorem 10 for the many-to-one CIC using the outer bound (174), can also be concluded from Theorem 11.

6. For the Gaussian network (3), using Lemma 2 and the general formula given in Theorem 11, one can derive explicit constraints on the network gains so that it lies in the strong interference regime as described before in several examples.

## V.B.4) Examples

In this subsection, we provide some additional examples to illustrate our general result in Theorem 11.

***Example 1:*** **An interference network with two transmitters and three receivers**

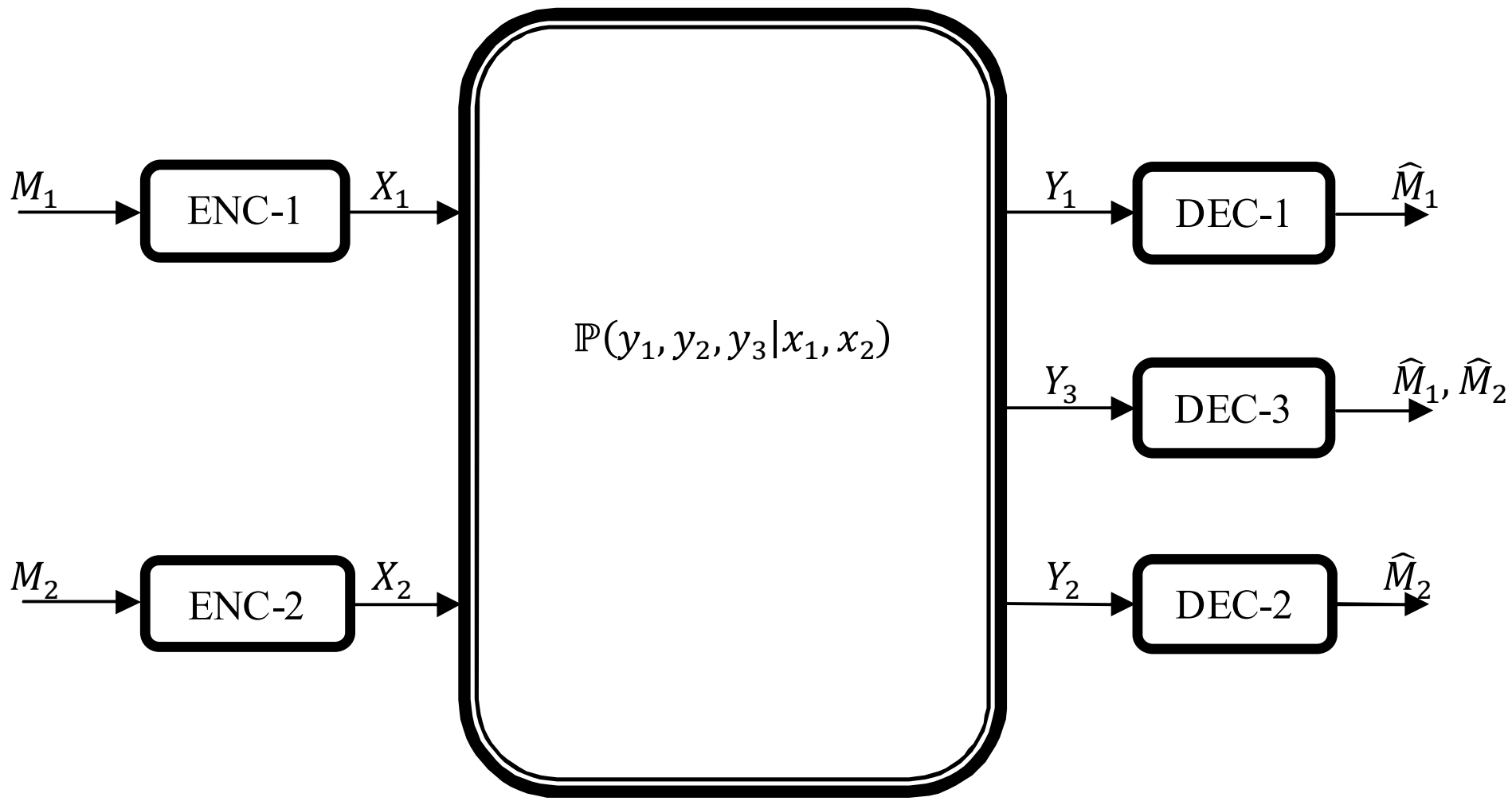

Figure 10. An interference network with two transmitters and three receivers.

Consider the inference network shown in Fig. 10. In this scenario, two transmitters $X_1$ and $X_2$ send the independent messages $M_1$ and $M_2$, respectively, to three receivers $Y_1$, $Y_2$ and $Y_3$. The receiver $Y_1$ decodes $M_1$, the receiver $Y_2$ decodes $M_2$ and the receiver $Y_3$ decodes both messages. It is clear that this network includes the two-user CIC as a special case. The receiver $Y_3$ experiences no interference because it decodes both messages. Similar to the two-user CIC, one can readily deduce that if the following conditions hold:

$$\begin{cases} I(X_1;Y_1|X_2) \leq I(X_1;Y_2|X_2) \\ I(X_2;Y_2|X_1) \leq I(X_2;Y_1|X_1) \end{cases} \quad \text{for all joint PDFs} \quad P_{X_1}P_{X_2} \tag{236}$$

then, the network lies in the strong interference regime. In fact, these conditions indicate that to settle the interference experienced by the receiver $Y_2$ at a strong level, the amount of information sent to the receiver $Y_1$ is taken as a reference, and vice versa. However, for the network in Fig. 10 other choices are also possible. Clearly, we can adapt the amount of information sent to the receiver $Y_3$ as a reference to measure the interference level at each of the receivers $Y_1$ and $Y_2$. In other words, using the general formula given in Theorem 11, one can derive the following strong interference regime, as well:

$$\begin{cases} I(X_1;Y_3|X_2) \leq I(X_1;Y_2|X_2) \\ I(X_2;Y_3|X_1) \leq I(X_2;Y_1|X_1) \end{cases} \quad \text{for all joint PDFs} \quad P_{X_1}P_{X_2} \tag{237}$$

As a combination of the above strategies, we can also present the following regimes:

$$\begin{cases} I(X_1;Y_3|X_2) \leq I(X_1;Y_2|X_2) \\ I(X_2;Y_2|X_1) \leq I(X_2;Y_1|X_1) \end{cases} \quad \text{for all joint PDFs} \quad P_{X_1}P_{X_2} \tag{238}$$

$$\begin{cases} I(X_1;Y_1|X_2) \leq I(X_1;Y_2|X_2) \\ I(X_2;Y_3|X_1) \leq I(X_2;Y_1|X_1) \end{cases} \quad \text{for all joint PDFs} \quad P_{X_1}P_{X_2} \tag{239}$$

Thus, we have obtained four different strong interference regimes for the network in Fig. 10. In general, for an arbitrary interference network, if there exists a receiver which decodes all messages, then one can adapt the amount of information sent to this receiver as a reference to settle the interference experienced by all other receivers at a strong level. This fact is formulated in the following corollary.

***Corollary 4)*** *Consider the general interference network with $K_1$ transmitters and $K_2$ receivers as shown in Fig. 1. Suppose that the receiver $Y_{K_2}$ decodes all messages, i.e., $\mathbb{M}_{Y_{K_2}} = \mathbb{M}$. Let the transition probability function of the network satisfies the following conditions:*

$$I\left(\mathbb{X}-\mathbb{X}_{\mathbb{M}_{Y_j}\cup\mathbb{M}_{c\nrightarrow Y_j}};Y_{K_2}\middle|\mathbb{X}_{\mathbb{M}_{Y_j}\cup\mathbb{M}_{c\nrightarrow Y_j}}\right)\leq I\left(\mathbb{X}-\mathbb{X}_{\mathbb{M}_{Y_j}\cup\mathbb{M}_{c\nrightarrow Y_j}};Y_j\middle|\mathbb{X}_{\mathbb{M}_{Y_j}\cup\mathbb{M}_{c\nrightarrow Y_j}}\right),\quad j=1,\dots,K_2-1 \tag{240}$$

*for all joint PDFs $P_{\mathbb{X}-\mathbb{X}_{\mathbb{M}_{Y_j}\cup\mathbb{M}_{c\nrightarrow Y_j}}}\prod_{X_i\in\mathbb{X}_{\mathbb{M}_{Y_j}\cup\mathbb{M}_{c\nrightarrow Y_j}}}P_{X_i}$. Under these conditions, the network lies in the strong interference regime.*

*Proof of Corollary 4)* Consider the general formula given in Theorem 11. For each $j\in[1:K_2]$, select $\lambda_j=1$ and $j(1)=\{K_2\}$. Since $\mathbb{M}_{Y_{K_2}}=\mathbb{M}$, the inclusions (220) are satisfied. By this choice, we derive (240) from (222). ■

***Example 2:*** **The Multiple-Access-Interference Networks (MAIN)**

Let us discuss the multi-receiver MAIN which is composed of multiple interfering MACs. These networks were introduced in Part II of our multi-part papers [2, Sec. IV]. Specifically, consider an interference network with $K_1=\sum_{l=1}^{K_2}\mu_l$ transmitters and $K_2$ receivers, where $\mu_1,\dots,\mu_{K_2}$ are arbitrary natural numbers. The transmitters are partitioned into $K_2$ sets labeled $\mathbb{X}_1,\dots,\mathbb{X}_{K_2}$ such that those in $\mathbb{X}_j=\{X_{j,1},\dots,X_{j,\mu_j}\}$ send the messages $\mathbb{M}_{Y_j}$ to the receiver $Y_j, j=1,\dots,K_2$, but they have no message for the other receivers; in other words, the message sets $\mathbb{M}_{Y_1},\dots,\mathbb{M}_{Y_{K_2}}$ are pairwise disjoint. Also, the distribution of messages $\mathbb{M}_{Y_j}$ among the transmitters $\mathbb{X}_j=\{X_{j,1},\dots,X_{j,\mu_j}\}$ is arbitrary. The network model has been shown in Fig. 11.

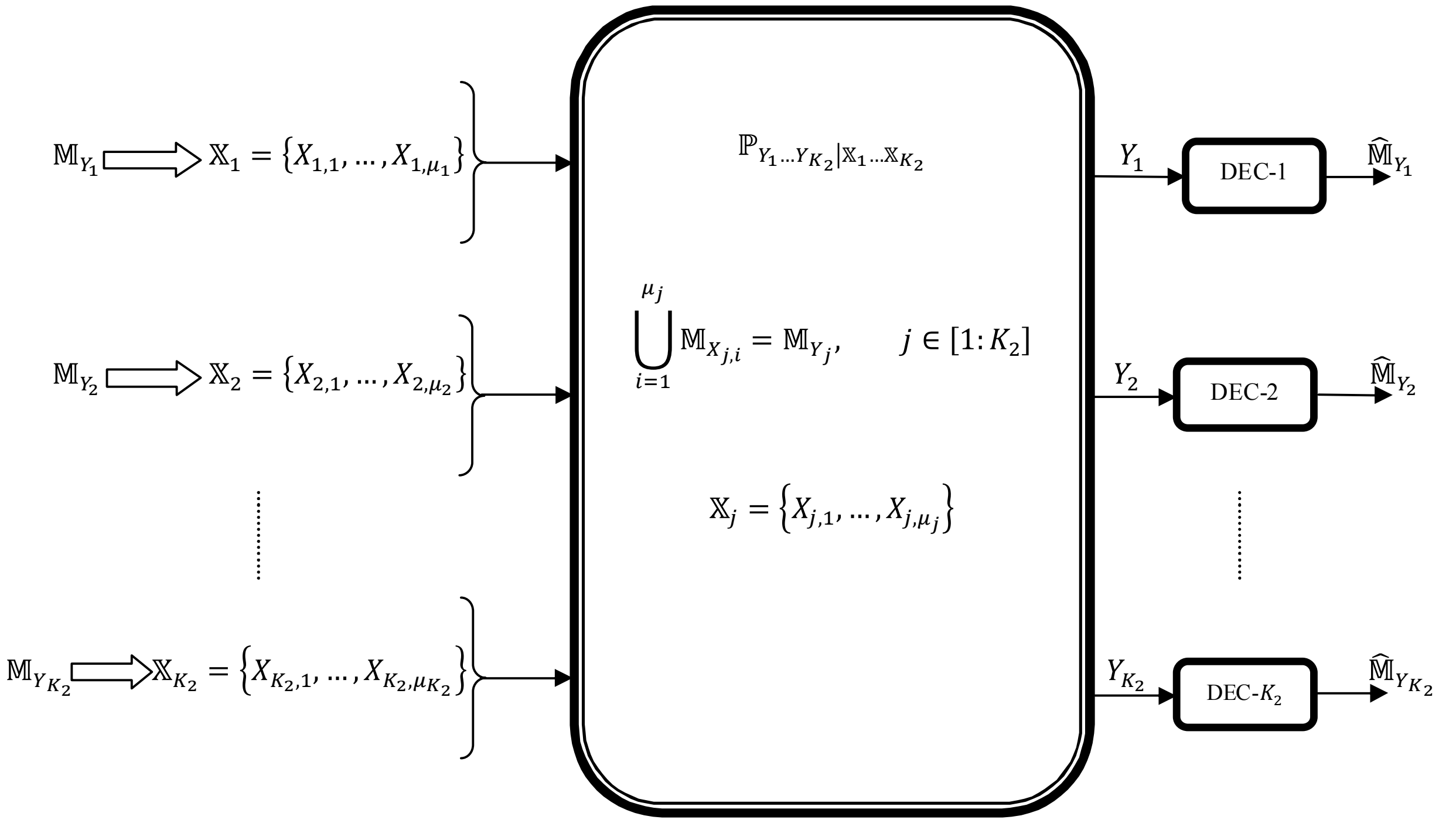

Figure 11. The general Multiple-Access-Interference Networks (MAIN). For $j=1,\dots,K_2$, $\mathbb{X}_j$ denotes a set of arbitrary transmitters which send (in an arbitrary order) the messages $\mathbb{M}_{Y_j}$ to the receiver $Y_j$.

These scenarios do not contain broadcasting messages to multiple receivers because each transmitter sends information to only one receiver. Note that when $\mu_1=\mu_2=\dots=\mu_{K_2}=1$, the MAIN is reduced to a $K_2$-user CIC. These networks indeed behave similarly to the multi-user CICs from many aspects. Specifically, one can readily derive strong interference regimes for such scenarios inspired by those for the CICs with the same number of receivers. For example, we can directly write the following strong interference regime for the fully connected MAIN in Fig. 11:

$$\begin{cases} I(\mathbb{X}_2,\mathbb{X}_3,\mathbb{X}_4,\dots,\mathbb{X}_{K_2};Y_{K_2}|\mathbb{X}_1) \leq I(\mathbb{X}_2,\mathbb{X}_3,\mathbb{X}_4,\dots,\mathbb{X}_{K_2};Y_1|\mathbb{X}_1), & \text{for all joint PDFs} \quad P_{\mathbb{X}_2\mathbb{X}_3\mathbb{X}_4\dots\mathbb{X}_{K_2}}\prod_{X_i\in\mathbb{X}_1}P_{X_i} \\ I(\mathbb{X}_1,\mathbb{X}_3,\mathbb{X}_4,\dots,\mathbb{X}_{K_2};Y_1|\mathbb{X}_2) \leq I(\mathbb{X}_1,\mathbb{X}_3,\mathbb{X}_4,\dots,\mathbb{X}_{K_2};Y_2|\mathbb{X}_2), & \text{for all joint PDFs} \quad P_{\mathbb{X}_1\mathbb{X}_3\mathbb{X}_4\dots\mathbb{X}_{K_2}}\prod_{X_i\in\mathbb{X}_2}P_{X_i} \\ \vdots \\ \vdots \\ I(\mathbb{X}_1,\mathbb{X}_2,\dots,\mathbb{X}_{K_2-1};Y_{K_2-1}|\mathbb{X}_{K_2}) \leq I(\mathbb{X}_1,\mathbb{X}_2,\dots,\mathbb{X}_{K_2-1};Y_{K_2}|\mathbb{X}_{K_2}), & \text{for all joint PDFs} \quad P_{\mathbb{X}_1\mathbb{X}_2\dots\mathbb{X}_{K_2-1}}\prod_{X_i\in\mathbb{X}_{K_2}}P_{X_i} \end{cases}$$

(241)

This regime is obtained inspired by that given in (216) for the multi-user CIC. In fact, similar to the CIC, the above conditions are sufficient to prove the desired constraints on the sum-rate, i.e.,

$$\boldsymbol{R}_{\Sigma\mathbb{M}_{Y_1}} + \boldsymbol{R}_{\Sigma\mathbb{M}_{Y_2}} + \dots + \boldsymbol{R}_{\Sigma\mathbb{M}_{Y_{K_2}}} \leq I\left(\mathbb{M}_{Y_1},\mathbb{M}_{Y_2},\dots,\mathbb{M}_{Y_{K_2}};Y_j\right), \qquad j = 1,\dots,K_2$$

Moreover, once we derived the necessary constraints on the sum-rate, then we are able to prove all desired partial sum-rates without being introduced any new condition, as discussed earlier. The strong interference regime (241) can be also extracted from the general formula given in Theorem 11. A remarkable point regarding the strong interference regime (241) for the MAIN is that it does not depend on the arrangement of the messages $\mathbb{M}_{Y_j}$ among the transmitters $\mathbb{X}_j$, $j = 1,2,\dots,K_2$. This is also the case for other strong interference regimes which are obtained for the MAIN using the general formula given in Theorem 11. To derive such strong interference regimes, we do not need to know the arrangement of messages among transmitters.

***Example 3:*** **Generalized Cyclic Z-Interference Network**

As a special case of the general MAIN in Fig. 11, consider a network for which the transition probability function is decomposed as follows:

$$\mathbb{P}_{Y_1\dots Y_{K_2}|X_1\dots X_{K_1}} = \mathbb{P}_{Y_1|\mathbb{X}_1\mathbb{X}_2}\mathbb{P}_{Y_2|\mathbb{X}_2\mathbb{X}_3}\dots\mathbb{P}_{Y_{K_2}|\mathbb{X}_{K_2}\mathbb{X}_1}$$

(242)

In this scenario, each MAC causes interference to only one receiver. Fig. 12 depicts the network model. In this figure, each receiver has been linked to its connected transmitters by a dashed line. These networks are indeed a natural generalization of the cyclic Z-interference channels previously studied in [29]; hence, we refer to them as the *generalized cyclic Z-interference networks*.

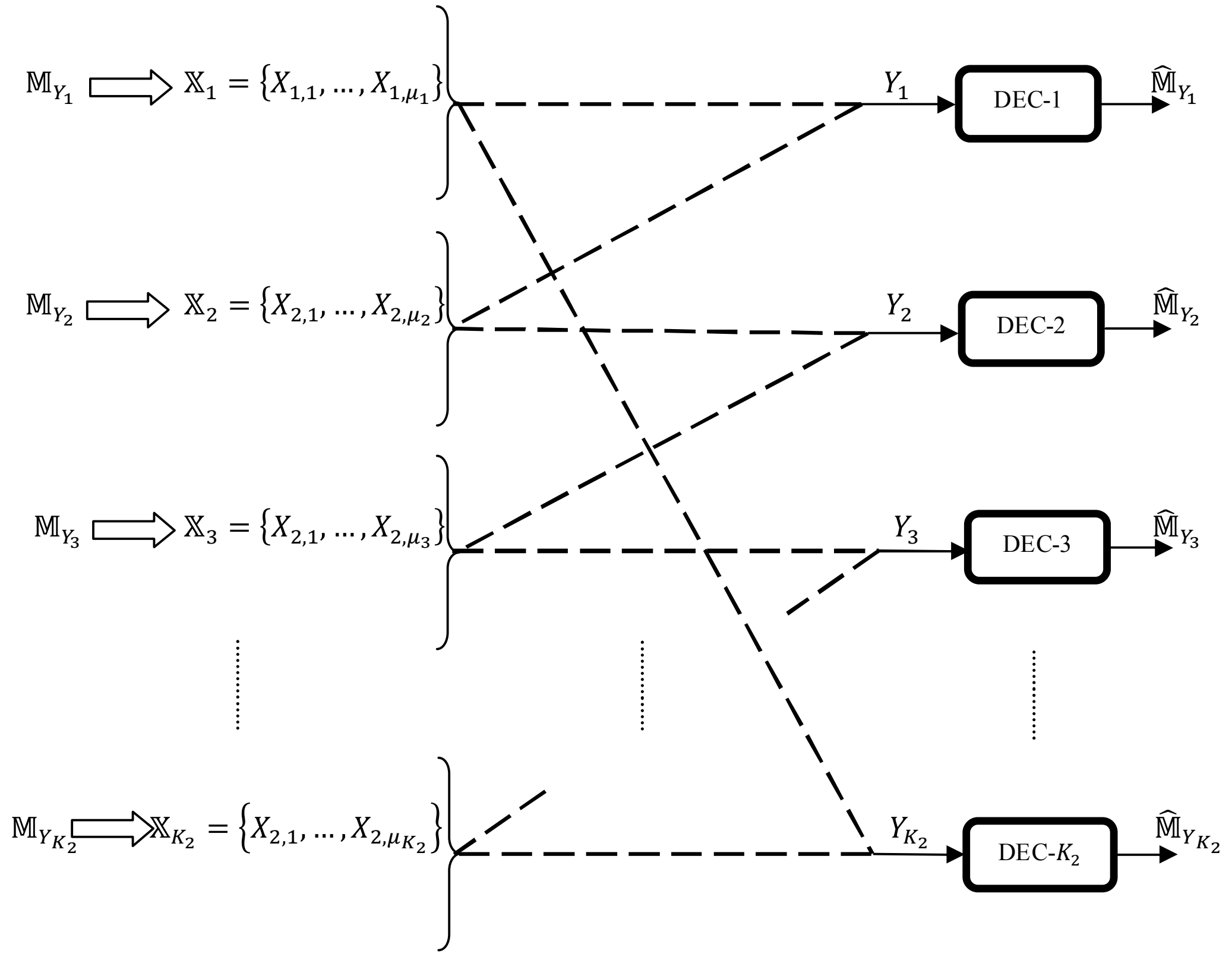

Figure 12. The Generalized Cyclic Z-Interference Network. For $i = 1, \dots, K_2$, $\mathbb{X}_i$ denotes a set of arbitrary transmitters which send (in an arbitrary order) the messages $\mathbb{M}_{Y_j}$ to the receiver $Y_j$.

For the generalized cyclic Z-interference network in Fig. 12, the strong interference conditions (241) cannot be satisfied. This point was discussed in details for the special case of cyclic Z-interference channel (204) in Subsection V.B.1. Nonetheless, using the general expression given in Theorem 11, one can readily derive the following strong interference regime for this network:

$$\begin{cases} I(\mathbb{X}_2; Y_2|\mathbb{X}_1, \mathbb{X}_3, \mathbb{X}_4, \dots, \mathbb{X}_{K_2}) \leq I(\mathbb{X}_2; Y_1|\mathbb{X}_1, \mathbb{X}_3, \mathbb{X}_4, \dots, \mathbb{X}_{K_2}), & \text{for all PDFs} \quad P_{\mathbb{X}_2} \prod_{X_i \in \mathbb{X} - \mathbb{X}_2} P_{X_i} \\ I(\mathbb{X}_3; Y_3|\mathbb{X}_1, \mathbb{X}_2, \mathbb{X}_4, \dots, \mathbb{X}_{K_2}) \leq I(\mathbb{X}_3; Y_2|\mathbb{X}_1, \mathbb{X}_2, \mathbb{X}_4, \dots, \mathbb{X}_{K_2}), & \text{for all PDFs} \quad P_{\mathbb{X}_3} \prod_{X_i \in \mathbb{X} - \mathbb{X}_3} P_{X_i} \\ \vdots & \\ I(\mathbb{X}_1; Y_1|\mathbb{X}_2, \mathbb{X}_3, \mathbb{X}_4, \dots, \mathbb{X}_{K_2}) \leq I(\mathbb{X}_1; Y_{K_2}|\mathbb{X}_2, \mathbb{X}_3, \mathbb{X}_4, \dots,, \mathbb{X}_{K_2}), & \text{for all PDFs} \quad P_{\mathbb{X}_1} \prod_{X_i \in \mathbb{X} - \mathbb{X}_1} P_{X_i} \end{cases} \tag{243}$$

Under the conditions (243), the optimal coding strategy is that the receiver $Y_j, j = 1, \dots, K_2 - 1$ jointly decodes the messages $\mathbb{M}_{Y_j}$ and $\mathbb{M}_{Y_{j+1}}$ and the receiver $Y_{K_2}$ decodes the messages $\mathbb{M}_{Y_{K_2}}$ and $\mathbb{M}_{Y_1}$.

***Example 3:*** **One-to-Many Interference Network**

Lastly, let discuss another special case of the MAINs in Fig. 11. Namely, consider the case where the transition probability function of the network is factorized as follows:

$$\mathbb{P}_{Y_1 \dots Y_{K_2}|X_1 \dots X_{K_1}} = \mathbb{P}_{Y_1|\mathbb{X}_1} \mathbb{P}_{Y_2|\mathbb{X}_1\mathbb{X}_2} \mathbb{P}_{Y_3|\mathbb{X}_1\mathbb{X}_3} \dots \mathbb{P}_{Y_{K_2}|\mathbb{X}_1\mathbb{X}_{K_2}} \tag{244}$$

In this special MAIN, the interference is caused by only one group of transmitters. Fig. 13 depicts the network model where each receiver has been linked to its connected transmitters by a dashed line. These networks are indeed a natural generalization of the so-called one-to-many interference channels [28]; hence, we refer to them as the *one-to-many interference networks*.

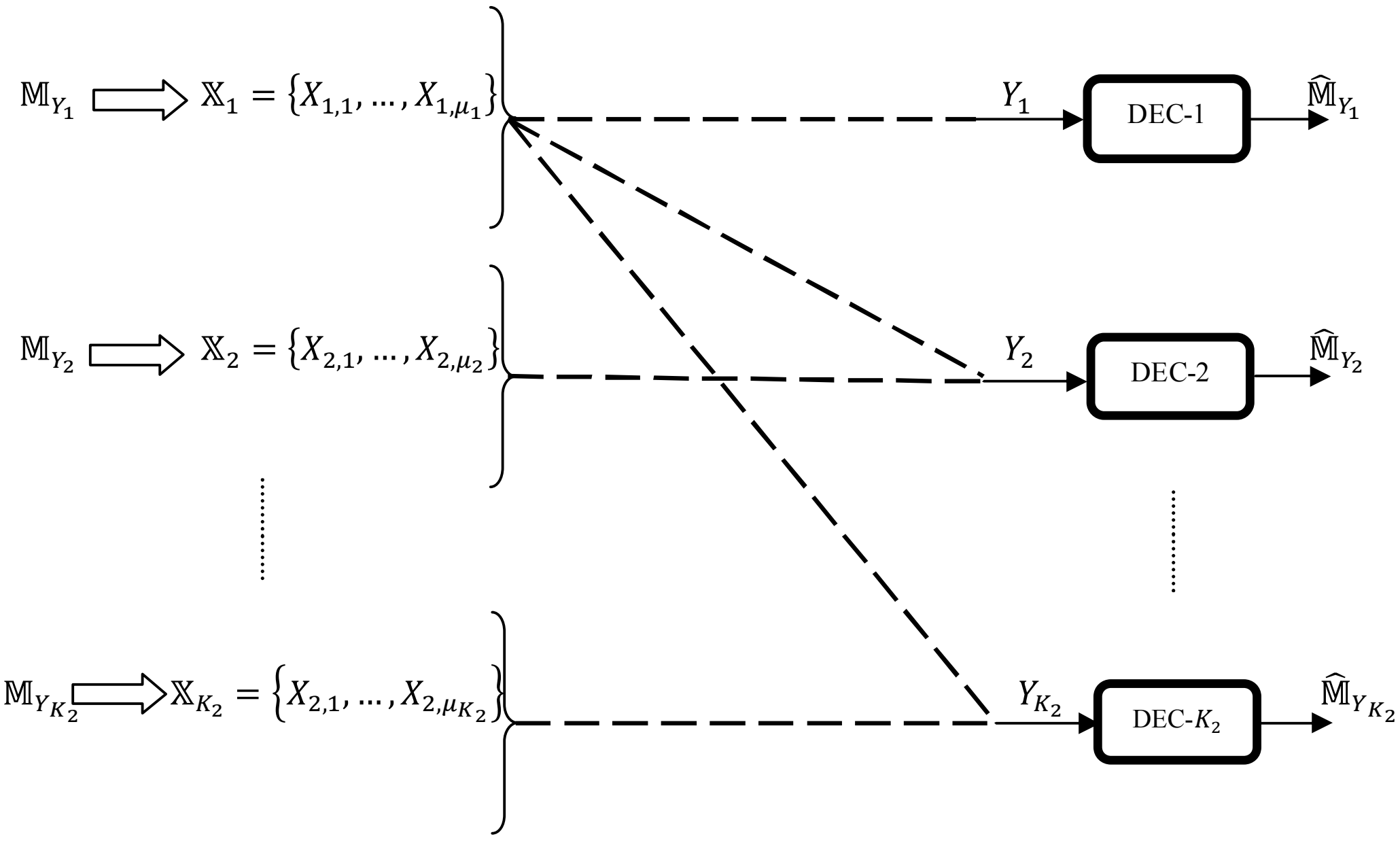

Figure 13. The One-to-Many Interference Network. For $i = 1, \dots, K_2$, $\mathbb{X}_i$ denotes a set of arbitrary transmitters which send the messages $\mathbb{M}_{Y_j}$ to the receiver $Y_j$.

By definition, in the strong interference regime for this network each receiver is required to decode its own messages as well as the messages $\mathbb{M}_{Y_1}$. Thus, to settle the interference experienced by the receivers at a strong level, the amount of information sent to the first receiver should be considered as a reference. Precisely, using the general formula given in Theorem 11, one can derive the following strong interference regime for this network:

$$I(\mathbb{X}_1; Y_1|\mathbb{X}_2, \mathbb{X}_3, \mathbb{X}_4, \dots, \mathbb{X}_{K_2}) \leq \min \begin{Bmatrix} I(\mathbb{X}_1; Y_2|\mathbb{X}_2, \mathbb{X}_3, \mathbb{X}_4, \dots, \mathbb{X}_{K_2}) \\ I(\mathbb{X}_1; Y_3|\mathbb{X}_2, \mathbb{X}_3, \mathbb{X}_4, \dots, \mathbb{X}_{K_2}) \\ \vdots \\ I(\mathbb{X}_1; Y_{K_2}|\mathbb{X}_2, \mathbb{X}_3, \mathbb{X}_4, \dots, \mathbb{X}_{K_2}) \end{Bmatrix} \quad \text{for all joint PDFs} \quad P_{\mathbb{X}_1} \prod_{X_i \in \mathbb{X} - \mathbb{X}_1} P_{X_i} \tag{245}$$

Note that, using the factorization (244), the above strong interference condition can be simplified as follows:

$$I(\mathbb{X}_1; Y_1) \leq \min \begin{Bmatrix} I(\mathbb{X}_1; Y_2|\mathbb{X}_2) \\ I(\mathbb{X}_1; Y_3|\mathbb{X}_3) \\ \vdots \\ I(\mathbb{X}_1; Y_{K_2}|\mathbb{X}_{K_2}) \end{Bmatrix} \quad \text{for all joint PDFs} \quad P_{\mathbb{X}_1} \prod_{X_i \in \mathbb{X} - \mathbb{X}_1} P_{X_i} \tag{246}$$

For the special case of the Gaussian one-to-many CIC, one can readily re-derive the result of [28, Sec. V] from the condition (246).

## V.B.5) Discussion

In the previous subsection, we derived strong interference conditions for the general interference network under which the optimal strategy is that each receiver decodes all its connected messages. The rate region due to this scheme is given by (223). As discussed before, such characterization for this rate region is useful to derive the general expressions (221)-(222); nonetheless, it involves numerous constraints many of which are redundant. The number of the constraints of the rate region (223) grows very rapidly with the size of the network.

Here, we intend to present an alternative characterization which is considerably simpler. The key idea is the use of the superposition coding technique. Recall that in Part II [2, Sec. II.A], we derived a simple characterization of the capacity region for the MAC with any arbitrary distribution of messages among transmitters based on the superposition coding scheme. To describe the achievability scheme, we presented the MACCM[6] graphs which illustrate the superposition structures among the generated codewords. We refer the reader to Part II [2, Sec. II.A] for the procedure of constructing the MACCM graphs, the corresponding probability law, and concepts such as cloud center and satellite messages. Inspired by the MACCM graphs, we also presented [2, Fig. 15] the MACCM plan of messages for any arbitrary interference network with a given distribution of messages. Let us briefly review the construction of these plans. Consider an arbitrary interference network with the message sets $\mathbb{M},\mathbb{M}_{X_i}, i=1,\dots,K_1$, and $\mathbb{M}_{Y_j}, j=1,\dots,K_2$ as shown in Fig. 1. Each subset of transmitters sends at most one message to each subset of receivers. There exist $K_1$ transmitters and $K_2$ receivers. Therefore, we can label each message by a nonempty subset of $\{1,\dots,K_1\}$ to denote which transmitters send the message, also a nonempty subset of $\{1,\dots,K_2\}$ to determine to which subset of receivers the message is sent. We represent each message of $\mathbb{M}$ as $M_\Delta^\nabla$, where $\Delta\subseteq\{1,\dots,K_1\}$ and $\nabla\subseteq\{1,\dots,K_2\}$. For example, $M_{\{1,2,3\}}^{\{2,4\}}$ indicates a message which is sent by transmitters 1, 2 and 3 to receivers 2 and 4.

Now, for each $\Delta\subseteq\{1,\dots,K_1\}$ we define:

$$\mathbb{M}_\Delta \triangleq \left\{M_\Delta^\nabla \in \mathbb{M} : \nabla\subseteq\{1,\dots,K_2\}\right\} \tag{247}$$

Using this representation, we arrange the messages into a MACCM graph-like illustration as shown in Fig. 14. This illustration is called the MACCM plan of messages. This plan includes $K_1$ columns so that the sets $\mathbb{M}_\Delta, \Delta\subseteq\{1,\dots,K_1\}$ with $\|\Delta\|=i$ are situated in its $i^{th}$ column, $i=1,\dots,K_1$. Also, the set $\mathbb{M}_{\Delta_1}$ in column $i, i=2,\dots,K_1$, is connected to the set $\mathbb{M}_{\Delta_2}$ in column $i-1$ provided that $\Delta_2\subseteq\Delta_1$. Please refer to Part II [2] for details.

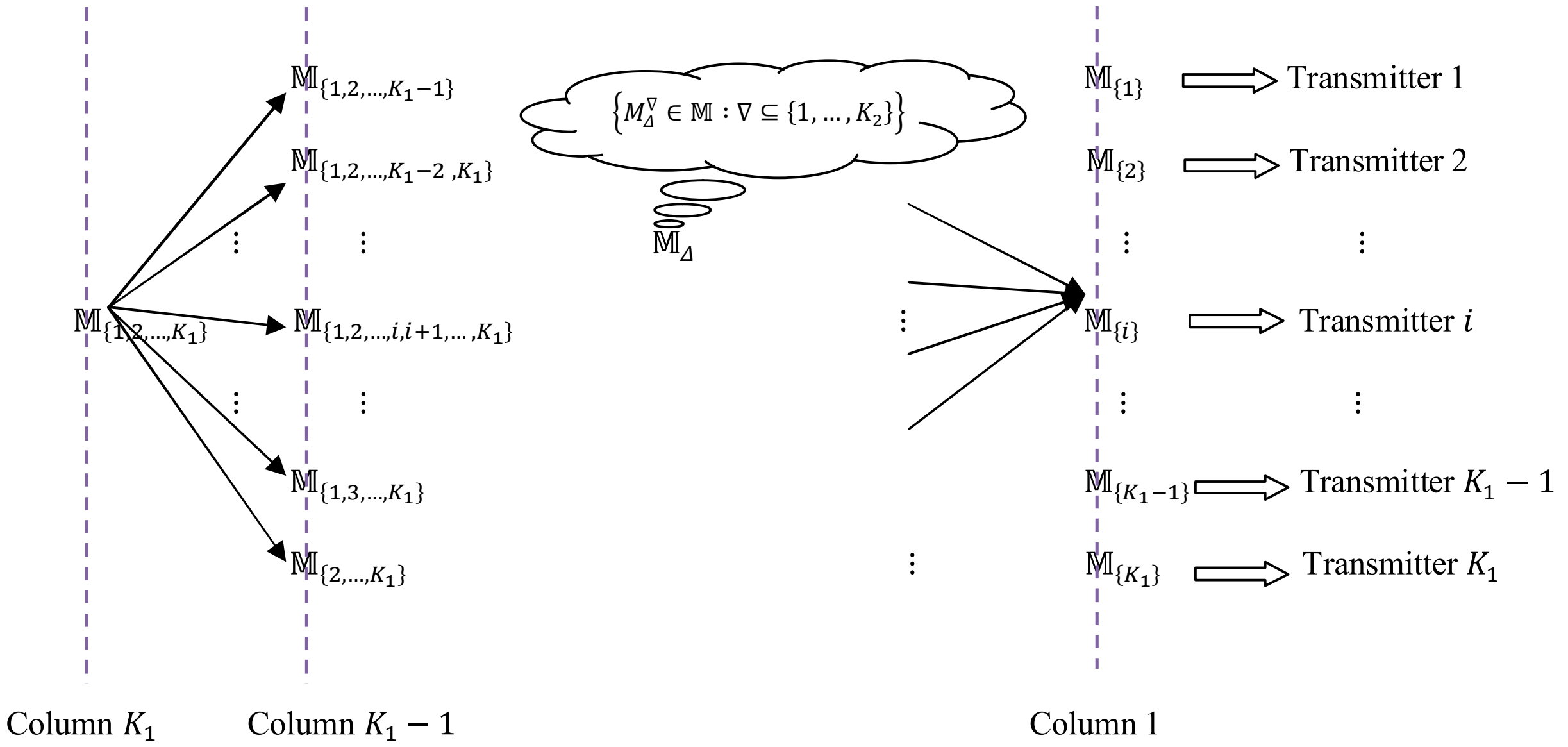

Figure 14. The MACCM plan of messages for an arbitrary interference Network.

Now, using the MACCM plan of messages we present a simple characterization for the rate region (223). Recall that in the achievability scheme leading to the rate region (223), all the messages are encoded separately using independent codewords and each receiver jointly decodes its connected messages. Let us present an alternative scheme based on superposition coding strategy. Assume that $\Delta$ is an arbitrary nonempty subset of $\{1,\dots,K_1\}$. According to the MACCM plan, the messages $\mathbb{M}_\Delta$ given in (247) are broadcasted by the transmitters $X_i, i\in\Delta$, meanwhile, no transmitter other than those in $\{X_i, i\in\Delta\}$ has access to these messages.

***Observation 2:*** Consider the general interference network in Fig. 1. If a message $M_\Delta^\nabla$ belonging to the set $\mathbb{M}_\Delta$ in (247) is connected to a certain receiver, then all other messages in $\mathbb{M}_\Delta$ are also connected to that receiver, as well. Moreover, if all the messages belonging to $\mathbb{M}_\Delta$ are unconnected to a receiver, then all the messages belonging to every $\mathbb{M}_{\bar{\bar{\Delta}}}$ with $\bar{\bar{\Delta}}\subseteq\Delta$ are also unconnected to that receiver.

[6] Multiple Access Channel with Common Messages

*Proof of Observation 2)* Let $M_\Delta^\nabla$ be connected to the receiver $Y_j, j \in [1:K_2]$. According to Definition 2, there is at least one transmitter $X_i, i \in [1:K_1]$ such that $X_i$ is connected to $Y_j$, i.e., $X_i \in \mathbb{X}_{c\to Y_j}$, and $M_\Delta^\nabla \in \mathbb{M}_{X_i}$. Moreover, since $M_\Delta^\nabla \in \mathbb{M}_{X_i}$, the index $i$ belongs to $\Delta$. Thus, $\mathbb{M}_\Delta$ is a subset of $\mathbb{M}_{X_i}$ and thereby, all the messages in $\mathbb{M}_\Delta$ are connected to the receiver $Y_j$.

Next, let all the messages in $\mathbb{M}_\Delta$ be unconnected to the receiver $Y_j$ and let $\bar{\bar{\Delta}}$ be a subset of $\Delta$. Consider a message $M_{\bar{\bar{\Delta}}}^\nabla$ belonging to $\mathbb{M}_{\bar{\bar{\Delta}}}$ and a transmitter $X_i, i \in [1:K_1]$ where $M_{\bar{\bar{\Delta}}}^\nabla \in \mathbb{M}_{X_i}$. Therefore, $i \in \bar{\bar{\Delta}}$ and since $\bar{\bar{\Delta}}$ is a subset of $\Delta$, we have $i \in \Delta$. Thus, $\mathbb{M}_\Delta$ is a subset of $\mathbb{M}_{X_i}$. This implies that the transmitter $X_i$ is unconnected to the receiver $Y_j$ because the messages $\mathbb{M}_\Delta$ are unconnected to this receiver. Therefore, $M_{\bar{\bar{\Delta}}}^\nabla$ does not belong to $\bigcup_{X_i \in \mathbb{X}_{c\to Y_j}} \mathbb{M}_{X_i}$ and hence it is unconnected to $Y_j$. ■

Observation 2 indicates that in the coding scheme leading to the rate region (223), for a given $\mathbb{M}_\Delta$, each receiver either decodes all the messages in $\mathbb{M}_\Delta$ or decodes no message belonging to this set. This leads us to a fact that to achieve the rate region (223), the messages in $\mathbb{M}_\Delta$ all can be embedded into a single codeword instead of being separately encoded using independent codeowords. Moreover, we can apply the superposition coding scheme. Specifically, consider the MACCM encoding graph shown in Fig. 15.

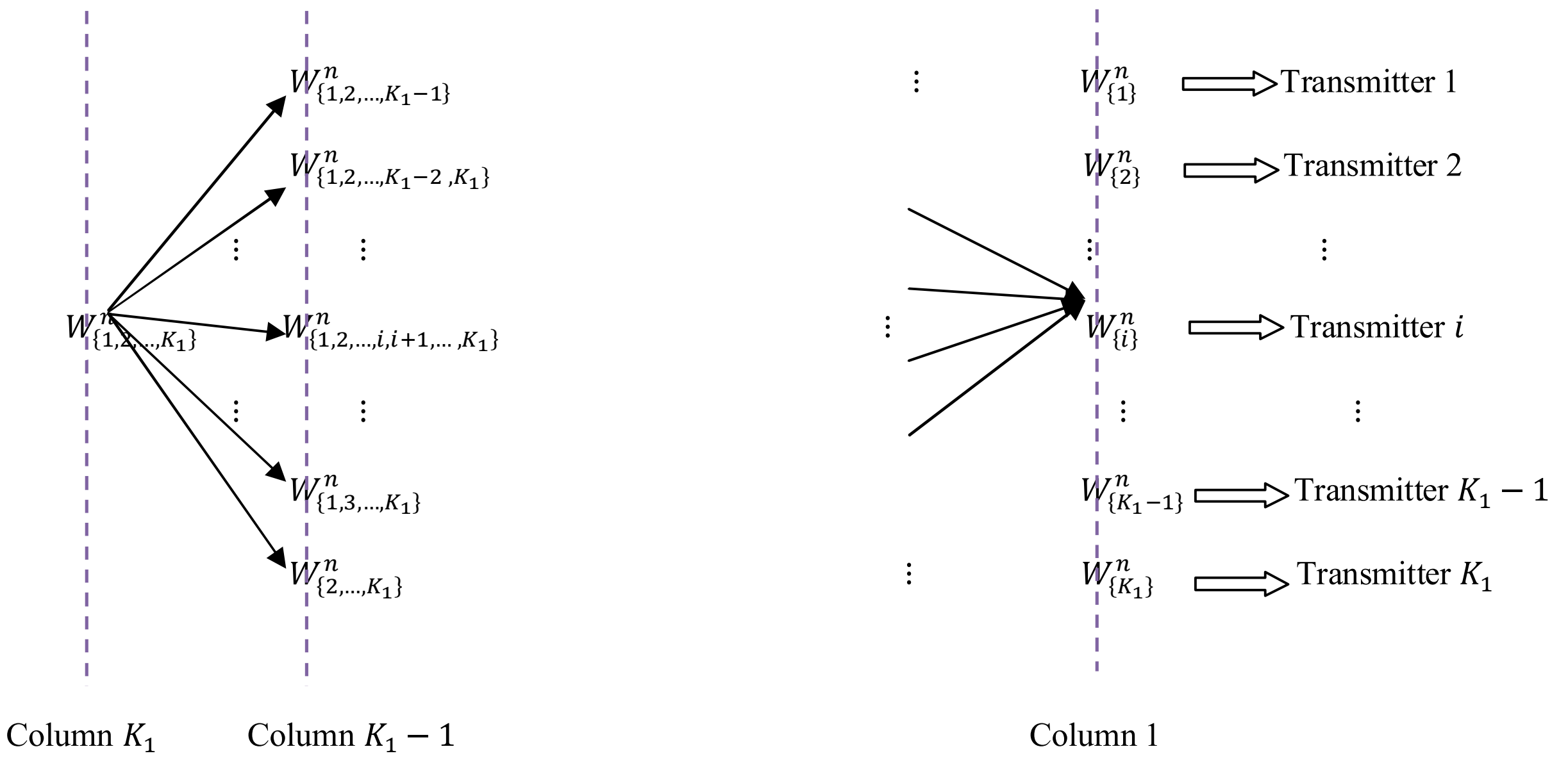

Figure 15. The encoding graph representing the superposition scheme for encoding the messages of the MACCM plan in Fig. 14: The messages in $\mathbb{M}_\Delta$ are embeded into the codeword $W_\Delta^n$. Every two codewords which are connected by a directed edge constitute a superposition structure.

This encoding graph is derived by replacing each set $\mathbb{M}_\Delta$ in the MACCM plan with a codeword $W_\Delta^n$, $\Delta \subseteq \{1,\ldots,K_1\}$. It represents a coding scheme (of length-$n$ where $n$ is a natural number) where its codewords $W_\Delta^n$, $\Delta \subseteq \{1,\ldots,K_1\}$ are generated based on the MACCM encoding probability law (see Part II [2, Sec. II.A]). Specifically, every two codewords in this graph which are connected by a direct edge build a superposition structure. This encoding graph can be directly translated into a scheme based on the superposition strategy to encode the messages in the MACCM plan. Clearly, all the messages $\mathbb{M}_\Delta$ are embedded into the codeword $W_\Delta^n$. Considering this point, the encoding scheme is completely determined based on the encoding graph. Note that in this scheme the rate with respect to the codeword $W_\Delta^n$, $\Delta \subseteq \{1,\ldots,K_1\}$, which is denoted by $\boldsymbol{R}_\Delta$, is given as follows:

$$\boldsymbol{R}_\Delta = \sum_{M_\Delta^\nabla \in \mathbb{M}_\Delta} R_{M_\Delta^\nabla} \tag{248}$$

where $R_{M_\Delta^\nabla}$ is the rate with respect to the message $M_\Delta^\nabla \in \mathbb{M}_\Delta$. Also, note that if the set $\mathbb{M}_\Delta$ is empty, then the respective codeword $W_\Delta^n$ in the encoding graph is nullified. On the other side, each receiver decodes all its connected codewords (a codeword $W_\Delta^n$ is connected to a receiver if $\mathbb{M}_\Delta$ in the MACCM plan is connected to that receiver) using a jointly typical decoder. Before presenting the resultant achievable rate region by this scheme, let us define a right-sided message set.

***Definition 7:*** *Consider the MACCM plan of messages in Fig. 14. A subset of* $\{\mathbb{M}_\Delta : \mathbb{M}_\Delta \subseteq \mathbb{M}\}$*, e.g.,* $\mathbb{Y}$ *is said to be right-sided provided that:*

$$\text{if} \quad \mathbb{M}_\Delta \in \mathbb{Y} \quad \Longrightarrow \quad \forall \; \bar{\bar{\Delta}} \subseteq \Delta : \quad \mathbb{M}_{\bar{\bar{\Delta}}} \in \mathbb{Y} \tag{249}$$

From the viewpoint of encoding graph, a subset of codewords is right-sided provided that all the codewords which are satellite for at least one codeword of that subset belong to the subset as well.

According to the above terminology, the rate region due to the superposition coding scheme shown in Fig. 15 is given by:

$$\bigcup_{\mathcal{P}^{MACCM}} \left\{ \begin{array}{l} \left(\dots, R_{M_\Delta^\nabla}, \dots\right) \in \mathbb{R}_+^{\|\mathbb{M}\|} : \\ \forall j \in [1:K_2] : \\ \forall \mathbb{Y} \subseteq \left\{\mathbb{M}_\Delta : \mathbb{M}_\Delta \subseteq \mathbb{M}_{c\to Y_j}\right\} : \qquad \mathbb{Y} \;\; \text{is} \;\; \text{right}-\text{sided} \\ \displaystyle\sum_{\Delta:\, \mathbb{M}_\Delta \in \mathbb{Y}} \boldsymbol{R}_\Delta \leq I\left(\{W_\Delta\}_{\Delta:\, \mathbb{M}_\Delta \in \mathbb{Y}}; Y_j \middle| \{W_\Delta\}_{\Delta:\, \mathbb{M}_\Delta \in \mathbb{M}_{c\to Y_j}} - \{W_\Delta\}_{\Delta:\, \mathbb{M}_\Delta \in \Omega}, Q\right) \end{array} \right\} \tag{250}$$

where $\mathcal{P}^{MACCM}$ denotes the set of all joint PDFs given by the MACCM encoding graph probability law (the auxiliary random variable $Q$ is the time-sharing parameter). Note that $\left(\dots, R_{M_\Delta^\nabla}, \dots\right)$ in (250) represents a $\|\mathbb{M}\|$-tuple belonging to $\mathbb{R}_+^{\|\mathbb{M}\|}$ each component of which stands for the rate with respect to one of the messages in $\mathbb{M}$. The rate region (250) is equivalent to that in (223) because both coincide with the capacity region of the network in which each receiver is required to decode all its connected messages. However, the characterization (250) is considerably simpler than (223) because it is described by a substantially fewer number of constraints.

Let us again consider the rate regions (223) and (250). These regions are derived by requiring that each receiver correctly decodes all its connected messages. For example, for the three-user CIC they lead to the achievable rate region (186). As discussed in Subsection V.B.1, such achievable rate regions are optimal in the strong interference regime; however, some of their constraints are still redundant. For example, for the three-user CIC in the strong interference regime, the rate region (186) is reduced to that in (198). In fact, some of the constraints in either (223) or (250) are actually required to *correctly decode* non-corresponding messages at the receivers. Such constrains can be removed from the achievable rate region because for the interference network the desired purpose is that each receiver decodes only its corresponding messages. By removing these unnecessary constraints, one may derive an achievable rate region for the general interference networks which is (but not in the strong interference regime) larger than those in (250) and (223). This result is given in the next theorem.

***Theorem 12)*** *Consider the general interference network in Fig. 1. The following rate region is achievable:*

$$\bigcup_{\mathcal{P}^{MACCM}} \left\{ \begin{array}{l} \left(\dots, R_{M_\Delta^\nabla}, \dots\right) \in \mathbb{R}_+^{\|\mathbb{M}\|} : \\ \forall j \in [1:K_2] : \\ \forall \mathbb{Y} \subseteq \left\{\mathbb{M}_\Delta : \mathbb{M}_\Delta \subseteq \mathbb{M}_{c\to Y_j}\right\} : \left\{ \begin{array}{l} \mathbb{Y} \;\; \text{is} \;\; \text{right}-\text{sided} \\ \exists \; \mathbb{M}_\Delta \in \mathbb{Y} \quad \text{s.t.} \;\; \mathbb{M}_\Delta \cap \mathbb{M}_{Y_j} \neq \emptyset \end{array} \right. \\ \displaystyle\sum_{\Delta:\, \mathbb{M}_\Delta \in \mathbb{Y}} \boldsymbol{R}_\Delta \leq I\left(\{W_\Delta\}_{\Delta:\, \mathbb{M}_\Delta \in \mathbb{Y}}; Y_j \middle| \{W_\Delta\}_{\Delta:\, \mathbb{M}_\Delta \in \mathbb{M}_{c\to Y_j}} - \{W_\Delta\}_{\Delta:\, \mathbb{M}_\Delta \in \Omega}, Q\right) \end{array} \right\} \tag{251}$$

*Proof of Theorem 12)* The achievability scheme again is derived using the superposition coding. The messages are encoded based on the MACCM encoding graph in Fig. 15. On the other side, each receiver jointly decodes its connected codewords; however, here each receiver is interested in correct decoding of only its corresponding messages. In fact, those constraints of (250) which are not given in (251) are regarded to correct decoding of non-corresponding messages at the receivers and thereby they can be removed. ■

***Remarks 18:***

1. For the three-user CIC, the rate region (251) is reduced to that in (198) discussed in details in Subsection V.B.1.
2. In general, the achievable rate region (251) is larger than that in (250). However, for the strong interference regime both coincide and yield the capacity region. Therefore, (251) is an alternative characterization of the capacity region in the strong interference regime which is described by a substantially fewer number of constraints than either (250) or (223).

In fact, both achievable rate regions (250) and (251) are derived by applying jointly decoding technique at the receivers. However, to achieve (250), each receiver correctly decodes all its connected messages; while to achieve (251) each receiver correctly decodes only its own messages. It should be noted that although in the strong interference regime both (250) and (251) coincide; nonetheless, in general (251) may be strictly larger than (250). For example, in Subsection V.B.1, we demonstrated that the rate region (198) for the three-user CIC satisfying the conditions (212) is optimal while (186) may not be strictly sub-optimal. According to Definition 4, a network is in the strong interference regime if the achievable rate region (250) achieves its capacity. For such networks the interference experienced by each receiver is indeed *decodable*. There exist scenarios for which the achievable rate region (250) is not optimal (and hence they do not lie in the strong interference regime) but (251) is optimal. In other words, in these scenarios the interference experienced by the receivers is not perfectly decodable; however, the capacity is still achieved by applying jointly decoding technique at each receiver with the purpose of correctly decoding only its corresponding messages. This motivates us to present the following general definition.

***Definition 8:*** **Almost Decodable Interference Regime**

*An interference network is said to be in the almost-decodable interference regime when the achievable rate region* (251) *derived in Theorem 12 is optimal and yields the capacity.*

According to Definitions 4 and 8 a network that is in the strong interference regime, is also in the almost-decodable interference regime; however, the inverse is not necessarily true. For example, the three-user CIC in (212) is in the almost decodable interference regime but it may not be in the strong interference regime.

In Theorem 11, we presented a general formula to identify networks with strong interference. How one can treat the almost-decodable interference regime? In response to this question, we must state that the procedure developed in Subsection V.B.1 to derive the almost-decodable interference regime (212) for the three-user could be adapted to all other interference networks. The essential features of this procedure are reviewed below:

1. We impose certain conditions on the achievable rate region (251) so that some of its constraints are redundant and can be relaxed. Usually these conditions include inequalities such as (207) which should be satisfied for all input distributions that achieve the boundary points of the rate region that is obtained by relaxing the non-desired constraints.
2. By following the same lines as the derivation of the strong interference conditions (221)-(222), we explore for those conditions under which one can derive the remaining constraints of the rate region (251) using Fano's inequality. The conditions due to this step are usually of the strong interference type conditions.
3. Under the collection of the conditions imposed in steps 1 and 2, we obtain the capacity region.

By the above approach, for any given network topology, one can identify several classes for which the almost-decodable achievable rate region (251) is optimal.

Let us again concentrate on the strong interference conditions given in Theorem 11. These conditions are are represented by inequalities of the following form:

$$I(\mathbb{X}-\mathbb{X}_B;Y_1|\mathbb{X}_B)\leq I(\mathbb{X}-\mathbb{X}_B;Y_2|\mathbb{X}_B) \tag{252}$$

which should be satisfied for all joint PDFs $P_{\mathbb{X}-\mathbb{X}_B}\prod_{X_i\in\mathbb{X}_B}P_{X_i}$. According to Lemma 1, these conditions extend to any arbitrary joint PDF on the inputs $\mathbb{X}$. For certain networks, actually it is not required that such conditions hold for all input distributions. For example, in Theorem 1, we proved that the CRC in Fig. 3 has strong interference provided that (45) and (46) hold. As discussed in Remark 4, to have the equality (46), it is sufficient that the inequality (50) is satisfied only for those input distributions that achieve the boundary points of the rate region in the right side of the equality (46). Many other networks could be identified such that if we restrict some of the conditions in Theorem 11 to hold for only specific input distributions, they still have strong interference.

In some scenarios, to some extent one may relax strong interference conditions such as (252) by introducing an auxiliary random variable. We conclude this subsection by providing an example on this issue. Consider the two-receiver MAIN shown in Fig. 16.

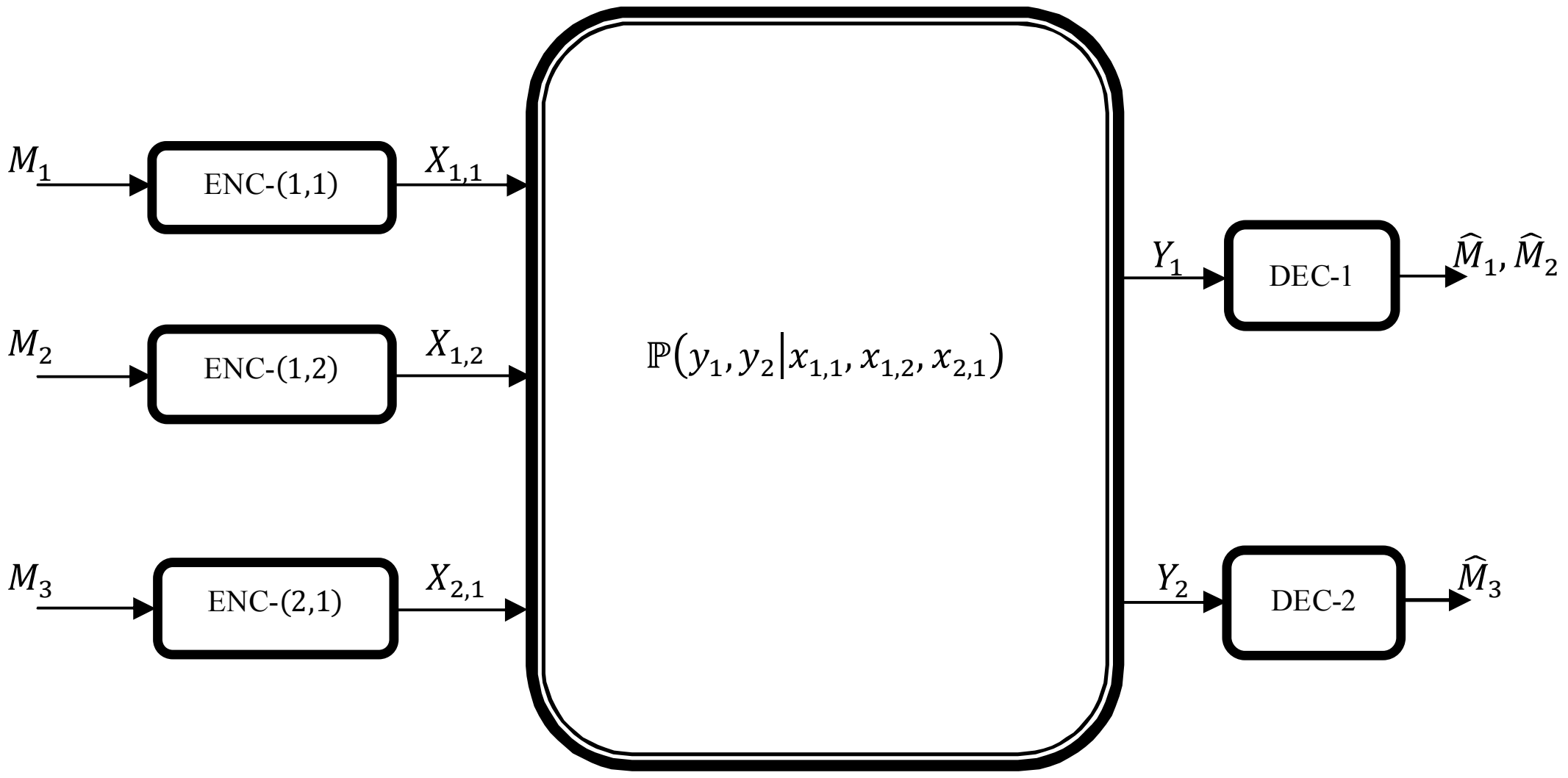

Figure 16. A two-receiver MAIN.

This network is derived by setting $\mu_1 = 2, \mu_1 = 1, \mathbb{M}_{Y_1} = \{M_1, M_2\}$ and $\mathbb{M}_{Y_2} = \{M_3\}$ in the two-receiver MAIN in Fig. 1. Considering (156), it is clear that a strong interference regime for this network is given by:

$$\begin{cases} I(X_{1,1},X_{1,2};Y_1|X_{2,1}) \leq I(X_{1,1},X_{1,2};Y_2|X_{2,1}) & \text{for all} \quad P_{X_{1,1}X_{1,2}}P_{X_{2,1}} \\ I(X_{2,1};Y_2|X_{1,1},X_{1,2}) \leq I(X_{2,1};Y_1|X_{1,1},X_{1,2}) & \text{for all} \quad P_{X_{1,1}}P_{X_{1,2}}P_{X_{2,1}} \end{cases} \tag{253}$$

According to Lemma 1, the conditions (253) imply that:

$$\begin{cases} I(X_{1,1},X_{1,2};Y_1|X_{2,1},D) \leq I(X_{1,1},X_{1,2};Y_2|X_{2,1},D) \\ I(X_{2,1};Y_2|X_{1,1},X_{1,2},D) \leq I(X_{2,1};Y_1|X_{1,1},X_{1,2},D) \end{cases}, \quad \text{for all joint PDFs} \quad P_{DX_{1,1}X_{1,2}X_{2,1}} \tag{254}$$

Now, we intend to show that under conditions weaker than (253), the network still lies in the strong interference regime. First, let present an outer bound on the capacity region of this network. Using $\mathfrak{R}_{o:(2)}^{GINTR}$ in (54), one can readily derive the following outer bound on the capacity region:

$$\bigcup_{\substack{P_Q P_{X_{1,1}|Q} P_{X_{1,2}|Q} P_{X_{2,1}|Q} \\ \times P_{U_1 U_2 V|X_{1,1}X_{1,2}X_{2,1}Q}}} \left\{ \begin{array}{l} (R_1,R_2,R_3) \in \mathbb{R}_+^3 : \\ R_1 \leq \min \left\{ \begin{array}{c} I(X_{1,1};Y_1|X_{1,2},X_{2,1},Q), \\ I(X_{1,1};Y_1|U_2,V,X_{1,2},X_{2,1},Q) + I(U_2,V;Y_2|X_{1,2},X_{2,1},Q) \end{array} \right\} \\ R_2 \leq \min \left\{ \begin{array}{c} I(X_{1,2};Y_1|X_{1,1},X_{2,1},Q), \\ I(X_{1,2};Y_1|U_1,V,X_{1,1},X_{2,1},Q) + I(U_1,V;Y_2|X_{1,1},X_{2,1},Q) \end{array} \right\} \\ R_3 \leq \min \left\{ \begin{array}{c} I(X_{2,1};Y_2|X_{1,1},X_{1,2},Q), \\ I(X_{2,1};Y_2|U_1,U_2,X_{1,1},X_{1,2},Q) + I(U_1,U_2;Y_1|X_{1,1},X_{1,2},Q) \end{array} \right\} \\ R_1 + R_2 \leq \min \left\{ \begin{array}{c} I(X_{1,1},X_{1,2};Y_1|X_{2,1},Q), \\ I(X_{1,1},X_{1,2};Y_1|V,X_{2,1},Q) + I(V;Y_2|X_{2,1},Q) \end{array} \right\} \\ R_2 + R_3 \leq \min \left\{ \begin{array}{c} I(X_{1,2};Y_1|U_1,V,X_{1,1},X_{2,1},Q) + I(U_1,V,X_{2,1};Y_2|X_{1,1},Q), \\ I(X_{2,1};Y_2|U_1,U_2,X_{1,1},X_{1,2},Q) + I(U_1,U_2,X_{1,2};Y_1|X_{1,1},Q) \end{array} \right\} \\ R_1 + R_3 \leq \min \left\{ \begin{array}{c} I(X_{1,1};Y_1|U_2,V,X_{1,2},X_{2,1},Q) + I(U_2,V,X_{2,1};Y_2|X_{1,2},Q), \\ I(X_{2,1};Y_2|U_1,U_2,X_{1,1},X_{1,2},Q) + I(U_1,U_2,X_{1,1};Y_1|X_{1,2},Q) \end{array} \right\} \\ R_1 + R_2 + R_3 \leq I(X_{1,1},X_{1,2};Y_1|V,X_{2,1},Q) + I(V,X_{2,1};Y_2|Q) \\ R_1 + R_2 + R_3 \leq I(X_{2,1};Y_2|U_1,U_2,X_{1,1},X_{1,2},Q) + I(U_1,U_2,X_{1,1},X_{1,2};Y_1|Q) \end{array} \right\} \tag{255}$$

This outer bound is derived by identifying $U_1 \equiv (Z, M_1), U_2 \equiv (Z, M_2)$ and $V \equiv (Z, M_3)$ in (54). Now consider the following conditions:

$$\begin{cases} I(X_{1,1}, X_{1,2}; Y_1|X_{2,1}, D) \leq I(X_{1,1}, X_{1,2}; Y_2|X_{2,1}, D) \\ I(X_{2,1}; Y_2|X_{1,1}, X_{1,2}, D) \leq I(X_{2,1}; Y_1|X_{1,1}, X_{1,2}, D) \end{cases}, \quad \text{for all joint PDFs} \quad P_{X_{1,1}} P_{X_{1,2}} P_{X_{2,1}} P_{D|X_{1,1}X_{1,2}X_{2,1}} \tag{256}$$

It is clear that the conditions (254) imply those in (256); however, the inverse is not true because (256) holds for only product PDFs on the inputs $X_{1,1}, X_{1,2}$, and $X_{2,1}$. On the one hand, since (253) and (254) are equivalent (according to Lemma 1), one may deduce that (253) also imply (256) and again the inverse is not true. Therefore, the conditions (256) in general are weaker than those in (253). Nonetheless, we now prove that under the conditions (256) the MAIN of Fig.16 still has strong interference, i.e., the decoding of all messages at each receiver is optimal. First, note that the condition (256) readily extends as:

$$\begin{cases} I(X_{1,1}, X_{1,2}; Y_1|X_{2,1}, D, Q) \overset{\text{(i)}}{\leq} I(X_{1,1}, X_{1,2}; Y_2|X_{2,1}, D, Q) \\ I(X_{2,1}; Y_2|X_{1,1}, X_{1,2}, D, Q) \overset{\text{(ii)}}{\leq} I(X_{2,1}; Y_1|X_{1,1}, X_{1,2}, D, Q) \end{cases}, \quad \text{for all joint PDFs} \quad P_Q P_{X_{1,1}|Q} P_{X_{1,2}|Q} P_{X_{2,1}|Q} P_{D|X_{1,1}X_{1,2}X_{2,1}Q} \tag{257}$$

Using the conditions (257), we can deduce that:

$$\begin{cases} I(X_{1,1}, X_{1,2}; Y_1|V, X_{2,1}, Q) \overset{(a)}{\leq} I(X_{1,1}, X_{1,2}; Y_2|V, X_{2,1}, Q) \\ I(X_{1,1}; Y_1|U_2, V, X_{1,2}, X_{2,1}, Q) \overset{(b)}{\leq} I(X_{1,1}; Y_2|U_2, V, X_{1,2}, X_{2,1}, Q) \\ I(X_{1,2}; Y_1|U_1, V, X_{1,1}, X_{2,1}, Q) \overset{(c)}{\leq} I(X_{1,2}; Y_2|U_1, V, X_{1,1}, X_{2,1}, Q) \\ I(X_{2,1}; Y_2|U_1, U_2, X_{1,1}, X_{1,2}, Q) \overset{(d)}{\leq} I(X_{2,1}; Y_1|U_1, U_2, X_{1,1}, X_{1,2}, Q) \end{cases}, \quad P_Q P_{X_{1,1}|Q} P_{X_{1,2}|Q} P_{X_{2,1}|Q} P_{U_1 U_2 V|X_{1,1}X_{1,2}X_{2,1}Q} \tag{258}$$

The inequality (a) is derived by setting $D \equiv V$ in (257~i), inequality (b) is derived by setting $D \equiv (U_2, V, X_{1,2})$ in (257~i), inequality (c) is derived by setting $D \equiv (U_1, V, X_{1,1})$ in (257~i), and inequality (d) is derived by setting $D \equiv (U_1, U_2)$ in (257~ii). Next, by substituting the inequalities (258) in (255), we obtain a rate region which is indeed the capacity region when both receivers are required to decode both messages. Therefore, under the conditions (256) the network has strong interference, as was claimed.

***Remarks 19:***

1. The strong interference conditions (253) hold also for the case where the transmitters $X_{1,1}$ and $X_{2,1}$ send information to the receiver $Y_1$ in any other order; for example, for the case where these transmitters cooperatively send a common message to $Y_1$. However, the same argument is not true regarding the conditions (256) because these conditions represents only the situation where $X_{1,1}$ and $X_{2,1}$ are independent. Thus, the conditions of the type (253) are more flexible to treat strong interference regime for large topologies. Another advantage of these conditions is that they do not include any auxiliary random variable so that for a given network it is more tractable to verify whether the conditions are satisfied or not.
2. Strong interference conditions such as (256) could be also identified for other networks. Specifically, one may derive similar conditions for the multi-user CICs.
3. The outer bound (255) for the network in Fig. 16 can be easily made tighter by introducing several other constraints. For example, one can also write:

   $$R_1 \leq I(U_1, X_{1,1}; Y_1|Q)$$

   However, here our main purpose was to determine the strong interference regime (256) for the network.

# VI. A Random Coding Scheme for the BCCR with Common Message

In Subsections IV.A, we established unified outer bounds on the capacity region of the general interference networks with two-receivers. We demonstrated the capability of these outer bounds to obtain explicit capacity results for the networks in the strong interference regime as well as the networks with less noisy receivers. But these are not all the results one can achieve. Our outer bounds are indeed very efficient to derive capacities for other scenarios such as more-capable networks and semi-deterministic networks. To demonstrate this fact, we concentrate on the BCCR with common message shown in Fig. 5. We present an achievability scheme for this network based on our systematic view developed in our previous paper [7] presented in [8]. Then, we specialize the derived achievable rate region for a sub-network of the BCCR, called as the *Cognitive Radio Channel with Common Message*, and show that it coincides with the outer bound given in (95) for a class of more-capable channels.

Let us once more look at the BCCR shown in Fig. 5. As we see, this network contains both the two-user BC with common message and the two-user CIC as special cases. It is clear that for such a large network, one can propose numerous achievability schemes. But the question is that what is the best coding? Indeed, here we intend to present a glimpse of our general random coding scheme in Part V [5] (also presented in [8]). We develop a unique achievability scheme for the interference networks with any arbitrary distribution of messages among transmitters and receivers. This achievability scheme is derived by a systematic combination of the best coding strategies for the MAC and the BC. Moreover, we provide a graphical illustration for our coding strategy using the MACCM graphs developed in Part II [2]. In fact, our graphical approach is a powerful tool not only to describe a given achievability scheme for a certain network but also to design achievability schemes with satisfactory performance for large networks. The BCCR in Fig. 5, which contains all the basic interference networks studied in Part I [1], is a suitable example to explain our approach in this regard. In the next theorem, we present the achievability scheme for the BCCR and then deal with about its capabilities. We note that to analyze the error probability of the coding scheme, we enjoy a novel application of a multivariate covering lemma proved in [15, page 15-40][7] by which the necessary conditions for vanishing error probability in the encoding steps are readily derived. Also, in the decoding steps, we make use of our general expression developed in [7, page. 12, Eq. 41-42] for analyzing the error probability of incorrect decoding of corresponding messages at a given receiver. This general expression holds for any memoryless network.

***Theorem 13)*** *Consider the BCCR with common message in Fig. 5. Define the rate region* $\mathfrak{R}_i^{BCCR_{cm}}$ *as follows:*

$$
\mathfrak{R}_i^{BCCR_{cm}} \triangleq \bigcup_{\mathcal{P}^{BCCR}} \left\{
\begin{array}{l}
(R_0,R_1,R_2)\in\mathbb{R}_+^3:\ \exists(B_0,B_1,B_2,R_0,R_{10},R_{11},R_{20},R_{22})\in\mathbb{R}_+^8:\\
R_1=R_{10}+R_{11},\qquad R_2=R_{20}+R_{22},\\
\qquad\qquad\qquad\qquad ,\\
\hfill B_0>I(U_1,V_2;W_B|W_1,W_2,Q)\triangleq\eta_0\\
\hfill B_0+B_1>\eta_0+I(V_2;U_B|U_1,W_1,W_2,W_B,Q)\triangleq\eta_0+\eta_1\\
\hfill B_0+B_2>\eta_0+I(U_1;V_B|V_2,W_1,W_2,W_B,Q)\triangleq\eta_0+\eta_2\\
\hfill B_0+B_1+B_2>\eta_0+\eta_1+\eta_2+\underbrace{I(U_B;V_B|U_1,V_2,W_1,W_2,W_B,Q)}_{\triangleq\eta_4}\\
\hfill \theta_1\triangleq I(U_1;W_B|W_1,W_2,Q),\\
\hfill R_{11}+B_1<I_{Y_1}^1+\theta_1\triangleq I(U_1,U_B;Y_1|W_1,W_2,W_B,Q)+\theta_1\\
\hfill R_0+B_0+B_1<I_{Y_1}^2+\theta_1\triangleq I(W_B,U_B;Y_1|W_1,W_2,U_1,Q)+\theta_1\\
\hfill R_{20}+R_0+B_0+B_1<I_{Y_1}^3+\theta_1\triangleq I(W_2,W_B,U_B;Y_1|W_1,U_1,Q)+\theta_1\\
\hfill R_0+B_0+R_{11}+B_1<I_{Y_1}^4+\theta_1\triangleq I(U_1,W_B,U_B;Y_1|W_1,W_2,Q)+\theta_1\\
\hfill R_{20}+R_0+B_0+R_{11}+B_1<I_{Y_1}^5+\theta_1\triangleq I(U_1,W_2,W_B,U_B;Y_1|W_1,Q)+\theta_1\\
\hfill R_{10}+R_0+B_0+R_{11}+B_1<I_{Y_1}^6+\theta_1\triangleq I(W_1,U_1,W_B,U_B;Y_1|W_2,Q)+\theta_1\\
\hfill R_{10}+R_{20}+R_0+B_0+R_{11}+B_1<I_{Y_1}^7+\theta_1\triangleq I(W_1,U_1,W_2,W_B,U_B;Y_1|Q)+\theta_1\\
\hfill \theta_2\triangleq I(V_2;W_B|W_1,W_2,Q)\\
\hfill R_{22}+B_2<I_{Y_2}^1+\theta_2\triangleq I(V_2,V_B;Y_2|W_1,W_2,W_B,Q)+\theta_2\\
\hfill R_0+B_0+B_2<I_{Y_2}^2+\theta_2\triangleq I(W_B,V_B;Y_2|W_1,W_2,V_2,Q)+\theta_2\\
\hfill R_{10}+R_0+B_0+B_2<I_{Y_2}^3+\theta_2\triangleq I(W_1,W_B,V_B;Y_2|W_2,V_2,Q)+\theta_2\\
\hfill R_0+B_0+R_{22}+B_2<I_{Y_2}^4+\theta_2\triangleq I(V_2,W_B,V_B;Y_2|W_1,W_2,Q)+\theta_2\\
\hfill R_{10}+R_0+B_0+R_{22}+B_2<I_{Y_2}^5+\theta_2\triangleq I(W_1,V_2,W_B,V_B;Y_2|W_2,Q)+\theta_2\\
\hfill R_{20}+R_0+B_0+R_{22}+B_2<I_{Y_2}^6+\theta_2\triangleq I(W_2,V_2,W_B,V_B;Y_2|W_1,Q)+\theta_2\\
\hfill R_{10}+R_{20}+R_0+B_0+R_{22}+B_2<I_{Y_2}^7+\theta_2\triangleq I(W_1,V_2,W_2,W_B,V_B;Y_2|Q)+\theta_2
\end{array}
\right\}
\tag{259}
$$

[7] It was proved in [15] to obtain an admissible source region for the two-user BC.

*where $\mathcal{P}^{BCCR}$ denotes the set of all joint PDFs $P_{QW_1U_1W_2V_2W_BU_BV_BX_1X_2X_B}(q,w_1,u_1,w_2,v_2,w_B,u_B,v_B,x_1,x_2,x_B)$ which are factorized as:*

$$P_Q P_{W_1U_1X_1|Q} P_{W_2V_2X_2|Q} P_{W_BU_BV_B|W_1U_1W_2V_2Q} P_{X_B|X_1X_2V_BU_BW_BU_1W_1V_2W_2Q} \tag{260}$$

*The set $\mathfrak{R}_i^{BCCR_{cm}}$ constitutes an inner bound on the capacity region.*

*Proof of Theorem 13)* We derive $\mathfrak{R}_i^{BCCR_{cm}}$ using a random coding technique. We outline key elements of our coding scheme here. The detailed analysis is given in Appendix. The achievability scheme includes a rate splitting technique. Each of the messages $M_1$ and $M_2$ and thereby their communication rates $R_1$ and $R_2$ are split into two parts:

$$\begin{aligned} M_1 &= (M_{10}, M_{11}), \qquad R_1 = R_{10} + R_{11} \\ M_2 &= (M_{20}, M_{22}), \qquad R_2 = R_{20} + R_{22} \end{aligned} \tag{261}$$

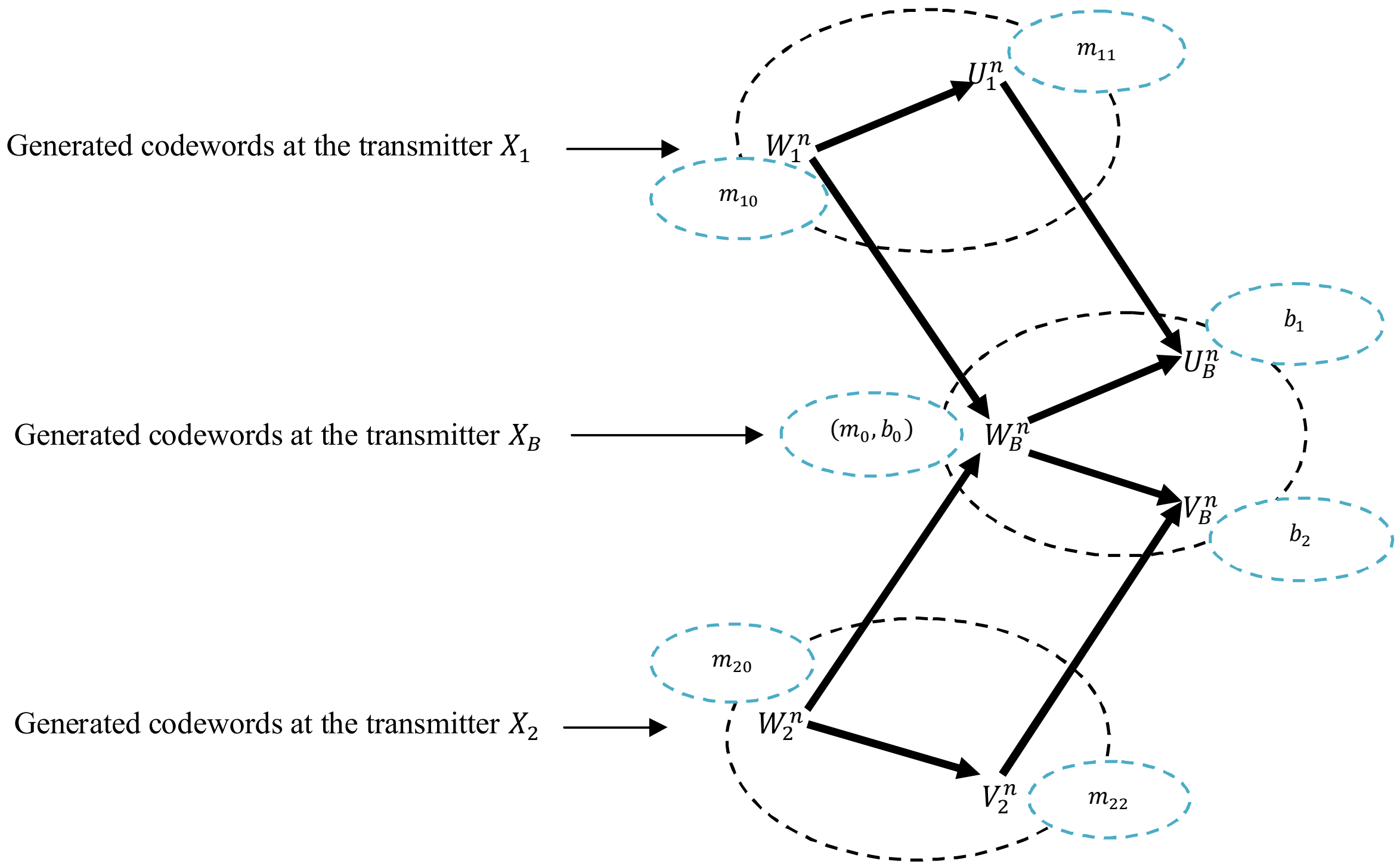

Figure 17. The graphical illustration of the achievability scheme for the BCCR with common message.

Similar to the Han-Kobayashi scheme [16], one part of each private message, i.e., $M_{10}$ of $M_1$ and $M_{20}$ of $M_2$ is used to contribute in building a joint decoding at the non-respective receiver. The split messages are then encoded by a random code of length-$n$. Roughly speaking, in our coding scheme the first transmitter encodes its corresponding messages $M_{10}$ and $M_{11}$ by the codewords $W_1^n$ and $U_1^n$ in a superposition style such that $W_1^n$ serves as the cloud center and $U_1^n$ as the satellite codeword. Similarly, the second transmitter (independent of the first one) encodes its corresponding messages $M_{20}$ and $M_{22}$ by the codewords $W_2^n$ and $V_2^n$ in a superposition style where $W_2^n$ serves as the cloud center and $V_2^n$ as the satellite. At the third transmitter, three bins of codewords are generated:

- $2^{nB_0}$ codewords $W_B^n$ conveying the common message $M_0$ which are superimposed upon $W_1^n, W_2^n$.
- $2^{nB_1}$ codewords $U_B^n$ which are superimposed upon $W_1^n, W_2^n, W_B^n, U_1^n$.
- $2^{nB_2}$ codewords $V_B^n$ which are superimposed upon $W_1^n, W_2^n, W_B^n, V_2^n$.

In fact, the codewords $U_B^n$ and $V_B^n$ do not convey any part of the messages beyond those conveyed by their cloud centers. These are satellite codewords (with the cloud centers $W_1^n, W_2^n, W_B^n, U_1^n$ for $U_B^n$, and $W_1^n, W_2^n, W_B^n, V_2^n$ for $V_B^n$) which are served just for building a Marton's type encoding (refer to our paper presented in [8]) at the third transmitter (broadcast node). Fig. 17 represents the generated codewords at the transmitters. Similar to our graphical illustrations in [8], in this figure every two codewords connected by a directed edge build a superposition structure: The codeword at the beginning of the edge is the cloud center and the one at the end of the edge serves as the satellite codeword. The sizes of the bins $B_0, B_1, B_2$ are selected large enough to guarantee the existence of a 7-tuple $(W_1^n, U_1^n, W_2^n, V_2^n, W_B^n, U_B^n, V_B^n)$ of jointly typical (with respect to the distribution (260)) codewords for each given 5- tuple $(M_0, M_{10}, M_{11}, M_{20}, M_{22})$ of messages. These typical codewords are designated for transmission. The first transmitter then generates a codeword $X_1^n$ superimposed upon its designated codewords $W_1^n, U_1^n$, and sends it over the network. Similarly, the second transmitter generates a codeword $X_2^n$ superimposed upon its corresponding codewords $W_2^n, U_2^n$, and transmit it. As the third transmitter has access to the codewords of the first and the second transmitters, it generates a codeword $X_B^n$ superimposed upon all the other designated codewords, i.e., $X_1^n, W_1^n, U_1^n, X_2^n, W_2^n, V_2^n, W_B^n, U_B^n, V_B^n$, and sends over the network. For decoding, each receiver uses of a jointly typical decoder. The receiver $Y_1$ explores within the codewords $W_1^n, W_2^n, W_B^n, U_1^n, U_B^n$ to find its intended messages, and the receiver $Y_2$ explores within the codewords $W_1^n, W_2^n, W_B^n, V_2^n, V_B^n$. The analysis of the coding scheme is found in Appendix. ■

But what is the philosophy behind this achievability design? We know that the best coding strategies for the two-user CIC is due to Han and Kobayashi [16] and that for the two-user BC is due to Marton [17] which were discussed in details in Parts I [1] and II [2]. Now it is reasonable that an efficient achievability scheme for the BCCR, which contains both the BC and the CIC as special cases, should include the benefits of the Han-Kobayashi scheme for the CIC and the Marton's coding for the BC simultaneously. It is clear that one may split the messages $M_1$ and $M_2$ into numerous sub-messages and then manipulate them in an order to build a combination of the Han-Kobayashi scheme and the Marton's scheme. In fact, such naive techniques are always available to combine coding schemes of simple channels and derive an achievable rate region for a given large network. Nonetheless, this approach is clearly never efficient. It should be noted that splitting messages into sub-messages yields achievable rate regions with complicated descriptions, specifically for large networks. In fact, both the number of parameters involved in the resultant achievable rate region and the number of constraints included in it (rapidly) increases in achievability schemes based on random coding techniques as the number of split messages increases. Accordingly, evaluation of such achievable rate regions is very difficult. Moreover, as we demonstrated in Part I [1, Sec. III.A.4], splitting a message into several sub-message does not necessarily lead to a larger achievable rate region. In some cases, it also causes some rate loss. For example, in [30] the authors derived an achievable rate region for the CRC where the message of the primary transmitter is split into two parts and the message of the cognitive transmitter is split into three parts. The primary transmitter then ignores one of its sub-messages and relegates its transmission to the cognitive transmitter. In Part I [1, Sec. III.A.4], we proved that such a type of message splitting yields a rate loss besides being superfluous. In fact, we showed that by splitting each message just into two parts, a larger achievable rate region is derived.

Now let us examine our achievability scheme designed in Theorem 13 for the BCCR. Refer to Fig. 17. In this scheme, we have split each message just into two parts, similar to the Han-Kobayashi achievability scheme. According to the description of our achievability scheme given in Theorem 13, it is clear that by setting $W_B \equiv U_B \equiv V_B \equiv \emptyset$ and also $M_0 \equiv \emptyset$, we obtain the Han-Kobayashi rate region for the two-user CIC. Now, consider the codewords generated at the broadcasting node, i.e., the transmitter $X_3$. As discussed in the description of the coding scheme, the codewords $U_B^n$ and $V_B^n$ do not contain any part of the messages beyond those conveyed by their cloud centers. But how does the coding scheme include the Marton's achievable rate region while we have split both private messages just into two parts? Let us set $W_1 \equiv U_1 \equiv \emptyset$ and also $W_2 \equiv V_2 \equiv \emptyset$ in our scheme. What happens by this choice? First note that the messages $M_{10}, M_{11}, M_{20}, M_{22}$ are conveyed by the codewords $W_1^n, U_1^n, W_2^n, V_2^n$, respectively; but this does not mean that by setting $W_1 \equiv U_1 \equiv W_2 \equiv V_2 \equiv \emptyset$, the messages $M_{10}, M_{11}, M_{20}, M_{22}$ have been withdrawn from the transmission scheme. These messages still are contained by the codewords $W_B^n, U_B^n, V_B^n$. When two codewords build a superposition structure, then the satellite codeword includes also those messages and bin indices which are conveyed by the cloud center. For example, in Fig. 17 the codeword $V_B^n$ actually conveys the triple $(m_{10}, m_{11}, b_1)$ and the ellipse beside each codeword only shows what is conveyed by that codeword in addition to the ones conveyed by its cloud centers. Now, when we set $W_1^n \equiv U_1^n \equiv W_2^n \equiv V_2^n \equiv \emptyset$, the task of the transmission of those messages conveyed by these codewords, i.e., $M_{10}, M_{11}, M_{20}, M_{22}$ automatically is transferred to their respective satellite codewords. In other words, the coding scheme is reduced to the one shown in Fig. 18:

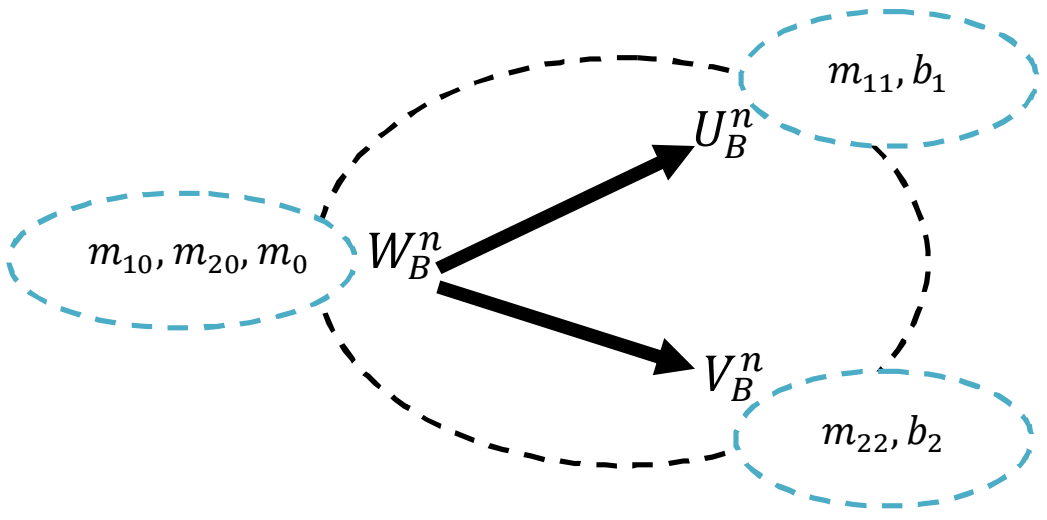

Figure 18. The coding scheme that is derived when we set $W_1^n \equiv U_1^n \equiv W_2^n \equiv V_2^n \equiv \emptyset$ in the scheme of Fig. 17.

Also, note that when we set $W_1 \equiv U_1 \equiv W_2 \equiv V_2 \equiv \emptyset$ in the achievable rate region $\mathfrak{R}_i^{BCCRcm}$ in (259), we have $B_0 = 0$; therefore, $b_0$ is not included in the ellipse beside the codeword $W_B^n$ in Fig. 18. Let us compare the scheme in Fig. 18 with the Marton's coding for the two-user BC with a common message as shown in the following illustration (see also our paper presented in [8]):

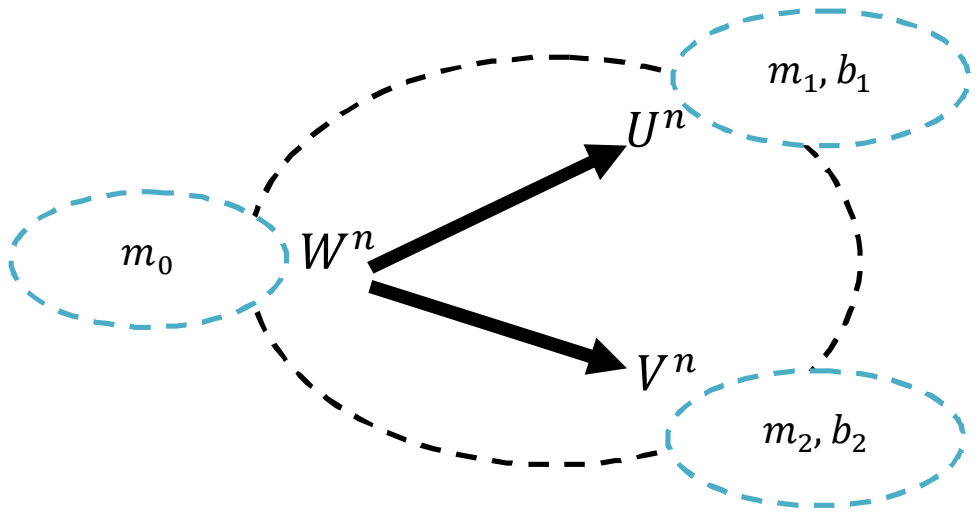

Figure 19. The Marton's coding scheme for the BC with a common message.

Here, we briefly review the Marton's coding scheme. Consider broadcasting the messages $M_0, M_1, M_2$ to two receivers where the first receiver is required to decode the messages $M_0, M_1$ and the second to decode the messages $M_0, M_2$. Roughly speaking, in the Marton's coding scheme (for a length-$n$ code) the common message $M_0$ is encoded by a codeword $W^n$ generated based on $P_W$. For each of the private messages, a bin of codewords is randomly generated which are superimposed upon the common message codeword $W^n$: The bin corresponding to $M_1$ contains the codewords $U^n$ generated based on $P_{U|W}$ and that one for $M_2$ contains the codeword $V^n$ generated based on $P_{V|W}$. The sizes of the bins are selected large enough to guarantee that for each triple $(M_0, M_1, M_2)$, there exists a triple $(W^n, U^n, V^n)$ jointly typical with respect to the PDF $P_{WUV}$. Superimposed on the designated jointly typical codewords $W^n, U^n, V^n$, the encoder then generates its codewords $X^n$ based on $P_{X|WUV}$, and sends it over the channel. The first receiver decodes the codewords $W^n, U^n$ and the second one decodes $W^n, V^n$, both using a jointly typical decoder. At the last step, the resulting achievable rate region is further enlarged by the fact that if the rate triple $(R_0, R_1, R_2) \in \mathbb{R}_+^3$ is achievable for the BC, then $(R_0 - \tau_1 - \tau_2, R_1 + \tau_1, R_2 + \tau_2) \in \mathbb{R}_+^3$ is also achievable. This latter technique can be re-interpreted as follows: Each of the messages $M_1$ and $M_2$ are split into two parts as in (261) and then the parts $M_{10}$ and $M_{20}$ are transferred to the common message codeword, i.e., $W_B^n$. In other words, we allow that each receiver decodes a part of its non-respective private message. Therefore, the scheme in Fig. 18 essentially does work similar to the Maton's coding.

The above discussion demonstrates the capability of our achievability scheme for the BCCR presented in Theorem 13: our design systematically combines the Han-Kobayashi scheme and the Marton's coding scheme; meanwhile, it requires a simple rate splitting (261) which is identically exploited also in the Han-Kobayashi scheme [16]. In Part V of our multi-part papers [5], we will extend the approach presented here and design a unique achievability scheme for any arbitrary interference network. The two key elements of our coding scheme are as follows:

1. Each message is split at most into $2^{K_2-1}$ parts, where $K_2$ is the number of receivers of the network: Each sub-message is designated to be transmitted to the desired receivers as well as a subset of non-desired receivers. For two-receiver networks each message is split at most into two parts. In fact, for the two-receiver networks our achievability design is derived by a simple generalization of the random coding scheme in Fig. 17 on the MACCM plan of messages. See also [7, 8].
2. Given a certain network, our systematic approach is such that when the designated coding scheme is specialized to the basic building blocks contained in the network, it does work essentially similar to the best known coding strategies.

As discussed in our paper [8], we in fact derive our achievability strategy by a systematic combination of the MAC capacity achieving scheme and the Matron's coding scheme for the two-user BC with common messages. Our design in Theorem 13 also falls in this framework.

Let now specialize our achievability scheme for some sub-networks of the BCCR with common message in Fig. 5. Specifically, we first consider the BCCR without common message which was previously studied in [25] and [31]. In the following, based on the achievability scheme in Fig. 17, we present an achievable rate region for this network, which is strictly larger than previous results.

***Corollary 5:*** *Consider the BCCR in Fig. 5 but without common message, i.e.,* $M_0 \equiv \emptyset$. *Define the rate region* $\mathfrak{R}_i^{BCCR}$ *as follows:*

$$\mathfrak{R}_i^{BCCR} \triangleq \bigcup_{\mathcal{P}^{BCCR}} \left\{ \begin{array}{r} (R_1,R_2) \in \mathbb{R}_+^2 : \ \exists (B_0,B_1,B_2,R_{10},R_{11},R_{20},R_{22}) \in \mathbb{R}_+^7 : \\ R_1 = R_{10} + R_{11}, \qquad R_2 = R_{20} + R_{22}, \\ , \\ B_0 > I(U_1,V_2;W_B|W_1,W_2,Q) \triangleq \eta_0 \\ B_0 + B_1 > \eta_0 + I(V_2;U_B|U_1,W_1,W_2,W_B,Q) \triangleq \eta_0 + \eta_1 \\ B_0 + B_2 > \eta_0 + I(U_1;V_B|V_2,W_1,W_2,W_B,Q) \triangleq \eta_0 + \eta_2 \\ B_0 + B_1 + B_2 > \eta_0 + \eta_1 + \eta_2 + \underbrace{I(U_B;V_B|U_1,V_2,W_1,W_2,W_B,Q)}_{\triangleq \eta_4} \\ \theta_1 \triangleq I(U_1;W_B|W_1,W_2,Q), \\ R_{11} + B_1 < I_{Y_1}^1 + \theta_1 \triangleq I(U_1,U_B;Y_1|W_1,W_2,W_B,Q) + \theta_1 \\ B_0 + R_{11} + B_1 < I_{Y_1}^4 + \theta_1 \triangleq I(U_1,W_B,U_B;Y_1|W_1,W_2,Q) + \theta_1 \\ R_{20} + B_0 + R_{11} + B_1 < I_{Y_1}^5 + \theta_1 \triangleq I(U_1,W_2,W_B,U_B;Y_1|W_1,Q) + \theta_1 \\ R_{10} + B_0 + R_{11} + B_1 < I_{Y_1}^6 + \theta_1 \triangleq I(W_1,U_1,W_B,U_B;Y_1|W_2,Q) + \theta_1 \\ R_{10} + R_{20} + B_0 + R_{11} + B_1 < I_{Y_1}^7 + \theta_1 \triangleq I(W_1,U_1,W_2,W_B,U_B;Y_1|Q) + \theta_1 \\ \theta_2 \triangleq I(V_2;W_B|W_1,W_2,Q) \\ R_{22} + B_2 < I_{Y_2}^1 + \theta_2 \triangleq I(V_2,V_B;Y_2|W_1,W_2,W_B,Q) + \theta_2 \\ B_0 + R_{22} + B_2 < I_{Y_2}^4 + \theta_2 \triangleq I(V_2,W_B,V_B;Y_2|W_1,W_2,Q) + \theta_2 \\ R_{10} + B_0 + R_{22} + B_2 < I_{Y_2}^5 + \theta_2 \triangleq I(W_1,V_2,W_B,V_B;Y_2|W_2,Q) + \theta_2 \\ R_{20} + B_0 + R_{22} + B_2 < I_{Y_2}^6 + \theta_2 \triangleq I(W_2,V_2,W_B,V_B;Y_2|W_1,Q) + \theta_2 \\ R_{10} + R_{20} + B_0 + R_{22} + B_2 < I_{Y_2}^7 + \theta_2 \triangleq I(W_1,V_2,W_2,W_B,V_B;Y_2|Q) + \theta_2 \end{array} \right\} \tag{262}$$

*where* $\mathcal{P}^{BCCR}$ *is given in* (260). *The set* $\mathfrak{R}_i^{BCCR}$ *constitutes an inner bound on the capacity region.*

*Proof of Corollary 5)* This achievable rate region is derived by setting $M_0 \equiv \emptyset$ in the achievability scheme of Fig. 17. Let us discuss the achievable rate region that is derived by this choice. Consider the rate region (259). When we set $R_0 = 0$, the constraints including $I_{Y_1}^2, I_{Y_2}^2, I_{Y_1}^3, I_{Y_2}^3$ are redundant because they do not correspond to correctly decoding of any sub-message at its true receiver. In other words, for the case of $M_0 \equiv \emptyset$, without the latter constraints, each receiver can still decode its respective sub-messages with small error probability. The achievable rate region (262) is derived by setting $R_0 = 0$ in (259) and removing redundant constraints. ■

***Remark 20:*** By setting $W_B \equiv \emptyset$ and subsequently $B_0 = 0$ in (262), our achievable rate region is directly reduced to that one previously given in [25, Th. 3.2] for the network. This can be verified by a simple comparison. In fact, our rate region can strictly include that of [25, Th. 3.2]. The reason is that our achievable rate region specialized to the two-user BC (see Fig. 18) includes the superposition random variable $W_B$ while this is not the case for the achievability scheme of [25, Th. 3.2]. On the one hand, according to [32], for the two-user BC, the Marton's rate region with superposition random variable strictly includes the one without this variable (the region which includes only binning variables, i.e., $U$ and $V$ in Fig. 19). Also, note that the achievable region of [25, Th. 3.2] includes that of [31] as shown in [25, Sec. B]. Thus, our region contains previous results.

Next, we consider the case of $X_2 \equiv \emptyset$, i.e., a cognitive interference network with a cognitive common message. Fig. 20 depicts this network. This scenario includes a primary transmitter $X_1$, a cognitive transmitter $X_B$ which has access to the message of the primary transmitter non-causally, and two receivers $Y_1$ and $Y_2$. The primary transmitter $X_1$ sends the message $M_1$ to its respective receiver $Y_1$ and the cognitive transmitter $X_B$ broadcasts the common message $M_0$ and private messages $M_1$ and $M_2$ to the receivers $Y_1$ and $Y_2$, correspondingly. Comparing with the cognitive radio channel [33], this model includes the additional feature that the cognitive transmitter sends also a common message to the receivers. We refer to this new scenario as "Cognitive Radio Channel with a Common Message (CRCCM)".

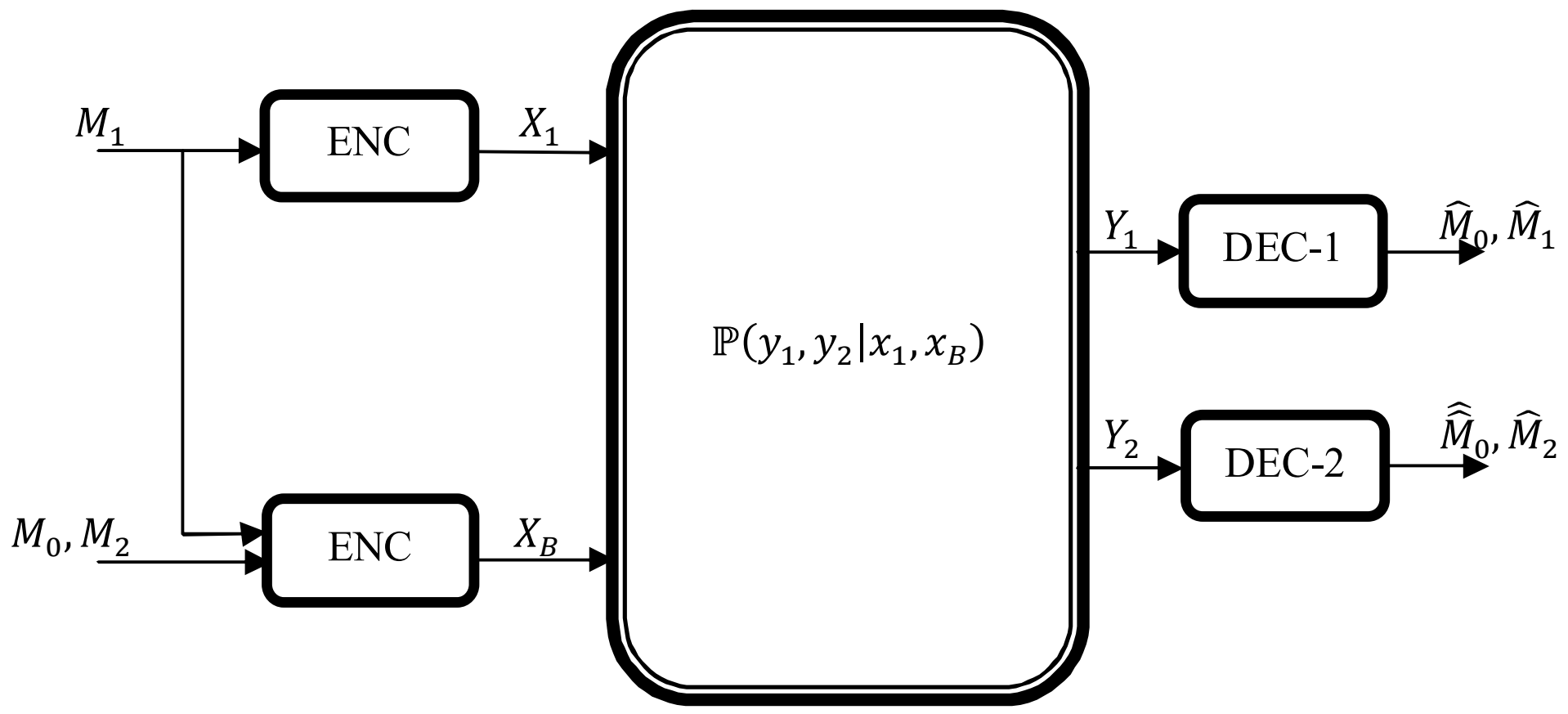

Figure 20. The Cognitive Radio Channel with a Common Message (CRCCM).

It is clear that an achievable rate region is directly derived for the CRCCM by setting $V_2 \equiv W_2 \equiv X_2 \equiv Q \equiv \emptyset$ in (259)[8]. The resulting achievability scheme is depicted in Fig. 21.

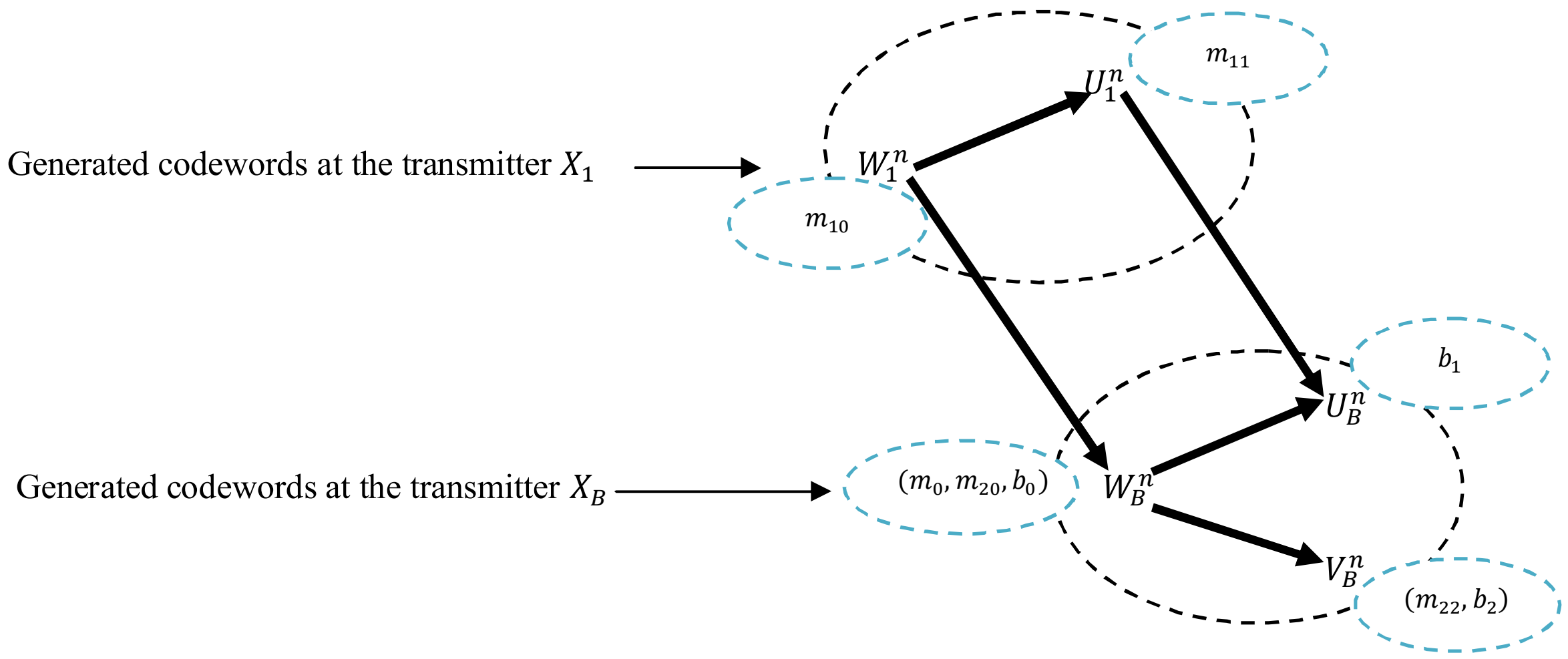

Figure 21. The graphical illustration of the achievability scheme for the CRCCM in Fig. 20.

We remark that by setting $V_2 \equiv W_2 \equiv \emptyset$ in the coding scheme of Fig. 17, the task of transmitting the messages $M_{20}$ and $M_{22}$ is automatically transferred to the codewords $W_B^n$ and $V_B^n$. This point can be perceived from the ellipses besides the latter codewords in Fig. 21.

Also, an outer bound is readily derived for the network by setting $X_2 \equiv \emptyset$ in (95). By this choice, the outer bound is simplified as follows:

[8] We remark that when we set $V_2 \equiv W_2 \equiv X_2 \equiv \emptyset$ in (259), there is no need to the time-sharing parameter "$Q$". In fact, by setting $Q \equiv \emptyset$, the resulting rate region is still convex.

$$\bigcup_{\substack{P_{UVWX_1X_B}(u,v,w,x_1,x_B)\\ P_{X_1|U},P_{X_B|UV}\in\{0,1\}}} \left\{\begin{array}{l} (R_0,R_1,R_2)\in\mathbb{R}_+^3: \\ R_0\leq \min\left\{\begin{array}{c} I(W;Y_1),I(W;Y_2),\\ I(U,W;Y_1|X_1),I(U,W;Y_2|X_1)\\ I(X_B;Y_1|X_1),I(X_B;Y_2|X_1)\end{array}\right\} \\ , \\ R_0+R_1\leq \min\left\{\begin{array}{c} I(X_1,X_B;Y_1),I(U,W,X_1;Y_1),\\ I(U,X_1;Y_1|W)+I(W;Y_2)\end{array}\right\} \\ , \\ R_0+R_2\leq \min\left\{\begin{array}{c} I(X_B;Y_2|X_1),I(V,W;Y_2),\\ I(V;Y_2|W)+I(W;Y_1)\\ I(X_B;Y_2|X_1,U,W)+I(U,W;Y_1|X_1)\end{array}\right\} \\ , \\ R_0+R_1+R_2\leq I(X_1,X_B;Y_1|V,W)+I(V,W;Y_2) \\ R_0+R_1+R_2\leq I(X_B;Y_2|X_1,U,W)+I(U,W,X_1;Y_1) \\ R_0+R_1+R_2\leq I(X_1,X_B;Y_1|V,W)+I(V;Y_2|W)+I(W;Y_1) \\ R_0+R_1+R_2\leq I(X_B;Y_2|X_1,U,W)+I(U,X_1;Y_1|W)+I(W;Y_2) \end{array}\right\} \tag{263}$$

Note that, for the rate region (95) if $X_2 \equiv \emptyset$, one can absorb the time-sharing parameter $Q$ into the auxiliary variables $U, V$ and $W$.

We now derive a special case where our achievability scheme in Fig. 21 and the outer bound in (263) coincide and yield the capacity region. Specifically, consider a CRCCM for which the following condition holds:

$$I(X_1,X_B;Y_1)\leq I(X_1,X_B;Y_2),\qquad \text{for all joint PDFs}\qquad P_{X_1X_B}(x_1,x_B) \tag{264}$$

In this case, the network is called to be *more-capable*. The capacity region of the more-capable CRC (without common message) in (264) was derived in Part I [1, Th. III.8]. For the case of $X_1 \equiv \emptyset$, the network with (264) is reduced to a more-capable BC where its capacity region was established in [18]. Note that one may present an alternative more-capable condition for the CRCCM by swapping the outputs $Y_1$ and $Y_2$ in (264). Due to asymmetry of the network, each of these conditions represents a situation different from the other. As remarked in Part I [1, Sec. III.C.2], we only derived the capacity region of the more-capable CRC (without common message) in the sense of (264) and for the alternative scenario the capacity is still unresolved. Therefore, here we also consider only the case of (264). In Part I [1, Th. III. 8], we proved that the superposition coding scheme is optimal for the more-capable CRC (264). Now, consider the following sub-scheme of the general achievability scheme given in Fig. 21 for the CRCCM.

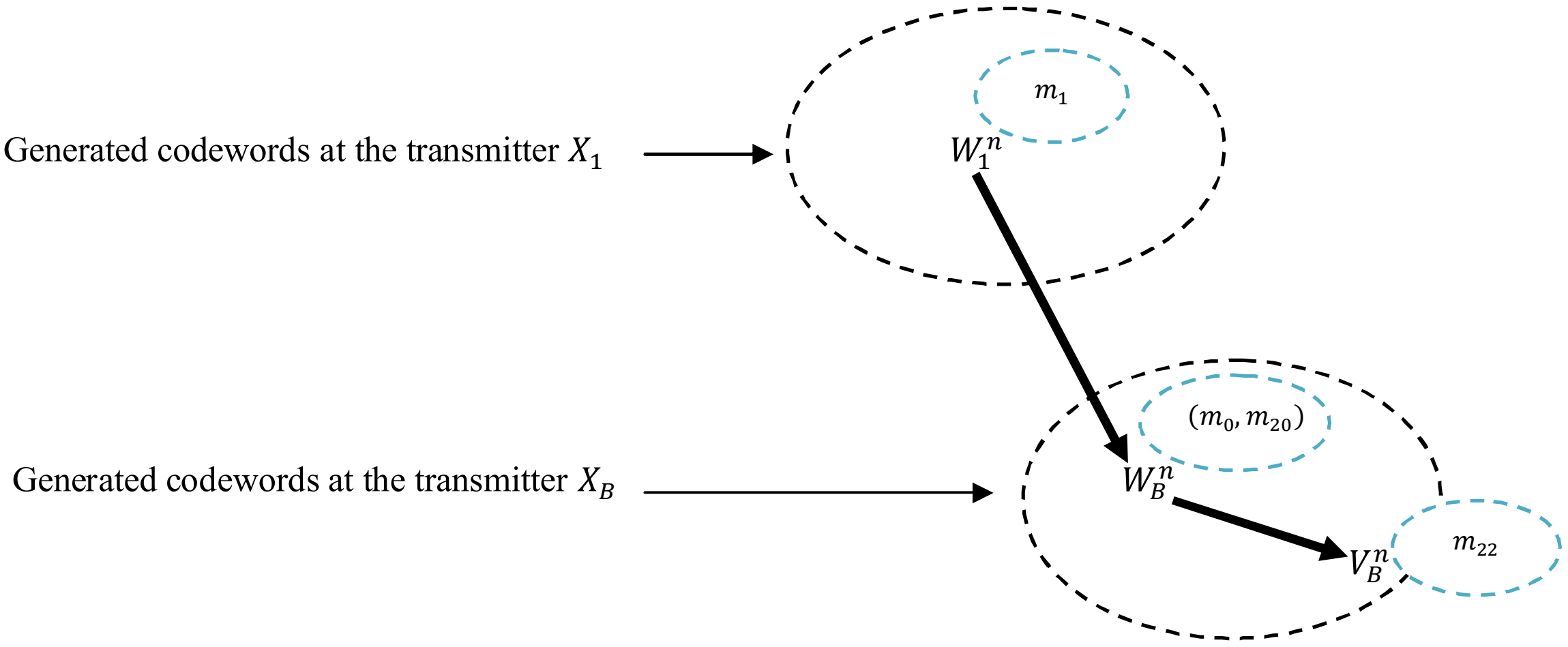

Figure 22. An achievability scheme based on superposition coding for the CRCCM.

This sub-scheme does not include any binning. It is derived only by superposition coding, i.e., by setting $U_1 \equiv U_B \equiv \emptyset$ in the scheme of Fig. 21. Therefore, $B_0 = B_1 = B_2 = 0$. The primary receiver $Y_1$ decodes the codewords $W_1^n$ and $W_B^n$ and the cognitive receiver $Y_2$ jointly decodes all the codewords $W_1^n, W_B^n$ and $V_B^n$. Note that in this scheme, we do not need not to split the primary message $M_1$ because this message is perfectly decoded at both receivers. Therefore, $R_1 = R_{10}$ and $R_{11} = 0$. We remark that a special case of the coding scheme in Fig. 22, that is given by $X_1 \equiv W_1 \equiv \emptyset$, achieves the capacity region of the more-capable BC with common message [18]. Also, for the case of no common message, i.e., $R_0 = 0$, this scheme achieves the capacity region for the more-capable CRC in (264), as derived in Part I [1, Th. III.8]. Now, as a generalization of the two latter cases, we intend to prove that the coding scheme is also optimal for the more-capable CRCCM (264) in Fig. 20. This result is given in the next theorem.

***Theorem 14)*** *The capacity region of the more-capable CRCCM (264) in Fig. 20 is given below:*

$$\bigcup_{P_{UX_1X_2}(u,x_1,x_2)} \left\{ \begin{array}{l} (R_0,R_1,R_2) \in \mathbb{R}_+^3 : \\ R_0 \leq I(U;Y_1|X_1) \\ R_0 + R_1 \leq I(U,X_1;Y_1) \\ R_0 + R_2 \leq I(X_B;Y_2|X_1) \\ R_0 + R_2 \leq I(X_B;Y_2|U,X_1) + I(U;Y_1|X_1) \\ R_0 + R_1 + R_2 \leq I(X_B;Y_2|U,X_1) + I(U,X_1;Y_1) \\ R_0 + R_1 + R_2 \leq I(X_1,X_B;Y_2) \end{array} \right\} \tag{265}$$

*Proof of Theorem 14) Consider the achievability scheme in Fig. 21. The resulting rate region is as follows:*

$$\bigcup_{P_{W_1}P_{X_1|W_1}P_{W_BV_B|W_1}P_{X_B|X_1V_BW_BW_1}} \left\{ \begin{array}{l} (R_0,R_1,R_2) \in \mathbb{R}_+^3 : \\ R_2 = R_{22} + R_{20} \\ R_0 + R_{20} \leq I(W_B;Y_1|W_1) \\ R_0 + R_{20} + R_1 \leq I(W_B,W_1;Y_1) \\ R_{22} \leq I(V_B;Y_2|W_B,W_1) \\ R_0 + R_2 \leq I(V_B,W_B;Y_2|W_1) \\ R_0 + R_1 + R_2 \leq I(W_1,W_B,V_B;Y_2) \end{array} \right\} \tag{266}$$

Now by setting $W_1 \equiv X_1$, $W_B \equiv U$ and $V_B \equiv X_B$ in (266) and then applying a simple Fourier-Motzkin elimination to remove the partial rates $R_{22}$ and $R_{20}$, we obtain (265). To derive the converse part, we prove that the outer bound (263) is optimal. To see this, by redefining $U \cong (U,W)$ in (263), we observe that all the constraints in (265) are directly found in (263) except the last bound on the sum-rate. To derive the latter constraint, we have:

$$\begin{aligned} R_0 + R_1 + R_2 &\leq I(X_1,X_B;Y_1|V,W) + I(V,W;Y_2) \\ &\overset{(a)}{\leq} I(X_1,X_B;Y_2|V,W) + I(V,W;Y_2) = I(X_1,X_B;Y_2) \end{aligned}$$

where inequality (a) is due to the more-capable condition (264) and its extension as in Lemma 1. The proof is thus complete. ■

***Remarks 21:***

1. By setting $X_1 \equiv \emptyset$ in (265), we re-derive the capacity region of the more-capable BC with common message [18].
2. By setting $R_0 = 0$ in (265), we re-derive the capacity region of the more-capable CRC without common message as established in Part I of our multi-part papers [1, Th. III.8].

Finally, we remark that the approach presented in this subsection can be followed to derive capacity results for many other multi-user/multi-message interference networks. Our methodology, through which large network topologies are systematically treated based on the basic building blocks, indeed constitutes a clear framework for this purpose.

# CONCLUSION

In this third part of our multi-part papers, we studied the information flow in interference networks with strong interference. First, we essentially revisited the concept of "strong interference". In our definition, a network is said to be in the strong interference regime when the coding scheme achieving the capacity region is that each receiver decodes all transmitted messages by its connected transmitters. Indeed, this characteristic deserves to be considered as a criterion while referring to the strong interference.

Next, we investigated conditions under which this regime holds. First, the two-receiver networks were considered. We established a unified outer bound on the capacity region of these networks. As was shown, our general outer bound can be systematically translated into simple and computable capacity outer bounds for specific networks such as the two-user Classical Interference Channel (CIC) and the Broadcast Channel with Cognitive Relays (BCCR) with common information. We also derived special cases for the latter scenarios where the outer bounds are optimal and yield the capacity. Our outer bounds are indeed efficient to derive strong interference regime for two-receiver interference networks with any topologies. By using these outer bounds, we indentified a strong interference regime for the general two-receiver interference networks with any arbitrary topology. A remarkable point is that our strong interference regime for the two-receiver networks is represented by only two conditions. Also, it subsumes all previously known results for simple topologies such as the two-user CIC and the cognitive interference channel.

Then, networks with arbitrary number of receivers were considered. We developed a new approach based on which one can obtain strong interference regimes not only for the multi-user CICs but also for any interference network of arbitrary large size. Our new technical lemmas given in Section II have a central role in this development. A general formula was also presented to derive strong interference conditions for any given network topology. As a result, in this part of our multi-part papers, by solving one of open problems in network information theory [15, page 6-68], we established the first non-trivial capacity result for the multi-user classical interference channels.

Finally, we presented an achievability scheme for the BCCR with common message based on the systematic approach developed in our previous paper [7] presented in [8]. Our achievability design systematically combines the Han-Kobayashi scheme for the two-user CIC [16] and the Marton's coding scheme for the two-user BC [17]; meanwhile, it enjoys only a simple rate splitting which is identically exploited in the Han-Kobayashi scheme. Clearly, in our scheme each message is split into only two parts. We also argued that the same coding scheme can be derived for any two-receiver interference network. Moreover, we developed a framework based on which one can intuitively obtain capacity results for certain interference networks such as more-capable networks. In our methodology, large multi-user/multi-message networks are systematically treated based on the basic building blocks.

Our systematic study of fundamental limits of communications in interference networks will be followed in Part IV [4] where we will introduce networks with a sequence of less-noisy receivers and discuss the behavior of information flow for them.

# APPENDIX

➢ ***Proof of Theorem 2***

Consider a code $\mathfrak{C}^n(R_1,\dots,R_K)$ with vanishing average error probability for the network. Let $\Omega_1^{Y_j},\Omega_2^{Y_j},\dots,\Omega_\mu^{Y_j}$ be a sequence of subsets of the messages $\mathbb{M}_{Y_j}$ , $j=1,2$ , where $\mu\in\mathbb{N}$ is an arbitrary number. Assume also that all the sets $\Omega_1^{Y_1},\Omega_2^{Y_1},\dots,\Omega_\mu^{Y_1},\Omega_1^{Y_2},\Omega_2^{Y_2}\dots,\Omega_\mu^{Y_2}$ are pairwise disjoint. Let $\Omega_s\subseteq\mathbb{M}-\left(\Omega_1^{Y_1}\cup\Omega_2^{Y_1}\cup\dots\cup\Omega_\mu^{Y_1}\cup\Omega_1^{Y_2}\cup\Omega_2^{Y_2}\cup\dots\cup\Omega_\mu^{Y_2}\right)$. In the following argument, the messages $\Omega_s$ are considered as the *virtual side-information* for the receivers. Since receiver $Y_j, j=1,2$ decodes all the messages $\mathbb{M}_{Y_j}$, using Fano's inequality, we have:

$$\frac{1}{n}H\left(\Omega_l^{Y_j}\middle|Y_j^n\right)\overset{(a)}{\leq}\frac{1}{n}H\left(\mathbb{M}_{Y_j}\middle|Y_j^n\right)\leq\epsilon_{j,n},\qquad l=1,\dots,\mu,\qquad j=1,2 \tag{A~1}$$

where $\epsilon_{j,n}\to 0$ as $n\to\infty$. Note that inequality (a) holds because $\Omega_l^{Y_j}$ is a subset of $\mathbb{M}_{Y_j}$. Define new random variables $Z_t, t=1,\dots,n$, as follows:

$$Z_t \triangleq \left(Y_1^{t-1}, Y_{2,t+1}^n\right)$$

(A~2)

Now using (A~1), we can write:

$$n\left(\sum_{l=1}^{\mu} \boldsymbol{R}_{\Sigma_{\Omega_l^{Y_1}}} + \boldsymbol{R}_{\Sigma_{\Omega_l^{Y_2}}}\right) - n\mu(\epsilon_{1,n} + \epsilon_{2,n}) = \sum_{l=1}^{\mu} H\left(\Omega_l^{Y_1}\right) + H\left(\Omega_l^{Y_2}\right) - n\mu(\epsilon_{1,n} + \epsilon_{2,n})$$

$$\leq \sum_{l=1}^{\mu} I\left(\Omega_l^{Y_1}; Y_1^n\right) + I\left(\Omega_l^{Y_2}; Y_2^n\right)$$

$$\leq \sum_{l=1}^{\mu} \begin{pmatrix} I\left(\Omega_l^{Y_1}; Y_1^n, \Omega_1^{Y_1}, \dots, \Omega_{l-1}^{Y_1}, \Omega_1^{Y_2}, \dots, \Omega_{l-1}^{Y_2}, \Omega_l^{Y_2}, \Omega_s\right) \\ +I\left(\Omega_l^{Y_2}; Y_2^n, \Omega_1^{Y_1}, \dots, \Omega_{l-1}^{Y_1}, \Omega_1^{Y_2}, \dots, \Omega_{l-1}^{Y_2}, \Omega_s\right) \end{pmatrix}$$

$$\overset{(a)}{\leq} \sum_{l=1}^{\mu} \begin{pmatrix} I\left(\Omega_l^{Y_1}; Y_1^n \middle| \Omega_1^{Y_1}, \dots, \Omega_{l-1}^{Y_1}, \Omega_1^{Y_2}, \dots, \Omega_{l-1}^{Y_2}, \Omega_l^{Y_2}, \Omega_s\right) \\ +I\left(\Omega_l^{Y_2}; Y_2^n \middle| \Omega_1^{Y_1}, \dots, \Omega_{l-1}^{Y_1}, \Omega_1^{Y_2}, \dots, \Omega_{l-1}^{Y_2}, \Omega_s\right) \end{pmatrix}$$

$$= \sum_{l=1}^{\mu} \sum_{t=1}^{n} \begin{pmatrix} I\left(\Omega_l^{Y_1}; Y_{1,t} \middle| Y_1^{t-1}, \Omega_1^{Y_1}, \dots, \Omega_{l-1}^{Y_1}, \Omega_1^{Y_2}, \dots, \Omega_{l-1}^{Y_2}, \Omega_l^{Y_2}, \Omega_s\right) \\ +I\left(\Omega_l^{Y_2}; Y_{2,t} \middle| Y_{2,t+1}^n, \Omega_1^{Y_1}, \dots, \Omega_{l-1}^{Y_1}, \Omega_1^{Y_2}, \dots, \Omega_{l-1}^{Y_2}, \Omega_s\right) \end{pmatrix}$$

(A~3)

where inequality (a) holds because the messages $\Omega_1^{Y_1}, \dots, \Omega_{l-1}^{Y_1}, \Omega_l^{Y_1}, \Omega_1^{Y_2}, \dots, \Omega_{l-1}^{Y_2}, \Omega_l^{Y_2}, \Omega_s$ are independent. Now, consider the summands in (A~3). For the first term we have:

$$\sum_{t=1}^{n} I\left(\Omega_l^{Y_1}; Y_{1,t} \middle| Y_1^{t-1}, \Omega_1^{Y_1}, \dots, \Omega_{l-1}^{Y_1}, \Omega_1^{Y_2}, \dots, \Omega_{l-1}^{Y_2}, \Omega_l^{Y_2}, \Omega_s\right)$$

$$= \sum_{t=1}^{n} I\left(\Omega_l^{Y_1}, Y_{2,t+1}^n; Y_{1,t} \middle| Y_1^{t-1}, \Omega_1^{Y_1}, \dots, \Omega_{l-1}^{Y_1}, \Omega_1^{Y_2}, \dots, \Omega_{l-1}^{Y_2}, \Omega_l^{Y_2}, \Omega_s\right)$$

$$- \sum_{t=1}^{n} I\left(Y_{2,t+1}^n; Y_{1,t} \middle| Y_1^{t-1}, \Omega_1^{Y_1}, \dots, \Omega_{l-1}^{Y_1}, \Omega_l^{Y_1}, \Omega_1^{Y_2}, \dots, \Omega_{l-1}^{Y_2}, \Omega_l^{Y_2}, \Omega_s\right)$$

$$= \sum_{t=1}^{n} I\left(\Omega_l^{Y_1}; Y_{1,t} \middle| Y_1^{t-1}, Y_{2,t+1}^n, \Omega_1^{Y_1}, \dots, \Omega_{l-1}^{Y_1}, \Omega_1^{Y_2}, \dots, \Omega_{l-1}^{Y_2}, \Omega_l^{Y_2}, \Omega_s\right)$$

$$+ \sum_{t=1}^{n} I\left(Y_{2,t+1}^n; Y_{1,t} \middle| Y_1^{t-1}, \Omega_1^{Y_1}, \dots, \Omega_{l-1}^{Y_1}, \Omega_1^{Y_2}, \dots, \Omega_{l-1}^{Y_2}, \Omega_l^{Y_2}, \Omega_s\right)$$

$$- \sum_{t=1}^{n} I\left(Y_{2,t+1}^n; Y_{1,t} \middle| Y_1^{t-1}, \Omega_1^{Y_1}, \dots, \Omega_{l-1}^{Y_1}, \Omega_l^{Y_1}, \Omega_1^{Y_2}, \dots, \Omega_{l-1}^{Y_2}, \Omega_l^{Y_2}, \Omega_s\right)$$

(A~4)

Also, for the second term we have:

$$\sum_{t=1}^{n} I\left(\Omega_l^{Y_2}; Y_{2,t} \middle| Y_{2,t+1}^n, \Omega_1^{Y_1}, \dots, \Omega_{l-1}^{Y_1}, \Omega_1^{Y_2}, \dots, \Omega_{l-1}^{Y_2}, \Omega_s\right)$$

$$= \sum_{t=1}^{n} I\left(\Omega_l^{Y_2}, Y_1^{t-1}; Y_{2,t} \middle| Y_{2,t+1}^n, \Omega_1^{Y_1}, \dots, \Omega_{l-1}^{Y_1}, \Omega_1^{Y_2}, \dots, \Omega_{l-1}^{Y_2}, \Omega_s\right)$$

$$- \sum_{t=1}^{n} I\left(Y_1^{t-1}; Y_{2,t} \middle| Y_{2,t+1}^n, \Omega_1^{Y_1}, \dots, \Omega_{l-1}^{Y_1}, \Omega_1^{Y_2}, \dots, \Omega_{l-1}^{Y_2}, \Omega_l^{Y_2}, \Omega_s\right)$$

$$= \sum_{t=1}^{n} I\left(\Omega_l^{Y_2}; Y_{2,t} \middle| Y_1^{t-1}, Y_{2,t+1}^n, \Omega_1^{Y_1}, \dots, \Omega_{l-1}^{Y_1}, \Omega_1^{Y_2}, \dots, \Omega_{l-1}^{Y_2}, \Omega_s\right)$$

$$+ \sum_{t=1}^{n} I\left(Y_1^{t-1}; Y_{2,t} \middle| Y_{2,t+1}^n, \Omega_1^{Y_1}, \dots, \Omega_{l-1}^{Y_1}, \Omega_1^{Y_2}, \dots, \Omega_{l-1}^{Y_2}, \Omega_s\right)$$

$$- \sum_{t=1}^{n} I\left(Y_1^{t-1}; Y_{2,t} \middle| Y_{2,t+1}^n, \Omega_1^{Y_1}, \dots, \Omega_{l-1}^{Y_1}, \Omega_1^{Y_2}, \dots, \Omega_{l-1}^{Y_2}, \Omega_l^{Y_2}, \Omega_s\right)$$

(A~5)

Note that using the Csiszar-Korner identity (see Part I [1, Lemma III.4]), the following equalities hold:

$$\begin{cases} \sum_{t=1}^{n} I\left(Y_1^{t-1}; Y_{2,t} \middle| Y_{2,t+1}^n, \Omega_1^{Y_1}, \dots, \Omega_{l-1}^{Y_1}, \Omega_1^{Y_2}, \dots, \Omega_{l-1}^{Y_2}, \Omega_s\right) = \sum_{t=1}^{n} I\left(Y_{2,t+1}^n; Y_{1,t} \middle| Y_1^{t-1}, \Omega_1^{Y_1}, \dots, \Omega_{l-1}^{Y_1}, \Omega_1^{Y_2}, \dots, \Omega_{l-1}^{Y_2}, \Omega_s\right) \\ \sum_{t=1}^{n} I\left(Y_1^{t-1}; Y_{2,t} \middle| Y_{2,t+1}^n, \Omega_1^{Y_1}, \dots, \Omega_{l-1}^{Y_1}, \Omega_1^{Y_2}, \dots, \Omega_{l-1}^{Y_2}, \Omega_l^{Y_2}, \Omega_s\right) = \sum_{t=1}^{n} I\left(Y_{2,t+1}^n; Y_{1,t} \middle| Y_1^{t-1}, \Omega_1^{Y_1}, \dots, \Omega_{l-1}^{Y_1}, \Omega_1^{Y_2}, \dots, \Omega_{l-1}^{Y_2}, \Omega_l^{Y_2}, \Omega_s\right) \end{cases}$$

(A~6)

By substituting (A~6) in (A~5) and then combining (A~4) and (A~5), we derive:

$$\sum_{t=1}^{n}\begin{pmatrix} I\left(\Omega_l^{Y_1};Y_{1,t}\middle|Y_1^{t-1},\Omega_1^{Y_1},\dots,\Omega_{l-1}^{Y_1},\Omega_1^{Y_2},\dots,\Omega_{l-1}^{Y_2},\Omega_l^{Y_2},\Omega_s\right)\\ +I\left(\Omega_l^{Y_2};Y_{2,t}\middle|Y_{2,t+1}^{n},\Omega_1^{Y_1},\dots,\Omega_{l-1}^{Y_1},\Omega_1^{Y_2},\dots,\Omega_{l-1}^{Y_2},\Omega_s\right)\end{pmatrix}$$

$$=\sum_{t=1}^{n}\begin{pmatrix} I\left(\Omega_l^{Y_1};Y_{1,t}\middle|Z_t,\Omega_1^{Y_1},\dots,\Omega_{l-1}^{Y_1},\Omega_1^{Y_2},\dots,\Omega_{l-1}^{Y_2},\Omega_l^{Y_2},\Omega_s\right)\\ +I\left(\Omega_l^{Y_2};Y_{2,t}\middle|Z_t,\Omega_1^{Y_1},\dots,\Omega_{l-1}^{Y_1},\Omega_1^{Y_2},\dots,\Omega_{l-1}^{Y_2},\Omega_s\right)\end{pmatrix}-\sum_{t=1}^{n}I\left(Y_{2,t+1}^{n};Y_{1,t}\middle|Y_1^{t-1},\Omega_1^{Y_1},\dots,\Omega_{l-1}^{Y_1},\Omega_l^{Y_1},\Omega_1^{Y_2},\dots,\Omega_{l-1}^{Y_2},\Omega_l^{Y_2},\Omega_s\right)$$
$$+\sum_{t=1}^{n}I\left(Y_{2,t+1}^{n};Y_{1,t}\middle|Y_1^{t-1},\Omega_1^{Y_1},\dots,\Omega_{l-1}^{Y_1},\Omega_1^{Y_2},\dots,\Omega_{l-1}^{Y_2},\Omega_s\right)$$

(A~7)

where $Z_t, t=1,\dots,n$, is defined in (A~2). Now by substituting (A~7) in (A~3), we have:

$$n\left(\sum_{l=1}^{\mu}\boldsymbol{R}_{\sum_{\Omega_l^{Y_1}}}+\boldsymbol{R}_{\sum_{\Omega_l^{Y_2}}}\right)-n\mu(\epsilon_{1,n}+\epsilon_{2,n})$$

$$\leq\sum_{l=1}^{\mu}\sum_{t=1}^{n}\begin{pmatrix} I\left(\Omega_l^{Y_1};Y_{1,t}\middle|Z_t,\Omega_1^{Y_1},\dots,\Omega_{l-1}^{Y_1},\Omega_1^{Y_2},\dots,\Omega_{l-1}^{Y_2},\Omega_l^{Y_2},\Omega_s\right)\\ +I\left(\Omega_l^{Y_2};Y_{2,t}\middle|Z_t,\Omega_1^{Y_1},\dots,\Omega_{l-1}^{Y_1},\Omega_1^{Y_2},\dots,\Omega_{l-1}^{Y_2},\Omega_s\right)\end{pmatrix}$$

$$-\sum_{l=1}^{\mu}\sum_{t=1}^{n}I\left(Y_{2,t+1}^{n};Y_{1,t}\middle|Y_1^{t-1},\Omega_1^{Y_1},\dots,\Omega_{l-1}^{Y_1},\Omega_l^{Y_1},\Omega_1^{Y_2},\dots,\Omega_{l-1}^{Y_2},\Omega_l^{Y_2},\Omega_s\right)$$

$$+\sum_{l=1}^{\mu}\sum_{t=1}^{n}I\left(Y_{2,t+1}^{n};Y_{1,t}\middle|Y_1^{t-1},\Omega_1^{Y_1},\dots,\Omega_{l-1}^{Y_1},\Omega_1^{Y_2},\dots,\Omega_{l-1}^{Y_2},\Omega_s\right)$$

$$=\sum_{l=1}^{\mu}\sum_{t=1}^{n}\begin{pmatrix} I\left(\Omega_l^{Y_1};Y_{1,t}\middle|Z_t,\Omega_1^{Y_1},\dots,\Omega_{l-1}^{Y_1},\Omega_1^{Y_2},\dots,\Omega_{l-1}^{Y_2},\Omega_l^{Y_2},\Omega_s\right)\\ +I\left(\Omega_l^{Y_2};Y_{2,t}\middle|Z_t,\Omega_1^{Y_1},\dots,\Omega_{l-1}^{Y_1},\Omega_1^{Y_2},\dots,\Omega_{l-1}^{Y_2},\Omega_s\right)\end{pmatrix}$$

$$-\sum_{t=1}^{n}I\left(Y_{2,t+1}^{n};Y_{1,t}\middle|Y_1^{t-1},\Omega_1^{Y_1},\dots,\Omega_{\mu-1}^{Y_1},\Omega_\mu^{Y_1},\Omega_1^{Y_2},\dots,\Omega_{\mu-1}^{Y_2},\Omega_\mu^{Y_2},\Omega_s\right)$$

$$+\sum_{t=1}^{n}I\left(Y_{2,t+1}^{n};Y_{1,t}\middle|Y_1^{t-1},\Omega_s\right)$$

$$\leq\sum_{l=1}^{\mu}\sum_{t=1}^{n}\begin{pmatrix} I\left(\Omega_l^{Y_1};Y_{1,t}\middle|Z_t,\Omega_1^{Y_1},\dots,\Omega_{l-1}^{Y_1},\Omega_1^{Y_2},\dots,\Omega_{l-1}^{Y_2},\Omega_l^{Y_2},\Omega_s\right)\\ +I\left(\Omega_l^{Y_2};Y_{2,t}\middle|Z_t,\Omega_1^{Y_1},\dots,\Omega_{l-1}^{Y_1},\Omega_1^{Y_2},\dots,\Omega_{l-1}^{Y_2},\Omega_s\right)\end{pmatrix}+\sum_{t=1}^{n}I\left(Y_{2,t+1}^{n};Y_{1,t}\middle|Y_1^{t-1},\Omega_s\right)$$

(A~8)

Now, note that due to Csiszar-Korner identity, we have the following equality:

$$\sum_{t=1}^{n}I\left(Y_{2,t+1}^{n};Y_{1,t}\middle|Y_1^{t-1},\Omega_s\right)=\sum_{t=1}^{n}I\left(Y_1^{t-1};Y_{2,t}\middle|Y_{2,t+1}^{n},\Omega_s\right)$$

(A~9)

Moreover, since adding information increases mutual information, we can write:

$$\begin{cases}\sum_{t=1}^{n}I\left(Y_{2,t+1}^{n};Y_{1,t}\middle|Y_1^{t-1},\Omega_s\right)\leq\sum_{t=1}^{n}I\left(Y_1^{t-1},Y_{2,t+1}^{n};Y_{1,t}\middle|\Omega_s\right)=\sum_{t=1}^{n}I\left(Z_t;Y_{1,t}\middle|\Omega_s\right)\\ \sum_{t=1}^{n}I\left(Y_1^{t-1};Y_{2,t}\middle|Y_{2,t+1}^{n},\Omega_s\right)\leq\sum_{t=1}^{n}I\left(Y_1^{t-1},Y_{2,t+1}^{n};Y_{2,t}\middle|\Omega_s\right)=\sum_{t=1}^{n}I\left(Z_t;Y_{2,t}\middle|\Omega_s\right)\end{cases}$$

(A~10)

Then based on (A~8), (A~9) and (A~10), we derive:

$$n\left(\sum_{l=1}^{\mu}\boldsymbol{R}_{\sum_{\Omega_l^{Y_1}}}+\boldsymbol{R}_{\sum_{\Omega_l^{Y_2}}}\right)-n\mu(\epsilon_{1,n}+\epsilon_{2,n})$$

$$\leq\sum_{l=1}^{\mu}\sum_{t=1}^{n}\begin{pmatrix} I\left(\Omega_l^{Y_1};Y_{1,t}\middle|Z_t,\Omega_1^{Y_1},\dots,\Omega_{l-1}^{Y_1},\Omega_1^{Y_2},\dots,\Omega_{l-1}^{Y_2},\Omega_l^{Y_2},\Omega_s\right)\\ +I\left(\Omega_l^{Y_2};Y_{2,t}\middle|Z_t,\Omega_1^{Y_1},\dots,\Omega_{l-1}^{Y_1},\Omega_1^{Y_2},\dots,\Omega_{l-1}^{Y_2},\Omega_s\right)\end{pmatrix}+\min\left\{\sum_{t=1}^{n}I\left(Z_t;Y_{1,t}\middle|\Omega_s\right),\sum_{t=1}^{n}I\left(Z_t;Y_{2,t}\middle|\Omega_s\right)\right\}$$

(A~11)

Now, define a time-sharing random variable $Q$ uniformly distributed over the set $\{1,\dots,n\}$ and independent of all other RVs. Also, define:

$$Y_1 \triangleq Y_{1,Q}, \qquad Y_2 \triangleq Y_{2,Q}, \qquad Z \triangleq Z_Q, \qquad X_i \triangleq X_{i,Q}, \qquad i = 1, \dots, K_1$$

(A~12)

We know that by code definition $X_{i,t}, t = 1, \dots, n$ is a deterministic function of $\mathbb{M}_{X_i}$ for $i = 1, \dots, K_1$; hence, according to (A~12), we have: $P_{X_i|\mathbb{M}_{X_i},Q} \in \{0,1\}$.

By using the time-sharing parameter $Q$, the inequality (A~11) can be re-written below:

$$\sum_{l=1}^{\mu} \boldsymbol{R}_{\Sigma_{\Omega_l^{Y_1}}} + \boldsymbol{R}_{\Sigma_{\Omega_l^{Y_2}}} - \mu(\epsilon_{1,n} + \epsilon_{2,n})$$

$$\leq \sum_{l=1}^{\mu} \begin{pmatrix} I(\Omega_l^{Y_1}; Y_{1,Q}|Z_Q, \Omega_1^{Y_1}, \dots, \Omega_{l-1}^{Y_1}, \Omega_1^{Y_2}, \dots, \Omega_{l-1}^{Y_2}, \Omega_l^{Y_2}, \Omega_s, Q) \\ +I(\Omega_l^{Y_2}; Y_{1,Q}|Z_Q, \Omega_1^{Y_1}, \dots, \Omega_{l-1}^{Y_1}, \Omega_1^{Y_2}, \dots, \Omega_{l-1}^{Y_2}, \Omega_s, Q) \end{pmatrix} + \min\{I(Z_Q; Y_{1,Q}|\Omega_s, Q), I(Z_Q; Y_{2,Q}|\Omega_s, Q)\}$$

(A~13)

Lastly, considering the definitions (A~12) and by letting $n$ tend to infinity in (A13), we derive the desired constraint given in (52). This completes the proof. ■

- ***Proof of Proposition 5***

Consider the outer bound $\mathfrak{R}_{o:(2)}^{GIN}$ given by (172) for the general interference network. For the three-user CIC, we have: $\mathbb{M} = \{M_1, M_2, M_3\}$, and $X_i$ is a deterministic function of $(M_i, Q)$, $i = 1,2,3$. Moreover, given the inputs $X_1, X_2, X_3$, the outputs $Y_1, Y_2, Y_3$ are independent of other random variables. Define:

$$U_1 \triangleq (Z_{1,2}, M_1, M_3), \qquad V_1 \triangleq (Z_{1,2}, M_2, M_3), \qquad U_{1,23} \triangleq (Z_{1,23}, M_1), \qquad V_{1,23} \triangleq (Z_{1,23}, M_2, M_3)$$

(A~14)

We have:

- ✓ *Select:* $\Omega_0 = \emptyset$, $\Omega_1 = \{M_1\}$, $\Omega_2 = \emptyset$, $\Omega_s = \{M_2, M_3\}$, $J_1 = \{1\}$, $J_2 = \{2,3\} \Rightarrow$

$$\mathfrak{R}_{o:(2)}^{GIN}\langle 1 \rangle: \; R_1 \leq I(Z_{1,23}, M_1; Y_1|M_2, M_3, Q) = I(Z_{1,23}, M_1, X_1; Y_1|X_2, X_3, M_2, M_3, Q)$$
$$\leq I(X_1; Y_1|X_2, X_3, Q)$$

(A~15)

- ✓ *Select:* $\Omega_0 = \emptyset$, $\Omega_1 = \{M_1\}$, $\Omega_2 = \emptyset$, $\Omega_s = \emptyset$, $J_1 = \{1\}$, $J_2 = \{2,3\} \Rightarrow$

$$\mathfrak{R}_{o:(2)}^{GIN}\langle 1 \rangle: \; R_1 \leq I(Z_{1,23}, M_1; Y_1|Q) = I(Z_{1,23}, M_1, X_1; Y_1|Q)$$
$$= I(U_{1,23}, X_1; Y_1|Q)$$

(A~16)

- ✓ *Select:* $\Omega_0 = \emptyset$, $\Omega_1 = \{M_1\}$, $\Omega_2 = \emptyset$, $\Omega_s = \{M_3\}$, $J_1 = \{1\}$, $J_2 = \{2\} \Rightarrow$

$$\mathfrak{R}_{o:(2)}^{GIN}\langle 1 \rangle: \; R_1 \leq I(Z_{1,2}, M_1; Y_1|M_3, Q) = I(Z_{1,2}, M_1, X_1; Y_1|X_3, M_3, Q)$$
$$\leq I(Z_{1,2}, M_1, M_3, X_1; Y_1|X_3, Q) = I(U_1, X_1; Y_1|X_3, Q)$$

(A~17)

- ✓ *Select:* $\Omega_0 = \emptyset$, $\Omega_1 = \emptyset$, $\Omega_2 = \{M_2\}$, $\Omega_s = \{M_3\}$, $J_1 = \{1\}$, $J_2 = \{2\} \Rightarrow$

$$\mathfrak{R}_{o:(2)}^{GIN}\langle 2 \rangle: \; R_2 \leq I(Z_{1,2}, M_2; Y_2|M_3, Q) = I(Z_{1,2}, M_2, X_2; Y_2|X_3, M_3, Q)$$
$$\leq I(Z_{1,2}, M_2, M_3, X_2; Y_2|X_3, Q) = I(V_1, X_2; Y_2|X_3, Q)$$

(A~18)

✓ *Select:* $\Omega_0 = \emptyset,\ \Omega_1 = \{M_1\},\ \Omega_2 = \{M_2\},\ \Omega_s = \{M_3\},\ J_1 = \{1\},\ J_2 = \{2\} \Rightarrow$

$$\begin{aligned}\mathfrak{R}_{o:(2)}^{GIN}\langle 3\rangle:\ R_1 + R_2 &\leq I(M_1;Y_1|Z_{1,2},M_2,M_3,Q) + I(Z_{1,2},M_2;Y_2|M_3,Q)\\ &= I(M_1,X_1;Y_1|X_2,X_3,Z_{12},M_2,M_3,Q) + I(Z_{1,2},M_2,X_2;Y_2|X_3,M_3,Q)\\ &\leq I(X_1;Y_1|X_2,X_3,Z_{1,2},M_2,M_3,Q) + I(Z_{1,2},M_2,M_3,X_2;Y_2|X_3,Q)\\ &= I(X_1;Y_1|X_2,X_3,V_1,Q) + I(V_1,X_2;Y_2|X_3,Q)\end{aligned} \tag{A~19}$$

$$\begin{aligned}\mathfrak{R}_{o:(2)}^{GIN}\langle 4\rangle:\ R_1 + R_2 &\leq I(M_2;Y_2|Z_{1,2},M_1,M_3,Q) + I(Z_{1,2},M_1;Y_1|M_3,Q)\\ &= I(M_2,X_2;Y_2|X_1,X_3,Z_{1,2},M_1,M_3,Q) + I(Z_{1,2},M_1,X_1;Y_1|X_3,M_3,Q)\\ &\leq I(M_2,X_2;Y_2|X_1,X_3,Z_{1,2},M_1,M_3,Q) + I(Z_{1,2},M_1,M_3,X_1;Y_1|X_3,Q)\\ &= I(X_2;Y_2|X_1,X_3,U_1,Q) + I(U_1,X_1;Y_1|X_3,Q)\end{aligned} \tag{A~20}$$

✓ *Select:* $\Omega_0 = \emptyset,\ \Omega_1 = \emptyset,\ \Omega_2 = \{M_2,M_3\},\ \Omega_s = \emptyset,\ J_1 = \{1\},\ J_2 = \{2,3\} \Rightarrow$

$$\begin{aligned}\mathfrak{R}_{o:(2)}^{GIN}\langle 2\rangle:\ R_2 + R_3 \leq I(Z_{1,23},M_2,M_3;Y_2,Y_3|Q) &= I(Z_{1,23},M_2,M_3,X_2,X_3;Y_2,Y_3|Q)\\ &= I(V_{1,23},X_2,X_3;Y_2,Y_3|Q)\end{aligned} \tag{A~21}$$

✓ *Select:* $\Omega_0 = \emptyset,\ \Omega_1 = \{M_1\},\ \Omega_2 = \{M_2,M_3\},\ \Omega_s = \emptyset,\ J_1 = \{1\},\ J_2 = \{2,3\} \Rightarrow$

$$\begin{aligned}\mathfrak{R}_{o:(2)}^{GIN}\langle 3\rangle:\ R_1 + R_2 + R_3 &\leq I(M_1;Y_1|Z_{1,23},M_2,M_3,Q) + I(Z_{1,23},M_2,M_3;Y_2,Y_3|Q)\\ &= I(M_1,X_1;Y_1|X_2,X_3,Z_{1,23},M_2,M_3,Q) + I(Z_{1,23},M_2,M_3,X_2,X_3;Y_2,Y_3|Q)\\ &= I(X_1;Y_1|X_2,X_3,V_{1,23},Q) + I(V_{1,23},X_2,X_3;Y_2,Y_3|Q)\end{aligned} \tag{A~22}$$

$$\begin{aligned}\mathfrak{R}_{o:(2)}^{GIN}\langle 4\rangle:\ R_1 + R_2 + R_3 &\leq I(M_2,M_3;Y_2,Y_3|Z_{1,23},M_1,Q) + I(Z_{1,23},M_1;Y_1|Q)\\ &= I(M_2,M_3,X_2,X_3;Y_2,Y_3|X_1,Z_{1,23},M_1,Q) + I(Z_{1,23},M_1,X_1;Y_1|Q)\\ &= I(X_2,X_3;Y_2,Y_3|X_1,U_{1,23},Q) + I(U_{1,23},X_1;Y_1|Q)\end{aligned} \tag{A~23}$$

The other constraints of the outer bound $\mathfrak{R}_o^{CIC\to 3-user}$ in (174) can be derived symmetrically. ■

➢ ***Proof of Achievable Rate Region in Theorem 13***

Consider the message and the rate splitting in (261). Without loss of generality, we prove the achievability of (259) for the case of no time-sharing, i.e., $Q \equiv \emptyset$. Fix a joint distribution of the form:

$$P_{W_1U_1X_1}P_{W_2V_2X_2}P_{W_BU_BV_B|W_1U_1W_2V_2}P_{X_3|X_1X_2V_BU_BW_BU_1W_1V_2W_2} \tag{A~24}$$

The messages are encoded by random codewords of length $n$ as follows. Let $(B_0, B_1, B_2)$ be a triple of non-negative real numbers.

*Encoding at the transmitter $X_1$:*

1. Generate at random $2^{nR_{10}}$ independent codewords $W_1^n$ according to $Pr(w_1^n) = \prod_{t=1}^n P_{W_1}(w_{1,t})$. Label these codewords as $W_1^n(m_{10})$, where $m_{10} \in [1:2^{nR_{10}}]$.
2. For each $W_1^n$, randomly generate $2^{nR_{11}}$ independent codewords $U_1^n$ according to $Pr\,(u_1^n) = \prod_{t=1}^n P_{U_1|W_1}(u_{1,t}|w_{1,t})$. Label these codewords as $U_1^n(m_{10},m_{11})$, where $m_{11} \in [1:2^{nR_{11}}]$.
3. For each pair $\left(W_1^n(m_{10}),U_1^n(m_{10},m_{11})\right)$, generate at random a codeword $X_1^n$ according to $Pr(x_1^n) = \prod_{t=1}^n P_{X_1|U_1W_1}(x_{1,t}|u_{1,t},w_{1,t})$. Label this codeword as $X_1^n(m_{10},m_{11})$.

Given the messages $(m_{10},m_{11})$, the first transmitter sends $X_1^n(m_{10},m_{11})$ over the channel.

*Encoding at the transmitter $X_2$:*

The random codebook generation at the transmitter $X_2$ is exactly similar to the one for the transmitter $X_1$, except that the indices 1 and 2, as well as the symbols $U$ and $V$ should be exchanged everywhere. Given the messages $(m_{20}, m_{22})$, the second transmitter sends $X_2^n(m_{20}, m_{22})$ over the channel.

*Encoding at the transmitter $X_3$:*

1. For each pair $\left(W_1^n(m_{10}), W_2^n(m_{20})\right)$, generate at random $2^{n(R_0+B_0)}$ codewords $W_B^n$ according to $Pr(w_B^n) = \prod_{t=1}^{n} Pr\left(w_{B,t}\middle|w_{1,t}, w_{2,t}\right)$. Label these codewords as $W_B^n(m_{10}, m_{20}, m_0, b_0)$, where $m_0 \in [1:2^{nR_0}]$ and $b_0 \in [1:2^{nB_0}]$.
2. For each 4-tuple $\left(W_1^n(m_{10}), U_1^n(m_{10}, m_{11}), W_2^n(m_{20}), W_B^n(m_{10}, m_{20}, m_0, b_0)\right)$, randomly generate $2^{nB_1}$ independent codewords $U_B^n$ according to $Pr(u_B^n) = \prod_{t=1}^{n} P_{U_B|W_BW_2U_1W_1}\left(u_{B,t}\middle|w_{B,t}, w_{2,t}, u_{1,t}, w_{1,t}\right)$. Label these codewords as $U_B^n(m_{10}, m_{11}, m_{20}, m_0, b_0, b_1)$, where $b_1 \in [1:2^{nB_1}]$.
3. For each 4-tuple $\left(W_1^n(m_{10}), W_2^n(m_{20}), V_2^n(m_{20}, m_{22}), W_B^n(m_{10}, m_{20}, m_0, b_0)\right)$, randomly generate $2^{nB_2}$ independent codewords $V_B^n$ according to $Pr(v_B^n) = \left(v_{B,t}\middle|w_{B,t}, v_{2,t}, w_{2,t}, w_{1,t}\right)$. Label these codewords as $V_B^n(m_{10}, m_{20}, m_{22}, m_0, b_0, b_2)$, where $b_2 \in [1:2^{nB_2}]$.

Let $\Lambda_3(.): \mathbb{N}^3 \to \mathbb{N}$ be an arbitrary bejection with the "min" operator $\min\Lambda_3$ defined in (8). Given the messages $(m_0, m_{10}, m_{11}, m_{20}, m_{22})$, define the triple $(b_0^T, b_1^T, b_2^T)$ as follows:

$$(b_0^T, b_1^T, b_2^T) \triangleq \min\Lambda_3 \left\{ \begin{array}{l} (b_0, b_1, b_2), \quad b_l \in [1:2^{nB_l}], l = 0,1,2 : \\ \begin{pmatrix} W_1^n(m_{10}), U_1^n(m_{10}, m_{11}), \\ W_2^n(m_{20}), V_2^n(m_{20}, m_{22}), \\ W_B^n(m_{10}, m_{20}, m_0, b_0), U_B^n(m_{10}, m_{11}, m_{20}, m_0, b_0, b_1), V_B^n(m_{10}, m_{20}, m_{22}, m_0, b_0, b_2) \end{pmatrix} \in \mathcal{T}_\epsilon^n \end{array} \right\}$$

(A~25)

In other words, $(b_0^T, b_1^T, b_2^T)$ is the minimum (with respect to $\Lambda_3(.)$) triple $(b_0, b_1, b_2)$ such that the respective codewords $W_1^n, U_1^n, W_2^n, V_2^n, W_B^n, U_B^n, V_B^n$ are jointly typical with respect to the PDF $P_{W_1U_1}P_{W_2V_2}P_{W_BU_BV_B|W_1U_1W_2V_2}$. If there is no such triple of codewords, then define $(b_0^T, b_1^T, b_2^T) \triangleq (1,1,1)$.

4. For each 9-tuple as follows:

$$\begin{pmatrix} W_1^n(m_{10}), U_1^n(m_{10}, m_{11}), X_1^n(m_{10}, m_{11}) \\ W_2^n(m_{20}), V_2^n(m_{20}, m_{22}), X_2^n(m_{20}, m_{22}) \\ W_B^n(m_{10}, m_{20}, m_0, b_0^T), U_B^n(m_{10}, m_{11}, m_{20}, m_0, b_0^T, b_1^T), V_B^n(m_{10}, m_{20}, m_{22}, m_0, b_0^T, b_2^T) \end{pmatrix}$$

(A~26)

randomly generate a codeword $X_B^n$ according to:

$$Pr(x_B^n) \triangleq \prod_{t=1}^{n} P_{X_B|X_1X_2V_BU_BW_BU_1W_1V_2W_2}\left(x_{B,t}\middle|x_{1,t}, x_{2,t}, v_{B,t}, u_{B,t}, w_{B,t}, u_{1,t}, w_{1,t}, v_{2,t}, w_{2,t}\right)$$

(A~27)

Label this codeword as $X_B^n(m_0, m_{10}, m_{11}, m_{20}, m_{22})$.

Given the messages $(m_0, m_{10}, m_{11}, m_{20}, m_{22})$ the third transmitter sends $X_B^n(m_0, m_{10}, m_{11}, m_{20}, m_{22})$ over the channel.

*Decoding at the receiver $Y_1$:* The decoder, assuming that the sequence $Y_1^n$ has been received, tries to find a unique $(\hat{m}_0, \hat{m}_{10}, \hat{m}_{11})$ for which there exists a triple $(\bar{m}_{20}, \bar{b}_0, \bar{b}_1)$ such that:

$$\begin{pmatrix} W_1^n(\hat{m}_{10}), W_2^n(\bar{m}_{20}), W_B^n(\hat{m}_{10}, \bar{m}_{20}, \hat{m}_0, \bar{b}_0) \\ U_1^n(\hat{m}_{10}, \hat{m}_{11}), U_B^n(\hat{m}_{10}, \hat{m}_{11}, \bar{m}_{20}, \hat{m}_0, \bar{b}_0, \bar{b}_1), \\ Y_1^n \end{pmatrix} \in \mathcal{T}_\epsilon^n(P_{W_1W_2W_BU_1U_BY_1})$$

(A~28)

If there exists such a unique triple, the decoder estimates its respective messages as $(\hat{m}_0, \hat{m}_{10}, \hat{m}_{11})$. Otherwise, it declares a decoding error.

*Decoding at the receiver $Y_2$:* The decoder, assuming that the sequence $Y_2^n$ has been received, tries to find a unique $(\hat{m}_0,\hat{m}_{20},\hat{m}_{22})$ for which there exists a triple $(\bar{m}_{10},\bar{b}_0,\bar{b}_2)$ such that:

$$\begin{pmatrix} W_1^n(\bar{m}_{10}), W_2^n(\hat{m}_{20}), W_B^n(\bar{m}_{10},\hat{m}_{20},\hat{m}_0,\bar{b}_0) \\ V_2^n(\hat{m}_{20},\hat{m}_{22}), V_B^n(\bar{m}_{10},\hat{m}_{20},\hat{m}_{22},\hat{m}_0,\bar{b}_0,\bar{b}_2), \\ Y_2^n \end{pmatrix} \in \mathcal{T}_\epsilon^n(P_{W_1W_2W_BV_2V_BY_2})$$

(A~29)

If there exists such a unique triple, the decoder estimates its respective messages as $(\hat{m}_0,\hat{m}_{20},\hat{m}_{22})$. Otherwise, it declares a decoding error.

*Analysis of probability of error:*

Let $0<\epsilon<p_{min}(P_{W_1U_1W_2V_2W_BU_BV_BX_1X_2X_B})$. Denote $P_{e\to Y_j}^n, j=1,2$, as the average probability of decoding at the receiver $Y_j$. Also, let $P_e^n$ be the total average probability of the code. Therefore, we have:

$$P_e^n \leq P_{e\to Y_1}^n + P_{e\to Y_2}^n$$

(A~30)

Given the messages $(m_0,m_{10},m_{11},m_{20},m_{22}) \in [1:2^{nR_0}]\times[1:2^{nR_{10}}]\times[1:2^{nR_{11}}]\times[1:2^{nR_{20}}]\times[1:2^{nR_{22}}]$, the encoding error events $E_1^e$ and $E_2^e$ , and also the decoding error events at the first receiver $E_0^{d\to Y_1},E_1^{d\to Y_1},\dots,E_7^{d\to Y_1}$ are defined as follows:

- Encoding errors:

$$E_1^e \triangleq \left\{\begin{matrix} \forall (b_0,b_1,b_2)\in[1:2^{nB_0}]\times[1:2^{nB_1}]\times[1:2^{nB_2}]: \\ \begin{pmatrix} W_1^n(m_{10}), U_1^n(m_{10},m_{11}), \\ W_2^n(m_{20}), V_2^n(m_{20},m_{22}), \\ W_B^n(m_{10},m_{20},m_0,b_0), U_B^n(m_{10},m_{11},m_{20},m_0,b_0,b_1), V_B^n(m_{10},m_{20},m_{22},m_0,b_0,b_2) \end{pmatrix} \notin \mathcal{T}_\epsilon^n \end{matrix}\right\}$$

(A~31)

$$E_2^e \triangleq \left\{\begin{pmatrix} W_1^n(m_{10}), U_1^n(m_{10},m_{11}), X_1^n(m_{10},m_{11}), \\ W_2^n(m_{20}), V_2^n(m_{20},m_{22}), X_2^n(m_{20},m_{22}), \\ W_B^n(m_{10},m_{20},m_0,b_0^{\mathcal{T}}), U_B^n(m_{10},m_{11},m_{20},m_0,b_0^{\mathcal{T}},b_1^{\mathcal{T}}), V_B^n(m_{10},m_{20},m_{22},m_0,b_0^{\mathcal{T}},b_2^{\mathcal{T}}), \\ X_B^n(m_0,m_{10},m_{11},m_{20},m_{22}) \end{pmatrix} \notin \mathcal{T}_\epsilon^n\right\}$$

(A~32)

- Decoding errors at receiver $Y_1$:

Two types of decoding error may occur at the receiver. The first one is that the transmitted codewords do not satisfy the decoding condition (A~28). This error event is given by:

$$E_0^{d\to Y_1} \triangleq \left\{\begin{pmatrix} W_1^n(m_{10}), W_2^n(m_{20}), W_B^n(m_{10},m_{20},m_0,b_0^{\mathcal{T}}) \\ U_1^n(m_{10},m_{11}), U_B^n(m_{10},m_{11},m_{20},m_0,b_0^{\mathcal{T}},b_1^{\mathcal{T}}), \\ Y_1^n \end{pmatrix} \notin \mathcal{T}_\epsilon^n(P_{W_1W_2W_BU_1U_BY_1})\right\}$$

(A~33)

The second type is that there exist some codewords other than the transmitted ones, which satisfy the decoding error condition (A~28). Precisely, there exist some 6-tuples $(m_{10}^*,m_{11}^*,m_{20}^*,m_0^*,b_0^*,b_1^*)$ with $(m_0^*,m_{10}^*,m_{11}^*)\neq(m_0,m_{10},m_{11})$ such that:

$$\begin{pmatrix} W_1^n(m_{10}^*), W_2^n(m_{20}^*), W_B^n(m_{10}^*,m_{20}^*,m_0^*,b_0^*) \\ U_1^n(m_{10}^*,m_{11}^*), U_B^n(m_{10}^*,m_{11}^*,m_{20}^*,m_0^*,b_0^*,b_1^*), \\ Y_1^n \end{pmatrix} \in \mathcal{T}_\epsilon^n(P_{W_1W_2W_BU_1U_BY_1})$$

(A~34)

Note that when two codewords construct a superposition structure, incorrect decoding of the cloud center codeword leads to incorrect decoding of the satellite one. Considering this fact and using the graphical illustration in Fig. 17, one can enumerate 7 different decoding error events of the second type in (A~34) at the receiver as described in table below.

| - | $W_1^n$ | $W_2^n$ | $W_B^n$ | $U_1^n$ | $U_B^n$ |
|---|---|---|---|---|---|
| - | $m_{10}$ | $m_{20}$ | $(m_0, b_0)$ | $m_{11}$ | $b_1$ |
| $E_1^{d\to Y_1}$ | ✓ | ✓ | ✓ | * | * |
| $E_2^{d\to Y_1}$ | ✓ | ✓ | * | ✓ | * |
| $E_3^{d\to Y_1}$ | ✓ | * | * | ✓ | * |
| $E_4^{d\to Y_1}$ | ✓ | ✓ | * | * | * |
| $E_5^{d\to Y_1}$ | ✓ | * | * | * | * |
| $E_6^{d\to Y_1}$ | * | ✓ | * | * | * |
| $E_7^{d\to Y_1}$ | * | * | * | * | * |

Table 1. The decoding error events at the first receiver.

In Table 1, the mark "*" indicates incorrect decoding of the corresponding codeword. Now, for the error probability of decoding at the first receiver, i.e., $P_{e\to Y_1}^n$, we can write:

$$P_{e\to Y_1}^n \leq \frac{1}{2^{n(R_0+R_{10}+R_{11}+R_{20}+R_{22})}} \sum_{\substack{m_0,m_{10},m_{11},\\ m_{20},m_{22}}} Pr^*\left(E_1^e \cup E_2^e \cup E_0^{d\to Y_1} \cup E_1^{d\to Y_1} \cup \ldots \cup E_7^{d\to Y_1}\right)$$

$$\leq \frac{1}{2^{n(R_0+R_{10}+R_{11}+R_{20}+R_{22})}} \sum_{\substack{m_0,m_{10},m_{11},\\ m_{20},m_{22}}} \left(Pr^*(E_1^e) + Pr^*(E_2^e|(E_1^e)^c) + Pr^*\left(E_0^{d\to Y_1}\middle|(E_2^e)^c\right) + \sum_{l=1}^{7} Pr^*\left(E_l^{d\to Y_1}\right)\right)$$

(A~35)

where $Pr^*(.) = Pr\left(.\middle|\begin{matrix} m_0, m_{10}, m_{11}, \\ m_{20}, m_{22}\end{matrix}\right)$, and $A^c$ denotes the complement of the event $A$. Then, we bound the summands in (A~35). In the following arguments $O(\epsilon)$ denotes a deterministic function of $\epsilon$, where $O(\epsilon) \to 0$ as $\epsilon \to 0$. First consider the encoding error event $E_1^e$. For bounding the probability of this event, we finely apply a multivariate covering lemma proved in [15, page 15-40]. We recall this lemma below.

***Lemma 6)*** [15, page 15-40] Consider a joint PDF $P_{U_1V_2W_BU_BV_B}(u_1, v_2, w_B, u_B, v_B)$ and its marginal PDFs $P_{U_1V_2}(u_1, v_2)$, $P_{W_B}(w_B)$, $P_{U_B|W_BU_1}(u_B|w_B, u_1)$, and $P_{V_B|W_BV_2}(v_B|w_B, v_2)$. Let $0 < \epsilon_1 < \epsilon_2 < p_{min}(P_{U_1V_2W_BU_BV_B})$. Also, let $(B_0, B_1, B_2) \in \mathbb{R}_+^3$ be a triple of non-negative real numbers. Given a pair of deterministic $n$-sequences $(u_1^n, v_2^n) \in \mathcal{T}_{\epsilon_1}^n(P_{U_1V_2})$, a random codebook is generated as follows:

1. Randomly generate $2^{nB_0}$ independent codewords $W_B^n$ according to $Pr(w_B^n) \triangleq \prod_{t=1}^n P_{W_B}(w_{B,t})$. Label these codewords as $W_3^n(b_0)$, where $b_0 \in [1:2^{nB_0}]$ .
2. For the given deterministic $n$-sequences $u_0^n$ and for each $W_B^n(b_0)$ where $b_0 \in [1:2^{nB_0}]$, randomly generate $2^{nB_1}$ independent codewords $U_B^n$ according to $Pr(u_B^n) = \prod_{t=1}^n P_{U_B|W_BU_1}(u_{B,t}|w_{B,t}, u_{1,t})$ . Label these codewords as $U_B^n(u_0^n, b_0, b_1)$ where $b_1 \in [1:2^{nB_1}]$.
3. For the given deterministic $n$-sequences $v_0^n$ and for each $W_B^n(b_0)$ where $b_0 \in [1:2^{nB_0}]$, randomly generate $2^{nB_2}$ independent codewords $V_B^n$ according to $Pr(v_B^n) = \prod_{t=1}^n P_{V_B|W_BV_2}(v_{B,t}|w_{B,t}, v_{2,t})$ . Label these codewords as $V_B^n(v_0^n, b_0, b_2)$ where $b_2 \in [1:2^{nB_2}]$.

Then, there exists $O(\epsilon) \to 0$ as $\epsilon \to 0$, where if:

$$\begin{cases} B_0 > I(U_1, V_2; W_B) + O(\epsilon) \\ B_0 + B_1 > I(U_1, V_2; W_B) + I(V_2; U_B|U_1, W_B) + O(\epsilon) \\ B_0 + B_2 > I(U_1, V_2; W_B) + I(U_1; V_B|V_2, W_B) + O(\epsilon) \\ B_0 + B_1 + B_2 > I(U_1, V_2; W_B) + I(V_2; U_B|U_1, W_B) + I(U_1, U_B; V_B|V_2, W_B) + O(\epsilon) \end{cases}$$

(A~36)

we have:

$$Pr\left(\bigcap_{\substack{b_l, l=0,1,2\\ b_l\in[1:2^{nB_l}]}} \left(W_B^n(b_0), U_B^n(u_1^n, b_0, b_1), V_B^n(v_2^n, b_0, b_2)\right) \notin \mathcal{T}_{\epsilon_2}^n\left(P_{U_1V_2W_BU_BV_B}\middle|u_1^n, v_2^n\right) \middle| u_1^n, v_2^n\right) \xrightarrow{n\to\infty} 0$$

(A~37)

In Fig. 23, we have depicted the structure of codeword generation for Lemma 6. This figure depicts the superposition structures among the generated codewrods.

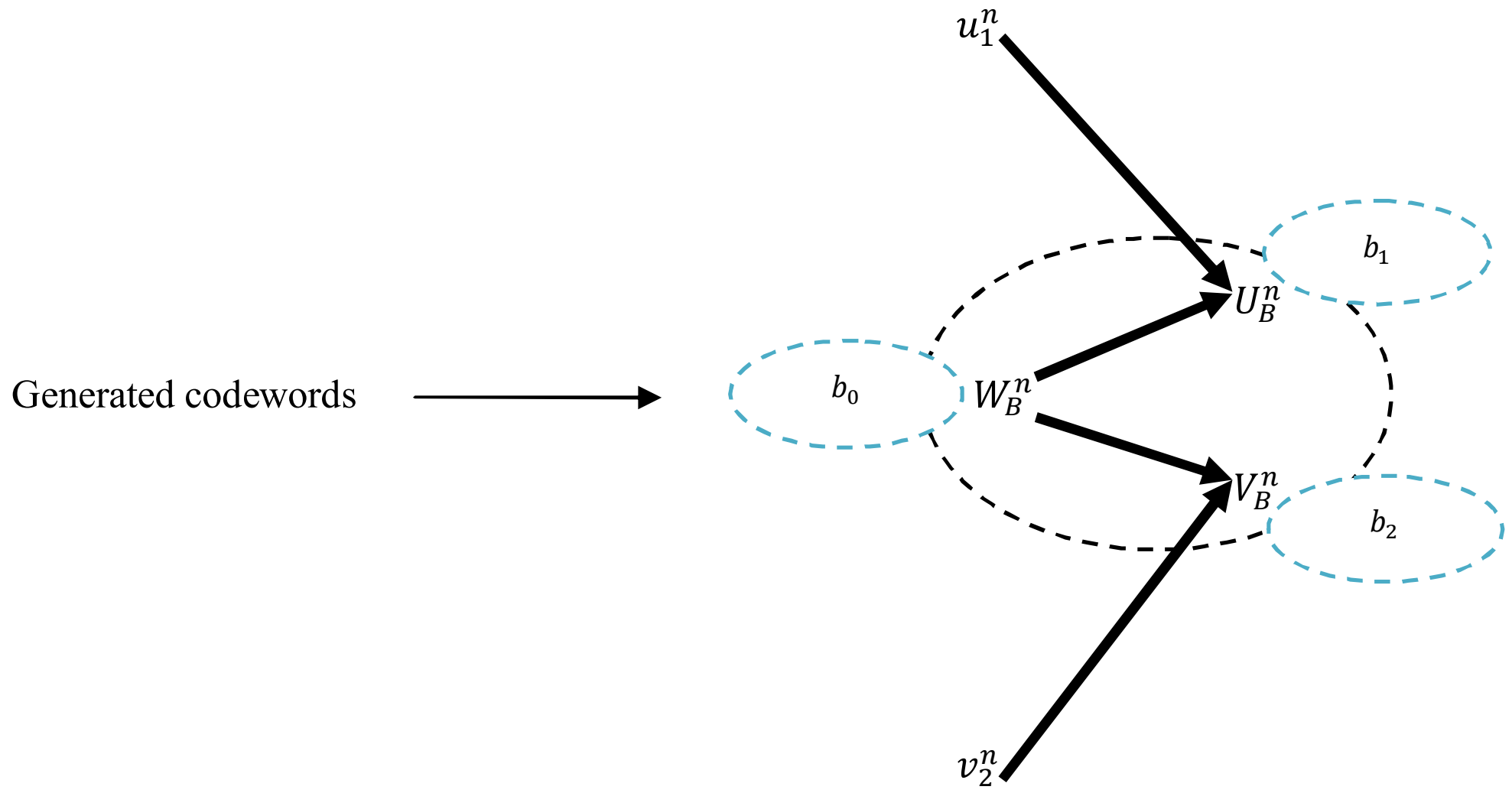

Figure 23. The graphical illustration for the generated codewords in Lemma 6.

Let us compare the graphical illustrations in Fig. 23 and Fig. 17. As we see, regardless of $W_1^n$ and $W_2^n$, the superposition structures among the codewords of the achievability scheme in Theorem 13 are exactly similar to those among the corresponding codewords in Lemma 6. Therefore, we can directly apply Lemma 6 to bound the probability of the error event $E_1^e$ in (A~31). Precisely, we can write:

$$Pr^*(E_1^e) \leq \sum_{(w_1^n,w_2^n,u_1^n,v_2^n)\in\mathcal{T}_\epsilon^n} Pr(w_1^n,w_2^n,u_1^n,v_2^n)Pr^*(E_1^e|w_1^n,w_2^n,u_1^n,v_2^n) + \sum_{(w_1^n,w_2^n,u_1^n,v_2^n)\notin\mathcal{T}_\epsilon^n} Pr(w_1^n,w_2^n,u_1^n,v_2^n)$$
$$\leq \sum_{(w_1^n,w_2^n,u_1^n,v_2^n)\in\mathcal{T}_\epsilon^n} Pr(w_1^n,w_2^n,u_1^n,v_2^n)Pr^*(E_1^e|w_1^n,w_2^n,u_1^n,v_2^n) + O(\epsilon)$$

(A~38)

Now consider $Pr^*(E_1^e|w_1^n,w_2^n,u_1^n,v_2^n)$ where $(w_1^n,w_2^n,u_1^n,v_2^n)$ are jointly typical. We have:

$$Pr^*(E_1^e|w_1^n,w_2^n,u_1^n,v_2^n)$$
$$= Pr\left(\bigcap_{\substack{b_l,l=0,1,2\\ b_l\in[1:2^{nB_l}]}} \left\{\begin{pmatrix} W_B^n(m_{10},m_{20},m_0,b_0), \\ U_B^n(m_{10},m_{11},m_{20},m_0,b_0,b_1), \\ V_B^n(m_{10},m_{20},m_{22},m_0,b_0,b_2)\end{pmatrix} \notin \mathcal{T}_\epsilon^n\left(P_{W_1U_1W_2V_2W_BU_BV_B}|w_1^n,w_2^n,u_1^n,v_2^n\right)\right\}\middle| w_1^n,w_2^n,u_1^n,v_2^n\right)$$

(A~39)

Note that the expression in the right hand side of (A~39) is similar to (A~37) except that here all the generated codewords $W_B^n, V_B^n, U_B^n$ are superimposed upon $(w_1^n,w_2^n)$. Thereby, based on Lemma 6, for vanishing the probability in (A~39), all the mutual information functions in (A~36) should be reformed to contain the conditioning on $W_1,W_2$. Therefore, we deduce that $Pr^*(E_1^e)\xrightarrow{n\to\infty} 0$ if:

$$\begin{cases} B_0 > I(U_1,V_2;W_B|W_1,W_2)+O(\epsilon) \\ B_0+B_1 > I(U_1,V_2;W_B|W_1,W_2)+I(V_2;U_B|U_1,W_1,W_2,W_B)+O(\epsilon) \\ B_0+B_2 > I(U_1,V_2;W_B|W_1,W_2)+I(U_1;V_B|V_2,W_1,W_2,W_B)+O(\epsilon) \\ B_0+B_1+B_2 > I(U_1,V_2;W_B|W_1,W_2)+I(V_2;U_B|U_1,W_1,W_2,W_B)+I(U_1,U_B;V_B|V_2,W_1,W_2,W_B)+O(\epsilon)\end{cases}$$

(A~40)

We remark that Lemma 6 always can be used to bound the probability of the encoding errors for any given achievability scheme for a two-receiver interference network (see Part V [5]).

Next we bound the other summands of (A~35). Based on the Markov relations which are implied by the joint PDF (A~24), it is readily derived that $Pr^*(E_2^e|(E_1^e)^c)\to 0$ . Also, since $W_1,U_1,W_2,V_2,W_B,U_B,V_B \to X_1,X_B,X_2 \to Y_1$ form a Markov chain, we have $Pr^*\left(E_0^{d\to Y_1}\middle|(E_2^e)^c\right) \to 0$. Lastly, consider the error events in Table 1. To bound the probability of these error events, we use the general formula derived in our previous paper [7, page. 12, Eq. 41-42]. Based on this general formula, we have:

$$\sum_{E_l^{d\to Y_1}} \begin{pmatrix}\text{Rates with respect to incorrect decoded} \\ \text{messages and bin indices}\end{pmatrix} < I_{E_l^{d\to Y_1}} - O(\epsilon), \qquad l = 1,\dots,7, \qquad \Rightarrow \qquad Pr^*\left(E_l^{d\to Y_1}\right) \to 0$$

(A~41)

where,

$$I_{E_l^{d\to Y_1}} =$$

$$\sum_{\substack{A^n \text{ is incorrctly} \\ \text{decoded in } E_l^{d\to Y_1}}} H\left(A\middle|\left\{\begin{matrix} B: \\ B^n \text{ is a cloud center for } A^n\end{matrix}\right\}\right) - H\left(\left\{\begin{matrix} C: \\ C^n \text{ is incorrectly decoded} \\ \text{in } E_l^{d\to Y_1}\end{matrix}\right\}\middle|\left\{\begin{matrix} D: \\ D^n \text{ is correctly decoded} \\ \text{in } E_l^{d\to Y_1}\end{matrix}\right\}, Y_1\right)$$

(A~42)

As mentioned in [7], this general formula can be used to evaluate the decoding error probabilities at the receivers for any given achievability scheme for a memoryless network. Using the error decoding table and also the graphical illustration in Fig. 17 which depicts the superposition structures among the generated codewords, one can easily check that for vanishing the probability of the events $E_1^{d\to Y_1},\dots,E_7^{d\to Y_1}$, it is required:

$$\begin{aligned}
R_{11} + B_1 &< I_{E_1^{d\to Y_1}} - O(\epsilon) = I(U_1, U_B; Y_1|W_1, W_2, W_B) + I(U_1; W_2, W_B|W_1) - O(\epsilon) \\
R_0 + B_0 + B_1 &< I_{E_2^{d\to Y_1}} - O(\epsilon) = I(W_B, U_B; Y_1|W_1, W_2, U_1) + I(U_1; W_B|W_1, W_2) - O(\epsilon) \\
R_{20} + R_0 + B_0 + B_1 &< I_{E_3^{d\to Y_1}} - O(\epsilon) = I(W_2, W_B, U_B; Y_1|W_1, U_1) + I(U_1; W_B|W_1, W_2) + I(W_2; W_1, U_1) - O(\epsilon) \\
R_0 + B_0 + R_{11} + B_1 &< I_{E_4^{d\to Y_1}} - O(\epsilon) = I(U_1, W_B, U_B; Y_1|W_1, W_2) + I(U_1; W_2, W_B|W_1) - O(\epsilon) \\
R_{20} + R_0 + B_0 + R_{11} + B_1 &< I_{E_5^{d\to Y_1}} - O(\epsilon) = I(U_1, W_2, W_B, U_B; Y_1|W_1) + I(U_1; W_2, W_B|W_1) + I(W_2; W_1) - O(\epsilon) \\
R_{10} + R_0 + B_0 + R_{11} + B_1 &< I_{E_6^{d\to Y_1}} - O(\epsilon) = I(W_1, U_1, W_B, U_B; Y_1|W_2) + I(U_1; W_2, W_B|W_1) + I(W_1; W_2) - O(\epsilon) \\
R_{10} + R_{20} + R_0 + B_0 + R_{11} + B_1 &< I_{E_7^{d\to Y_1}} - O(\epsilon) = I(W_1, U_1, W_2, W_B, U_B; Y_1) + I(U_1; W_2, W_B|W_1) + I(W_1; W_2) - O(\epsilon)
\end{aligned}$$

(A~43)

Also, note that the distribution (A~24) implies:

$$I(U_1; W_2, W_B|W_1) = I(U_1; W_B|W_1, W_2), \qquad I(W_2; W_1, U_1) = 0, \qquad I(W_2; W_1) = 0$$

(A~44)

Substituting (A~44) in (A~43), we derive the desired constraints with respect to the receiver $Y_1$. The constraints concerning the receiver $Y_2$ are symmetrically obtained by the following exchanges in (A~43):

$$Y_1 \to Y_2, \quad W_1 \to W_2, \quad U_1 \to V_2, \quad U_B \to V_B, \quad R_{10} \to R_{20}, \quad R_{11} \to R_{22}, \quad R_{20} \to R_{10}, \quad B_1 \to B_2$$

(A~45)

The proof is thus complete. ■

## ACKNOWLEDGEMENT

The author would like to appreciate Dr. H. Estekey, head of School of Cognitive Sciences, IPM, Tehran, Iran, and Dr. R. Ebrahimpour, faculty member at School of Cognitive Sciences, for their kindly support of the first author during this research. He also greatly thanks his mother whose love made this research possible for him. Lastly, F. Marvasti is acknowledged whose editing comments improved the language of this work.